\documentclass[letterpaper,twocolumn,notitlepage,final,superscriptaddress,balancelastpage,rmp,nobibnotes]{revtex4}%
\usepackage{amssymb}
\usepackage{amsmath}
\usepackage{amsfonts}
\usepackage{graphicx}
\usepackage{bm}%
\begin{document}

\title{Multi-photon entanglement and interferometry}
\author{Jian-Wei Pan}
\email{pan@ustc.edu.cn}
\affiliation{Hefei National Laboratory for Physical Sciences at Microscale and Department
of Modern Physics, University of Science and Technology of China, Hefei, Anhui
230026, China}
\author{Zeng-Bing Chen}
\email{zbchen@ustc.edu.cn}
\affiliation{Hefei National Laboratory for Physical Sciences at Microscale and Department
of Modern Physics, University of Science and Technology of China, Hefei, Anhui
230026, China}
\author{Chao-Yang Lu}
\email{cylu@ustc.edu.cn}
\affiliation{Hefei National Laboratory for Physical Sciences at Microscale and Department
of Modern Physics, University of Science and Technology of China, Hefei, Anhui
230026, China}
\author{Harald Weinfurter}
\email{h.w@lmu.de}
\affiliation{Fakult\"{a}t f\"{u}r Physik, Ludwig-Maximilians-Universit\"{a}t, D-80799
M\"{u}nchen, Germany}
\affiliation{Max-Planck-Institut f\"{u}r Quantenoptik, D-85748 Garching, Germany}
\author{Anton Zeilinger}
\email{anton.zeilinger@univie.ac.at}
\affiliation{Vienna Center of Quantum Science and Technology (VCQ), Faculty of Physics, Universit\"{a}t Wien, Boltzmanngasse 5,
A-1090 Wien, Austria}
\affiliation{Institut f\"{u}r Quantenoptik und Quanteninformation (IQOQI),
\"{O}sterreichische Akademie der Wissenschaften, Boltzmanngasse 3, A-1090
Wien, Austria}
\author{Marek \.{Z}ukowski ~}
\email{marek.zukowski@univie.ac.at}
\affiliation{Instytut Fizyki Teoretycznej i Astrofizyki, Uniwersytet Gda\'{n}ski, PL-80-952
Gda\'{n}sk, Poland}
\affiliation{Institut f\"{u}r Experimentalphysik, Universit\"{a}t Wien, Boltzmanngasse 5,
A-1090 Wien, Austria}
\affiliation{Hefei National Laboratory for Physical Sciences at Microscale and Department
of Modern Physics, University of Science and Technology of China, Hefei, Anhui
230026, China}

\date{Accepted for publication in Rev. Mod. Phys. on 20 Sep. 2011}

\begin{abstract}

Multi-photon interference reveals strictly non-classical phenomena.
Its applications range from fundamental tests of quantum mechanics to
photonic quantum information processing, where a significant fraction
of key experiments achieved so far comes from multi-photon state manipulation.
We review the progress, both theoretical
and experimental, of this rapidly advancing research. The emphasis is given to the creation
of photonic entanglement of various forms, tests of the completeness of quantum mechanics (in
particular, violations of local realism), quantum information protocols for quantum communication
(e.g., quantum teleportation, entanglement purification and quantum repeater), and quantum
computation with linear optics. We shall limit the scope of our review to ``few photon'' phenomena
involving measurements of discrete observables.

\end{abstract}

\maketitle

\tableofcontents

\section{Introduction}

In his 1704 treatise \textit{Opticks} Newton claimed that light is composed of
particles, and strongly opposed Huygens wave picture. Later on with Young's double slit interference experiments,
the wave picture seemed to be correct and sufficient. This view was
further strengthened by Maxwell's electrodynamics. Yet, 201 years after Newton, during his \textit{annus mirabilis}
Einstein re-introduced \textit{lichtquanten} (light particles) and in this way
explained the photoelectric effect Einstein (1905).\footnote{The current term
\textit{photon} was introduced by Lewis (1926).} The ultimate consequences of Einstein's ideas, after fundamental works of Bohr, Heisenberg and Schroedinger, gave birth to quantum mechanics in 1925. Quantum electrodynamics, the final theory of light, in which
photons are elementary excitations of the quantized electromagnetic field
interacting with charges, was given by Dirac (1927), and its internal
consistency was proved by Dyson, Feynman, Schwinger and Tonomaga around twenty
years later. According to these theories, photons, as all quantum particles, reveal both wave-like and
particle-like properties - a phenomenon known as wave-particle duality. The wave nature is revealed
by interference, while the particle nature can be recognized in absorption and detection
events, or more generally in the statistics of counts. The interference patterns involving single photons or, equally well, the light intensity
does not reveal strictly non-classical phenomena.
Some of the most counterintuitive  effects begin with
two or more photon interference and in intensity correlation measurements: a plethora of
classically impossible phenomena occurs - most of them completely incomprehensible with
any classical concepts, neither particle nor wave.
 As always in the history of
human scientific endeavor, harnessing of new phenomena leads to new applications.
The aim of this review is to describe the recent theoretical and experimental advances in multi-photon
interference, entanglement, manipulation, and their applications in quantum
communication and computation.

\subsection{Quantum optics}

An intensive research of the quantum properties of light started around half a
century ago. Its advances allow one to gain a coherent control of quantum
optical systems, enabling true quantum engineering. As a result, quantum
optical methods made possible to actually perform \textit{gedanken}experiments
concerning the foundations of quantum theory. This control of quantum
phenomena further allows one to search for novel information processing protocols,
which now promise new technologies based on quantum information science.

Soon after Einstein's introduction of light quanta, Taylor (1909) tried to
find some new effects in a two-slit Young-type experiment using extremely
faint light, so faint that on average only one photon at a time was inside the
apparatus. No deviation from the classical interference was observed. Now,
with a fully developed theory of quantized light we know that experiments of
this type cannot differentiate between the classical explanation (based on the
interference of electric field waves) and the quantum explanation (based on
the interference of probability amplitudes for photons passing through either
of the two slits). The inherently \textquotedblleft quantum nature of the
electromagnetic field\textquotedblright, as we know now, is revealed directly
in multi-photon experiments which were not possible earlier.

Still, quantum interference of truly individual photons is certainly a fascinating
phenomenon. The first precise experiment aimed at exactly this was performed by
Grangier, Roger, and Aspect (1986). They used photon pairs emitted in atomic
cascades, one of the photons was used as a trigger, and the other was fed into
a Mach-Zehnder interferometer.
When detectors are placed in the two arms of the interferometer, besides background noise, no simultaneous detection (i.e., coincidence) in both detectors was observed\footnote{As a matter of
fact some coincidence was observed, however it was below the case that was to be
expected if photon were treated as classical wave-packets}, i.e., the photon was found only in one of the two arms - a typical particle-like behavior. However, after overlapping the two arms by the output beamsplitter of the interferometer the usual (wave-like) interference pattern was observed. Recently Braig et al. (2002) demonstrated that, also when observing interference depending on the phase difference between both arms,  the light from the interferometer output exhibited the characteristic single-photon antibunching.

Modern quantum optics was effectively born in 1956 when
Hanbury Brown and Twiss (1956) introduced intensity interferometry. It was
the first serious attempt to study the correlations between intensities
recorded at two separated detectors. It motivated more sophisticated
photon-counting and correlation experiments. The quantum theory of optical
coherence of Glauber (1963) gave theoretical clues to search for unambiguously
quantum optical phenomena. Carmichael and Walls (1976) predicted photon
antibunching in a resonance-fluorescence, which was observed experimentally by
Kimble, Dagenais and Mandel (1977, 1978). The early experiments used atomic
beams as sources. Thus, atomic number (and thus the emission statistics)
fluctuations were unavoidable. Later, Diedrich and Walther (1987) realized
such experiments using single atoms in traps and observed photon antibunching
as well as sub-Poissonian statistics in the system. Squeezed states of light were
experimentally generated by using a four-wave mixing in atomic sodium (Slusher
\textit{et al}., 1985) and in optical fibers (Shelby \textit{et al}., 1986),
or by using an optical parametric oscillator (Wu \textit{et al}., 1986). Interested readers may find excellent reviews in, e.g., Walls and Milburn
(1994), Mandel and Wolf (1995), Scully and Zubairy (1997), and Lounis and
Orrit (2005).

The study on photon statistics and photon counting techniques enables a direct
examination of some of the fundamental distinctions between quantum and
classical concepts of light. Parallel developments in neutron, atomic, and
molecular interferometry, as well as modern methods of cooling and trapping
ions, etc., allowed one to probe ever deeper the properties of
\textit{individual} quantum systems and to realize many of the
\textit{gedanken} (thought) experiments testing the foundations quantum
physics.

\subsection{The essence of the quantum world: entanglement}
Entanglement is a property of more than one quantum system such that the state
of one system cannot be seen independent of the other's. It forms the basis for
the most remarkable, purely quantum effects and is the main resource for the many
applications of quantum information processing. Initially, it was used by Einstein,
Podolsky, and Rosen (EPR, 1935) to show that quantum mechanics is incomplete. The
trio, EPR, argued that the outcome of a measurement on any physical system
is determined prior to and independent of the
measurement (realism) and that the outcome cannot depend on any actions in
space-like separated regions (Einstein's locality).\footnote{A more detailed
discussion of the EPR paradox is in Section V.} This EPR criterion of, what is now called,
\textquotedblleft local realism\textquotedblright\
should be fullfilled by every physical description of nature, and, indeed, it looks quite reasonable to
us, particularly as all our Classical World and experience fully adhere to it.
They used the perfect correlations of entangled states (thus often called EPR states) to
define ``elements of reality'', a notion which according to them was missing in quantum theory.
Elements of reality are deterministic predictions for a
measurement result, which can be established without actually performing the
measurement, and without physically disturbing the (sub-)system to which
they pertain. As elements of reality in the studied case were argued to exist
necessarily even for pairs of non-commuting observables, they claimed  they are
contradicting the Heisenberg uncertainty relation.

The EPR paradox and with it the entanglement of quantum states remained a philosophical issue (Feynman \textit{et al}. 1963), and seemed experimentally untestable for almost 30 years.
The breakthrough happened when Bell (1964) derived his remarkable inequalities which revealed that two-particle correlations for the two
spin-$\frac{1}{2}$ singlet disagree with any local realistic model.  The
pioneering \textquotedblleft Bell experiment\textquotedblright was done by Freedman and Clauser (1972), followed by the famous ones by Aspect \textit{et al}. (1981, 1982a,
1982b) and many others.\footnote{See Aspect, (1999); Tittel and
Weihs (2001) for a recent survey.} The early experiments used polarization
entangled photon pairs from atomic cascades (Clauser and Shimony, 1978). In
late 1980's, parametric down conversion was discovered as a convenient and robust
method to produce entangled photons (see Section~\ref{section:nonlocality}).

A quarter of century after Bell's paper, it turned out that the conflict of
local realism with quantum mechanics is even more striking for certain three or
more particle entangled states. The Greenberger-Horne-Zeilinger (GHZ) theorem
(Greenberger, Horne, and Zeilinger, 1989; Greenberger \textit{et al}., 1990;
Mermin, 1990a) showed that the concept of EPR's elements of reality is
self-contradictory.
That is, there are situations for which local realism and quantum mechanics
make completely opposite predictions, even for perfectly correlated
results - which were the starting point of the EPR argumentation. The GHZ paper showed that three or more particle interferometry is
a rich untested area, full of exciting classically paradoxical phenomena. However, at that
time no effective sources of three or four photon entanglement were present.
Thus, a new chapter in experimental multi-photon quantum optics was opened, but had to wait
for new ideas and experimental techniques.

\subsection{Sources of photonic entanglement}

The standard source of entangled photon pairs is nowadays the nonlinear
optical process of spontaneous parametric down-conversion (SPDC; for a pioneering observation see Burnham
and Weinberg, 1970), the inverse of frequency doubling and up-conversion. In SPDC photons
from a pump laser beam, within a nonlinear crystal, can spontaneously be
converted into pairs that are momentum and frequency entangled, and in the
so-called type-II process can be also polarization entangled (Kwiat \textit{et
al}., 1995). Today, SPDC sources of entangled photon pairs of a high
quality and brightness can be routinely realized using various methods.

However, since the GHZ paper, and even more after the birth of quantum
information, three or more photon entanglement was in demand. It turned out
that using the primary two-photon entanglement, by a procedure which is called
entanglement swapping, one can entangle
without any direct interaction particles which were independent of each other,
or what is more important for us, \emph{construct} entanglement of higher
order (\.{Z}ukowski \textit{et al}, 1993). Since photons basically do not
interact with each other this method is of special importance for schemes
aimed at creating multi-photon entanglements. Practical versions
of this technique (Zukowski et al, 1995; Rarity, 1995; Zeilinger et al, 1997) are thus the basis of all experiments with three or more entangled photons, as well as of many realizations of quantum information protocols, up to measurement based quantum computation.

\subsection{Applications in quantum information}

Experimental quantum information processing was started right after new
experimental techniques allowed to control individual or compound quantum systems like atoms
in traps, pairs of entangled photons, etc, and to observe a new set of classically impossible phenomena.
As always, new controllable phenomena lead to new practical applications, especially in information transfer and processing.
Quantum information processing harnesses the superposition principle and
non-classical correlations of quantum mechanics and employs them in communication and computation.
In quantum cryptography [Wiesner, 1983; Bennett and Brassard, 1984; Ekert, 1991;
for a review see (Gisin \textit{et al}., 2002)] complementary measurements on quantum systems are used to establish a secret key
shared by two partners, thus enabling for the first time a provably secure
communication. Quantum teleportation (Bennett \textit{et al}., 1993) enables
a faithful transfer of an unknown quantum state from one location to another,
using entangled states as a quantum channel. Quantum computers promise
to increase greatly the efficiency of solving problems such as factoring large
integers, database search and simulation of some quantum phenomena.

Photons are the fastest information carriers, have a very weak coupling to
the environment, and are thus best suited for quantum communication tasks. Thanks to this property,
quantum key distribution with photons has now went long beyond first few-meter laboratory
demonstrations, to free-space or fiber-based distributions over hundred
kilometers (see e.g. Ursin \textit{et al}, 2007; Rosenberg \textit{et at}., 2007),
and is rapidly commercialized into real life inter-city cryptographic networks.
To ultimately extend the range of quantum communication to a global scale,
a collection of quantum toolkits still has to be developed. A
quantum  repeater (Briegel \textit{et al.} 1998) would allow in principle an efficient long
distance high-fidelity transmission of entanglement. Several ingredients of
this scheme  have been already demonstrated: entanglement swapping (Pan \textit{et al}. 1998),
purification (Pan \textit{et al}. 2003), quantum memory (e.g., Yuan
\textit{et al}. 2008), {\em etc.}

Despite the difficulty to ``localize'' photons, there has been
a considerable interest in linear optical quantum computing.
This is motivated by the photon's robustness
against decoherence and the relative ease with which it
can be manipulated with a high precision. Remarkably, by exploiting
the nonlinearity induced by measurement, Knill,
Laflamme, and Milburn (KLM) showed that scalable
quantum computation is in principle possible with linear optics, single photon sources and
detectors. A new and probably more practical approach is the
concept of a \textquotedblleft one-way quantum computer\textquotedblright%
\ (Raussendorf and Briegel, 2001; see section~\ref{sec:LO-oneway}). In this approach, one starts with
the so-called \textquotedblleft cluster states\textquotedblright%
(Briegel and Raussendorf, 2001). The computation algorithm is then
performed by applying a sequence of one-qubit measurements. Optical quantum
computing proposals (Browne and Rudolph, 2005) based on the one-way model reduce the computational resource
overhead by three orders of magnitudes compared to the KLM scheme.
Cluster states up to six entangled photons have been realized
(Walther \textit{et al}., 2005b; Kiesel \textit{et al}., 2005a; Lu \textit{et al}.,
2007), and applied to demonstrate elementary quantum gates and algorithms
(Walther \textit{et al}., 2005b).

The rapid growth of literature on photonic realizations of quantum information processing
still continues. One can expect much more exciting new developments.

\subsection{Related reviews}

The earlier stages of the research of photonic entanglement
have been reviewed in Clauser and Shimony (1978), Mandel and Wolf (1995) and Mandel (1999). They contain
a collection of descriptions of beautiful experiments demonstrating the very nature of quantum mechanics. Because of a limit of space,
we shall not discuss these experiments in this review. We start our description more or less at the stage of developments at which
these earlier reviews ended. For detailed discussions on quantum entanglement, we refer to Alber \textit{et al}. (2001) and Horodecki
\textit{et al}. (2009). Reviews on Bell's theorem can be found in Lalo\"{e} (2001), Werner and Wolf (2001),
and Genovese (2005). For an introduction to quantum information and quantum
computation see a short survey by Bennett and DiVincenzo (2000) and
textbooks by Preskill (1998), Nielsen and Chuang (2000), and Bouwmeester \textit{et
al}. (2001). Quantum cryptography has been reviewed by Gisin \textit{et al}. (2002).
Linear optical quantum computing with photonic qubits has been reviewed by (Kok \textit{et al}., 2007),
O'Brien (2007), O'Brien \textit{et al}. (2009) and Ralph \textit{et al}. (2009).
Zeilinger \textit{et al}. (2005) have given a concise review on experimental progress
on photon interference and quantum information applications.

\subsection{Our aims}

In this review, we wish to describe the progress in the last two decades or so, both theoretical
and experimental, in multi-photon interferometry, and its applications ranging
from  fundamental tests of quantum mechanics to photonic quantum information
processing. Emphasis will be put on creation and control of photonic entanglement
with linear optics, and its application in quantum communication and computation.
We shall limit the scope of our review to ``few photon'' phenomena involving measurements of discrete observables, thus many fascinating experiments involving continuous variables will be not discussed here.

\section{Interference and quantum entanglement}

Classical interference is a macroscopic expression of the quantum one i.e.,
the coherent or thermal states of the electro-magnetic fields can also be described with Maxwell's laws. The
interference phenomena in the quantum realm are richer and more pronounced. We
discuss here the basic differences between the classical interference
understood as interference of electro-magnetic waves in space, and the quantum one which is
interference of various operationally indistinguishable processes.

\subsection{Classical interference}

In classical physics interference results from the superposition of waves. It
may express itself in the form of intensity variations or intensity correlations.

%\subsubsection{Interference expressed by intensity variations}

Consider two quasi-monochromatic plane waves linearly polarized in the same
direction, described by
\begin{equation}
E_{j}(r,t)=E_{j}e^{i[\mathbf{k}_{j}\cdot\mathbf{r}-\omega t-\phi_{j}(t)]}+c.c.
\end{equation}
where $E_{j}$ is the real amplitude of one of the fields, $\mathbf{k}_{j}$ the
wave vector, $\omega$ the frequency of both waves, $j=1,2$ the index numbering
the fields, and finally $c.c.$ denotes the complex conjugate of the previous
expression. The intensity of the superposed fields at a certain point in space
is given by
\begin{equation}
I(\mathbf{r},t)=E_{1}^{2}+E_{2}^{2}+2E_{1}E_{2}\cos{[{\Delta_{12}\mathbf{k}%
}\cdot\mathbf{r}-\Delta_{12}\phi(t)]},
\end{equation}
where $\Delta_{12}$ is the difference of the respective parameters for the two
fields, e.g., $\Delta_{12}\phi(t)=\phi_{1}(t)-\phi_{2}(t)$. For $\Delta
_{12}\phi(t)$ constant in time, or of values varying much less than
$\pi$, this formula (after averaging over time) describes a Young-type
interference pattern. In the opposite case, of widely fluctuating $\Delta
_{12}\phi(t)$ no interference can be observed, because the pattern is washed out.
In the case of $E_{1}=E_{2}$ one has maximal possible interference. This can
be quantified in terms of the interferometric contrast, or visibility,
$V=({I_{max}-I_{min}})/({I_{max}+I_{min}}),$ which in the aforementioned case
equals $1$.

%\subsubsection{Interference in intensity correlations}

The Hanbury Brown and Twiss experiment introduced intensity correlation
measurements to optics. Such correlations between two points in space and two moments of
time, for two classical fields are described by an intensity correlation
function
\begin{equation}
G^{(2)}(\mathbf{r_{1}},t_{1};\mathbf{r_{2}},t_{2})=\langle I(\mathbf{r_{1}%
},t_{1})I(\mathbf{r_{2}},t_{2})\rangle_{av}\ .
\end{equation}
The average is taken over an ensemble, and for stationary fields this is
equivalent to the temporal average. Even when no intensity variations are
observable (i.e., for averaged intensity constant in space), the intensity
correlations can reveal interference effects. Assume that the phases of the
two fields fluctuate independently of one another. Then for $t_{1}=t_{2}$, the
$G^{(2)}$ function still exhibits a spatial modulation or maximal visibility
of $50\%$ as exhibited by the formula:
\begin{align}
G^{(2)}(\mathbf{r_{1}},t;\mathbf{r_{2}},t)  &  =(I_{1}+I_{2})^{2}\nonumber\\
&  +2I_{1}I_{2}\cos[(\Delta_{12}\mathbf{k})(\mathbf{r_{1}}-\mathbf{r_{2}})]\ ,
\end{align}
where $I_{i}=E_{i}^{2}$, $i=1,2$. This formula can be easily reached by noting
that the temporal average of
\begin{equation}
\cos[\alpha+\Delta_{12}\phi(t)]\cos[\alpha^{\prime}+\Delta_{12}\phi(t)],
\end{equation}
where $\Delta_{12}\phi(t)=\phi_{1}(t)-\phi_{2}(t)$, is given by
\begin{align}
&  \cos\alpha\cos\alpha^{\prime}\langle\cos^{2}\Delta_{12}\phi(t)\rangle
_{av}+\nonumber\\
&  \sin\alpha\sin\alpha^{\prime}\langle\sin^{2}\Delta_{12}\phi(t)\rangle
_{av}-\nonumber\\
&  \frac{1}{2}\sin(\alpha+\alpha^{\prime})\langle\sin2\Delta_{12}%
\phi(t)\rangle_{av},
\end{align}
and due to the random nature of $\Delta_{12}\phi(t)$ only the first two terms
survive because both $\langle\cos^{2}\Delta_{12}\phi(t)\rangle_{av}$ and
$\langle\sin^{2}\Delta_{12}\phi(t)\rangle_{av}$ give $\frac{1}{2}$, whereas
$\langle\sin2\Delta_{12}\phi(t)\rangle_{av}=0.$

In addition to the phase fluctuations one can take into account also amplitude
fluctuations. Nevertheless, the basic features of the earlier formula must be
retained. Amplitude fluctuations tend to lower the
visibility of the intensity correlations patterns even further. Thus, the visibility of
intensity correlations for fields with fluctuating phase differences is never full, maximally $50\%$. As we shall see, there
is no bound on visibility in the quantum case. For a broader treatment of
these matters see a review by Paul (1986), and Belinskii and Klyshko (1993).

\subsection{Quantum interference}

Quantum interference rests on the concept of superposition of probability amplitudes of indistinguishable processes that contribute to the given phenomenon.

\subsubsection{Single-particle quantum interference}

Single-particle interference looks almost identical to the classical one.
We replace the fields (waves) by amplitudes, $A(\mathbf{x},t)$, which differ
only by the fact that they must be suitably normalized, if one wants to compute the probabilities. Suppose that the (not normalized) amplitude
to detect a photon at $\mathbf{x}$, is given by
\begin{equation}
A_{\mathbf{b}_{1}}(\mathbf{x},t)=e^{i[\mathbf{k}{_{1}}\cdot(\mathbf{x}%
-\mathbf{b}_{1})+\Phi_{\mathbf{x},\mathbf{b}_{1}}(t)]}, \label{AMPLITUDE1}%
\end{equation}
if it originates from point $\mathbf{b}_{1}$, and by
\begin{equation}
A_{\mathbf{b}_{2}}(\mathbf{x},t)=e^{i[\mathbf{k}{_{1}^{\prime}}\cdot
(\mathbf{x}-\mathbf{b}_{2})+\Phi_{\mathbf{x},\mathbf{b}_{2}}(t)]},
\label{AMPLITUDE2}%
\end{equation}
if it originates from $\mathbf{b}_{2}$. The quantum mechanical probability
density that a particle is detected at $\mathbf{x}$ is given by
\begin{align}
P(\mathbf{x},t)  &  \sim\left\vert A_{\mathbf{b}_{1}}(\mathbf{x}%
,t)+A_{\mathbf{b}_{2}}(\mathbf{x},t)\right\vert ^{2}\nonumber\\
&  \sim1+\cos[\Delta\mathbf{k}_{1}\cdot\mathbf{x}+\Phi_{0}+\Phi_{\mathbf{x}%
,\mathbf{b}_{1}}(t)-\Phi_{\mathbf{x},\mathbf{b}_{2}}(t)],
\end{align}
where $\Delta\mathbf{k}_{1}=\mathbf{k}_{1}^{\prime}-\mathbf{k}_{1},$
and $\Phi_{0}$ is an irrelevant constant phase. Thus if the phase
difference,
$\Phi_{\mathbf{x},\mathbf{b}_{1}}(t)-\Phi_{\mathbf{x},\mathbf{b}_{2}}(t)$,
is stable, one can have the Young-type interference patterns of up
to $100\%$ visibility. Such a stable phase difference can be observed with single photons
in, e.g., a double-slit experiment. Also in the case of a classical wave description and classical-like fields the observed intensity is
proportional to the probability density $P(\mathbf{x},t).$, that is in this respect nothing changes.

Nevertheless, the above description differs drastically from the classical \emph{particle}
picture, in which one would expect that a process originating with state $A$
and with possible intermediate stages $B_{1},\ldots B_{N}$, leading to an
event $C$, would be described by
\begin{equation}
P(C|A)=\sum_{j=1}^{N}P(C|B_{j})P(B_{j}|A). \label{eq4}%
\end{equation}
In the quantum case $P(C|A)=|\langle C|A\rangle|^{2}$, where
\begin{equation}
\langle C|A\rangle=\sum_{j=1}^{N}\langle C|B_{j}\rangle\langle B_{j}|A\rangle,
\label{eq5}%
\end{equation}
this means, one sums over {\em amplitudes}, not probabilities. For a double slit we have $N=2$, and  $\langle A|B_{1}\rangle\approx\langle A|B_{2}\rangle$ give the
amplitudes to reach the slits. Finally  one has $\langle C|B_{j}\rangle=A_{\mathbf{b}_{j}}.$ Please note that, for classical particles the terms of the summation,
$P(C|B_{j})$ etc., are real numbers, while in the quantum case the amplitudes
$\langle C|B_{j}\rangle$ are complex numbers. Thanks to that interference effects can be predicted.

The difference between (\ref{eq4}) and (\ref{eq5}) is in the
assumption, inherent in (\ref{eq4}), that the particle had to be in
\emph{one} of the intermediate situations (states) $B_{i}$. In the
quantum case any attempt to verify by measurement\footnote{Or by securing the possibility of
a postponed measurement (which could be made by correlating our
particle, while it is at the intermediate stage $B_{i}$, with
another system, which could be measured later).}
which of the situations actually took place puts one back to
the classical formula (\ref{eq4}). The formula (\ref{eq5}) leads to
interference phenomena, and may be thought as a manifestation of a
wave nature of quantum particles, whereas, if we make measurements
discriminating  events $B_i$, we learn by which way (\emph{welcher weg}) the particles
travel. The \textquotedblleft which-way
information\textquotedblright is a clear signature of the particle nature.

\subsubsection{Two-particle quantum interference}

All this becomes much more puzzling once we consider a two-particle
experiment. Already Einstein, Podolsky and Rosen (1935) pointed out some
strange features of such a case. Schr\"{o}dinger noticed that these features
are associated with what he called \textquotedblleft entangled
states\textquotedblright, which will be discussed in detail later. At the
moment consider these as superpositions of fully distinguishable products of
single-particle states, i.e., that there is a specific pair of local measurements for which the two subsystems are perfectly correlated (a
result of a measurement on one subsystem reveals the unique value of the corresponding
observable for the second subsystem).

Consider such a correlation: assume that if particle $1$ is at $b_{1}$, then
particle $2$ is also at $b_{1}$, and, whenever $1$ is at $b_{2}$ then particle
$2$ is at $b_{2}$. Later on the particles are detected at two different points,
$\mathbf{x}_{1}$ and $\mathbf{x}_{2}$. Then, according to the rules given
above
\begin{align}
P(\mathbf{x}_{1},\mathbf{x}_{2},t)  &  \sim|A_{b_{1}}(\mathbf{x}%
_{1},t)A_{b_{1}}(\mathbf{x}_{2},t)+A_{b_{2}}(\mathbf{x}_{1},t)A_{b_{2}%
}(\mathbf{x}_{2},t)|^{2}\nonumber\\
&  \sim1+\cos(\Delta\mathbf{k}_{1}\cdot\mathbf{x}_{1}+\Delta\mathbf{k}%
_{2}\cdot\mathbf{x}_{2}+\Delta\Phi_{b_{1},b_{2}}),
\end{align}
with
\begin{equation}
\Delta\Phi_{b_{1},b_{2}}=\Delta\Phi_{\mathbf{x}_{1},b}(t)+\Delta
\Phi_{\mathbf{x}_{2},b}(t)+\Phi_{o}^{\prime},
\end{equation}
where the amplitudes for the second particle are given by formulas
(\ref{AMPLITUDE1}) and (\ref{AMPLITUDE2}), with $\mathbf{x}_{2}$ replacing
$\mathbf{x}_{1}$ and $\mathbf{k}_{2}$ replacing $\mathbf{k}_{1}$, and
$\Delta\Phi_{\mathbf{x}_{i},b}(t)=\Phi_{\mathbf{x}_{i},b_{1}}(t)-\Phi
_{\mathbf{x}_{i},b_{2}}(t)$, with $i=1,2$. Thus, if the phase relation between
the two amplitudes is stable one can have absolutely noiseless interference
with $100\%$ visibility, while there is no single particle interference:
\begin{equation}
P(\mathbf{x}_{1},t)=\int d\mathbf{x}_{2}P(\mathbf{x}_{1},\mathbf{x}%
_{2},t)=const.
\end{equation}
As we shall see, the unbounded visibility is not the only feature by which two-particle
interference differs from classical one.

\subsection{Quantum entanglement}

Entanglement, according to Erwin Schr\"{o}dinger (1935a, 1935b) contains \textquotedblleft\emph{the essence of quantum mechanics}%
\textquotedblright. Consider a spin-$0$ particle which decays into two
spin-$1/2$ particles (Bohm, 1951). The quantum state is such that along any
chosen direction, say the $z$-axis, the spin of particle 1 when measured can either be up or
down, which in turn, by angular momentum conservation, implies that for
particle 2 it is respectively down or up. The state of the two spins is the rotationally invariant
singlet
\begin{equation}
\left\vert \psi\right\rangle _{12}=\frac{1}{\sqrt{2}}\left(  \left\vert
\uparrow\right\rangle _{1}\left\vert \downarrow\right\rangle _{2}-\left\vert
\downarrow\right\rangle _{1}\left\vert \uparrow\right\rangle _{2}\right)  ,
\label{singlet1}%
\end{equation}
where, e.g., $\left\vert \uparrow\right\rangle _{1}$ ($\left\vert
\downarrow\right\rangle _{1}$) describes the state of particle 1 with its spin
up (down) along the $z$-direction. The minus sign is necessary to get the
rotational invariance. The state describes a coherent superposition of the two
product states: there is no information in \emph{the whole Universe} on which
of the two possibilities will be detected at the measurement stage. {None of
those two possibilities is the actual case. Actualization can happen only via
a measurement. This superposition, like any other, e.g., in the double-slit experiment, survives as long as
no measurement actualizing one of those possibilities is performed, and any possible interaction of the particles with an environment does not destroy it. While none of the two possibilities actually can be assigned without measurement, both of
them affect the predictions for all measurements.} Another property of the
state (\ref{singlet1}) is that it does not make \emph{any} prediction about the result of
spin measurement on only one of the two particles: the result is
random. The spin state of one of the particles is described by a reduced
density operator\footnote{Reduced density matrices (fully) describe states of subsystems. Take a two subsystems $1$ and $2$. The average of any observable  of the composite system, say $\hat{A}_{12}$, is given by $Tr_{12}(\hat{A}_{12}\varrho_{12})$, where $\varrho_{12}$ is the density matrix of the full system $12$. If the observable acts {\em only} on system $2$, that is, it is of the form $\hat{A}_{12}=\hat{I}_1\otimes \hat{B}_2$, where $\hat{I}_1$ is the unit operator for system $1$ , one has $Tr_{12}(\varrho_{12}\hat{I}_1\otimes \hat{B}_2)=Tr_2(\hat{B}_2 Tr_1\varrho_{12}) $. This relation holds because,  to calculate the trace, one can always use a basis which consists of tensor products of basis states of the two subsystems,
$|a_{i}\rangle_1|b_{j}\rangle_2$, with ranges of the indices $i,j$ defined by the dimensions of the subsystems (trace is basis independent). Thus it is evident that the average is effectively defined by $\varrho_{2}=Tr_1\varrho_{12}$, which is the density matrix describing all predictions concerning system $2$ alone (disregarding its possible correlations with system $1$). This is the reduced density matrix of system $2$.},
which is a totally random state $\frac{1}{2}I_{k}$, where $k=1,2$ is indexing the subsystems, and $I_k=|\uparrow\rangle_{kk}\langle\uparrow|+|\downarrow\rangle_{kk}\langle\downarrow|$ is
the unit operator for a given subsystem. All information contained in the state in Eq.
(\ref{singlet1}) defines only joint properties (Schr\"{o}dinger, 1935). The joint property
can be put as follows:
\emph{the two spins, if measured with respect to the same direction, will be
found opposite}. As a matter of fact this property for any pair of complementary measurement settings fully defines the singlet state.

Imagine that the two particles can be separated far apart, one in the laboratory of Alice
and the other one in Bob's. As soon as Alice measures the value of a spin
projection along some axis, new information is gained, and for {\em her} the state
of Bob's particle is a well defined pure one. This is independent of the spatial
separation between them. Thus a state like (\ref{singlet1}) is a perfect
case to study --- and to reveal --- the EPR paradox.\footnote{Following Bohm (1951), one could apply the EPR
reality criterion to the singlet state (\ref{singlet1}): \textquotedblleft If,
without in any way disturbing a system, we can predict with certainty (i.e.,
with probability equal to unity) the value of a physical quantity, then there
exists an element of physical reality corresponding to this physical
quantity\textquotedblright. This would imply that to any possible spin
measurement on any one of our particles we can assign such an element of
physical reality on the basis of a corresponding measurement on the other
particle. Whether or not we can assign an element of reality to a specific
spin component of one of the systems must be independent of which measurement
we actually perform on the other system and even independent of whether we
care to perform any measurement at all on that system. This approach was shown to be leading to a class of theories incompatible with quantum mechanics (Bell, 1964). The concept of elements-of-reality was shown to be strictly self-contradictory via the GHZ theorem (see further on).} Basically, all earlier studies of entanglement concentrated
on entangled states of spins $1/2$ or photonic polarizations, for a review see Clauser and Shimony (1978).
Much later, we saw an emergence of research on entanglement of three or more
particles, practically together with the advent of quantum information.

\subsubsection{Theoretical methods for entanglement analysis}

The most important tool for the analysis of pure states of two subsystems is
the so-called Schmidt decomposition. For any pure state, $\left\vert
\Psi\right\rangle $, of a \textit{pair} of subsystems, one described by a
Hilbert space of dimension $N$, the other by a space of dimension $M$, say
$N\leq M$, one can find preferred bases, one for the first system, the other
one for the second, such that the state is a sum of bi-orthogonal terms,
i.e.,
\begin{equation}
\left\vert \Psi\right\rangle =\sum_{i=1}^{N}r_{i}\left\vert a_{i}\right\rangle
_{1}\left\vert b_{i}\right\rangle _{2} \label{Schmidt}%
\end{equation}
with $_{n}\langle x_{i}|x_{j}\rangle_{n}=\delta_{ij}$, for $x=a,b$ and
$n=1,2$. The coefficients $r_{i}$ are real and positive. The appropriate single subsystem
bases, here $\left\vert a_{i}\right\rangle _{1}$ and $\left\vert
b_{j}\right\rangle _{2}$,
depend upon the state.
A proof\footnote{The crux of the proof is that the greatest of the coefficients is given by
$Max_{|a\rangle_1|b\rangle_2}|\langle \Psi |a\rangle_1|b\rangle_2|$ and after finding it and
the states that give the maximization, say $|a_1\rangle_1$ and $|b_1\rangle_2$, one
searches for the second greatest coefficient by performing maximization
over the linear subspace to which $|a_1\rangle_1$ and $|b_1\rangle_2$ do
not belong. This is  recursively continued to get next coefficients and
basis states, till the procedure halts.} of Schmidt
decomposition can be found in, e.g., the book by Peres (2002). A generalization of
Schmidt decomposition to more than two subsystems is not straightforward, see e.g.
Carteret, Higuchi, and Sudbery, (2000). It is easy to show that the two
reduced density matrices of Eq.$\,$(\ref{Schmidt}) are endowed with the same
spectrum. This does not hold for three or more particle subsystems.

Every pure state of two spins $1/2$ can be put into the following form:
\[
\cos{\alpha}\left\vert \uparrow\right\rangle _{1}\left\vert
\uparrow \right\rangle _{2}+\sin{\alpha}\left\vert
\downarrow\right\rangle _{1}\left\vert \downarrow\right\rangle _{2},
\]
where the states $\left\vert \uparrow\right\rangle _{n}$ and $\left\vert
\downarrow\right\rangle _{n}$, $n=1,2$, are the eigenstates of the
$\mathbf{{z}^{(n)}}\cdot{\sigma}^{(n)}$ operator. The unit vectors
$\mathbf{{z}^{(n)}}$ are individually defined for each of the observers' particles. They
define the basis for the Schmidt decomposition for each of the subsystems.

More complicated is the theory of entanglement of mixed states. A state (pure
or mixed) described by a density matrix $\rho_{AB}$ of a composite quantum
system consisting of two  sub-systems A and B is separable if and only
if $\rho_{AB}$ is a convex combination of products density matrices, $\rho
_{A}^{\lambda}$ and $\rho_{B}^{\lambda}$, of the two sub-systems, namely,
$\rho_{AB}=\sum_{\lambda}p_{\lambda}\rho_{A}^{\lambda}\otimes\rho_{B}%
^{\lambda}, $ where $p_{\lambda}\geq0$ and $\sum_{\lambda}p_{\lambda}=1$.
Otherwise, $\rho_{AB}$ is entangled (Werner, 1989). For composite systems of
more than two sub-systems this definition can be generalized in a
straightforward way.

Basic structural criteria, which decide whether a given density operator
represents an entangled state, were first given by Peres (1996) and in the full form by Horodecki
\textit{et al}. (1996). The full set of separable mixed states is a bounded convex set
in a multidimensional real space of hermitian operators. Thus, any entangled
state is separated from the set of separable states by a hyperplane. The
equation of such a hyperplane is defined by an element of the space, namely, a
hermitian operator $\hat{W}$, which is called \textquotedblleft entanglement
witness\textquotedblright\ (Horodecki \textit{et al}., 1996; Terhal, 2000;
Lewenstein \textit{et al}., 2000; Bru\ss \ \textit{et al}., 2002, Bourennane \textit{et al}., 2004a). Since the
scalar product in the operator space is given by $\mathrm{Tr}(\hat{A}%
^{\dagger}\hat{B})$, the equation of a hyperplane in the space is given by
$\mathrm{Tr}(\hat{W}\varrho)=\mathrm{const}$. A hermitian operator $\hat{W}$ is an entanglement witness if for all
separable states one has  $\mathrm{Tr}(\hat{W}\varrho_{sep})\geq0$, whereas there exists an entangled state for which one has $\mathrm{Tr}(\hat{W}\varrho_{ent}%
)<0$. Thus, via measurement of a suitably chosen witness operator one can
detect entanglement.

The original idea of Peres was the observation that positivity of a partial transposition (PPT) of a density matrix (i.e. its transposition for just one subsystem) is a necessary condition for a state to be separable. This was extended by Horodecki Family to a fully general necessary and sufficient condition for separability, which is that a density matrix after {\em any} positive transformation (map) on one of the subsystems should remain a density operator\footnote{Partial transposition is a positive operation but is not ``completely positive'', while e.g. the most general quantum evolution of a subsystem is always represented by a ``completely positive map'', as such map leads from one density matrix to another one for the compound system even if the subsystem is entangled with another one}. The spin-offs of such considerations are measures of entanglement  via the negativity of the eigenvalues of a partial transpose of the density matrix, etc.
Other methods that give a quantitative measure of the degree
of entanglement of bipartite entangled states include the entanglement of
formation (Bennett et al. (1996c)), concurrence (Wootters (1998) and
Coffman et al. (2000)), and tangle (White et al. (2001)). For more
details on the entanglement theory, we refer the readers to
comprehensive reviews by Alber et al. (2001), Mintert et al. (2005),
Horodecki et al. (2009), and G\"{u}hne and Toth (2009).

An interesting feature of the theory of entanglement of mixed states is that for two three dimensional systems, or for more complicated ones, one can find states which are entangled, but from which no maximally entangled state can be distilled\footnote{Distillation is process in which two or more parties obtain some amount of  maximally  entangled states out of a more numerous set of copies of less entangled states by making only local operations and classical communication (LOCC) (Bennett et al. 1996a, 1996c).} (Horodecki, Horodecki and Horodecki, 1998). Such states are called bound entangled.

\subsection{Interferometry with entangled two- and multi-photon states}

\label{sec:MP-IF}

Entanglement can manifest itself in strictly quantum interference phenomena, that is, these phenomena can neither be explained by a classical wave nor by a classical particle picture.
To show this, below we present the basics of multi-particle interferometry.

\subsubsection{EPR interferometry}

\label{sec:interfero2}

\begin{figure}[ptb]
\begin{center}
\includegraphics[width=0.45\textwidth]{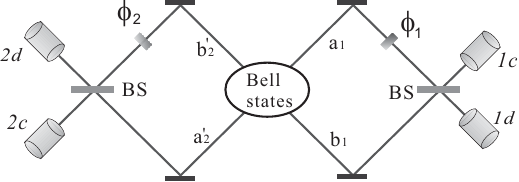}
\caption{A two-photon interferometer with variable phase shifts $\phi_1$ and
$\phi_2$. Before being combined at the 50:50 beamsplitter (BS) and then subject to
single-photon detections, the two paths of each photon acquire a relative
phase shift. For experimental realization of such a two-photon interferometer
see Fig.~\ref{dc_rarityscheme}.}%
\label{int2}%
\end{center}
\end{figure}

Recall that in a single-particle interferometer, as in Young's double-slit
experiment, the interference pattern appears only if the particle's two
paths are indistinguishable. However, for interferometers involving  two or more
particles, dramatically new features arise. Figure \ref{int2} is a
sketch of the generalization of the concept of a Mach-Zehnder interferometer to two-photon interferometry (Horne and Zeilinger, 1986; \.{Z}ukowski
and Pykacz, 1988; Horne \textit{et al}., 1989, 1990; Rarity and Tapster, 1990a; Greenberger \textit{et
al}., 1993). We assume that  a central source emits two photons in an entangled state
\begin{equation}
\left\vert \psi\right\rangle _{12}=\frac{1}{\sqrt{2}}(\left\vert
a\right\rangle _{1}\left\vert a^{\prime}\right\rangle _{2}+\left\vert
b\right\rangle _{1}\left\vert b^{\prime}\right\rangle _{2}). \label{aabb}%
\end{equation}
Here $\left\vert a\right\rangle $ and $\left\vert b\right\rangle $
($\left\vert a^{\prime}\right\rangle $ and $\left\vert b^{\prime}\right\rangle
$) are two different spatial modes of photon-1 (photon-2). The entanglement of
$\left\vert \psi\right\rangle _{12}$ is actually called momentum entanglement
(Horne and Zeilinger, 1985, Rarity and Tapster, 1990a), whose creation will be described in
section~\ref{sec:creation}. Before being combined at a 50:50 beamsplitter (BS) and then
subject to single-photon detections, the two paths of each photon acquire a
relative phase shift.%

%EndExpansion

By taking into account the phase shifts $\phi_1$\ and $\phi_2$ and the action
of the two beamsplitters, the probabilities of the \textit{coincidence detections} of
two photons at the detector pairs (D$_{1c/d},$D$_{2c/d}$) read
\begin{align}
p_{1c,2d}(\phi_1,\phi_2)  &  =p_{1d,2c}(\phi_1,\phi_2)=\frac{1}{4}[1+\cos
(\phi_1-\phi_2)],\nonumber\\
p_{1c,2c}(\phi_1,\phi_2)  &  =p_{1d,2d}(\alpha,\phi_2)=\frac{1}{4}[1-\cos
(\phi_1-\phi_2)]. \label{sin}%
\end{align}
Thus, by simultaneously monitoring the detectors on both sides of the
interferometer, while varying the phase shifts $\phi_1$\ and$\phi_2$, the
interference fringes will be observed as shown by the sinusoidal terms. In contrast, for any single detector
the count rate shows no interference at all. For example, $p_{1c}=p_{1c,2c}%
(\phi_1,\phi_2)+p_{1c,2d}(\phi_1,\phi_2)=\frac{1}{2}$, independent of $\phi_1
$\ and$\phi_2$.

\subsubsection{GHZ interferometry}

\begin{figure}
[ptb]
\begin{center}
\includegraphics[
width=0.36\textwidth
]%
{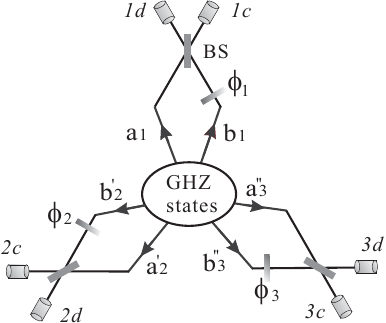}%
\caption{A three-photon interferometer with variable phase shifts $\phi_{1}$,
$\phi_{2}$ and $\phi_{3}$.}%
\label{int3}%
\end{center}
\end{figure}

After many years of studying only two-particle entanglements, in 1989 a generalization of the EPR interferometry to  three-photons
was proposed (Greenberger \textit{et al}., 1989, later refereed to as GHZ). The most elementary case is shown in
Fig.~\ref{int3}. Though such a step from 2 to 3 seems to be small, it
nevertheless leads to profound implications, one of which is the GHZ
theorem (Greenberger \textit{et al}., 1989, 1990; Mermin, 1990a).

At the
center of the interferometer is a source emitting three photons in a so-called
GHZ-entangled state
\begin{equation}
\left\vert \mathrm{GHZ}\right\rangle _{123}=\frac{1}{\sqrt{2}}(\left\vert
a\right\rangle _{1}\left\vert a^{\prime}\right\rangle _{2}\left\vert
a^{\prime\prime}\right\rangle _{3}+\left\vert b\right\rangle _{1}\left\vert
b^{\prime}\right\rangle _{2}\left\vert b^{\prime\prime}\right\rangle _{3}). \label{ghzint}%
\end{equation}
Here each photon has two different spatial modes, which are, e.g., for
photon-1 $\left\vert a\right\rangle $ and $\left\vert b\right\rangle
$.%
%EndExpansion
By taking into account the actions of the relative phase shifts and the 50:50 beamsplitters (for their properties see the next Section),
one can deduce novel features of three-particle interference
(Greenberger \textit{et al}., 1990). First, the respective three-fold
coincidence detection probability for the output modes
[$(1c,2c,3c), (1d,2d,3d), etc.$] reads
\begin{align}
p_{1c,2c,3c}(\phi_{1},\phi_{2},\phi_{3})  &  =\frac{1}{8}[1+\sin
(\phi_{1}+\phi_{2}+\phi_{3})],\nonumber\\
p_{1d,2c,3c}(\phi_{1},\phi_{2},\phi_{3})  &  =\frac{1}{8}[1-\sin
(\phi_{1}+\phi_{2}+\phi_{3})], ~ \mathrm{etc.} \label{singhz}%
\end{align}
The three-fold coincidence rates given in Eq.~(\ref{singhz})
display sinusoidal oscillations depending on the sum $\phi_{1}+\phi_{2}+\phi_{3}$. Second, this
three-particle interferometer does not exhibit any two-particle fringes. For
example, if only two-particle coincidences $2c$-$3c$ are detected, while
the modes $1c$ and $1d$ are ignored, the observed rate will be constant
$p_{1c,2c,3c}(\phi_{1},\phi_{2},\phi_{3})+p_{1d,2c,3c}(\phi
_{1},\phi_{2},\phi_{3})=\frac{1}{4}$, and completely independent of the
phases\footnote{This holds only for observables dependent on $\phi_i$, like those shown in the Fig.~\ref{int3}}. Finally, a similar argument shows that, of course, no single-particle fringes can
be observed. Actually, an $n$-particle
interferometer generalized along the above reasoning will only exhibit
$n$-particle fringes, but no ($n-1$)-, ($n-2$)-,..., single-particle fringes
(Greenberger \textit{et al}., 1990).

The above two- and three-photon interferometry was described here using photon's path degree of freedom. However, similar
interference effects can  be observed using any of the photon's degrees of
freedom, e.g. polarization. Moreover, the above argument
should be understood as a special case of a wider concept indicating that  entangled massive particles (e.g,
electrons and atoms) could also display multi-particle interference.

\subsection{Interferometry with multiport beamsplitters}

Novel interferometric effects can be obtained with N-port
beamsplitters, which are  devices which split light into more than two output
beams  (for a general theory of such devices see Reck et al. (1994)). Such
devices can be utilized in multiparticle interferometry (Zeilinger et al.
1993). With current technology  such experiments are becoming feasible.
%%%%%%%%%%%%%%%%%%%%%%%%%%%%%%%%%%%%%%%%%%%%%%%%%%%%%%%%%%%%%%%%%%%%

\section{Photonic qubits and linear optics}

\label{sec2-lo}
The possibility of performing quantum information-processing
tasks with photons is based on the fact that quantum
information can be encoded in quantum states of certain degrees of freedom
(e.g., polarizations) of individual photons, and that
individual photons can be manipulated either by simple optical elements (e.g.,
wave plates and interferometers), or by letting them
interact with matter (trapped ions, atoms and so on) at a atom-photon
interface. Here we shall show to what extent a photon can carry a qubit,
and the simplest elements that are used to manipulate it. To this end, let us
begin with a formal definition of photons and their quantum states.

%\subsection{Photon-number states}

%\label{sec:Fock}

The formal theory of quantization of electromagnetic fields was formulated by
Dirac (1927). Here we only give its basic mathematical devices for
completeness; for detailed treatment, we refer to standard textbooks on
quantum optics, e.g., Walls and Milburn (1994), Mandel and Wolf (1995), Scully
and Zubairy (1997), and especially Bialynicki-Birula and Bialynicka-Birula (1975).

A single-photon pure state can be characterized by a specific wave packet
profile $g_{\lambda\mathbf{k}}$, i.e., by the quantum amplitudes for a given
momentum $\mathbf{k}$ and polarization $\mathbf{\lambda}$. According to the
Born rule, $|g_{\lambda\mathbf{k}}|^{2}$ gives the probability density of
having the single photon with the momentum $\hbar\mathbf{k}$ and
polarization $\lambda$. Thus, one has $\sum_{\lambda=1,2}\int d\mathbf{k}%
|g_{\lambda\mathbf{k}}|^{2}=1$. The wave packet profiles are vectors in a
Hilbert space with a scalar product given by
\begin{equation}
\langle g|h\rangle=\sum_{\lambda=1,2}\int d\mathbf{k}g_{\lambda\mathbf{k}%
}^{\ast}h_{\lambda\mathbf{k}}.
\end{equation}

One can introduce an arbitrary orthonormal basis
$g^{l}_{\lambda\mathbf{k}}$,
where $l$ are natural numbers and $\langle g^{n}|g^{m}\rangle=\delta_{nm}$.
Two different orthonormal bases, to be denoted respectively as primed, and
unprimed, are related by a unitary operation: ${g^{\prime}}^{m}=\sum
_{l=1}^{\infty}U_{ml}g^{l}$. The complex numbers $U_{lm}$ satisfy $\sum
_{l=1}^{\infty}U_{ml}{U^{\ast}}_{kl}=\sum_{l=1}^{\infty}{U^{\ast}}_{lm}%
U_{lk}=\delta_{mk}$. One can choose a specific basis of the wave packet
profiles of the single photon, say $g^{l}$, and with each element of such a
basis one associates a quantum oscillator-like construction to introduce
number states, namely, the Fock states. One introduces the vacuum state
$|\Omega\rangle\equiv|0,0,0,...\rangle$, the state with no photons at all for
any modes. Next, for a chosen basis one
associates a pair of operators satisfying the usual relations for creation and
annihilation operators, namely, $[\hat{a}_{l},\hat{a}_{l}^{\dagger}]=1$ and
requires that for all $l$ $\hat{a}_{l}|\Omega\rangle=0$. Using the standard oscillator
algebra one constructs states like $\frac{\hat{a}_{l}^{\dagger n_{l}}}%
{\sqrt{n_{l}!}}|\Omega\rangle$, which is a state of the electromagnetic
field in which one has $n_{l}$ identical photons of the same wave packet
profile $g^{l}$, and no other photons whatsoever. This is denoted by
$|0,...,0,n_{l},0,0,...\rangle$. Finally, one assumes that $[\hat{a}_{n}%
,\hat{a}_{m}]=0$ and $[\hat{a}_{n},\hat{a}_{m}^{\dagger}]=\delta_{nm}$, that
is, creation and annihilation operators of photons with orthogonal wave packet
profiles always commute. A general (normalized) basis state of the Fock space
is therefore of the following form:
\begin{equation}
|n_{1},n_{2},n_{3},...\rangle=\prod_{l=1}^{\infty}\frac{\hat{a}_{l}^{\dagger
n_{l}}}{\sqrt{n!}}|\Omega\rangle.
\end{equation}
All vectors of the Fock space are linear combinations of the above basis
states, which have a finite total number of photons. It is easy to see that if one defines the
creation operators with respect to an alternative basis of wave packet
profiles (here the primed ones) one has:
\begin{equation}
{\hat{a}}_{m}^{\prime\dag}=\sum_{n=1}^{\infty}U_{mn}\hat{a}_{n}^{\dagger}.
\label{LINEAR}%
\end{equation}
The vacuum state is invariant with respect to such transformation, i.e., one
still has $\hat{a}_{m}^{\prime}|\Omega\rangle=0$. For more details see, e.g.,
Bialynicki-Birula and Bialynicka-Birula (1975).

\subsection{Photonic qubits}

A quantum bit, or \emph{qubit}, is the most elementary unit of quantum
information. It is a generalization of the classical bit, which has two distinguishable states
\textquotedblleft0\textquotedblright\ and \textquotedblleft1\textquotedblright. Similarly,
we can have a qubit in two distinguishable, i.e., orthogonal states $|0\rangle$ and $|1\rangle$. However, in
contradistinction to its classical counterpart, a qubit
 can  be prepared as, or transformed to, any superposition of these
two states (normalization requires $\alpha_{0}^2+\alpha_{1}^2=1$):
\begin{equation}
|\Psi_{\mathrm{qubit}}\rangle=\alpha_{0}|0\rangle+\alpha_{1}|1\rangle. \label{QUBIT}%
\end{equation}

Any isolated two-level system consisting of a pair of orthogonal
quantum states represents a qubit. Photons, massless spin 1
particles, have only two eigenvalues of their spin along the direction of
their propagation (wave vector), $\pm\hbar$. These two spin values correspond to
right-handed and left-handed circular polarization.  Thus this property makes the photon an ideal candidate for a qubit. However, there are other degrees of freedom of a
photon that can be used to encode qubit information.

\textit{Polarization qubits.--}The most commonly used photonic qubits are
realized using polarization. In this case arbitrary qubit states can be
$\alpha_{0}\left\vert H\right\rangle +\alpha_{1}\left\vert V\right\rangle $,
where $H$ and $V$ stand for horizontal and vertical polarizations,
respectively. The advantage of using polarization qubits stems from the fact
that they can easily be created and manipulated with high precision by simple linear-optical elements such as polarizing beam splitters
(PBS), polarizers and wave plates. Photon polarization states, and spin-states of a spin $1/2$ particle are perfect qubits given by Nature, no human invention is required.

\textit{Spatial qubits.-}A single photon can also appear in two different
spatial modes or paths, $a$ and $b$: the general state reads $\alpha
_{0}|a\rangle+\alpha_{1}|b\rangle$. This may occur, e.g., if a single photon
exits a beam splitter (BS), with two output modes $a$ and $b$. Any state of
spatial qubits can be prepared by using suitable phase shifters and BSs. A
disadvantage of using spatial qubits is that the coherence between $|a\rangle$
and $|b\rangle$ is sensitive to the relative phase between the paths $a$ and $b$,
and this is difficult to control in long-distance cases.

\textit{Time-bin qubits.-}For a more robust long-distance
transmission of quantum information, one may use \emph{time-bin}
qubits. The computational basis\footnote{A basis of a qubit is called
computational if one associates logical 0 and 1 to its two orthogonal
states} consists of two states which are of
the same spectral shape, but time shifted by much more than the
coherence time\footnote{The coherence time is the time over which
the relative phase of a propagating wave remains stable. It can be
approximately estimated as $\tau \leq  \frac {\lambda^2}{c\, \Delta
\lambda}$ , where $\lambda$ is the central wavelength of the source,
$\Delta \lambda$ is the spectral width of the source, and $c$ is the
speed of light in vacuum.}. Time-bin qubits can be realized by sending a
single photon through an unbalanced Mach-Zehnder
interferometer. Its wavepacket is split by the first BS, with
transmission coefficient $T=|\alpha_{0}|^{2}$ and reflection
coefficient $R=|\alpha_{1}|^{2}$ into two pulses. The transmitted
one propagates along the short arm, and the reflected one along the
long arm. If the wave packet extension  is shorter than the arm length
difference, the output from the ports of the second,
50:50 BS is two wavepackets well separated in time. If no
photon is registered in, say, output port I, in the other output one
has a single photon in a coherent superposition of two time-bin
states $\alpha
_{0}|\mathrm{early}\rangle+\alpha_{1}|\mathrm{late}\rangle$. The
phase relation can be controlled with a phase shifter in one of the
arms of the interferometer. For more details, see a
review by Gisin \textit{et al}. (2002).

While in this review we are mainly concerned with the above three
implementations of photonic qubits, one should keep in mind that
other implementations are possible. Frequency qubits have been
first implemented in quantum cryptography (Sun \textit{et al}., 1995; Mazurenko
\textit{et al}.,1997) and more recently also in entangled atom-photon systems (Madsen
\textit{et al}., 2006).

Quantum $d$-level (high-dimensional) systems (\textquotedblleft
qudit\textquotedblright) can also be realized using, e.g., orbital
angular momentum states of photons (Mair \textit{et al}., 2001), or
using simultaneously two or more degrees of freedom listed above.
For instance, for the latter case a polarized single photon in a
coherent superposition of two spatial modes can be thought as
a quantum system in a four-dimensional Hilbert space
(Boschi \textit{et al}. 1996; Michler \textit{et al}., 2000a; Chen
\textit{et al}., 2003; Kim, 2003).

\textit{Two-photon polarization entangled states.--}
The so-called Bell states
\footnote{Please note, that the term ``Bell state'' was earlier used with a
completely opposite meaning. E.g., Mann, Revzen and Schleich (1992) define
a  Bell state as ``a pure state which when split in any way cannot violate
Bell's inequality''.} form a basis in the four dimensional
two-qubit Hilbert space. Bell states of photonic polarization
qubits can be for example:
\begin{eqnarray}%
|\psi^{\pm}\rangle _{12} & = & \frac{1}{\sqrt{2}}(\left\vert H\right\rangle _{1}\left\vert V\right\rangle _{2} \pm \left\vert V\right\rangle _{1}\left\vert H\right\rangle _{2}), \\
|\phi^{\pm}\rangle _{12} & = & \frac{1}{\sqrt{2}}(\left\vert H\right\rangle _{1}\left\vert H\right\rangle _{2} \pm \left\vert V\right\rangle _{1}\left\vert V\right\rangle _{2}).
\label{Bell_s}
\end{eqnarray}

 As we shall see in the following chapters,
such entangled states serve as a central physical resource in
various quantum information protocols like quantum cryptography, quantum teleportation,
entanglement swapping and in tests aimed at excluding hidden variable models of quantum mechanics.

\subsection{Simple linear-optical elements}

\label{sec:sec2-LOE}

In the photonic domain, quantum states of photons can be easily, and with a high precision, manipulated by
simple passive linear-optical devices. These linear-optical elements
include BS, polarizing beamsplitter (PBS), wave plates and
phase shifters. Classically, such devices conserve energy: The total input energy
equals the total output energy, and there is no energy transfer between
different frequencies. A passive linear optical device is described by a
unitary transformation of annihilation operators for the same frequency:
\begin{equation}
\hat{a}_{m}^{out}=\sum_{m}U_{mn}\hat{a}_{n}^{in},
\end{equation}
where $U$ is a unitary matrix, and the indices denote a basis of orthogonal
modes.
\begin{figure}
[ptb]
\begin{center}
\includegraphics[width=0.45\textwidth
]%
{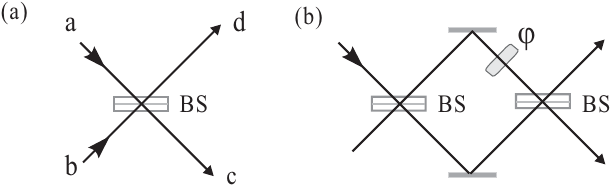}%
\caption{The function of a BS. (a) The BS coherently transforms two input
spatial modes (a, b) into two output spatial modes (c, d). (b) A Mach-Zehnder
interferometer consists of two BSs, mirrors and phase shift (as a whole it forms
a universal tunable beamsplitter).}%
\label{bs}%
\end{center}
\end{figure}

The BS is one of the most important optical elements. It has two
spatial input modes $a$ and $b$ and two output modes $c$ and $d$
(Fig.~\ref{bs}). The theory of the lossless BS was developed by
Zeilinger (1981), and Fearn and Loudon (1987), for lossy BS, see
Barnett \textit{et al}.\ (1989). In a simplified theory of BS, one
assumes identical action for every relevant frequency. The most
commonly used BS is the symmetric 50:50 BS characterized by the
following transformation
\begin{align}
&  \hat{a}\longrightarrow\frac{1}{\sqrt{2}}\hat{c}+\frac{i}{\sqrt{2}}\hat{d},\nonumber\\
&  \hat{b}\longrightarrow\frac{i}{\sqrt{2}}\hat{c}+\frac{1}{\sqrt{2}}\hat{d}. \label{bs50}%
\end{align} \
%\begin{align}
%&  \hat{a}^{\dagger}\longrightarrow\frac{1}{\sqrt{2}}\hat{c}^{\dagger}+\frac{i}{\sqrt{2}}\hat{d}^{\dagger},\nonumber\\
%&  \hat{b}^{\dagger}\longrightarrow\frac{i}{\sqrt{2}}\hat{c}^{\dagger}+\frac{1}{\sqrt{2}}\hat{d}^{\dagger}. \label{bs50}%
%\end{align}
In such a case, an outgoing particle can be found with equal probability
($50\%$) in either of the output modes $c$ and $d$, no matter
through which single input beam it came. The factor $i$ in
Eq.~(\ref{bs50}) is a consequence of unitarity. It describes a phase
jump upon reflection (Zeilinger, 1981).

Using two 50:50 BSs and a phase shifter one can build a Mach-Zehnder
interferometer, which in fact can be used as a universal BS of a variable transitivity and reflectivity (see Fig. 1(b)). The phase shifter
can be some glass plate, a birefringent optical crystal, or a path-length tuner (e.g., an optical trombone). Together with additional phase shifters in the input and output ports, the Mach-Zehnder
interferometer  can perform an arbitrary SU(2) unitary transformation
on a qubit encoded in the two  spatial modes (see, e.g., Englert
et al. (2001)).

\begin{figure}
[ptb]
\begin{center}
\includegraphics[
width=0.45\textwidth
]%
{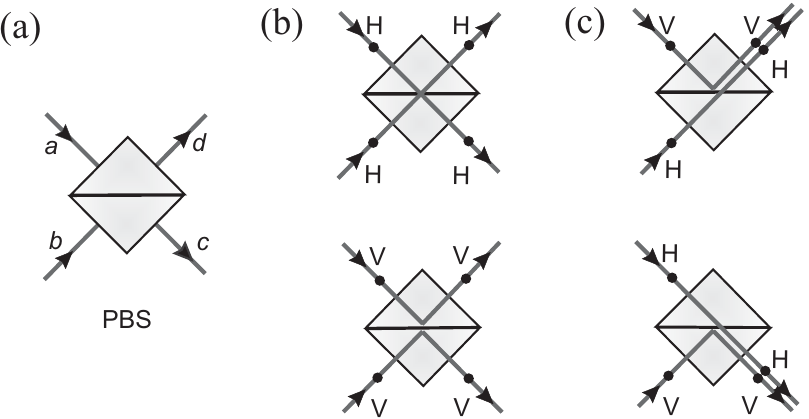}%
\caption{The operation of a polarizing beamsplitter (PBS). (a) In the usual
configuration, the PBS transmits horizontal, and reflects
vertical, polarization. (b) Two photons, each entering via a different input: If the two photons incident onto the PBS have
identical polarization, then they will always go out along different
directions, so there will be one photon in each of the two output modes. (c)
If the two incident photons have opposite polarizations,
they will always go out along the same direction, so there will be two
photons in one of the two outputs and none in the other. In essence, a PBS can
thus be used as a polarization parity checker (Pan and Zeilinger, 1998; Pan
\textit{et al}., 2001b).}%
\label{pbs-fig}%
\end{center}
\end{figure}

Another important component is the polarizing beamsplitter (PBS) (see Fig.~\ref{pbs-fig}). A standard
PBS transmits the horizontal and reflects the vertical polarization. The
transformations between the incoming modes ($a$ and $b$) and the outgoing
modes ($c$ and $d$) are as follows
\begin{equation}
\hat{a}_{H}\rightarrow\hat{c}_{H}~\text{and}~\hat{a}_{V}\rightarrow i\hat{d}
_{V}\ ;\ \ \hat{b}_{H}\rightarrow\hat{d}_{H}~\text{and}~\hat{b}_{V}
\rightarrow i\hat{c}_{V}. \label{pbs}%
\end{equation}

%EndExpansion

For polarization qubits, any single-qubit operation can be accomplished by
using a sequence of suitably oriented quarter- and half-wave plates. Simply, a
half-wave plate retarding the $45^{\circ}$ polarization acts on the $H$ and
$V$ modes exactly as a certain 50:50 BS, etc. Now, the input modes are two
orthogonal polarizations, instead of two  spatial input modes of a
Mach-Zehnder interferometer.

With these simple optical elements, large optical networks can be
constructed, mapping an input state onto an output state via a
\textit{linear} transformation determined by the networks' unitary transformation. As a
possible generalization of the usual interferometers with two ports,
an $N$-port interferometer was proposed by Reck \textit{et al}.
(1994), which can realize any $U(N)$ transformation for $N$ optical
spatial modes by using an arrangement of BS, phase shifters and
mirrors. Weihs \textit{et al}. (1996) realized an all-fiber
three-path Mach-Zehnder interferometer, which is based on the idea
of symmetric unbiased multiport BSs (Zeilinger \textit{et al}.,
1993).

\subsection{Two-photon interference due to indistinguishability of photons}

\label{sec:HOM}

Quantum interference\footnote{See Mandel (1999) for a review on a series of
pioneering parametric down-conversion experiment revealing quantum effects in
one-photon and two-photon interference. Our present review can be treated as
a direct continuation of Mandel's one.} may occur also entirely due to indistinguishability of
particles. We shall describe this phenomenon in the case of photons, i.e., bosons. One can obtain all
effects due to indistinguishability by a suitable symmetrization of the
amplitudes for elementary processes, for which we do not know which particle
ended up in which final state (for bosons amplitudes do not change sign when
particles are interchanged). However it is much more convenient to use the
formalism of bosonic creation and annihilation operators, as its algebra directly takes into account the
symmetrization. Here we shall present the most elementary optical effect due
to the indistinguishability of photons, the Hong-Ou-Mandel interference (Hong,
Ou and Mandel, 1987) behind a 50:50 BS.

\begin{figure}
[ptb]
\begin{center}
\includegraphics[width=0.41\textwidth]%
{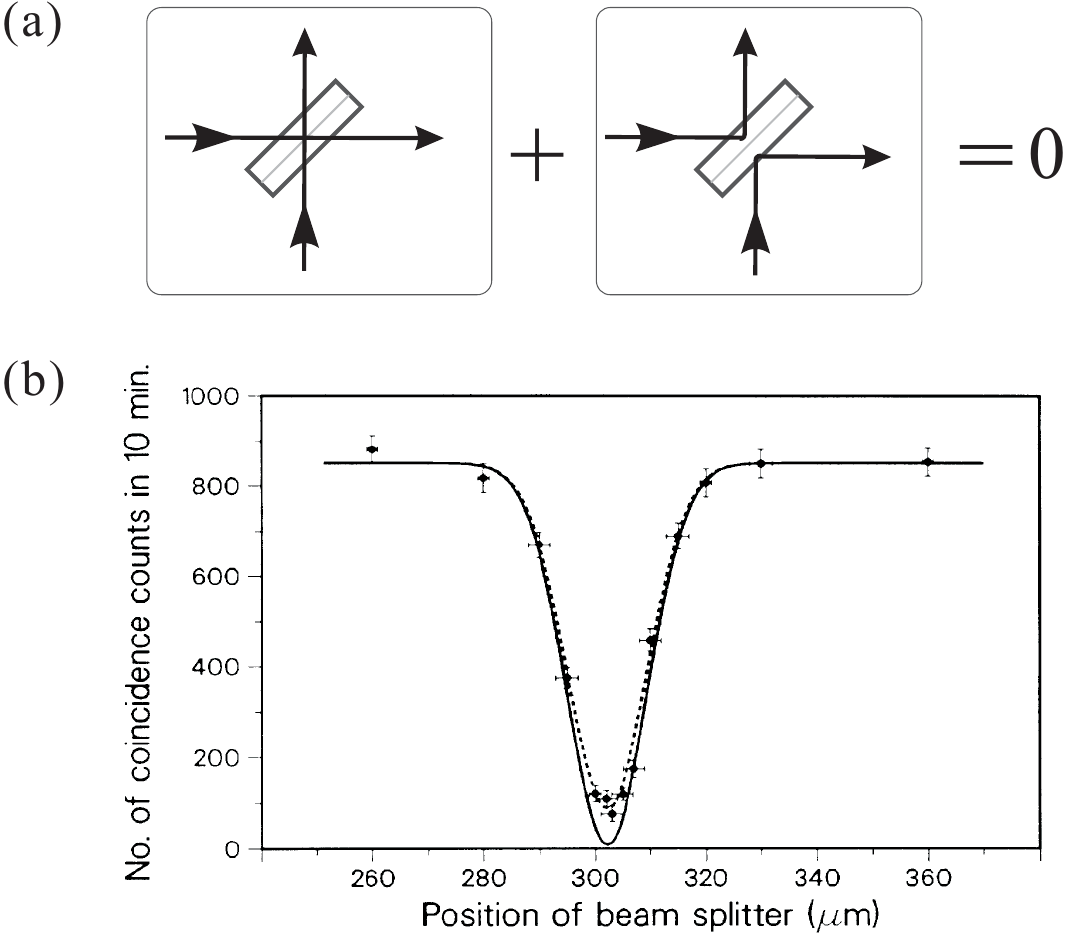}%
\caption{(a) Individually incoming photons can be transmitted or reflected. However, since the
two photons are identical, we cannot distinguish between the two cases when both are transmitted or both reflected. The BS
introduces a phase difference of $\pi$ between the two  amplitudes describing such possibilities.
This
leads to a destructive interference. Thus no coincidence detection can be found.
(b) Data  taken from Hong, Ou and Mandel, (1987): Pairs of photons impinging on a BS are produced by spontaneous parametric down-conversion
(see Section~\ref{sec:spdc}) and have the
same polarization and frequency distributions. The measured number of coincidence counts as a
function of relative path displacement (temporal distinguishability) shows a ``Hong-Ou-Mandel" dip, for equal optical paths.
}%
\label{homdata}%
\end{center}
\end{figure}

As we have seen previously, upon reflection off a symmetric 50:50 BS  a photon
picks a phase shift $i$. Denote the input modes  as $a$ and $b$ and output
modes as $c$ and $d$. If we have two spectrally identical photons (of the same polarization)
each entering at exactly the same moment an opposite input port of the BS, the initial
state $\hat{a}^{\dagger}\hat{b}^{\dagger}|\Omega\rangle$, is transformed into
\begin{equation}
\frac{1}{\sqrt{2}}(\hat{c}^{\dagger}+i\hat{d}^{\dagger})\frac{1}%
{\sqrt{2}}(\hat{d}^{\dagger}+i\hat{c}^{\dagger})|\Omega\rangle
\rightarrow\frac{i}{{2}}(\hat{c}^{\dagger2}+\hat{d}^{\dagger2})|\Omega\rangle.
\label{eq:hom}
\end{equation}
We have a cancellation of the two terms, $\hat{c}\hat{d}-\hat{d}\hat{c}=0$, which describe the cases in which each photon exits by a different exit port. This cancellation occurs if the two photons are perfectly indistinguishable in terms of all their other degrees of freedom such as frequency, time, or polarization. The two
photons exit the BS paired via only one (random) output port. This is a
bunching effect due to the bosonic character of photons. Thus, there are no
coincidences between the output ports\footnote{If the two incident photons are in an antisymmetric
polarization entangled state $|\Psi^-\rangle$, the amplitudes for photons to exit via different ports
interfere \emph{constructively}, as in this case their spatial wave function has to be antisymmetric, too (Weinfurter 1994). For proper path length, one has a coincidence peak (instead of a dip) (Mattle \textit{et al}. 1996). This observation is crucial for discriminating the
$|\Psi^-\rangle$ using a BS (see the ``Bell-state analyzer'' Section~\ref{sec:analyzer}). For a color (at frequency $\omega_1$ and $\omega_2$) entangled
state, one can further observe ``spatial quantum beating'', where the two-photon detection exhibits a modulation of the form $\cos(\omega_1-\omega_2)\tau$
(see e.g. Ou and Mandel, 1988b).}. Another (and more graphical) way to
look at this is shown in Fig.~\ref{homdata}(a).

If the photons are at least partially distinguishable (in this case we shall label the annihilation operators of the photons with different
subscripts as $\hat{a}_{1}$ and $\hat{b}_{2}$) the initial state $\hat
{a}_{1}^{\dagger}\hat{b}_{2}^{\dagger}|\Omega\rangle$ is transformed by a perfect 50:50 BS via the relations in Eqn.\ref{bs50}
into
\begin{equation}
\frac{1}{\sqrt{2}}(\hat{c}_{1}^{\dagger}+i\hat{d}_{1}^{\dagger})\frac{1}%
{\sqrt{2}}(\hat{d}_{2}^{\dagger}+i\hat{c}_{2}^{\dagger})|\Omega\rangle.
\end{equation}
Since $\hat{c}_1 \neq \hat{c}_2$ and $\hat{d}_1 \neq \hat{d}_2$, the  terms that
contribute to the cases in which each photon exits by a different exit port, namely
$\frac{1}{{2}}(\hat{c}_{1}^{\dagger}\hat{d}_{2}^{\dagger}-\hat{d}_{1}^{\dagger}\hat{c}_{2}^{\dagger})|\Omega\rangle$, do not cancel with each other.

Let us use $\beta$ to denote the degree of distinguishability between the photon 1 and 2. The probability of finding a
coincidence count at exits $c$ and $d$, which is given by the square of the norm of $\frac{1}{{2}}(\hat{c}_{1}^{\dagger}\hat{d}_{2}^{\dagger}-\hat{d}_{1}^{\dagger}\hat{c}_{2}^{\dagger})|\Omega\rangle$,
is $(1/2)|\beta|^{2}.$ Thus, if $|\beta|=1$ (the photons $\hat{a}$ and $\hat{b}$ are
fully distinguishable), this probability reads $1/2$; if $\beta=0$
(the photons are indistinguishable), it vanishes. Therefore, the
Hong-Ou-Mandel effect depends critically on the distinguishability of photons.
The distinguishability was tuned in the original experiments with the
temporal delay between the two photons (Fig.~\ref{homdata}(b)).

The original Hong-Ou-Mandel experiment used the two photons of the same signal-idler
pair from parametric down-conversion (see Section~\ref{sec:spdc}).
Later experiments of this kind evolved into observations of a  Hong-Ou-Mandel dip for photons
originating from two sources, which were  progressively more and more independent of each other.
This was motivated both by fundamental issues, such as: whether independent photons indeed interfere,
as well as practical ones: interference of photons emerging from different sources must be harnessed if
one wants to build complicated schemes realizing quantum protocols, e.g. quantum repeaters  (see Section~\ref{sec:repeater}) and distributed quantum computing
(see Section~\ref{computing}). Rarity \textit{et al}.~(1996) observed the interference between
independent photons, one of which was a triggered single photon from a down-converted pair,
and the other one was in an attenuated beam\footnote{The beam was in a pulsed weak coherent
state: for each pulse there was only a very small probability for it to contain a single photon.}
derived from the pumping laser light. Interference of two triggered single photons created via
parametric down-conversion by the same pump pulse passing twice through a nonlinear
crystal was achieved in the Innsbruck teleportation experiment (Bouwmeester \textit{et al}., (1997),
for more details see Section~\ref{sec:teleportation}). With a similar method of triggering, Keller \textit{et al}., (1998) used photons generated by two mutually
coherent but time-separated pulses from the same mode-locked laser (Keller \textit{et al}., 1998).
Experiments aimed at observing the Hong-Ou--Mandel dip for fully independently emitted photons, like the one of Kaltenbaek et al. (2006),
will be described in Section~\ref{sec:swapping}.

The Hong-Ou-Mandel interference
 provides a powerful tool to estimate the degree of
indistinguishability of  two separately emitted photons. For instance, two single
photons successively emitted from a single quantum dot were overlapped on a BS and
the Hong-Ou-Mandel  effect was observed  (Santori \textit{et al}.
2002). Other examples include single photons from independent trapped
atoms (Beugnon \textit{et al}. 2006; Maunz \textit{et al}. 2007),
and from remote cold atomic ensembles (Yuan \textit{et al}. 2007) or
independent, tunable quantum dots (Patel et al. 2010). The interference of
 indistinguishable photons enables a process called entanglement swapping and teleportation
(Zukowski et al. 1993, see Section~\ref{sec:swapping}),
which in turns opens up prospects of distributing of  entanglement between distant matter qubits
such as ions (Moehring \textit{et al.} 2007) and atoms (Yuan \textit{et al.} 2008).

A characteristic feature of the Hong-Ou-Mandel interference is that it is sensitive
to path length changes on the order of the coherence length of the photons,
while in a Mach-Zehnder interferometer one has a sub-wavelength sensitivity.
The photon's coherence length can range from a few hundred micrometers, in pulsed
parametric down-conversion, to a few meters for trapped ions. This  makes
the Hong-Ou-Mandel interferometers very stable.

\subsection{Post-selection of entanglement and quantum erasure}

\label{sec:inter-more}

\begin{figure}
[ptb]
\begin{center}
\includegraphics[
width=0.45\textwidth
]%
{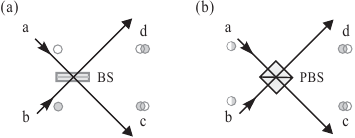}%
\caption{(a) Shih-Alley type polarization entanglement. One combines two orthogonally polarized
photons at a BS and registers the coincidence events behind it (Shih and Alley, 1988; Ou and Mandel, 1998a).
(b) In a similar way, by superposing two $(H+V)$ polarized photons at a PBS, one can  observe an entangled state
$|H\rangle_c|H\rangle_d+|V\rangle_c|V\rangle_d$ in the post-selected coincidence counts.}%
\label{SA}%
\end{center}
\end{figure}

Let us now discuss other tricks and states obtainable with a single BS on which a pair of photons impinges.

Note first, that
according to Eq. (\ref{eq:hom}),
in a Hong-Ou-Mandel interferometer,
the electromagnetic field emerging from the BS associated with the pair of photons
 is in an entangled state $[(\hat{c}%
^{\dagger})^{2}+(\hat{d}^{\dagger})^{2}]|\Omega\rangle=|2_c0_d\rangle+|0_c2_d\rangle$,
a so-called NOON state (see Section~\ref{N00N}), with $N=2$. This state has interesting interferometric properties which will be discussed later.

With the simple device one can also produce effects characteristic for a maximally entangled polarization state.
Shih and Alley (1988), and Ou and Mandel (1988a) were the first
to achieve this. In their experiments, the two photons entering the BS, each by a different input port, were  indistinguishable in all degrees of freedom,
except that they were oppositely polarized
(say, one  horizontally polarized, $H$, and the other vertically polarized,
$V$, see Fig.~\ref{SA}(a)). After being superposed on the BS and leaving the
two output ports \textit{c} and \textit{d}, the photons emerged in a
polarization-tagged two-photon state: $$\frac{1}{2}[\sqrt{2}\left\vert \psi^{-}\right\rangle
_{cd}+i(\left\vert H\right\rangle _{c}\left\vert V\right\rangle _{c}%
+\left\vert H\right\rangle _{d}\left\vert V\right\rangle _{d})].$$ A
coincident detection at the two outputs, \textit{c} and \textit{d}, can occur  only due the entangled
$\left\vert \psi^{-}\right\rangle_{cd}$ component of the total state. As in the case of the Hong-Ou-Mandel
interferometer, the visibility of this polarization interference
is reduced once the photons are partially distinguishable. E.g.,  by varying the relative optical paths before the photons reach the BS the overlap of the
photon wave packets behind the beam splitter may be controlled, up to a total distinguishability. The full effect occurs for perfect overlap.
This kind of post-selected entanglement was used to violate
Bell's inequality\footnote{There was a controversy whether such a kind of experiment can constitute
a true Bell test. It was positively resolved in Popescu et al. (1997).}, both in experiments of Shih and Alley (1988),
and Ou and Mandel (1988a).  The effect was later  demonstrated for two indistinguishable
photons from quasi-independent sources (Pittman and Franson, 2003; Fatal \textit{et al}.,
2004).

Figure~\ref{SA}(b) shows another setup for generating entanglement, in
a similar spirit, however using a PBS (such an effect was used in a proposal for obtaining multi-photon entanglement
by Zeilinger \textit{et al}. 1997, see Section~\ref{sec:multi}).
As a PBS customarily transmits H and reflects V polarization, a coincidence detection
between the two outputs can originate only if either  both
photons are transmitted (resulting in a $|HH\rangle$ case) or both are reflected
($|VV\rangle$ case). The two cases are quantum mechanically indistinguishable,
if the photons are indistinguishable. Thus again,
an entangled state $\left\vert \phi^{-}\right\rangle_{cd}$ can be generated
using postselection. As in the case of the Hong-Ou-Mandel interferometer, the entanglement
quality is  sensitive to optical path length changes of the order of the photons'
coherence length. Such interferometers play a
crucial role in creation, manipulation and projection of various multi-photon
entangled states.

When analyzing the above experiments, one could be mislead to suppose that the interference arises due to the fact that the wave packets of the two photons overlap at the BS. Indeed the Hong-Ou-Mandel dip is presented as a function of the temporal delay between the two photons, i.e. effectively in terms of the overlap  (Fig.~\ref{homdata}(b)). However, it is important to
note that essentially, the origin of this interference is due to
the indistinguishability of  two-photon amplitudes describing the various
alternatives leading to a coincidence count. To dispel a misconception that
the photons must arrive at the BS at the same
time for some type of classical local ``agreement'', Pittman \textit{et al}. (1996) performed
a ``postponed compensation'' two-photon Shih-Alley type experiment. Interference
is observed, even though the photons were arriving at the BS at different times.
In the experiment, the optical path of one input light beam was much longer than of
the other one with a difference exceeding the photon's coherence length.
However, after the interferometer, this delay was compensated  (using a polarization-selective
unbalanced Mach-Zehnder interferometer) in such a way that the firing times of the two detectors
did not provide any information whatsoever concerning which of the two-photon  processes led to
the coincidence detection. The experimental results confirmed that the quantum
interference can be indeed revived. A discussion of related ``quantum erasure'' experiments
can be found in Kwiat \textit{et al}. (1992); Michler \textit{et al}. (1996); and Scully \textit{et al}. (1991).

%%%%%%%%%%%%%%%%%%%%%%%%%%%%OCT-08-7:55 AM%%%%%%%%%%%%%%%%%%%%%%%%%%%%%%%%%%%%%%%

\subsection{Entangled-state analyzers}

\label{sec:analyzer}

The projection of two photons into a Bell state lies at the heart for many quantum
information processing protocols, such as quantum dense coding (Bennett and
Wiesner, 1992), quantum teleportation (Bennett
\textit{et al}., 1993), and entanglement swapping (\.{Z}ukowski \textit{et al}., 1993).
A deterministic {\small CNOT} gate\footnote{The quantum CNOT gate, a fundamental quantum circuit, is a two-qubit gate acting on a target qubit $|\alpha\rangle_t$ and a control qubit $|\beta\rangle_c$. It flips the target qubit ($|0\rangle_t\rightarrow|1\rangle_t$, $|1\rangle_t\rightarrow|0\rangle_t$) if the control qubit is in logic $|1\rangle_c$, and does nothing if the control qubit is $|0\rangle_c$. Note that if the control qubit is in a superposition $(1/\sqrt{2})(|0\rangle+|1\rangle)$, the action of a CNOT gate produces a maximally entangled state of the target and control qubits.} would make such a the Bell-state measurement possible. However, {\small CNOT} gates are difficult to realize
with linear optics and single photons (see section~\ref{computing}).
Nevertheless, by exploiting  quantum interference effects due to the bosonic nature of photons discussed in the above section,
 photonic Bell-state and GHZ-state analyzers can be realized
in a probabilistic way.

%%%%%%%%%%%%%%%%%%%%%%%%%OCT-08-8:06 AM %%%%%%%%%%%%%%%%%%%%%%%%%%%%%%%%%%%%%%%%

\subsubsection{Bell-state analyzer}

\label{sec:bsm}
\begin{figure}
[ptb]
\begin{center}
\includegraphics[width=0.45\textwidth
]%
{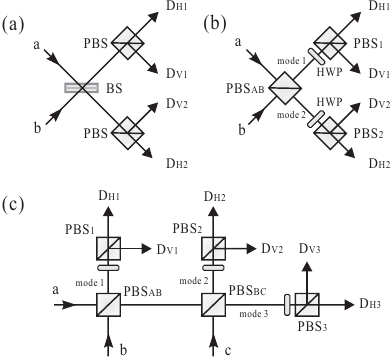}%
\caption{(a). A Bell-state analyzer using a BS.
(b). A modified Bell-state analyzer.
The angle between the half-wave plate axis and the horizontal
direction is $22.5^{\circ}$. It corresponds to a $45^{\circ}$
rotation of the polarization. (c). A GHZ-state analyzer
(Pan and Zeilinger, 1998).}%
\label{PBS2}%
\end{center}
\end{figure}
%EndExpansion

A linear optical Bell-state analyzer was  suggested by Weinfurter
(1994) and by Braunstein and Mann (1995). It is based on the two-photon interference
effect at a 50:50 BS and via a coincidence analysis can distinguish two of
the four Bell states. As shown in Fig.~\ref{PBS2}(a),
the setup consists of a BS followed by two-channel polarizers in each of its
output beams. As the polarization  state $|\Psi^-\rangle$ is antisymmetric,
it results in a  coincidence detection
at the outputs of the BS (i.e.,
a coincidence at detectors DH1 and DV2 or at
DH2 and DV1). In fact, the state $|\Psi^-\rangle$ can be encoded with
any degree of freedom (e.g., color encoding, see Moehring \textit{et al}. 2007) and can be pinpointed by a two-channel coincidence behind a BS.
This property can be easily checked by reversing the action of the Shih-Alley interferometer. The minus sign
in $|\Psi^-\rangle$ leads to a cancellation of the photon bunching
amplitudes, that is to  a ``fermionic''-like behavior. For the remaining three states, both
photons exit via the same output port of the BS.
The state $|\Psi^+\rangle$ can be distinguished from the other
two by the fact that the emerging photons have different polarizations
\footnote{In addition to the ``fermionic''-like antibunching for the state $|\Psi^-\rangle$ and ``boson''-like
bunching behavior of the state $|\Psi^+\rangle$, there are also intermediate behaviors
observed by tuning the phase between the two Bell states $|\Psi^-\rangle$, which could be used
to simulate anyons (Michler \textit{et al}. 1996)}. This results in
 a coincidence counts at detectors DH1 and DV1 or at DH2 and DV2.
The two states $|\Phi^+\rangle$ and $|\Phi^-\rangle$ both lead to  a two-photon
event at a single detector, and thus cannot
be distinguished.

A modified version of a Bell-state analyzer, which can directly be generalized
to the $N$-particle case was  introduced by Pan and Zeilinger (1998).
Consider the arrangement of Fig.~\ref{PBS2}(b). Two spectrally identical photons enter
the Bell-state analyzer by modes $a$ and $b$. Assume that they
 arrive at PBS$_{ab}$ simultaneously, so that their
wavefunctions overlap behind it. The properties of a PBS depicted in
Fig.~\ref{pbs-fig}, and a coincidence detection  in modes 1 and
 2, allow one to distinguish  $|\phi^{+}\rangle_{12}$
and $|\phi^{-}\rangle_{12}$ polarization Bell states. Specifically, for the incident state $\frac{1}{\sqrt{2}}(|H\rangle_{a}
|H\rangle_{b}+|V\rangle_{a}|V\rangle_{b})$ we always observe a
coincidence either between detectors D$_{H1}$ and D$_{H2}$ or D$_{V1}$ and
D$_{V2}$. On the other hand, if the incident state is $\frac{1}{\sqrt{2}}(|H\rangle_{a}|H%
\rangle_{b}-|V\rangle_{a}|V\rangle_{b})$ we observe
coincidence at detectors D$_{H1}$ and D$_{V2}$ or D$_{V1}$ and D$_{H2}$. The other two  Bell states would
lead to no coincidence at detectors in modes 1 and 2.

Finally, let us mention that by taking the advantage of the properties of hyper-entanglement one can implement a  complete Bell-state analysis (Kwiat and
Weinfurter, 1998; Walborn \textit{et al}., 2003b). Such a  scheme was
experimentally realized (Schuck \textit{et al}., 2006) and was used to beat
the channel capacity limit for linear photonic superdense coding (Barreiro,
Wei, and Kwiat, 2008, see Section~\ref{sec:DCoding} for more details on dense coding).

%%%%%%%%%%%%%%%%%%%%%%%%%%%%%OCT-08-8:45AM%%%%%%%%%%%%%%%%%%%%%%%%%%%%%%%%%%%%%

\subsubsection{GHZ-state analyzer}

\label{sec:GHZ-analyzer}

Bell state measurement schemes can be generalized to the $N$-particle case. One can  construct a  GHZ-state analyzer
(Pan and Zeilinger, 1998),  with which one can  identify two out of the
$2^{N}$ maximally entangled GHZ states.%

In the case of three spectrally indistinguishable identical photons,  eight maximally
entangled polarization GHZ states could be given by
\begin{align}
|{\Phi_{0}}^{\pm}\rangle &  =\frac{1}{\sqrt{2}}(|H\rangle|H\rangle|H\rangle
\pm|V\rangle|V\rangle|V\rangle),\label{nghz}\\
|{\Psi_{1}}^{\pm}\rangle &  =\frac{1}{\sqrt{2}}(|V\rangle|H\rangle|H\rangle
\pm|H\rangle|V\rangle|V\rangle),\label{nghz1}\\
|{\Psi_{2}}^{\pm}\rangle &  =\frac{1}{\sqrt{2}}(|H\rangle|V\rangle|H\rangle
\pm|V\rangle|H\rangle|V\rangle),\label{nghz2}\\
|{\Psi_{3}}^{\pm}\rangle &  =\frac{1}{\sqrt{2}}(|H\rangle|H\rangle|V\rangle
\pm|V\rangle|V\rangle|H\rangle). \label{nghz3}%
\end{align}
The notation used is such that the index $i$ in $|{\Psi_{i}}^{\pm}\rangle$ designates a GHZ state with the property that the
polarization of photon $i$, in each of the terms of the superposition, is different from the polarization of the other two. Consider now the
setup of Fig.~\ref{PBS2} and suppose that three photons enter the GHZ
analyzer by modes a, b and c.
The polarization beam splitters transmit $H$ and reflect $V$
polarizations, thus a
coincidence detection at the three outputs can only originate from
either the case that all photons are transmitted (this corresponds to the
input state $|H\rangle|H\rangle|H\rangle$) or all reflected
($|V\rangle|V\rangle|V\rangle$). The two cases are fully
indistinguishable if the photons  perfectly overlap spatially
and temporally. Thus, two GHZ states, namely
$|\Phi_0^{\pm}\rangle=(|H\rangle|H\rangle|H\rangle\pm|V\rangle|V\rangle|V\rangle)$,
can be filtered out of the  eight. One can further tell apart
the states $|\Phi^{\pm}\rangle$  by
placing a polarizer after each PBS, setting it to distinguish the $+/-$ polarization basis. In such a case,  the state $|\Phi_0^{+}\rangle$ leads to coincidences
 $+++$, $+--$, $-+-$, and $--+$,  while $|\Phi_0^{-}\rangle$ causes totally
different events: $++-$, $+-+$, $-++$, and $---$. The success
probability of the GHZ analyzer is thus $1/4$.

\section{Experimental realizations of photonic entanglement}

%%%%%%%%%%%%%%%%%%%%%%%%%%%%%%%%%%%%OCT-08-9:16 AM%%%%%%%%%%%%%%%%%%%%%%%%%%%%%%%%%%%%%%%%%%%

\label{sec:creation}

Sources of entangled photons play a central role in the experimental study of
 foundations of quantum mechanics and are an essential resource in optical quantum
information processing. The early Bell-test experiments used entangled photons
from atomic cascades, see Clauser and Shimony (1978). Such a source has some drawbacks.
The directions of entangled-photon emissions are uncorrelated. This causes very low collection efficiency.
Moreover, the entanglement is only perfect for photons that are emitted back to back, a loophole that could
allow local hidden-variable model to explain the experimental data (Santos, 1991, 1992).
Meanwhile it was discovered  that the
process of spontaneous parametric down-conversion allows pairs of entangled photons to be collected
in clearly specified directions, with reasonable intensity and with very high purity.

Today, essentially all entangled photon sources employ the second order optical nonlinearity
leading to SPDC or more recently also the third order Kerr nonlinearity in four-wave mixing (FWM) in optical
fibers. Such processes can be realized with an increasing quality and  brightness. For instance,
Altepeter, Jeffrey, and Kwiat (2005) have reported an entangled photon
pair source with an impressive count rate of over one million per second and a fidelity
of $97.7\%$. In this section, we shall focus on the creation of photonic entanglement of
various forms.

\subsection{Spontaneous parametric down-conversion}

\label{sec:spdc}

If one shines strong laser light on a nonlinear crystal, the pump photons have some
probability to split into correlated pairs of lower energy. This is called
spontaneous parametric down conversion (SPDC). The new photons, customarily
called \textquotedblleft signal\textquotedblright\ and \textquotedblleft
idler\textquotedblright, satisfy the following relations: for the wave vectors within the crystal one has $\mathbf{k}_{0}%
\approx\mathbf{k}_{i}+\mathbf{k}_{s}$ where subscripts $0$, ${s}$ and $i$
denote, respectively, pump, signal and idler wave vectors, and the respective
frequencies satisfy $\omega_{o}\approx\omega_{i}+\omega_{s}.$ This is usually called phase matching. It governs the directional correlations of the emissions. It also implies the emerging photon pairs have entangled  frequencies and  linear momenta. There are two different types of the process: either signal and idler photons share the same polarization (type I) or they have perpendicular polarizations (type II).

The quantum nature of SPDC was first studied by Klyshko and Zel'dovich
(Klyshko, 1967, 1988; Zel'dovich and Klyshko, 1969). With the works of Mollow
(1973) and Hong and Mandel (1985) the theory reached its final form (see Appendix). The
predicted strong quantum correlations between the photon pairs created in
SPDC were first experimentally observed by Burnham and Weinberg (1970).
Quantum interference of (type-I) SPDC photons was  used to violate Bell's
inequality first by Shih and Alley (1988), and Hong, Ou, and
Mandel (1987). The process was shown to be a ready source of (path) entangled
pairs by Horne, Shimony and Zeilinger (1989). This was demonstrated independently by Rarity
and Tapster (1990a). Polarization entanglement in
type-II process was discovered by Kwiat \textit{et al}. (1995). For a  survey of
SPDC, we refer to (Shih, 2003).

We give a brief description of the SPDC process in the Appendix and show below
how to create photons entangled in various degrees of freedom.

\subsubsection{Types of entanglement}

\begin{figure}
[ptb]
\begin{center}
\includegraphics[
height=1.2989in,
width=3.0436in
]%
{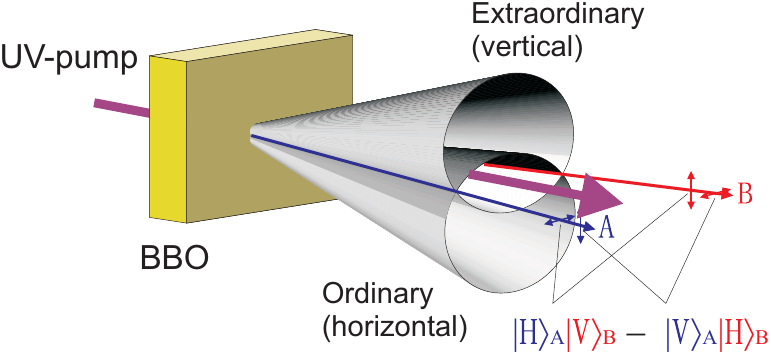}%
\caption{Type-II parametric down-conversion. Inside a specially cut (birefringent) nonlinear
crystal (e.g., BBO), a  pump photon can convert spontaneously into a pair of photons of lower frequencies.
The polarization of signals is orthogonal with respect to the one of idlers. One can attempt to pick only photons
of frequency which is one half of the frequency of the pumping field. In such a case, if the crystal is suitably oriented,
$H$ polarized photons are emitted into one cone (in the figure, the lower one), while $V$ polarized photons are emitted into
the other cone,  and the two cones intersect. As due to the phase matching pairs of photons pop up only along an intersection
of a plane containing  the pump beam with the cones, emissions along the directions at which the two cones intersect are phase matched
and  have undefined polarizations. However, as one of the photons is from the upper cone and the other one is from the lower one, and due to the
indistinguishability of the two processes we have no way to establish which is which (except from a direct polarization measurement),
photons in beams A and B are polarization entangled. This holds under condition that one erases all possible distinguishing features of upper and
lower cone photons (which arise because one of them is an ordinary ray and the other one an extraordinary one which can easily be done via compensation methods described in the text.)}%
\label{typeII}%
\end{center}
\end{figure}

\emph{Polarization entanglement.--}Currently the standard method to produce
polarization-entangled photons is  the noncollinear type-II SPDC process
(Kwiat \textit{et al}., 1995). Its principle is described in
Fig.~\ref{typeII} and the caption.%

%EndExpansion
The state emerging through the two beams A and B is a superposition of
$|H\rangle|V\rangle$ and $|V\rangle|H\rangle$, namely,
\begin{equation}
\frac{1}{\sqrt{2}}(|H\rangle_{A}|V\rangle_{B}+e^{i\alpha}|V\rangle
_{A}|H\rangle_{B}) \label{EPRS}%
\end{equation}
where the relative phase $\alpha$ is due to the birefringence. Using an
additional birefringent phase shifter (or even slightly rotating the
down-conversion crystal itself), the value of $\alpha$ can be set as desired,
e.g., to 0 or $\pi$. A net phase shift of $\pi$ may be obtained by a
$90^{\circ}$ rotation of a quarter wave plate in one of the paths. A half wave
plate in one path can be used to change horizontal polarization to vertical
and \textit{vice versa}. One can thus produce any of the four Bell states in
Eq.~(\ref{Bell_s}) (Mattle \textit{et al.} 1996).

The birefringence of the nonlinear crystal introduces complications. Since the
ordinary and extraordinary photons have different velocities, and propagate
along different directions, the resulting longitudinal and transverse walk-offs between the
two terms in the state (\ref{EPRS}) are maximal for pair creation near the
entrance face. This results in a relative time delay $\delta T=L(1/u_{o}%
-1/u_{e})$ ($L$ is the crystal length, and $u_{o}$ and $u_{e}$ are the
ordinary and extraordinary group velocities, respectively) and a relative
lateral displacement $d=L\tan\rho$ ($\rho$ is the angle between the ordinary
and extraordinary beams inside the crystal). If the coherence time,
$\tau_{c}$, of the down-converted light is shorter than $\delta T$,
the terms in Eq.~(\ref{EPRS}) become, in principle, distinguishable, and no
two-photon polarization interference, and thus no entanglement is observable.
Similarly, if $d$ is larger than the coherence width, the terms can become partially
labeled by their spatial location. Fortunately, because the photons are produced
coherently along the entire length of the crystal, one can completely
compensate for the longitudinal walk-off (Rubin \textit{et al}., 1994). After
the compensation, interference occurs pairwise between processes in which the
photon pair is created at distances $\pm x$ from the middle of the crystal.
The ideal compensation is therefore to use two crystals, one in each path,
which are identical with the down-conversion crystal, but only half as long.
If the polarizations are rotated by $90^{\circ}$ (e.g., with a half wave
plate), the retardations of the $o$ and $e$ components are exchanged and
temporal indistinguishability is restored. The method
also provides reasonable compensation for the transverse walk-off effect.

An alternative method is using type-I  SPDC in two orthogonally oriented crystals.
Two-photon states of tunable purity and degree of entanglement can be produced
(White \textit{et al}., 1999; Peters \textit{et al}., 2004; Cinelli \textit{et al}., 2004).
Periodic poling of nonlinear crystals  greatly relaxes the phase-matching
conditions for SPDC and thus allows to exploit the material's nonlinear properties
more efficiently compared to bulk crystals. Kim \textit{et al}. (2006) proposed
and implemented a polarization Sagnac interferometric configuration with
bidirectional pumping of a type-II phase-matched periodically poled
KTiOPO$_4$ (PPKTP) crystal.
A pulsed (Kuzucu and Wong, 2008),
and narrow-band (0.15nm), wavelength tunable entangled photon source based on such a configuration with a spectral brightness up to 273000 pairs  (s$\cdot$mW$\cdot$nm)$^{-1}$ has been
reported (Fedrizzi \textit{et al}. 2007).
In a further development, via four-wave mixing in a photonic crystal fiber Fulconis \textit{et al}. (2007) have developed a bright, pulsed source of photon pairs.

\begin{figure}
[ptb]
\begin{center}
\includegraphics[width=0.495\textwidth]
{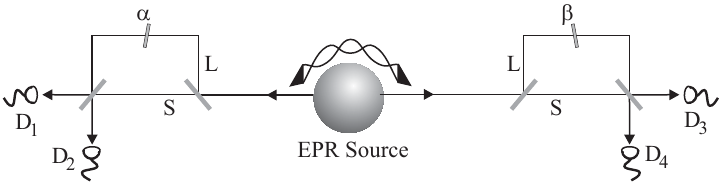}%
\caption{A Franson-type experiment allowing  interference of time-entangled
photon pairs (under the condition of a strict coincidence). The main feature of the scheme is a pair of two identical unbalanced
Mach-Zehnder interferometers.}%
\label{dc_fransonschematic}%
\end{center}
\end{figure}
%EndExpansion

\emph{Temporal entanglement.--}When the crystal is pumped by a CW laser with a coherence time of $\Delta L_c$,
the time at which pair emission happens is undefined and phase stable within $\Delta L_c$. Now
imagine that each of the photons is sent to a different unbalanced Mach-Zehnder interferometer.
We assume that the interferometers have an {\em identical} arm-length difference, $\Delta L$ much longer than
the coherence length of the photons, but shorter than the coherence length of the pump laser
(see Fig.~\ref{dc_fransonschematic}). As it was shown
by Franson (1989), who put forward such an interferometric configuration, the coincidence count rates at the outputs of both
interferometers show a sinusoidal interference pattern, which depends on the
sum of the local phase shifts. If the coincidence gate is much shorter than
$\Delta L_c$ only the following coherent processes are selected and contribute to the interference:
\begin{equation}
\frac{1}{\sqrt{2}}\left(  |\mathrm{long}\rangle_{1}|\mathrm{long}\rangle
_{2}+e^{i\phi}|\mathrm{short}\rangle_{1}|\mathrm{short}\rangle_{2}\right)  ,
\end{equation}
where $|\mathrm{short}\rangle$ and $|\mathrm{long}\rangle$ denote the photon in short or long
arm of the given local interferometer. This is the principle of Franson-type
interferometry (Franson, 1989)\footnote{The original motivation for this type of interferometry was to have an alternative method of obtaining correlations that lead to violations of Bell inequalities. However, as shown by Aerts et al. (1999), the post selection inherent in this type of  experiment makes this connection much more complicated: the original scheme in its ideal form has an explicit local hidden variable model. One has to modify the experiment, and use non-standard Bell inequalities to make it a valid Bell test.}. The predicted interference phenomena were observed by Brendel \textit{et al}. (1992) and Kwiat \textit{et al}. (1993), many other experiments have followed.

\begin{figure}
[ptb]
\begin{center}
\includegraphics[width=0.43\textwidth
]%
{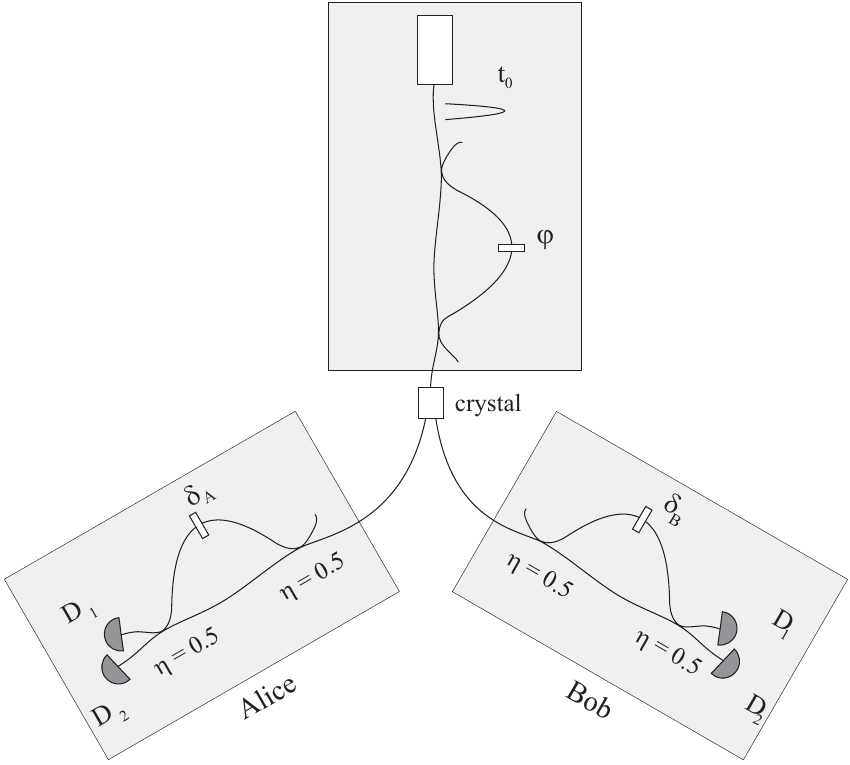}%
\caption{Schematic of the pulsed time-entangled twin-photon source (Gisin
\textit{et al}., 2002). The interferometers are represented by curvy lines. This is so to stress that practical realizations of this type of interferometers are usually built out of optical fibers and couplers.}%
\label{dc_pulsedc}%
\end{center}
\end{figure}

A pulsed-pump version of Franson-interferometry uses so-called time-bin
entanglement (Brendel \textit{et al}., 1999; Marcikic \textit{et al}., 2002;
Gisin \textit{et al}., 2002) (see Fig.~\ref{dc_pulsedc}). A short, ultraviolet pump pulse is sent first through
an unbalanced Mach-Zehnder interferometer (the pump interferometer) creating two mutually coherent pump pulses $\alpha|\mathrm{short}%
\rangle_{p}+\beta|\mathrm{long}\rangle_{p}$, and next
through a BBO crystal. Then, if via SPDC a pump photon
is converted into a pair, the latter is in the state
\begin{equation}
\alpha|\mathrm{short}\rangle_{1}|\mathrm{short}\rangle_{2}+\beta
|\mathrm{long}\rangle_{1}|\mathrm{long}\rangle_{2}.
\end{equation}
The two entangled photons can be separated and subjected to local measurements
in unbalanced interferometers with $\Delta L$ identical as in the pump
interferometer (For more details, see Gisin \textit{et al}., 2002). By varying
the parameters of the beamsplitters and phases in the pump interferometer, all
two-qubit entangled time-bin states can be generated. An advantage of the
time-bin entanglement is that it is insensitive to polarization fluctuations.
Using reference laser pulses to actively lock the phase, it can be robustly
distributed over long distances in optical fibers.  Note that because the pulses
are only separated on the order of a few nanoseconds, and this is much shorter
than the timescale of any phase drifts in the fiber, the drifts do not affect the quality of entanglement. An experimental by Marcikic
\textit{et al}., in 2004 has demonstrated the distribution of time-bin entanglement
over 50 km in optical fibers.

\begin{figure}
[ptb]
\begin{center}
\includegraphics[width=0.5\textwidth]
{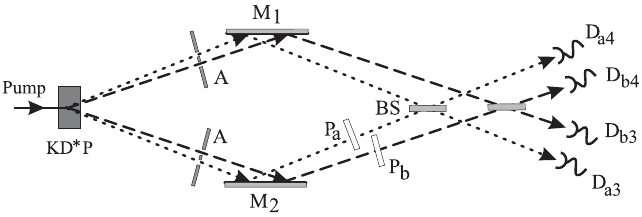}%
\caption{The Rarity-Tapster experiment with momentum entanglement from a type-I
SPDC (Rarity and Tapster, 1990). Using double apertures A two correlated pairs of modes are selected
from the emission spectrum of a type-I down-conversion source. On each of the  two BS beams of the same wavelength are superimposed. The detectors
$D_{a3}$, $D_{b3}$, $D_{a4}$, and $D_{b4}$ register two-photon coincidences between the $a$ and $b$ outputs.}%
\label{dc_rarityscheme}%
\end{center}
\end{figure}

\emph{Path entanglement.--}Entanglement experiments involving
path (momentum) entanglement were proposed by Horne and Zeilinger (1986), and their
feasible version by \.{Z}ukowski and Pykacz (1988). Finally Horne, Shimony and
Zeilinger (1989) proposed that SPDC is an ideal source in case of such
experiments. This was realized by Rarity and Tapster (1990a). Due to the
phase-matching relation, idler and signal photons of given frequencies are
correlated in emission directions. One can use
apertures, see Fig.~\ref{dc_rarityscheme}, to select only two pairs of spatially
conjugate modes (directions). The photon pairs then emerge via the
apertures such that they are either in the upper $a$-mode ($a1$) and lower
$b$-mode ($b2$), or in the lower $a$-mode ($a2$) and upper $b$-mode ($b1$). The
resulting state is thus
\begin{equation}
|\Psi\rangle=\frac{1}{\sqrt{2}}(e^{i\phi_{b}}|a1\rangle|b2\rangle+e^{i\phi
_{a}}|a2\rangle|b1\rangle).
\end{equation}
The $a$-modes enter a BS via opposite inputs, so do $b$-modes. Behind the BSs
upper and lower paths cannot be distinguished, leading to two-photon
interference, which depends on the difference of the relative phase shifts in $a$ and $b$ modes.

%EndExpansion

%%%%%%%%%%%%%%%%%%%%%%%%%%%%%%%%%%%%OCT08-16:10pm%%%%%%%%%%%%%%%%%%%%%%%%%%%%%%%%%%%%%%%%%%%%%

\subsection{Photonic entanglement in higher dimensions}

\subsubsection{Entangled qudits}

Photonic entanglement in higher dimensions can in principle be generated by SPDC
processes in a form of generalization of path or temporal entanglement into more than two
conjugate pairs of beams or time-bins, respectively (Zeilinger \textit{et al}., 1993), and analyzed with
$N$-port beamsplitters (Reck \textit{et al}.,
1994). As it was shown by \.{Z}ukowski, Zeilinger and Horne (1997) such
configuration can lead to new types of EPR correlations, and can be used for
tests of local realism (which are more discriminating than two qubit tests,
see [Kaszlikowski \emph{et al}. (2000)]).

Another route is to use the photons' orbital angular momentum. Orbital
angular momentum eigenstates of photons are states of the electromagnetic
field with phase singularities. They can be utilized for observation of higher dimensional entanglement (Mair
\textit{et al}., 2001; Vaziri \textit{et al}., 2002, 2003). This approach has
advantages in certain quantum communication protocols (Vaziri \textit{et al}.,
2002, Molina-Terriza \textit{et al}., 2004; Gr\"{o}blacher \textit{et al}., 2006).
High dimensional entangled qudits have also been created by transverse spatial
correlations of two SPDC photons (Neves \textit{et al}., 2005), or using transverse momentum
and position entanglement of photons emitted in SPDC, in a form called pixel entanglement
(O'Sullivan-Hale \textit{et al}., 2005).

\subsubsection{Hyper-entanglement}

\label{sec:hyper}

As was shown earlier the SPDC photons are entangled in energy and momentum,
and if suitably selected, can be also entangled in polarization or path. If
one selects pairs which are entangled not only in polarization but also in
some other degree(s) of freedom, this specific entanglement is called
hyper-entanglement (Kwiat, 1997). Hyper-entanglement may have interesting
applications such as Bell-state analysis
(Kwiat and Weinfurter, 1998; Walborn \textit{et al}., 2003b; Schuck \textit{et
al}., 2006), two-particle
GHZ-type test of local realism (Michler et a., 2000a, Chen \textit{et al}., 2003), implementations
of single-photon two-qubit {\small CNOT} gate (Fiorentino \textit{et al}.,
2004), two-qubit {\small swap} gate (Fiorentino \textit{et al}., 2005), and
quantum cryptography (Chen \textit{et al}., 2006d).

\emph{Polarization-path entanglement}:  Polarization-path
entanglement can be generated by a double pass of a pump laser through a BBO
crystal (Chen \textit{et al}., 2003). The pump passes the crystal and is reflected to pass it again. While the polarization state in each of the two possible
emission processes is given by the respective  SPDC setting, the path state of the
pairs is $\left\vert \psi^{-}(\phi)\right\rangle _{\mathrm{path}}=\frac
{1}{\sqrt{2}}(\left\vert u\right\rangle _{A}\left\vert d\right\rangle
_{B}-e^{i\phi}\left\vert d\right\rangle _{A}\left\vert u\right\rangle _{B})$,
where the two orthonormal kets $\left\vert d\right\rangle $ and $\left\vert
u\right\rangle $ denote the two path states of photons. By properly adjusting
the distance between the mirror and the crystal such that $\phi=0$, one gets
\begin{equation}
\left\vert \Psi\right\rangle =\left\vert \psi^{-}\right\rangle _{\mathrm{pol}%
}\otimes\left\vert \psi^{-}(0)\right\rangle _{\mathrm{path}}, \label{hyper}%
\end{equation}
which is a two-photon state maximally entangled in both polarization and path.
It was independently realized by
Barbieri \textit{et al} (2005) using type-I nonlinear crystals, and by Yang
\textit{et al} (2005) using a type-II nonlinear crystal.

\emph{Polarization-time entanglement}: A more robust distribution of hyper-entanglement is possible with photon pairs
which are entangled both in time (i.e., time-bin entanglement) and in
polarization (Genovese and Novero, 2002; Schuck \textit{et al}., 2006; Chen \textit{et al}., 2006d). To
create such a polarization-time entanglement, we similarly combine creation of polarization entanglement with
the method to obtain temporal entanglement.
That is,  either by using a short, ultraviolet laser pulse sent first through an unbalanced Mach-Zehnder interferometer (the pump
interferometer), to have two pulses well separated in time, or  by taking advantage of the long coherence time of a cw-pump laser. Using a \textquotedblleft
time-path transmitter\textquotedblright\thinspace introduced by Chen
\textit{et al.} (2006d), one can realize a transformation between
polarization-path and polarization-time hyper-entanglement.

\emph{Entanglement in multiple degrees of freedom} In an experiment by Barreiro \textit{et al} (2005), besides the entanglement
in polarization and in energy, photon pairs from a single nonlinear crystal
were also entangled in orbital angular momentum.
By pumping two BBO crystals with optical axes aligned in
perpendicular planes, a two-photon $(2\otimes2)\otimes(3\otimes3)\otimes
(2\otimes2)$-dimensional hyper-entangled state was produced, approximately
described by
\begin{align}
&  \underbrace{\left(  |HH\rangle+|VV\rangle\right)  }_{\text{polarization}%
}\otimes\underbrace{\left(  |1,-1\rangle_{LG}+\alpha|0,0\rangle_{LG}%
+|-1,1\rangle_{LG}\right)  }_{\text{spatial}}\nonumber\\
&  \otimes\underbrace{\left(  |EE\rangle+|LL\rangle\right)  }%
_{\text{energy-time}}\,.
\end{align}
Here $\alpha$ describes the orbital-angular-momentum spatial mode balance
which is due the properties of the source~(Torres \textit{et al}., 2003) and
the selection via the mode-matching conditions.

\begin{figure}
[ptb]
\begin{center}
\includegraphics[width=0.5\textwidth]
{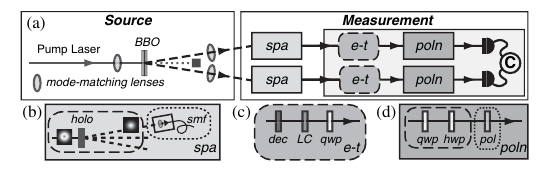}%
\caption{(a) Experimental setup for the creation and analysis of
hyper-entangled photons (Barreiro \textit{et al}. (2005)). (b) Spatial filtration
(spa): hologram (holo) and single-mode fiber (smf). (c) Energy time
transformation (e-t): thick quartz decoherer (dec) and liquid
crystal (LC). (d) Polarization filtration (poln): quarter-wave
plate (qwp), half-wave plate (hwp), and polarizer (pol).}%
\label{fig:hyper}%
\end{center}
\end{figure}

\subsection{Twin-beam multi-photon entanglement}

\label{N00N}

It is also possible to produce entangled states involving large numbers of
photons, approaching the macroscopic domain. Such entanglement is related to
experiments on twin beams (Smithey \textit{et al}., 1992) and could be called
a twin-beam multi-photon entanglement. It is different from
multi-photon entanglement in which each spatially separated photon represents a qubit, and can be
individually manipulated and observed.

The twin-beam multi-photon entanglement can be generated via standard SPDC,
but with a strong pump pulse. Stimulated
SPDC (Lamas-Linares \textit{et al}., 2001; Simon and Bouwmeester, 2003) can be seen as an extension of interferometrically enhanced SPDC (Herzog \textit{et al.} 1994)  and may
show an onset of a laser-like emission of entangled photons, i.e. we can have an (\textquotedblleft
entanglement laser\textquotedblright) in the sense that a (spontaneously
created) photon pair in two polarization-entangled modes stimulates, inside a
nonlinear gain medium,  emission of additional pairs.

A simplified Hamiltonian\footnote{It is interesting to note that this
simplified Hamiltonian $H_{\mathrm{SPDC}}^{sim}$ and the state generated
thereby [Eq.~(\ref{elaser})] are closely related to squeezing and
continuous-variable entanglement, see Braunstein and van Loock (2005) for a
review.} responsible for generation of polarization entangled SPDC photons can be put as
$H_{0}=i\kappa(a_{H}^{\dag}b_{V}^{\dag}-a_{V}^{\dag}%
b_{H}^{\dag})+h.c.$ Horizontally ($H$) and vertically ($V$) polarized photons
occupy two spatial modes (\textit{a} and \textit{b}); $\kappa$ is a coupling
constant that depends on the nonlinearity of the crystal and the intensity of
the pump pulse. After the interaction time $t$ the resulting photon state is
given by $|\psi\rangle=e^{-itH_{0}}|0\rangle$ (Kok
and Braunstein, 2000; Lamas-Linares \textit{et al}., 2001; Simon and
Bouwmeester, 2003), and in the number state representation reads
\begin{align}
|\psi\rangle &  =\frac{1}{\cosh^{2}\tau}\sum_{n=0}^{\infty}\sqrt{n+1}\tanh
^{n}\tau|\psi_{n}^{-}\rangle\,,\nonumber\\
|\psi_{n}^{-}\rangle &  =\sum_{m=0}^{n}\frac{(-1)^{m}}{\sqrt{n+1}%
}|n-m,m;m,n-m\rangle . \label{elaser}%
\end{align}
The ket $|n-m,m;m,n-m\rangle$ denotes a number state in the respective modes
${a_{H}}$, ${a_{V}}$, ${b_{H}}$ and ${b_{V}}$, and $\tau=\kappa t$ is the
interaction parameter. To avoid multi-pair emission events most SPDC
experiments are restricted to $\tau\ll0.1$. By going to higher values, bipartite
entangled states constituting of large numbers of photons are generated. The
state $|\psi\rangle$ is a superposition of the states $|\psi_{n}^{-}\rangle$
of $n$ indistinguishable photon pairs. Each $|\psi_{n}^{-}\rangle$ is an
analog of a singlet state of two spin-$n/2$ particles, thus $|\psi\rangle$ is
invariant under joint rotations of the polarization bases of both modes. The
polarization of each beam is completely undetermined, but the polarizations of
the two beams are always anti-correlated. The average photon pair number is
$\langle n\rangle=2\sinh^{2}\tau$.

Out of such states one can extract for example the following 2-pair term of
Eq.~(\ref{elaser}):
\begin{equation}
|\psi_{2}^{-}\rangle=\frac{1}{\sqrt{3}}(|2,0;0,2\rangle-|1,1,1,1\rangle
+|0,2;2,0\rangle),
\end{equation}
which can be treated as a singlet state of two (composite) spin-1 systems
[see Howell \textit{et al}. (2002) for a test of Bell's inequality by
entangled states of spin-$1$-like systems].

The theory of entanglement laser was developed by Simon and Bouwmeester
(2003). The basic principle of a stimulated entanglement creation was first
experimentally demonstrated in the few-photon regime (Lamas-Linares \textit{et
al}., 2001). Later, twin-beam entanglement of up to 12 photons (Eisenberg
\textit{et al}., 2004) was experimentally observed.

\begin{figure}
[ptb]
\begin{center}
\includegraphics[width=0.5\textwidth]%
{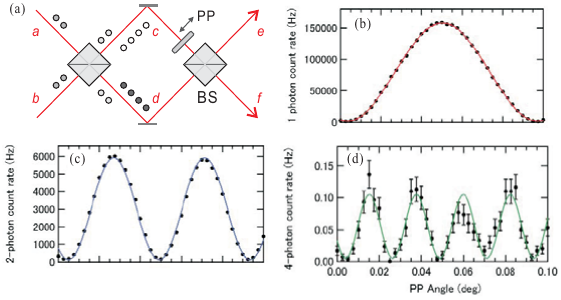}%
\caption{(a) A schematic of the experimental setup of Nagata et al. (2007): a Mach-Zehnder interferometer
consisting of two 50:50 beam splitters. (b) Single photon
count rate in mode e as a function of phase plate (PP) angle
with single-photon input $|10\rangle_{ab}$. (c) Two-photon
count rate in modes \textit{e} and \textit{f} for input state $|11\rangle_{ab}$. (d)
Four-photon count rate of three
photons in mode e and one photon
in mode f for the input state $|11\rangle_{ab}$. The visibility
of the fringes is 0.91(6), greater than
the threshold to beat the standard quantum limit.}%
\label{fig:noon}%
\end{center}
\end{figure}

A special twin-beam
entanglement is the so-called \textquotedblleft high NOON\textquotedblright%
\ type (Bouwmeester, 2004) state of two beams (Dowling, 1998; Kok
\textit{et al}., 2001, 2002):
\begin{equation}
|\mathrm{NOON}\rangle=|N,0;0,N\rangle=\frac{1}{\sqrt{2}}(|N\rangle
_{a}|0\rangle_{b}+|0\rangle_{a}|N\rangle_{b}).
\end{equation}
It was experimentally realized for $N=3$ (Mitchell, Lundeen, and Steinberg,
2004) and $N=4$ (Walther \textit{et al}., 2004) (see also other related experiments,
Rarity \textit{et al}., 1990b; Edamatsu \textit{et al}., 2002; Sun \textit{et al}., 2006;
Nagata \textit{et al}., 2007). These experiments demonstrated an interesting feature of
{NOON} states: the effective de Broglie wavelength of the multiphoton state is by $1/N$ shorter than the
wavelength of the single photon (Jacobson \textit{et al}., 1995). Nagata \textit{et al.} have not only
measured the reduced de Broglie wavelength of four-entangled photons, but also shown
a visibility that exceeds the threshold to beat the standard quantum limit (see the original paper for details). Let us see Fig.~\ref{fig:noon}(a) for more details.
If we put two single photons in each input of the Mach-Zehnder interferometer ($|11\rangle_{ab}$), the state after the first BS is,
due to Hong-Ou-Mandel effect,  $(|20\rangle_{cd}+|02\rangle_{cd})/\sqrt{2}$, which then evolves to $(|20\rangle+e^{i2\varphi}|02\rangle)/\sqrt{2}$.
After the second BS, the probability of detecting two photons in the output modes \textit{e} and \textit{f} is $P=(1-\cos2\varphi)/2$ which shows
a phase super-resolution (for the experimental data see Fig.~\ref{fig:noon}(c)). If two photons are fed into in each input of the interferometer ($|22\rangle_{ab}$),
after the first BS we get $\sqrt{3/8}(|40\rangle_{cd}+|04\rangle_{cd})+(1/2)|22\rangle_{cd}$ --- a generalized multi-photon Hong-Ou-Mandel interference phenomenon.
After the second BS, the probability of detecting three photons in one output \textit{e} and one in \textit{f} is $P=1-(3/8)\cos4\varphi$ (see Fig.~\ref{fig:noon}(d) for data).
The high-precision optical phase measurements have many important applications, e.g.,
overcoming the diffraction limit for classical light (Boto et al. 2000; Kok et al. 2001).

\subsection{Multi-photon entanglement}

\label{sec:multi}

The original motivation to observe entanglement of more than two particles,
with measurements on the particles performed at spatially separated stations,
stems from the observation by GHZ that three-particle entanglement leads to a
dramatic conflict between local realism and EPR's ideas with predictions of
quantum mechanics (GHZ, 1989; Greenberger \textit{et al}., 1990; Mermin,
1990a), see section~\ref{sec:GHZ-t}. However, in 1989 no ready sources of
three or more particle entanglement were available. Yurke and Stoler (1992a,
1992b) showed that in theory multiparticle entanglement effects should be in
principle observable for particles originating from independent sources. A
general method for making such an interference observable, and also to swap
entanglement, was given by \.{Z}ukowski \textit{et al}. (1993), \.{Z}ukowski
\textit{et al}. (1995), Rarity (1995) and Zeilinger \textit{et al}. (1997).
In the following, we will first present the basic methods followed by numerous experiments
in which multi-photon entanglement was observed. Once one is able to entangle
two photons that never interacted, one can construct very many types of
entanglement (Zeilinger \textit{et al}., 1997), which in turn can be utilized in
many ways (Bose \textit{et al}., 1998).

\subsubsection{Entanglement construction}

\label{sec:SWAP}

We have at hand only photon sources of two-particle entanglement. We shall
show in detail an operational method to swap entanglement of two pairs of particles
(\.{Z}ukowski, Zeilinger, and Weinfurter, 1995), which has been used in many
experiments (the pioneering one was the Innsbruck teleportation, Bouwmeester \textit{et al.} 1997). The technique of essentially
erasing which-source information, can be applied in many other configurations,
e.g., in the case of a double pair emission from a single source, etc. It
works even for totally independent emissions (provided they are synchronized).

\begin{figure}
[ptb]
\begin{center}
\includegraphics[
height=1.7974in,
width=2.984in
]%
{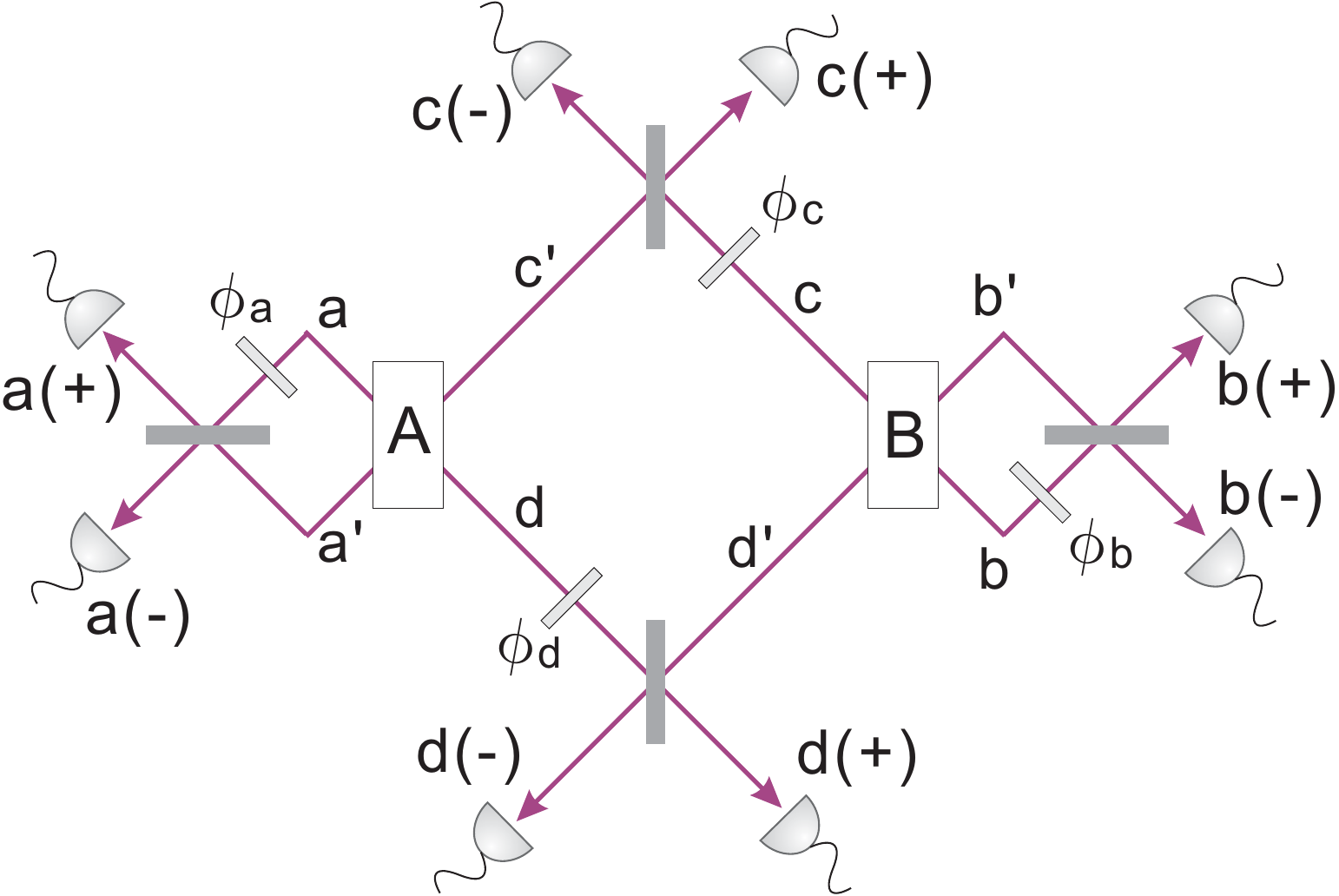}%
\caption{Four-particle interference effects for two pairs of particles
originating from two independent sources (\.{Z}ukowski \textit{et al}.,
1999).}%
\label{eswap93}%
\end{center}
\end{figure}

\emph{Entangling two independent particles: the principle. -- }%
Figure~\ref{eswap93} shows a configuration for obtaining interference effects
for two pairs of particles originating from two independent sources. Assume
that the sources of path entangled states in Fig.~\ref{eswap93}, A and B, each spontaneously emits a
pair of particles in an entangled state (all particles are supposed to be
identical) at nearly the same moment of time, and the states of the pairs are
$|\Psi^{A}\rangle=\frac{1}{\sqrt{2}}\left(  |a\rangle|d\rangle+|a^{\prime
}\rangle|c^{\prime}\rangle\right)  $ for source A, and $|\Psi^{B}\rangle
=\frac{1}{\sqrt{2}}\left(  |b^{\prime}\rangle|d^{\prime}\rangle+|b\rangle
|c\rangle\right)  $ for source B (the letters represent beams taken by the
particles in Fig.~\ref{eswap93}, and $\langle e|f\rangle=\delta_{ef}$). The
beams $x$ and $x^{\prime}$, where $x=a,b,c$ or $d$, are superposed at 50:50
BSs. Behind the BSs we place detectors in the output ports $x(\pm)$. In all
unprimed beams one can introduce a phase shift of $\phi_{x}$. The detector
stations differ in their role: $a(\pm)$ and $b(\pm)$ observe radiation coming
from one source only, but this is not so for stations $d(\pm)$ and $c(\pm)$.
For instance, if a \textit{single} particle is detected by $d(+)$, its origin may be,
under suitable conditions, completely unknown. If it cannot be determined, even in principle, which source produced the
particle which activated the detectors, say $d(+)$ and $c(+)$, then
four-particle interference effects may occur.

Assume that detectors $a(+)$ and $b(+)$ also fired. Simultaneous firings of
the four detectors can exhibit interference effects provided the two
contributing processes \textit{are indistinguishable}: detection of the particles from source A in $d(+)$ and
$a(+)$, and detection of the particles from B in $c(+)$ and $b(+)$; detection
of the particles from source A in $c(+)$ and $a(+)$, and detection of the
particles form B in $d(+)$ and $b(+)$. Note
that depending on the phase shifts the detection at, e.g., $c(+)$ and $d(+)$,
acts like a Bell-state measurement, projecting the two photons
into the state $\frac{1}{\sqrt{2}}\left(  |c^{\prime}\rangle|d^{\prime}%
\rangle+e^{i(\phi_{c}+\phi_{d})}|c\rangle|d\rangle\right)  .$ The
other two photons are, due to this event, in state: $\frac{1}{\sqrt{2}}\left(
|b^{\prime}\rangle|a^{\prime}\rangle+e^{-i(\phi_{c}+\phi_{d})}|b\rangle
|a\rangle\right)  .$ This process is called \textit{entanglement swapping}.

\emph{Enforcing source indistinguishability.--}Imagine now that the sources of
entangled states are two crystals pumped by independent, pulsed lasers
operating synchronously.
Assume that the time separation between two pulses is much larger than all other time scales of the experiment, i.e., we study the
radiation generated in each crystal by a single pulse. We omit retardation
effects by assuming equal optical paths. We assume that we pick the SPDC
radiation with frequencies close to $\frac{1}{2}\omega_{p}^{o}$, where
$\omega_{p}^{o}$ is the central frequency of the pump pulse. Suppose that the four
SPDC photons are detected in almost the same moment of time (up to a few nanosecond window), one
in each of the detectors $a(+)$, $b(+)$, $c(+)$, and $d(+)$. One could
determine that the photon detected at $d(+)$ came from crystal A (B) by noting the
near simultaneity of the detection of photon $d(+)$ and one of the photons at
$a(+)$ or $b(+)$ (the detection times of a true SPDC pair are extremely tightly
correlated, see the Appendix, formula (\ref{pro15}) and the discussion after it).
To ensure that the source of photons is unknowable, the (initially spectrally broadband) photons
should be detected behind a narrow band filtering system (to be called later simply a
filter) whose inverse of the bandwidth (coherence time) clearly exceeds the pulse
duration $\tau$  (e.g., by an order of magnitude). Then, the temporal separation of true SPDC
pairs, all created within $\tau$, spreads over times much longer than $\tau$ and thereby prevents identifying the source
of the photon by comparison of the arrival times. Note that filtering is necessary only in modes $c$ and $d$, while \emph{no filtering is required} in front of the
detectors $a(\pm)$ and $b(\pm)$.\footnote{This method also precludes the
possibility of inferring the source of the photon from the frequency
correlations. The frequency of the photons reaching $d(+)$ is better defined
than the pumping pulse frequency, and it is the spread of the latter one that
limits the frequency correlations of the idler-signal pair from one source.}

One can estimate the maximal
visibility expected for the interference process, using the results
presented in the Appendix, where the basic properties of the SPDC
radiation are derived.
The amplitude of the four-photon detections at, say, detectors $a(+)$, $b(+)$,
$c(+)$ and $d(+)$ at times $t_{a}$, $t_{b}$, $t_{c}$, and $t_{d}$, is
proportional to
\begin{align}
&  e^{i(\phi_{a}+\phi_{b}+\phi_{c}+\phi_{d})}A_{ad}(t_{a},t_{d})A_{cb}%
(t_{c},t_{b})\nonumber\\
&  +A_{b^{\prime}d^{\prime}}(t_{b},t_{d})A_{a^{\prime}c^{\prime}}(t_{a}%
,t_{c}),
\end{align}
where $\phi_{i}$, $i=a,b,c,d$ is the local phase shift in the given beam.
The probability amplitude $A(t,t')$ to detect one
photon of a SPDC pair at $t$ and the other one at $t'$,  is the one given
by equation (A20). To
get an overall probability of the process one has to integrate the square of
the modulus of the amplitude over the detection times. Since typical time
resolution of the detectors is of the order of nanoseconds, which is much
longer than the coherence times of typical filters and the width of fs pump
pulses, the integrations over time can be extended to infinity. The joint
probability to have counts in the four detectors behaves as $$1-V_{(4)}
\cos{(\sum_{x=a,b,c,d}\phi_{x})}$$ and the visibility $V_{(4)}$ is given by
\begin{equation}
V_{(4)}=\frac{\int d^{4}t|A_{ad}(t_{a},t_{d})A_{bc}(t_{b},t_{c})A_{b^{\prime
}d^{\prime}}(t_{b},t_{d})A_{a^{\prime}c^{\prime}}(t_{a},t_{c})|}{\int
d^{4}t|A_{ad}(t_{a},t_{d})A_{bc}(t_{b},t_{c})|^{2}},
\end{equation}
where $d^{4}t=dt_{a}dt_{b}dt_{c}dt_{d}$. Assume that the filter functions in
all beams are of identical Gaussian form: $F_{f}(t)=e^{-\frac{i}{2}\omega
_{p}^{o}t}|F_{f}(t)|$, whereas the pump beam is described by $G(t)=e^{-i\omega
_{p}^{o}t}|G(t)|$. Here $\omega_{p}^{o}$ is the central frequency of the
pulse, and $\left\vert F\right\vert $ and $\left\vert G\right\vert $ functions
are given by Fourier transforms of $\exp{\left[  -\frac{1}{2}(\omega-\Omega)^{2}/\sigma^{2}\right]
}$, where $\Omega=\frac{1}{2}\omega_{p}^{o}$ (for $\left\vert F\right\vert $)
or $\omega_{p}^{o}$ (for $\left\vert G\right\vert $), and $\sigma$ is the
respective spectral width. One gets:
\begin{equation}
V_{(4)}=\left[  \frac{\sigma_{p}^{2}}{\sigma_{p}^{2}+\sigma_{F}^{2}\sigma
_{f}^{2}/(\sigma_{p}^{2}+\sigma_{F}^{2}+\sigma_{f}^{2})}\right]  ^{1/2},
\end{equation}
where $\sigma_{p}$ is the spectral width of the pulse, $\sigma_{f}$ is the
spectral width of the filters in beams $a,a^{\prime},b,b^{\prime}$, and the
spectral width of the filters in $c$ and $d$ is $\sigma_{F}$. If one removes
the filters in beams $a$, $a^{\prime}$, $b$ and $b^{\prime}$, the formula
simplifies to $V_{(4)}=\left(  \frac{\sigma_{p}^{2}}{\sigma_{p}^{2}+\sigma
_{F}^{2}}\right)  ^{1/2}$, see \.Zukowski {\em et al.} (1995). Namely, narrow filters in paths $a$, $a^{\prime}$
and $b$, $b^{\prime}$ are not necessary to obtain high visibility. The other filters, for detectors which
observe radiation from both sources, should be always sufficiently narrow.

%%%%%%%%%%%%%%%%%%%%%%%%%%OCT-08-20:47%%%%%%%%%%%%%%%%%%%%%%%%%%%%

\emph{The influence of photon statistics.}--The visibility of the
four-particle fringes in the set-up of Fig.~\ref{eswap93} can be impaired by
the statistical properties of the emission process. The statistics of a single
beam of a down converter is thermal-like. The state of idler-signal pairs
emerging via a pair of (perfectly phase matched) pinholes is given by
\begin{equation}
|\psi\rangle=N^{-1}\Sigma_{m=0}^{\infty}z^{m}|m,s\rangle|m,i\rangle
\label{PDCfull}%
\end{equation}
where $z$ is a number dependent on the strength of the pump, $|m,s\rangle$
($|m,i\rangle$) denotes an $m$-photon state in the signal (idler) mode, and
$N$ is the normalization constant. It can be shown (\.{Z}ukowski \textit{et
al}., 1999) that the visibility is reduced below $50\%$ if $|z|^{2}>(\sqrt
{17}-3)/8\approx0.140$. Thus, to have high visibility the ratio of the
probability of \textit{each pulse} to produce a single down converted pair to
the probability of not producing anything must be less than $14\%$. This
threshold is at quite high pump powers. Nevertheless, this puts a strong
limitation as how many particles can be entangled using such methods. Simply,
creation of entanglement for many particles requires more and more initial
entangled pairs, thus one pumps stronger. However, strong pump leads to lower
visibility of quantum interference, which may prohibit to observe the
correlations due to the desired multi-photon entanglement (Laskowski \textit{et al.} 2009). Recently,
several experiments were performed to identify and quantify the
experimental imperfections that contribute error to the produced multi-photon
states (see Barbieri  \textit{et al}. 2009; Weinhold \textit{et al}. 2008).

\emph{Remarks.--}Note that source indistinguishability in principle can also
be achieved with an ultra-coincidence technique, which does not require
a pulsed pump, but an extremely good detection time resolution, $\Delta T$, much
sharper that the coherence time of the filtered SPDC
radiation, and rejection of all events at $c(+)$ and $d(+)$ which are detected
with time difference higher than, say, $2\Delta T$ [see \.{Z}ukowski
\textit{et al}. (1993)]. In such a case the pumping lasers may be cw
ones.\footnote{See e.g. an experiment reported by Halder \textit{et al}. (2007) who
use single-photon detectors with a time resolution of
$\sim70$ ps, which is much shorter than the coherence length ($\sim350$ ps) of
the tightly filtered photons in the experiment (see section~\ref{otherswapping}). Using an atom-cavity system,
Legero et al. (2004) generated single photons with coherence time of $\sim500ns$ exceeding the time resolution
of employed photon detectors by three orders of magnitude, and observed quantum beat between
photons of different frequencies with a near-unity visibility.}

\subsubsection{New methods}

The methods described above require a femtosecond pulsed laser pump.
Unfortunately, femtosecond pulse pumped SPDC shows relatively poor quantum
interference visibilities (Keller and Rubin, 1997). The following methods are used to increase the quantum interference visibility: (i) a thin nonlinear crystal (Sergienko
\textit{et al}., 1999), (ii) narrow-band spectral filters in front of
detectors, as shown above (Grice and Walmsley, 1997; Grice \textit{et al}.,
1998; Di Giuseppe \textit{et al}., 1997), or (iii) an interferometric
technique (Branning \textit{et al}., 1999, 2000) without spectral and
amplitude post-selection, which was making the spectral wave function of the two photons much more symmetric.\footnote{The method rest on two
distinct processes for emission of a down-converted pair
which are coherently overlapped. The axes of polarization
are switched between the two processes. This gives a symmetrization of the spectral properties.} The first two methods
reduce the intensity of the entangled photon pairs significantly
and cannot achieve perfect overlap of the two-photon amplitudes.
For the theoretical and experimental details of the last method, see (Kim
\textit{et al}., 2001).

A method gaining great importance is tuning the properties of the downconversion source and the pump such that
one obtains frequency uncorrelated pairs of photons, see, e.g.,  Grice,  U'Ren, and Walmsley (2001),
U'Ren, Banaszek and Walmsley (2003),
Walton \textit{et al}. (2003, 2004), Torres \textit{et al}. (2005),  Mosley
\textit{et al}. (2008) and Garay-Palmett et al. (2007).
Halder et al. (2009) demonstrated a source of photon pairs based on four-wave mixing
in photonic crystal fibers (see also Fulconis et al. 2007). Engineering of the phase matching
conditions in the fibers allowed  creation of photon pairs at 597 nm and
860 nm in an intrinsically factorable state of frequencies. Thus there were almost no spectral correlations. The source is narrow band and bright.
Two separate sources of such a kind were used to generate a Hong-Ou-Mandel interference.
The idlers were used to herald the singles. The observed interference, conditioned on a joint detection event
of two idlers, had a raw visibility of $76.1\%$. Since narrowband filtering is unnecessary in case of such sources, one can achieve a higher collection efficiency than in the case of using passive filtering (see Kim \textit{et al}., 2001, and Yao \textit{et al}. 2011 where the collection efficiency was reported to be about twice as high than in measurements using $3$nm filters).

With mastering these phase-matching tune-up techniques one can expect that in future they may replace
the ones based on passive filtering, as a new method for entanglement swapping and related processes.

\subsubsection{First proposals}

\label{sec:GHZ-creation}

In 1990's many proposals were made for observations of multi-photon
entanglement (\.{Z}ukowski \emph{et al}., 1993; \.{Z}ukowski, Zeilinger and
Weinfurter (1995); Rarity, 1995; Zeilinger \textit{et al}., 1997; Pan and
Zeilinger, 1998; see also section~\ref{sec:GHZ-analyzer}), or involving atoms
(Cirac and Zoller, 1994; Sleator and Weinfurter 1995; Haroche, 1995).
%BeginExpansion
\begin{figure}
[ptb]
\begin{center}
\includegraphics[
height=2.1249in,
width=2.0696in
]%
{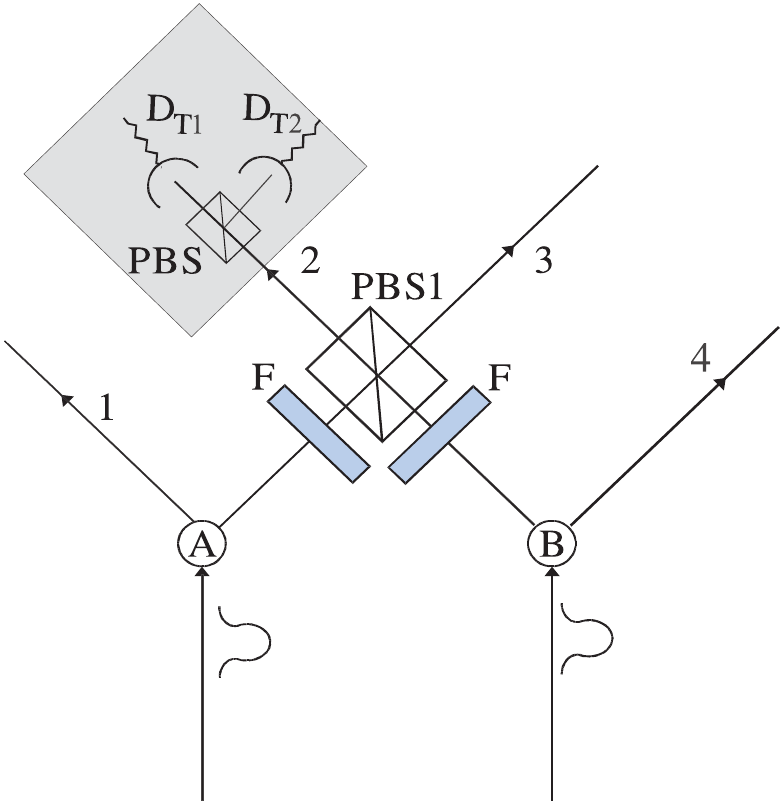}%
\caption{A three-photon polarization-entanglement source. The photon sources,
$A$ and $B$, are pumped by short pulses. The PBS1 transmits $45^{\circ}$ polarization and reflects $-45^{\circ}$
polarization, $F$ is a narrow filter, and D$_{T1}$ and D$_{T2}$ are two
single-photon detectors. A single-photon trigger event in one of the detectors
signals that coincident detections  in channels 1, 2 and 3 would result in GHZ correlations.
The setup can also be used for observation of four photon GHZ interference.}%
\label{pbs3}%
\end{center}
\end{figure}
%EndExpansion

The generic idea of observing photonic GHZ entanglement, later put into practice, was
given by Zeilinger \textit{et al}. (1997)\footnote{Earlier  proposals, \.Zukowski {\em at al.} (1993) and \.Zukowski {\em at al.} (1995), are essentially showing an explicit operational method to realize the ideas of Yurke and Stoler (1992a), (1992b). They involved techniques which required more complicated optical setups and more sources, however the basic principles were the same.}, see
Fig.~\ref{pbs3}. Assume that sources A and B simultaneously emit a photon pair
each. Pairs in beams $x,y$ ($1$, $3$, and $2$, $4$) are in identical
polarization states $\frac{1}{\sqrt{2}}(|H_{x},H_{y}\rangle+|V_{x}%
,V_{y}\rangle)$. The state of the four particles, after the passage of $2$ and $3$
via PBS1, and provided the photons are indistinguishable (which can be
secured using the methods as described earlier), reads
\begin{equation}%
\begin{array}
[c]{l}%
\frac{1}{2}(|H_{1},H_{2},H_{3},H_{4}\rangle+|V_{1},V_{2},V_{3},V_{4}\rangle\\
+|H_{1},H_{3},V_{3},V_{4}\rangle+|V_{1},V_{2},H_{2},H_{4}\rangle).
\end{array}
\label{4ghz}%
\end{equation}
Only the superposition $|H_{1},H_{2},H_{3},H_{4}\rangle+|V_{1},V_{2}%
,V_{3},V_{4}\rangle$, which is a GHZ state, leads to four-fold coincidence.
Therefore, four-fold coincidences can reveal four-particle GHZ correlations.

The scheme in Fig.~\ref{pbs3} also allows one to generate
\textit{unconditional} three-particle GHZ states\footnote{The original
proposal for the realization of three-photon GHZ states in Zeilinger \textit{et al}. (1997)
makes use of a slightly different interferometric setup.} via a method based
on the notion of entangled entanglement (Krenn and Zeilinger, 1996). For
example, one could analyze the polarization state of photon 2 by
passing it through a PBS selecting $45^{\circ}$ and $-45^{\circ}$
polarizations. Then the polarization state of the remaining three
photons 1, 3 and 4 will be projected into
$\frac{1}{\sqrt{2}}(|H_{1},H_{3},H_{4}\rangle+|V_{1},V_{3},V_{4}\rangle$,
if, and only if, detector D$_{T1}$ detects a \emph{single} photon. A
similar superposition, however with a minus sign, is obtained once
detector D$_{T2}$ detects a single photon. The detection of photon 2
excludes the last two terms in Eq.~(\ref{4ghz}), and projects the
remaining three photons into a spatially separated and freely
propagating GHZ state. However, the scheme works only with
photon-number discriminating detectors, and if both EPR sources emit
only a pair each without double pair (or more) emission events.

Unfortunately, this is not the case in the actual SPDC experiments. Due to
the absence of perfect pair sources and perfect single photon
detectors, in the experiments both three- and four-photon entanglement
(Bouwmeester \textit{et al., }1999a; Pan \textit{et al}.,
2001a; Zhao \textit{et al}., 2003a; Eibl \textit{et al}., 2003; see
also section~\ref{sec:GHZ-exp}) is observed only under the
condition that there is one and only one photon in each of the four
outputs. As there are other detection events where, e.g., two
photons appear in the same output port, this condition might raise
doubts about whether such a source can be used for a valid GHZ
test of local realism (section~\ref{sec:GHZ theorem}). By further developing the ideas of Yurke and Stoler (1992a),
 \.{Z}ukowski (2000) showed that
the above procedure indeed permits a valid GHZ test.

\subsubsection{Experimental realizations}

\label{sec:GHZ-exp}

\begin{figure}
[ptb]
\begin{center}
\includegraphics[
width=0.45\textwidth
]%
{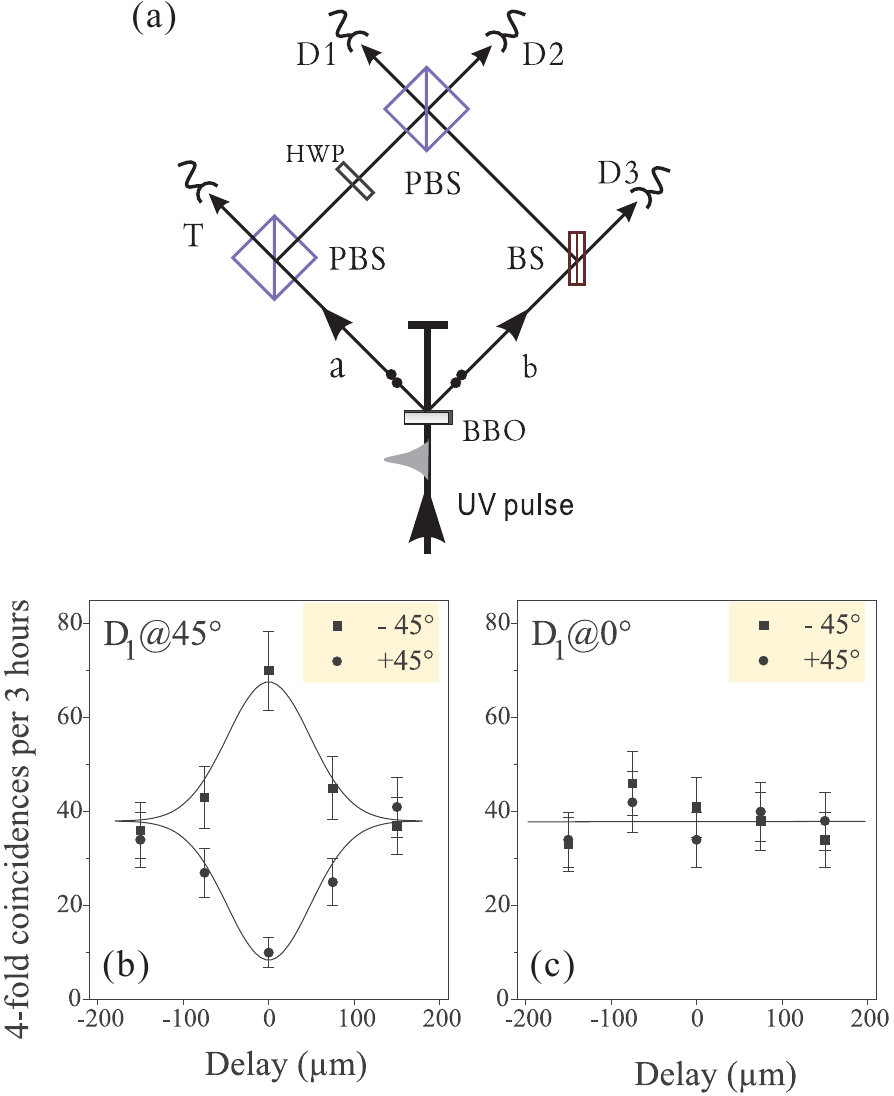}%
\caption{(a). Experimental setup for observing three-photon GHZ
entanglement (Bouwmeester \textit{et al.}, 1999a). The UV pulse incident
on the BBO crystal produces four photons, two in each mode.
Conditioned on the registration of one photon at the
trigger detector T, the three photons registered at D$_{1}$, D$_{2}$, and
D$_{3}$ exhibit the desired GHZ correlations. (b). Experimental data.
The photon at D1 polarized at $45^\circ$ and the photon at
detector D2 polarized at $-45^\circ$. The two curves show the
fourfold coincidences for a polarizer in front of detector D3 oriented at $-45^\circ$ and $45^\circ$,
respectively, as a function of the spatial
delay in path a. (c). Experimental data for the case of  the polarizer in front of detector D1  set to $0^\circ$}.%
\label{3ghz}%
\end{center}
\end{figure}

The first experiment involving techniques of forcing indistinguishability of photons from
different PDC pairs was the teleportation experiment by Bouwmeester \textit{et al.} (1997).
This however will be discussed later in the context of quantum communication (chapter VI).
A GHZ-type entanglement among three spatially
distributed photons, using the above methods of entanglement construction, was first observed
by Bouwmeester \textit{et al.} (1999a). The main idea behind this
experiment, as was put forward in Zeilinger et al. (1997), is to transform two pairs
of polarization entangled photons into three entangled photons. The fourth photon
served the role of a trigger. Figure~\ref{3ghz} illustrates the experimental setup.
Two pairs of polarization-entangled photons are generated via a pulsed SPDC.
The probability per pulse to create a single pair
in the desired modes was on the order of a few $10^{-4}$ with a correspondingly smaller probability to create four photons and negligible
for three pair events. The source was aligned to emit photon pairs
in the state $\frac{1}{\sqrt{2}}\left(  |H_{a},V_{b}\rangle-|V_{a},H_{b}\rangle\right)$.

%EndExpansion

The experimental arrangement was such that the GHZ entanglement could be observed
under the condition that both the trigger photon and the three entangled photons
are detected. This is usually called a ``post-selection'' (sometimes also called ``conditional
detection'', or ``observation in coincidence basis''). Let us go into more details
on how this works. As initially there were four photons, provided that each of them
is detected in a different output, one can infer that the two photons in mode $a$ were
horizontally and vertically polarized, respectively.
The photon of polarization $H$ was transmitted through the PBS
and detected by the trigger detector T, and the other one, $V$ polarized, must have been
reflected by the PBS and consequently rotated by the HWP into state $\frac{1}{\sqrt{2}}\left(|H\rangle+|V\rangle\right)$.
Concerning the two photons in mode $b$, they must have been one of polarization $H$ and
the other of $V$, in order to match the polarizations in mode a. After a random distribution by the BS, one
photon can be detected by the detector D3 while the other one can be combined on a PBS with the photon
from mode $a$. Therefore, if each of the three detectors, D1, D2 and D3, detects a photon, there are
only two possible polarization combinations: $|H_{1}H_{2}V_{3}\rangle$ and $|V_{1}V_{2}H_{3}\rangle$.
Bouwmeester \textit{et al}. adjusted the two path lengths such that the two photons arrived simultaneously at the last PBS.
The photons were spectrally filtered, with $\triangle\lambda=4.6$nm, and monitored
by single-mode fiber-coupled single-photon detectors. The filtering
process stretched the coherence time so that it was substantially
larger than the pump pulse duration ($\sim200$fs). These processes erase the possibility of distinguishing
the photons from different pairs by their arrival time or spatial modes (see previous Subsections). Thus the resulting counts originate from  a superposition between $|H_{1}H_{2}V_{3}\rangle$
and $|V_{1}V_{2}H_{3}\rangle$, that is, the three-photon GHZ state:
\begin{equation}\label{GHZstate}
    \frac{1}{\sqrt{2}}(|H\rangle_1|H\rangle_2|V\rangle_3+
|V\rangle_1|V\rangle_2|H\rangle_3).
\end{equation}

To test whether one indeed dealt with a the three-photon GHZ state, Bouwmeester \textit{et al}. performed a polarization analysis
(with polarizing filters in front of each detector, not shown in Fig. \ref{3ghz}). They first
compared the counts of all eight possible polarization
combinations $HHH,\cdots, VVV$. The results showed that the intensity ratio between
the desired events ($HHV$ and $VVH$)
and the six other undesired ones was about
$12:1$. The dominance  of the two terms is a necessary
but not yet sufficient condition for demonstrating a GHZ
entanglement, as there could, in principle, be just a
statistical mixture of the two states. Next, to test whether  the one indeed dealt with a coherent superposition of the two terms,
Bouwmeester \textit{et al}. performed measurements in the ``diagonal'' basis
$+/-=H\pm V$. A result $+$ at detector D1 should  effectively project the state of the other photons to $\frac{1}{\sqrt{2}}(|+\rangle_2|+\rangle_3-
|-\rangle_2|-\rangle_3)$. The obtained data are consistent with this  prediction, as can be seen from Fig. \ref{3ghz}(b). This is so only within a short interval
of delay between the  photons, while for path length differences larger than
the coherence length of the detected photons the coherence between the two GHZ-terms vanishes.
Further control experiments were done by setting
the polarizer in front of detector D1 at $0^\circ$ such that the GHZ state
(\ref{GHZstate}) is projected into a separable state $|V\rangle_2|H\rangle_3$.
In this case, the results in Fig.~\ref{3ghz}(c) show no correlation.

By extending the above technique, GHZ-type entanglement among more particles was observed.
Four-photon GHZ state was first observed by Pan \textit{et al}., (2001a) and its
genuine entanglement was confirmed by Zhao \textit{et al}., (2003a). Later on five-photon
entanglement was demonstrated by interfering a four-photon GHZ state
(Pan \textit{et al}., 2001a) with a source of pseudo-single photons from an attenuated
laser light (Zhao \textit{et al}. 2004). With further improvements in high-power pump source
and photon collection efficiency, GHZ entanglement among six photons (Lu \textit{et al}. (2007))
and eight photons (Yao \textit{et al}. (2011)) were also observed.
The number of GHZ-like-entangled qubits climbed up to ten by entangling
both the polarization and momentum degrees of freedom of five photons
(Gao \textit{et al}. 2010a).

SPDC is quite versatile, as it enables to observe a number of further, genuinely multipartite entangled states. Emission of two photons into each of two modes gives already a highly entangled state and is significantly different from simply emitting two pairs (Weinfurter and \.{Z}ukowski 2001). After beam-splitting of these two modes into  four modes and again conditioning on detection of one photon in each of the four modes, the observed state can be written as a superposition of a four-photon
GHZ state and a product of two EPR pairs. This state, first observed by Eibl et al. (2003), is the extension of the singlet state $|\Psi^-\rangle$ to four photons and is thus invariant under equal unitary transformations in the four output modes. This enables a decoherence free communication of a qubit encoded in four-photon states (Bourennane \textit{et al.} 2004b). By splitting the two emission modes into three output modes also the six-photon singlet was observed recently by Radmark \textit{et al}. (2009), with visibility of the six photon interference as high as $85\%$. The high visibility is possible  because the distinctive feature of this scheme is that it does not  involve interferometric overlaps; only beamsplitting is necessary.

Another important multipartite entangled state, the symmetric Dicke state, can be obtained by using collinear type-II SPDC and splitting the four (six) photons into four (six) output modes (Kiesel \textit{et al.} 2007; Wieczorek \textit{et al}. 2009a; Prevedel \textit{et
al}. 2009). The high symmetry of this state makes it an ideal resource to synthesize a number of different multipartite entangled states by projection as shown above for the GHZ-states (Wieczorek \textit{et al}. 2009b) but also for entanglement enhanced sub shot-noise measurements. Finally, all the above states can be observed also by a single, tunable setup. There, a waveplate rotating the polarization in one of the two emission modes of SPDC, followed by a PBS combining these two emission modes, serves to set any superposition between a GHZ state and the product of two entangled pairs (Wieczorek \textit{et al.}, 2008, see also Lanyon \textit{et al.}, 2009b).

To characterize the created multi-photon
entangled states, various methods have been introduced.
Quantum state tomography (for an introduction see James \textit{et al}. 2001)
uses projective measurements on an ensemble of
identically prepared quantum states each probing the state
from a different ``perspective''. It is a tool to fully
reconstruct the density matrix of a quantum system\footnote{A similar technique,
called quantum process tomography, has been used to fully characterize the
quantum controlled-NOT gates (O'Brien \textit{et al}. 2004).}. Experimentally, this
technique was used for two-, three-, four-photon polarization states
and also hyper-entangled photon pairs (White \textit{et al}. 1999; Resch \textit{et al}.
2005; Walther \textit{et al}. 2005b; Barreiro \textit{et al}. 2005).
A disadvantage of such method, however, is that
the number of measurements grows exponentially with the
number of photons, thus the reconstruction of a $n$-photon state
necessitates $4^n$ $n$-fold coincidence measurements which is experimentally demanding.

{\em Entanglement witnesses.---}The method of entanglement witnesses (Horodecki \textit{et al}., 1996; Terhal, 2000;
Lewenstein \textit{et al}., 2000; Bru\ss \ \textit{et al}., 2002) allows to detect the
entanglement in a given state. One speaks of genuine
multipartite entanglement (T\"oth and G\"uhne, (2005)) whenever
the state involves quantum correlations of all subsystems, such that there is no subsystem which is just classically correlated with the other particles. Detection of a genuine multiparticle entanglement with appropriate witnesses usually requires an experimental effort that increases
only polynomially with the number of qubits. A toolbox for efficient witness
operator construction has been created for some multi-particle states (such as GHZ, cluster, and W state, see \textit{e.g.}
G\"uhne \textit{et al}. 2007), and applied in a number of multi-photon
experiments (see \textit{e.g.} Bourennane \textit{et al}. 2004a; Kiesel \textit{et al}. 2005a; Lu  \textit{et al}. 2007a;
Wieczorek \textit{et al}. 2009a). We refer the readers to the  reviews by Horodecki \textit{et al.}
(2009) and by G\"uhne and Toth (2009) for more details.

%%%%%%%%%%%%%%%%%%%%%%%%%OCT-09-10:08%%%%%%%%%%%%%%%%%%%%%%%%%%%%%%%%%%%%%%%%%%%%%%%%

\section{Falsification of a realistic world view}

\label{section:nonlocality}

With a detailed analysis of the work of EPR and its extension by Bohm (1951), in a trailblazing paper Bell (1964) proved, that despite the hopes of Einstein {\em at al.}, there is a deep conflict between quantum mechanics and \textit{local realistic theories}. Not only a conceptual one, which was stressed by EPR is their claims concerning incompleteness of quantum mechanics, but one which straightforwardly leads to drastically different predictions concerning two-particle interference phenomena.

Realism, the cornerstone
of classical physics, is a view that for any physical system (also a subsystem of
a compound system), one can find a theoretical description (deterministic or probabilistic)
which involves all results of \textit{all} possible experiments that can be performed upon it
no matter which experiment actually was performed. Evidently,
this directly contradicts  the Bohr's complementarity principle.
A theory is local if it assumes that information, and influences, cannot
travel faster than light. By assuming that local measurement events are determined
by other events (i.e, causes) in their backward light cone, the term local realism could be replaced by local causality.

Bell's famous theorem is of profound scientific and philosophical consequences.
It also showed that the previously ignored class of entangled states is very important in experimentally
distinguishing between the classical and the quantum. Below we
 first present the formal aspects of Bell's theorem as well as other forms
of \textquotedblleft no-go theorems\textquotedblright\ for hidden-variable
theories.\footnote{Hidden variables are those hypothetical parameters that supposedly influence the  results of individual measurement acts, but are not present in the standard mathematical structure of quantum mechanics. This is why they are called hidden. If one introduces to the theory expressions containing algebraic functions of results of, e.g. pairs of, non commeasurable variables, this is tantamount to the introduction of  hidden variables (such operations are impossible in quantum mechanics). Also, as causes of individual measurement events do not appear in the quantum formalism, they are hidden variables as well.} Next, we present the most important photonic tests invalidating classes of such theories.

An important line of research was opened with
the discovery of the GHZ theorem, which pertains to three or more particle systems, and reveals a contradiction between
quantum mechanics and local realistic theories, even
for definite predictions. This result was the initial motivation for experimental efforts to produce entanglement of more
than two particles. With the advances in multi-photon entangled state preparation, discussed in the previous
Section, a new class of tests of the validity of local realistic theories became possible. Note that, as any classical information processing, or communication, has a local realistic model, the theorems and experiments that reveal phenomena impossible to describe by such formalism, clearly indicate existence of a different method of processing and transferring information. That is why quantum information processing is able to perform tasks impossible with the classical methods.

%%%%%%%%%%%%%%%%%%%%%%%%%OCT-09-13:35pm%%%%%%%%%%%%%%%%%%%%%%%%%%%%%%%%%%%%%

\subsection{Bell's inequality}

Consider pairs of photons simultaneously emitted in opposite directions. They
arrive at two very distant measuring devices A and B, operated by Alice and Bob, respectively.
Their apparatuses have a "knob", which specifies
which dichotomic (i.e., two-valued, yes-no) observable they
measure.\footnote{E.g., for a device consisting of a polarizing beamsplitter
and two detectors behind its outputs, this knob specifies the orientation of
the polarizer, etc.} One can assign to the two possible results (eigen-)
values $\pm1$ (for yes/no).\footnote{We assume, that we have a perfect
situation in which the detectors never fail to register a photon.} Alice and
Bob are at any time  \emph{free} to choose the
knob settings (also in a \textquotedblleft delayed
choice\textquotedblright\ mode, after an emission).

Assuming {\em realism} allows one to introduce, and compare values of results of all possible experiments that
can be performed on an individual system, without caring which measurement would be actually done on the system. According to \textit{locality},
random choices and the consecutive observations made by Alice and Bob, which can be simultaneous in certain reference
frames, cannot influence each other, and the choice made on one side cannot
influence the results on the other side, and \emph{vice versa}.

To test local realism Alice chooses randomly, with equal probability,
to measure either observable $\hat{A_{1}}$ or $\hat{A_{2}}$, and Bob either
$\hat{B_{1}}$ or $\hat{B_{2}}$.  Let us denote the hypothetical results that
they may get for the $j$-th pair by $A_{1}^{j}$ and $A_{2}^{j}$ for Alice's
two possible choices,\footnote{Note that if the results were depending also on the settings a Bob's side, one would  use a two index notation $A_{a,b}^{j}$, with $a,b=1,2$, where $a $ and $b$ index Alice's and Bob's choices of the settings. As index $b$ is missing, this is tantamount to the locality assumption.} and $B_{1}^{j}$ and $B_{1}^{j}$ for Bob's. The numerical
values of these can be $\pm1$. The assumption of local realism allows one to
treat $A_{1}^{j}$ and $A_{2}^{j}$ on equal footing as two numbers, one of them
revealed in the experiment, the second one unknown, but still either $\pm1$. Thus,
their sum and difference always exist, and are algebraic expressions with two unknowns.
For the dichotomic values for all the possible  measurement results one obtains
either the combination $|A_{1}^{j}-A_{2}^{j}|=2$
and $|A_{1}^{j}+A_{2}^{j}|=0$, or $|A_{1}^{j}-A_{2}^{j}|=0$ and $|A_{1}%
^{j}+A_{2}^{j}|=2$, and similarly for Bob's values. Thus, the following trivial
relation holds
\begin{equation}
\sum_{s_{1}=\pm1}\sum_{s_{2}=\pm1}S(s_{1},s_{2})(A_{1}^{j}+s_{1}A_{1}%
^{j})(B_{1}^{j}+s_{2}B_{2}^{j})=\pm4, \label{trivialrelation}
\end{equation}
where $S(s_{1},s_{2})$ is an arbitrary \textquotedblleft
sign\textquotedblright\ function of $s_{1}$ and $s_{2}$, that is, one always has
$|S(s_{1},s_{2})|=1.$ Imagine now that $N$ pairs of photons are emitted, pair
by pair, and {$N$ is sufficiently large, $\sqrt{1/N}\ll1$}. The average value
of the products of the local values for a joint test (the Bell correlation
function), during which, for all photon pairs, only {\em one} pair of observables
(say, $\hat{A_{n}}$ and $\hat{B_{m}}$) is chosen, is given by
%For a joint measurement of only one pair of observables (say, $\hat{A_{n}}$ and $\hat{B_{m}}$) the average value of the products of %the local results is given by
\begin{equation}
E(A_{n},B_{m})=\frac{1}{N}\sum_{j=1}^{j=N}A_{n}^{j}B_{m}^{j},\label{BELLEQ-0}%
\end{equation}
where $n=1,2$ and $m=1,2$. The relation \ref{trivialrelation} after averaging implies, together with \ref{BELLEQ-0}, that for the four possible choices the following inequalities\footnote{All such inequalities boil down to just
a single one
$\sum_{s_{1}=\pm1}\sum_{s_{2}=\pm1}\big|E(A_{1},B_{1})+s_{2}E(A_{1},B_{2})+s_{1}E(A_{2},B_{1})$\\
$+s_{1}s_{2}E(A_{2},B_{2})\big|\leq4.$\label{BELLEQ-1}%
}
must be satisfied for local realistic descriptions, see (Weinfurter and \.{Z}ukowski, 2001; Werner and Wolf (2001),
\begin{eqnarray}
&  -4\leq\sum_{s_{1}=\pm1,s_{2}=\pm1}S(s_{1},s_{2})[E(A_{1},B_{1}%
)+s_{2}E(A_{1},B_{2})\nonumber\\
&  +s_{1}E(A_{2},B_{1})+s_{1}s_{2}E(A_{2},B_{2})]\leq4.\label{BELLEQ}%
\end{eqnarray}
If one chooses a non-factorizable function $S(s_{1},s_{2})$, say, $\frac{1}{2}%
(1+s_{1}+s_{2}-s_{1}s_{2})$, the famous CHSH (Clauser, Horne, Shimony, and
Holt, 1969) inequality is recovered\footnote{The simple algebra to reach this
result rests on the fact that $\sum_{{s_{j}}=\pm1}s_{j}=0$, while
$\sum_{{s_{j}}=\pm1}s_{j}^{2}=2$.}
\begin{eqnarray}
S_{\mathrm{Bell}} &  \equiv\left\vert E(A_{1},B_{1})+E(A_{1},B_{2})\right.
\nonumber\\
&  \left.  \left.  +E(A_{2},B_{1})-E(A_{2},B_{2})\right\vert \leq2.\right.
\label{BELLINEQ}%
\end{eqnarray}

Let us discuss one more assumption, sometimes provocatively called of  {\em free will}.
For the actual experiment, we assume that choices of actual observables are random and
independent from all other processes in the experiment. Note that, only in a part of the
cases (around $1/4$) the given pair of observables [see Eq.~(\ref{BELLEQ-0})]
would be measured. However, as $N$ is large, the correlation function obtained
on a randomly pre-selected subensemble\footnote{Note that the subensemble is
selected by the choice of observables made by Alice and Bob \emph{before} the
actual measurements. If the choices are statistically independent of any other
processes in the experiment, which is equivalent to Alice and Bob having free
will, expectation values of correlation functions conditioned on a particular
choice of local settings do not differ from the unconditional ones, like
Eq.~(\ref{BELLEQ-0}). For more details see (Gill \textit{at al.}, 2002).} of
pairs, due to the aforementioned randomness and independence, cannot differ too much from the one that would have been obtained for
the full ensemble (say, by $\pm 2/\sqrt{N}$), as we deal with two statistically independent processes. Therefore, the results of the \textit{actually} chosen
measurements, under all the three assumptions, must satisfy a {\em Bell inequality} in the form of (\ref{BELLINEQ}), up to minor statistical fluctuations of the order of $1/\sqrt{N}$. {Note that the
presented Bell-type argument avoids any explicit introduction of hidden
variables, other than the hypothetical local realistic values.}

Some quantum processes involving entangled states violate the inequality.\footnote{The CHSH inequality was the first experimentally testable Bell inequality.
The original Bell (1964) inequality, since it assumes perfect correlations of the
singlet state, cannot be tested experimentally, as in such a case correlations
are never perfect. A generalization of the original inequality to the
imperfect case leads to the CHSH inequality. Nevertheless, the original inequality clearly reveals the conflict between local realism and quantum {\em theory}.} For example the
predictions for the spin-$1/2$ singlet give correlations for which
$S_{\mathrm{Bell}}$ can acquire the maximal value $2\sqrt{2}$ allowed by
quantum mechanics, known as the Cirel'son\footnote{In the meantime the transliteration of this surname was changed to Tsirelson.} bound (Cirel'son, 1980; Landau,
1987). In fact, predictions for any pure, non-factorisable (i.e., not
necessarily maximally entangled) two-system state lead to violations (Gisin
and Peres, 1992). This is also the case for a wide range of mixed states
(Horodecki, Horodecki and Horodecki, 1995).

%%%%%%%%%%%%%%%%%%%%%%%%%OCT-09-15:42%%%%%%%%%%%%%%%%%%%%%%%%%%%%%%%%%%%%%%%%%%%%%%%%%%%%%%%%%%%%%%%%%%%%
%%%%%%%%%%%%%%%%%%%%%%%%%%%%%%%%%%%%%%%%%%%%%%%%%%%%%%%%%%%%%%%%%%%%%%%%%%%%%%%%%%%%%%%%%%%%%%%%%%%%%%%%

\subsubsection{Experimental tests of Bell's inequality}

The initial experiments using photon pairs produced in
an atomic cascade to test Bell's inequalities (Freedman
and Clauser, 1972; Aspect \textit{et al}., 1982a, 1982b, see also Clauser and Shimony 1978  for more experiments) falsified Bell's
inequalities, and thus challenged the local realistic world-view. However, this falsification was up to certain loopholes, which are due to experimental imperfections, and still
allow to build local realistic models for the measurement results obtained in the experiments.

The locality loophole is present in experiments which do not have independent, i.e. space-time separated measurement settings, as can be guaranteed only with random and fast
switching of the local measurement settings. In such a case one of the
assumptions behind Bell inequalities, full certainty of the independence of Alice's results on Bob's
settings, or \emph{vice versa}, is not enforced. The efficiency loophole emerges
due to low collection and detection efficiency of the particles. For an efficiency lower than about $83\%$ one can show that one cannot derive
a (generalized) CHSH-type inequality that is violated by quantum predictions. For example see, e.g., Garg
and Mermin, (1987). Eberhard (1993) managed to lower this threshold to $67\%$ by employing, effectively, the Clauser and Horne (1974) inequalities,
albeit for non-maximally entangled states, only.\footnote{ All these results are derived under the usual assumption that
efficiency of all detectors is independent of the locally measured
observable, and equal for all detectors. E..g., a local model explaining
correlations of a maximally entangled state in which detector efficiency
is dependent on the measured observables was proposed by Selleri and
Zeilinger  (1988) . For a more recent discussion of the efficiency
loophole see e.g. Vertesi {\em et al.} (2010) } In the analysis of experiments with efficiency loophole, many authors usually use the so-called {\em fair sampling assumption}, which states that one expects that the fact whether a detector registers a particle or not is statistically independent of all other processes in the experiment.\footnote{When discussing experiments which were not directly aimed at closing the detection assumption we shall give an analysis of the experimental data which always would tacitly assume the fair sampling assumption. However, we shall not repeat this statement {\em ad nauseam}. This pertains also to the locality loophole.} Of course such an assumption in highly questionable. One can find  many {\em ad hoc} local realistic models that violate it,  see e.g. Cabello and Santos (1996). For example, for qubits one could assign three possible local outcomes: +/-1 and no count, or in the case of polarization experiments, one could expect the response of the detection systems might depend on the photon's polarization, even without turning to hidden variable approaches. In case of some experiments, especially the early cascade ones, the polarization state of the photons depends on the direction of emission, {\em etc}.

The famous Aspect \textit{et al}. experiments were the pioneering attempt to address the locality loophole.
To close the locality loophole, one must freely and rapidly choose the
directions of local analyzers and register the particles, in such way that
it is impossible for any information about the setting and the detection to
travel via any (possibly unknown) causal channel to the other observer before
he or she, in turn, chooses the setting and finishes the measurement.
Thus the selection of analyzer directions has to be completely unpredictable, which
necessitates a (quantum) physical random number generator. A numerical pseudo-random
number generator would not do: its state at any time is predetermined.
Furthermore, to achieve a complete independence of both observers, one should
avoid any common context, as would be a conventional use of coincidence
circuits. Individual events should be registered on both sides completely
independently, and compared only long after the measurements are finished.
This requires independent and highly accurate time bases on both sides.

\begin{figure}
[ptb]
\begin{center}
\includegraphics[width=0.425\textwidth]
{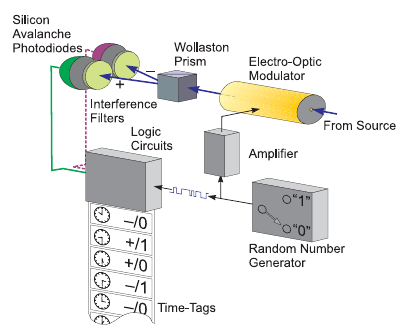}%
\caption{One of the two observer stations. A random number generator is
driving the electro-optic modulator. Silicon avalanche photodiodes are used as
detectors. A \textquotedblleft time tag\textquotedblright\ is stored for each
detected photon together with the corresponding random number
\textquotedblleft0\textquotedblright\ or \textquotedblleft1\textquotedblright%
\ and the code for the detector \textquotedblleft+\textquotedblright\ or
\textquotedblleft-\textquotedblright\ corresponding to the two outputs of the
polarizer. See Weihs \textit{et al}. (1998).}%
\label{weihs2}%
\end{center}
\end{figure}
%EndExpansion

Aspect's experiments were the firs to use fast but periodic switching of the local polarization analyzers. Although the settings  were quickly varying, they were derived from independent function generators with certain correlation times and thus not fully random as it is assumed in the derivation of Bell inequalities. This independent randomness was experimentally realized by Weihs \textit{et al}.
(1998). The observers
\textquotedblleft Alice\textquotedblright\ and \textquotedblleft
Bob\textquotedblright\ were spatially separated by 400 m across the Innsbruck
university science campus. The individual measurements were finished within a
time much shorter than 1.3 $\mu$s, which is the distance of the two observation
stations in light seconds. The actual orientation for local polarization
analysis was determined independently by a quantum physical random number generator (Jennewein et al. 2000).
The employed generator had a light-emitting diode illuminating a BS whose two outputs
were monitored by a pair of photomultipliers.
 Upon receiving a pulse from one photomultiplier a
\textquotedblleft0\textquotedblright\ was produced, whereas the pulse  coming form the other one was giving a \textquotedblleft1\textquotedblright.
This results in a set of binary random numbers\ (Fig.~\ref{weihs2}). The
polarization entangled photon pairs were distributed to the observers through long
optical fibers. A typical observed value of $S_{\mathrm{Bell}}$,   the right side of
inequality (\ref{BELLINEQ}), was as high as $2.73\pm0.02$. In 10 s 14700 coincidence
events were collected. This corresponds to a violation of the local realistic
threshold of $2$ by 30 standard deviations. Still the experiment had a detection efficiency well below the required minimum allowing to avoid the fair sampling assumption.

Meanwhile, there were many ideas to close the detection loophole, see, e.g., (Eberhard, 1993; Kwiat
\textit{et al}., 1994). It was first
closed in an ion-trap experiment by Rowe \emph{et al.} (2001) utilizing the nearly perfect
detection efficiency of fluorescence detection of single ions. However, as the two
entangled ions were separated by approximately $3\mu$m, the locality loophole was left widely open.
A recent experiment by Matsukevich et al. (2008) involved two separate ion traps (one meter distance) and
an entanglement-swapping procedure involving the photons emitted by the ions. The detection loophole
was again perfectly closed. This method gives high hopes for a future experiment simultaneously closing both loopholes (Simon and Irvine, 2003; Rosenfeld \emph{et al.} 2009). Violation of Bell's inequality was also
demonstrated using Josephson phase qubits with deterministic entangled-state preparation and single-shot readout
(Ansmann, 2009).

Other aspects of entanglement have been demonstrated distributing entanglement over long distances.
For example, over 10.5 km free space in Hefei by Peng \textit{et al}. (2005), or in an asymmetric
arrangement, in the case of which one photon is sent over 144 km between the islands of La Palma and Teneriffe (Ursin \textit{et al} 2007).
The Bell experiments were also performed using fiber-based entangled photon
sources, from which two photons were distributed over a distance of more than
10 km apart, see, e.g., (Tittel \textit{et al}., 1998; Zbinden \textit{et
al}., 2001). The later experiment and the subsequent ones (Stefanov \textit{et
al}., 2002; 2003) were done with moving reference frames. These Bell tests in
\textquotedblleft relativistic configurations\textquotedblright\ stress the
oddness of quantum correlations.\footnote{As in a properly performed Bell experiment the measurement events have to be spatially separated, in the relativistic meaning of this word (space time interval between them has to be of the spatial type), they are simultaneous in a certain reference frame. However if the  detectors move with respect to each other and the source, the detection events cease to be simultaneous in the rest reference frames of respective detectors, {\em etc.}. One can have various temporal sequences. The experiments showed that even with moving detectors the expected quantum correlations occur.}

\subsection{GHZ theorem}

\label{sec:GHZ-t}

\subsubsection{Impossibility of deriving realism via perfect quantum correlations and locality}

\label{sec:GHZ-t}

If there are $N>2$ maximally entangled quantum systems (qubits), and if
measurements on them are performed in $N$ mutually spatially separated regions
by $N$ independent observers, the correlations in such an experiment violate
bounds imposed by local realism much stronger than in the two-particle case.
More remarkably, instead of purely statistical reasoning for deriving
Bell's inequality, one can fully follow the spirit of the EPR paper, and first try to
define \textquotedblleft elements of reality\textquotedblright\ based on
specific perfect correlations of the entangled state. In a further step one then can show contradicting
predictions between local realistic theories and quantum mechanics with precisely those
quantum predictions which were used to define the ``elements of reality'' (Greenberger, Horne, and
Zeilinger, 1989).

Take the GHZ experiment,  Fig.~\ref{int3}. Assume that a click at the local
detector $D_{x_1}$, where $x=d,e,f$ is described as a result of value $+1$,
whereas clicks at $D_{x_2}$ are ascribed $-1$. According to the quantum probabilities (\ref{singhz})
the average values of the product of local results reads

\begin{eqnarray}
&E(\phi_{a},\phi_{b},\phi_{c})&\nonumber\\=&\sum_{i,j,k=1,2} (-1)^{i+j+k+1} p_{d_{i}e_{j}f_{k}}(\phi_{a},\phi_{b},\phi_{c})  & \nonumber\\ &
  =\sin(\phi_{A}+\phi_{B}+\phi_{c}).& \label{GHZ-corr-funct}
\end{eqnarray}

(Here $p_{d_i e_j f_k}$$(\phi\cdots)$ is the probability for a detection of one photon by detectors $d_i$, $e_j$, and $f_k$, given the phase settings $\phi_a$$\cdots$). Therefore, if $\phi_{a}+\phi_{b}+\phi_{c}=\pi/2+k\pi$ (where $k$ is an integer), one has
perfect correlations and perfect predictability of the common measurement result. For instance, for $\phi_{a}=\pi/2$, $\phi_{b}=0$ and
$\phi_{c}=0$, whatever may be the results of local measurements of the
observables for, say, the particles belonging to the $n$-th triple represented
by the GHZ quantum state, they have to satisfy
\begin{equation}
A^{n}(\pi/2)B^{n}(0)C^{n}(0)=1,\label{ghz-4}%
\end{equation}
where $X^{n}(\phi)$ ($X=A,B$ or $C$) is the value of a local measurement,
by Alice, Bob and Cecil, respectively, that \textit{would have been} obtained for the
$n$-th particle triple, if the setting of the measuring device is $\phi$.
Locality assumption forces one to assume that $X^{n}(\phi)$ depends solely on
the local phase. Equation (\ref{ghz-4}) indicates that we can predict
with certainty the result of measuring the observable pertaining to one of the
particles (say, $C$) by choosing to measure suitable observables for the other
two. Hence, in a local realistic model the values $X^{n}(\phi)$ for the angles specified in the equality
are EPR's elements of reality.

Similarly, when measuring different settings, according to (\ref{GHZ-corr-funct}), one would obtain
\begin{align}
&  A^{n}(0)B^{n}(0)C^{n}(\pi/2)=1,\label{ghz-6}\\
&  A^{n}(0)B^{n}(\pi/2)C^{n}(0)=1,\label{ghz-6b}
\end{align}
Now, in a local realistic model, from these results we can deduce a further correlation by simply multiplying Eqs.(\ref{ghz-4}-\ref{ghz-6b}). Since $X^n (0)^2 = +1$,
 regardless of whether $X^n (0) = +1$ or $-1$, we obtain
\begin{align}
&  A^{n}(\pi/2)B^{n}(\pi/2)C^{n}(\pi/2)=1.\label{ghz-7-lr}%
\end{align}
This, however, is in a full contradiction with the quantum mechanical prediction obtained from
(\ref{GHZ-corr-funct}) which reads:
\begin{align}
&  A^{n}(\pi/2)B^{n}(\pi/2)C^{n}(\pi/2)=-1.\label{ghz-7}%
\end{align}

Thus the EPR elements of reality program
breaks down, because it leads to a $1=-1$ contradiction. Introduction of EPR's
elements of reality leads to a prediction concerning one of the perfect
correlations, (\ref{ghz-7-lr}), which is \emph{opposite} with respect to the
quantum prediction. We have a \textquotedblleft Bell theorem without
inequalities\textquotedblright\ (Greenberger \textit{et al}., 1990), which
thrashes any hopes to {\em derive} realism from locality and perfect
correlations of the EPR type, as a necessary condition for {\em any} reasoning to be logically valid is that it does not lead to a $1=-1$ contradiction.

\emph{Multiparticle Bell inequalities. -- } In a laboratory one cannot expect
perfect correlations, and even if they were possible any necessarily finite sample would be endowed with
a finite, non-zero uncertainty. Thus, any test of local realism based on the GHZ
correlations has to resort to some Bell-type inequalities. Series of such
inequalities were discovered by Mermin (1990b), Ardehali (1992) and Belinskii
and Klyshko (1993). To get the full set of such inequalities, for correlation
functions involving the product of the result of all parties, it is enough to
generalize the relation (\ref{BELLEQ-0}) to the situation in question. E.g.,
extending (\ref{trivialrelation}) for three partners one has
\begin{eqnarray}
&  \sum_{s_{1},s_{2},s_{3}=\pm1}S(s_{1},s_{2},s_{3})(A_{1}^{n}+s_{1}A_{2}^{n})
(B_{1}^{n}+s_{2}B_{2}^{n})&\nonumber\\
&  \times(C_{1}^{n}+s_{3}C_{2}^{n})=\pm8.& \label{BELLEQ-4}\\ \nonumber
\end{eqnarray}
This leads to the following Bell inequality [Werner and Wolf (2001);
Weinfurter and \.{Z}ukowski (2001)]:
\begin{equation}
\sum_{s_{1},s_{2},s_{3}=\pm1}|\sum_{k,l,m=1,2}s_{1}s_{2}s_{3}E(A_{k}%
,B_{l},C_{m})|\leq8,
\end{equation}
which is the \emph{necessary and sufficient} condition for correlation
functions involved in it, $E(A_{k},B_{l},C_{m})$, to have a local realistic
model [for proofs see Werner and Wolf (2001), \.{Z}ukowski and Brukner
(2002)]. The reasoning is trivially generalizable to an arbitrary number of
parties.\footnote{As a matter of fact this single inequality either implies
all earlier derived tight inequalities, e.g., those of Mermin (1990b), or is
tighter.} The noise resistance\footnote{As indicated by the decreasing
threshold visibility for mixed states of the form $(1-V)\frac{1}{2^{N}}\hat
{I}+V|\mathrm{GHZ}_{N}\rangle\langle\mathrm{GHZ}_{N}|$ which is sufficient to
violate the inequalities.} of violations of such inequalities by $N$-qubit
GHZ states is growing exponentially with $N$. This
is an important fact, because usually one expects noise to increase with the number of photons
involved in an interferometric experiment (due to the increasing
complications, and alignment problems). Thus non-classicality of the GHZ
correlations can be significant even for many particles (Mermin, 1990b;
Klyshko, 1993; Roy and Singh, 1991; \.{Z}ukowski 1993; Gisin and
Bechmann-Pasquinucci, 1998).

\subsubsection{GHZ theorem for laboratory measurement}

\label{sec:GHZ theorem}

The first laboratory test of the GHZ-type paradox  was
done by Pan \textit{et al}. (2000). The experiment used a three-photon GHZ state
\begin{equation}
\left\vert \mathrm{\Delta}\right\rangle =\frac{1}{\sqrt{2}}(\left\vert
H\right\rangle _{1}\left\vert H\right\rangle _{2}\left\vert H\right\rangle
_{3}+\left\vert V\right\rangle _{1}\left\vert V\right\rangle _{2}\left\vert
V\right\rangle _{3}). \label{start}%
\end{equation}
The obtained data in the form of the (necessarily imperfect) GHZ correlations was shown to violate
local realism.

The GHZ state (\ref{start}) satisfies the following eigen-equations
\begin{equation}%
\begin{array}
[c]{rr}%
\hat{x}_{1}\hat{y}_{2}\hat{y}_{3}\left\vert \mathrm{\Delta}\right\rangle
=-\left\vert \mathrm{\Delta}\right\rangle , & \hat{y}_{1}\hat{x}_{2}\hat
{y}_{3}\left\vert \mathrm{\Delta}\right\rangle =-\left\vert \mathrm{\Delta
}\right\rangle ,\\
\hat{y}_{1}\hat{y}_{2}\hat{x}_{3}\left\vert \mathrm{\Delta}\right\rangle
=-\left\vert \mathrm{\Delta}\right\rangle , & \hat{x}_{1}\hat{x}_{2}\hat
{x}_{3}\left\vert \mathrm{\Delta}\right\rangle =\left\vert \mathrm{\Delta
}\right\rangle ,
\end{array}
\label{ghz4}%
\end{equation}
where $\hat{x}$ denotes the observable discriminating between $|45^\circ\rangle$
and $|135^\circ\rangle$ polarizations, whereas $\hat{y}$ discriminates between
left and right circular polarizations. In both cases the ascribed eigenvalues
are, respectively, $1$ and $-1$. With these settings one can get a GHZ paradox falsifying the elements of reality
of the form described in the previous subsection.
%%%%%%%%%%%%%%%%%%%%%OCT-10-9:56%%%%%%%%%%%%%%%%%%%

\begin{figure}
[ptb]
\begin{center}
\includegraphics[
height=4.4429in,
width=1.8944in
]%
{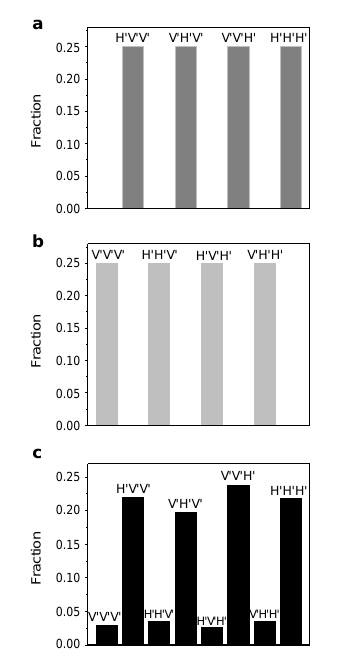}%
\caption{Predictions of quantum mechanics (a) and of local realism (b), and
observed results (c) for the  final $\hat{x}_{1}\hat{x}_{2}\hat{x}_{3}$
measurement in the first GHZ-type experiment of
Pan \emph{et al.} (2000). $H'$ and $V'$ denote here diagonal and anti-diagonal linear polarizations.}%
\label{ghzdata}%
\end{center}
\end{figure}
%EndExpansion

A demonstration of the conflict between local realism and quantum
mechanics for GHZ entanglement consists  of four experiments. In the experiment by Pan \textit{et al.} (2000)
the measured values for $\hat{x}_{1}\hat{y}_{2}\hat{y}_{3}$, $\hat{y}_{1}\hat{x}_{2}\hat
{y}_{3}$, and $\hat{y}_{1}\hat{y}_{2}\hat{x}_{3}$ followed
the values predicted by quantum physics in a fraction of $0.85\pm0.04$ of all
cases. The fourth experiment,  measuring $\hat{x}_{1}\hat{x}_{2}\hat{x}_{3}$,  also was performed, and
yielded results as shown in Fig. \ref{ghzdata}. The data again agree with quantum mechanics for the same fraction
of events. The results are in a clear disagreement with a prediction range that can be made with the data of the three first measurements using a local realistic model.

%%%%%%%%%%%%%%%%%%%%%%%%%%%%%%%%%%%%%%%%%%%%
 The experimental results confirmed
the quantum predictions, within an experimental uncertainty. The obtained average
visibility\footnote{The imperfect visibilities, which are typical in
multi-photon experiments, are mainly caused by two reasons (see also
section~\ref{sec:SWAP}). First, higher-order emissions of
entangled photons give rise to the undesired components in the multi-photon
states. Second, the partial distinguishability of
photons from different emissions or sources causes some degree of incoherence. There may be also alignment problems.} of $(71\pm4)\%$
clearly surpasses the threshold of $50\%$, necessary for a violation of local
realism in three-particle GHZ experiments (Mermin, 1990b; Roy and Singh, 1991;
\.{Z}ukowski and Kaszlikowski, 1997; Ryff, 1997).

Four-photon entanglement was later demonstrated by Pan \textit{et al}., 2001a,
and Eibl \textit{et al}., 2003), and used for a corresponding multi-photon
falsification of local realism (see also Zhao \textit{et al}. (2003a)).

\subsubsection{Two-observer GHZ-like correlations}

Interestingly, the GHZ reasoning can be reduced to a two-party (thus
two space-like separated regions) case while its all-versus-nothing feature is
still retained. One option is to encode three two-state quantum systems in distinct
degrees of freedom of only two photons. Thereby a GHZ-type argument, now also necessarily
involving non-contextuality\footnote{See section V.D for a discussion of this concept.}, can be applied in this two-particle scheme (Zukowski 1991,
Michler \textit{et al.} 2000a). The second option is to find suitable EPR elements of reality in the two
particle case, and to show that they are internally inconsistent. Such an
approach has been taken by Bernstein {\em et al.} (1993) for a (spin-less)
two particle interferometer. Later, after a considerable debate (Cabello, 2001a, 2001b, 2003; Lvovsky, 2002; Marinatto, 2003; Chen \textit{et al}., 2003),
it was shown that an all versus nothing violation of local realism can be
shown for two particle four-dimensional entangled systems. In this new refutation of
local realism, one recovers EPR's original situation of two-party perfect
correlations, but with much less complexity. This becomes possible
with a new approach for defining elements of reality, which nevertheless
strictly follows the EPR criteria.

A third protocol of the two-observer GHZ-like theorem uses a
two-photon hyper-entanglement (Chen \textit{et al}., 2003).
Due to the specific properties of the
hyper-entanglement, nine variables for each party can be regarded as
simultaneous EPR elements of reality. The nine variables can be arranged in
three groups of three, and the three variables of each group can be measured
by one and the same apparatus. This eliminates the necessity of an argument
based on non-contextuality as it is not necessary to assume any of these
variables to be independent of local experimental context. Experimental demonstrations of such a protocol
were done by Yang \textit{et al}. 2005 and Cinelli \textit{et al}. (2005)\footnote{Unfortunately the experimental configuration in the latter experiment did not eliminate the necessity of the non-contextuality assumption. Nevertheless, upon a permutation in the setup one could get an arrangement avoiding this problem. }
using a two-photon hyper-entangled state in Eq.~(\ref{hyper})
(for hyper-entanglement see section~\ref{sec:hyper}).

\subsubsection{Hardy's paradoxes}

Hardy's thought experiment (Hardy 1993) provides an alternative way
to demonstrate a violation of local realism without inequalities for two spin-half
particles, or equivalently for polarizations of two photons. A crucial distinguishing feature in Hardy's thought experiment is that
the two particles are nonmaximally entangled. In such a case, in the ideal situation, for a specific set of measurements quantum mechanics predicts that approximately
$9\%$ of the pairs of photons must give measurement results
 in a form of coincidence counts absolutely not allowed by local realism.\footnote{Note that in the GHZ scenario, one has situations in which $100\%$ coincidences are not allowed by local realism.} The original proposal was
 experimentally demonstrated  by Torgerson et al. (1995) and  White et al.(1999));
 as in the GHZ-type experiments, the results of the experiments were fed into specific inequalities, specially derived to take into account experimental imperfections (under the fair sampling assumption, of course). Their violation indicates underlying  Hardy's contradiction between local realism and quantum mechanics.  Hardy's scheme was later scaled up, so that
 $(50-h)\%$ of photon pairs lead to coincidence events prohibited by local realism
(where $h$ is any small finite number). The effect was demonstrated in an
experiment by Boschi et al. 1997.

Another proposal suggested by
Hardy (1992) was implemented using a pair of Mach-Zehnder
interferometers that couple via the bosonic photon bunching effect at a beam
splitter (Irvine et al. 2005). The original idea was based on a double Mach-Zehnder interferometer that involved particles that annihilate each other (say, electron and positron). The, say,  right internal path of the electron interferometer is at some place partially overlapping with the left internal path of the positron interferometer. The individual interferometers are tuned such that if only one of the particles is fed to its interferometer, it would always exit by, e.g., the left exit port. However, if both electron and positron are simultaneously fed into their interferometers, there is a non-zero probability amplitude of their annihilation. This is related to both particles being in the overlapping arms of the interferometers  (within the story, in such a case the annihilation is assumed to happen with probability one). The two particles act on each other like (two) bombs of the Elitzur and Vaidman (1993) paradox.\footnote{Imagine again a Mach-Zehnder interferometer, tuned in such a way that all photons emerge via its `left' exit. Just a single photon enters it. In the meantime somebody may put an ultrasensitive light detector into one of the paths, a bomb triggered by light in the Elizur-Vaidman anecdote. If the bomb is there, then we either have an explosion, or the photon emerges with equal probability from both outputs of the interferometer (if the bomb does not register the photon it must be propagating in the other internal path). In the case it exits via the `right' exit, we have detected the bomb, using light (photon!), without igniting it.  This is often called `interaction free measurement'. If there is no bomb, nothing changes, and all photons emerge via the left exit.} Thus, if one treats in a realistic way the presence, or the absence, of the particles in the internal paths of the interferometers, they can never be registered {\em both} in the right exit ports of their interferometers - because this would mean for each  of  them that the other particle was traveling via the overlapping arm. That is, both of them were there, thus annihilation must occur. Still, quantum prediction gives a probability of $1/16$ of both particles emerging via the right exit ports.  Of course all that is a {\em gedankenexperiment per se}. To formulate a feasible version of it, Irvine {\em at al.} resorted to photons, and a pair of Mach-Zehnder interferometers with one of their internal mirrors replaced by a  {\em shared} 50-50 beamsplitter. Instead of annihilation, they relied on the Hong-Ou-Mandel bunching effect: if two indistinguishable photons enter the shared beamsplitter, such that each enters by a different input port, they exit randomly via just one port (see section~\ref{sec:HOM}). Thus, if two photons meet (instead of annihilation) one of the interferometers looses the light traveling through it, and no detection is possible in its exit ports. Using a similar realistic reasoning as before this leads to the conclusion that no joint detection in the right exit ports of the two interferometers is possible. The quantum prediction is different, and the experiment gave results agreeing with it.

\subsection{Refutation of a class of  nonlocal realistic theories}

\label{sec:nonlr-t}

Violation of local realism implies that either locality, or
realism, or both, cannot provide a foundational basis for quantum theory
(provided the {\em freedom} assumption of randomness and independence of setting choices holds).
In a novel approach, Leggett (2003)
discussed a broad class of \emph{nonlocal} hidden-variable theories, which, based on
a very plausible type of realism, provide an explanation for all existing
Bell-type experiments. Nevertheless, they are in conflict with quantum
predictions as shown theoretically by Leggett (2003), and experimentally by G\"{o}blacher et all (2007). Subsequently, a reformulation enabled to reduce  the dependence on auxiliary assumptions as shown independently by Paterek et al. (2007) and by Branciard et al (2007) (see also Romero et al. 2010).

Let us discuss the description of the polarization of photons within such theories. The following assumptions are made: ($L_1$)
{\em realism}, ($L_2$) physical states are
statistical mixtures of subensembles with {\em definite} polarizations, ($L_3$)
local expectation values for polarization measurements taken for each
subensemble obey Malus' law.

 Importantly, locality is not assumed. Measurement outcomes may depend
 on whatever parameters in space-like separated regions. It can be shown that
such theories can explain some important features of
entangled  states of two particles: first, by assumption
($L_3$), they do not allow information to be transmitted faster than at the speed of
light; second, they are capable to reproduce perfect anti-correlations, a fundamental feature of the Bell singlet state; and
third, they provide a model for all Bell type experiments in which the
CHSH inequality was violated. Nevertheless, theories based on assumptions
($L_1$)-($L_3$) deliver predictions different from the quantum ones for certain other measurement outcomes.

%%%%%%%%%%%%%%%%%%%%%%

Let us discuss a general mathematical structure of such models. We shall concentrate
on the description of events at Alice's side, events at Bob's side must follow a similar model.
Assumption ($L_1$) allows
an individual binary measurement outcome $A$ for any possible polarization
measurement along any direction $\mathbf{a}$ (that is, whether a single photon is
transmitted or absorbed by a polarizer set at a specific angle) to be
a well defined function of some set of hidden-variables $\lambda$, and, by ($L_2 $), of a
three-dimensional vector\footnote{We use here the Bloch sphere, or spin $1/2$ like,
parametrization of polarization states and measurement settings.} $\mathbf{u}$. As locality is not assumed, A can depend on some set of other
possibly non-local parameters $\eta$ and the remote setting of Bob, ${\mathbf{b}}$. That is, the measurement outcome A depends on these five variables $A=A(\lambda,\mathbf{u}
,\mathbf{a},\eta, {\mathbf{b}}),$ and can take values $\pm 1$ (two distinct measurement outcomes). According to assumption ($L_3$), particles with the same
$\mathbf{u}$ but with different $\lambda$'s and $\eta$'s build up subensembles of
\textquotedblleft definite polarizations\textquotedblright\ described by a
probability distribution $\rho_{\mathbf{u}}(\lambda, \eta)$, and the local expectation
value $\overline{A}(\mathbf{u})$, obtained by averaging
over $\lambda$ and $\eta$, fulfills Malus' law, that is, $\overline{A}(\mathbf{u})=\int
d\lambda d\eta \rho_{\mathbf{u}}(\lambda, \eta )A(\lambda,\mathbf{u},\mathbf{a}%
,\eta, {\mathbf{b}})=\mathbf{u}\cdot\mathbf{a}$. Finally, with assumption ($L_2$), the measured
expectation value for a general physical state is given by averaging over a
distribution $F(\mathbf{u})$ of the subensembles, that is, $\langle A\rangle=\int
d\mathbf{u}F(\mathbf{u})\overline{A}(\mathbf{u})$. Of course one introduces a similar
dependence for Bob's measurement outcomes,
$B=B(\lambda,\mathbf{v}%
,\mathbf{b},\eta', {\mathbf{a}}),$ now depending on Bob's vector $\mathbf{v}$.

The correlation function of measurement results for a source emitting
well-polarized photons is defined as the average of the products of the
individual measurement outcomes:
\begin{eqnarray}
&\overline{AB}(\mathbf{u},\mathbf{v})&\nonumber\\
&=\int d\lambda d\eta d\eta'\rho_{\mathbf{u},\mathbf{v}
}(\lambda, \eta,\eta')A(\lambda,\mathbf{u}
,\mathbf{a},\eta, {\mathbf{b}})B(\lambda,\mathbf{v}
,\mathbf{b},\eta', {\mathbf{a}}).&\nonumber\\
\end{eqnarray}
For a general source producing mixtures of polarized photons the observable
correlations are averaged over a distribution of the polarizations
$F(\mathbf{u},\mathbf{v})$, and the general correlation function $E$ is given
by:
\begin{equation}
E=\langle AB\rangle=\int d\mathbf{u}d\mathbf{v}F(\mathbf{u},\mathbf{v}%
)\overline{AB}(\mathbf{u},\mathbf{v})
\end{equation}
It is a very important trait of this model that there exist subensembles of
definite polarizations (independent of measurements) and that the predictions
for the subensembles agree with Malus' law. It is clear that other classes of
non-local theories, possibly even fully compliant with all quantum mechanical
predictions, might exist that do not have this property when reproducing
entangled states. Such theories may, for example, include additional
communication or dimensions. A specific case deserving comment is Bohm's theory (Bohm, 1951). There the
non-local correlations are a consequence of the non-local quantum potential,
which exerts suitable torque on the particles leading to experimental results
compliant with quantum mechanics (Dewdney \textit{et al}., 1987).

%%%%%%%%%%%%%%%%%%%%%%%%%OCT-10-11:38am%%%%%%%%%%%%%%%%%%%%%%%%%%%%%%%

Following Leggett (2003) one can use the following
identity, which holds for any numbers $A=\pm1$ and $B=\pm1$:
\begin{equation}
-1+|A+B|=AB=1-|A-B|.
\end{equation}
This, plus the above assumptions, implies a Leggett-type inequality
(for details of the derivation see Groblacher \textit{et al}., 2007) of the following form:
\begin{align}
  & S_{NLHV}\equiv |E_{11}(\varphi)+E_{23}(0)|+|E_{22}(\varphi)+E_{23}%
(0)|\nonumber\\
&  \leq4-\frac{4}{\pi}|\sin\frac{\varphi}{2}|, \label{NLHV}%
\end{align}
where $E_{kl}(\varphi)$ is a uniform average of all correlation functions,
defined in the plane of $\mathbf{a}_{k}$ and $\mathbf{b}_{l}$, with the same
relative angle $\varphi$. {Inequalities avoiding the averaging were also derived and tested (Branciard
\textit{et al}., 2007; Paterek \textit{et al}., 2007).} For the inequality to be applicable, the
vectors $\mathbf{a}_{1}$ and $\mathbf{b}_{1}$ necessarily have to lie in a
plane orthogonal to the one defined by $\mathbf{a}_{2}$ and $\mathbf{b}_{2}$.
This contrasts with the
CHSH inequality. Thus, if, as it is experimentally
most easy, $\vec{a}_1$, $\vec{a}_2$ and $\vec{b}_1$ correspond to linear
polarizations, then $\vec{b}_2$ must correspond to an elliptical
polarization

\begin{figure}
[ptb]
\begin{center}
\includegraphics[width=0.44\textwidth]
{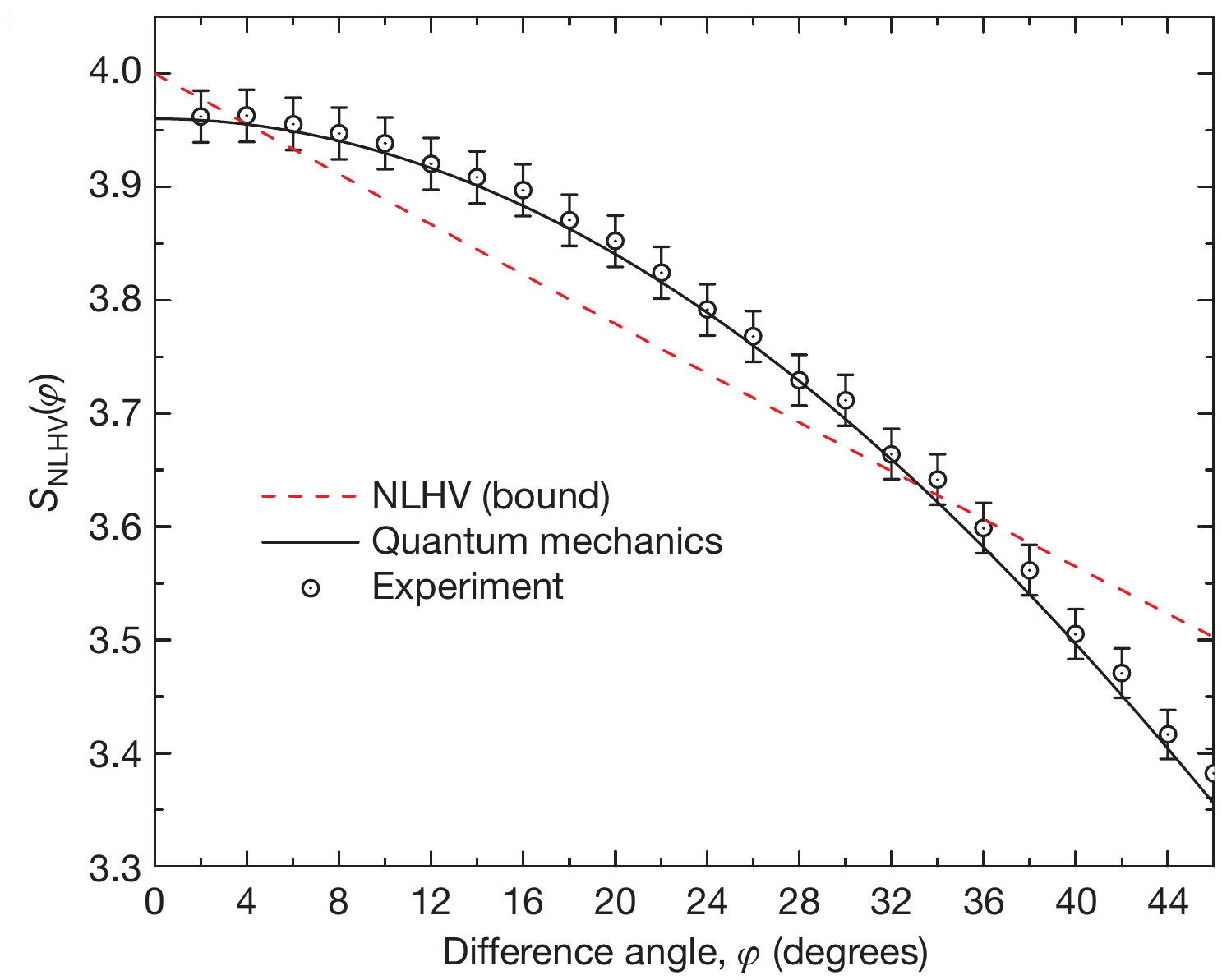}%
\caption{The experimental violation of non-local realism in Groeblacher {\em et al.} (2007).
The dashed line shows the bound of inequality (\ref{NLHV}) for the particular class of nonlocal realistic theories. The solid line is the quantum
 prediction reduced to fit the experimentally achieved visibility. The
experimental data were taken for various difference angles  of local measurement settings
(parameterized as on the Poincare sphere).}%
\label{coding}%
\end{center}
\end{figure}

Quantum theory violates inequality (\ref{NLHV}). Consider quantum predictions
for the polarization singlet state, $|\psi^{-}\rangle
_{AB}=\frac{1}{\sqrt{2}}\left[  |H\rangle_{A}|V\rangle_{B}-|V\rangle
_{A}|H\rangle_{B}\right]  $. The quantum correlation function for the  local
measurements defined by $\mathbf{a}_{k}$ and $\mathbf{b}_{l}$
depends only on the relative angles between the vectors:
$E_{kl}=-\mathbf{a}_{k}\cdot\mathbf{b}_{l}=-\cos\varphi$. Thus, the left hand
side of inequality (\ref{NLHV}), for quantum predictions, reads $|2(\cos
\varphi+1)|$. The maximal violation of inequality (\ref{NLHV}) is for
$\varphi_{max}=18.8^{\circ}$: the bound given by
inequality (\ref{NLHV}) equals 3.792 whereas the quantum value is 3.893.

The Leggett-type inequality valid for
nonlocal realistic theories of the discussed type  was experimentally tested  by Gr\"{o}blacher
\textit{et al}. (2007). In the experiment a SPDC source of high-fidelity two-photon entangled
states was used, with visibilities  $\sim99.0\pm1.2\%$ in the $H/V$
basis, $\sim99.2\pm1.6\%$ in the $\pm45^{\circ}$ basis and $\sim98.9\pm1.7\%$ in the
$R/L$ basis. At that time it was  the highest reported visibility for a pulsed SPDC
entanglement source, and was well beyond the
threshold of 97.4\% which is required for testing the Leggett-type inequality.
The observed $S_{NLHV}=3.8521\pm0.0227$ violates inequality (\ref{NLHV}) by $3.2$
standard deviations. At the same time measurements gave
$S_{\mathrm{Bell}}=2.178\pm0.0199$, which violates the CHSH
inequality by $\sim$9 standard deviations. This way, Gr\"{o}blacher
\textit{et al}. experiment excluded a broad class of
nonlocal hidden-variable theories.

\subsection{Non-contextual hidden variable theories}

Another class of theorems, which show the drastic difference between the
classical and the quantum, are the no-go theorems for non-contextual hidden
variables (NCHV) interpretations of quantum mechanics [Specker, 1960; Bell, 1966; Kochen and
Specker, 1967; Mermin, 1990c; for a survey, see (Mermin, 1993)]. Such
realistic theories assume that hidden variables fix the values of measurement
results of all possible observables for the given system, and that such values
are independent of the measurement context. That is, they do not depend on
which other observables are measured together with them.\footnote{ The measurement context
is defined by a maximal observable. An observable is maximal if it
has a fully non-degenerate spectrum, that is for a $d$ dimensional system it is of the form $\hat{M}=\sum_{i=1}^{d}
\lambda_i|b_i\rangle\langle b_i|,$ where $|b_i\rangle$ are the eigenstates and $\lambda_i$ the eigenvalues of the observable, which are such that $\lambda_i\neq\lambda_j$ for all pairs of the indices. A degenerate observable $\hat{D}$ does not satisfy this last requirement. That is one has at least one pair of indices, say $i=1,2$, such that $\lambda_1=\lambda_2$. Thus for our example $\hat{D}=\lambda_1\sum_{i=1,2}|b_i\rangle\langle b_i|+\sum_{i=3}^{d}
\lambda_i|b_i\rangle\langle b_i|$ . The observable commutes with an infinite set of different maximal observables. This is due to the fact that $\sum_{i=1,2}|b_i\rangle\langle b_i|)=\sum_{i=1,2}|b'_i\rangle\langle b'_i|)$, where the primed eigenvectors are any pair of orthogonal normalized states in the two dimensional subspace spanned by the pair $|b_1\rangle$ and $|b_2\rangle$. } It is interesting that already in the lowest
dimension of $3$, for which a degenerate observable can exist (only such
observables can be measurable in different contexts) non-contextual hidden variable models of quantum mechanics are impossible
(Kochen and Specker, 1967).\footnote{For dimension $2$ a degenerate observable
is just a constant.}

Bell's theorem is a case of a no-go theorem for NCHV in which non-contextuality
is given ``for free'' by the locality assumption. As locality forbids the result on Alice's
side to depend on the actual observable chosen to be measured by Bob, etc.,
the required non-contextuality is enforced by the relativistic causality. This is very
appealing, because relativity is generally assumed to be a principle setting theory for
causal links. Non-contextuality, without the help from relativistic
principles, seems to be a much stronger assumption, as it is difficult to argue why nature has to obey it.\footnote{However, please note that {\em expectation} values of all quantum observables are non-contextual.} Nevertheless, both NCHV
theories and local realistic ones can be reduced to the assumption of the
existence of a joint probability distribution for noncommuting
observables. Note that such distributions are impossible in the quantum
formalism.

Let us present an example of a Kochen-Specker type problem.
Recently Cabello (2008) showed that if  nine observables
$A$, $B$, $C$, $a$, $b$, $c$, $\alpha$, $\beta$, and
$\gamma$ have predefined noncontextual
outcomes $\pm 1$, they must satisfy the following
inequality:
\begin{equation}
S=\langle ABC\rangle+\langle abc\rangle+\langle \alpha\beta\gamma\rangle+\langle Aa\alpha\rangle+\langle Bb\beta\rangle-\langle Cc\gamma\rangle\leq 4
\label{CABELLO}
\end{equation}
where e.g. $\langle ABC\rangle$ denotes the ensemble average of the product
of the three outcomes of measuring the mutually compatible
observables $A$, $B$, and $C$. For any
four-dimensional quantum system, one can find a set of observables for
which the prediction of quantum mechanics is $S = 6$,
irrespectively of the actual state.
This inequality was violated in an experiment by Amselem et al. (2009), where the
chosen observables had the form
\begin{equation}
\begin{array}{ccc}
A=\sigma_z^{(1)} & B=\sigma_z^{(2)} & C=\sigma_z^{(1)}\otimes\sigma_z^{(2)}\\
a =\sigma_x^{(2)}& b=\sigma_x^{(1)} & c=\sigma_x^{(1)}\otimes\sigma_x^{(2)}\\
\alpha=\sigma_z^{(1)}\otimes\sigma_x^{(2)} & \beta=\sigma_x^{(1)}\otimes\sigma_z^{(2)} & \gamma=\sigma_y^{(1)}\otimes\sigma_y^{(2)}\\
\end{array}
\label{KS-STD}
\end{equation}
The above operators $\sigma_i  (i=x,y,z)$ are the usual Pauli operators, for two subsystems $1$ and $2$, respectively.
This set has the following lovely properties (Peres, 2002): all operators have spectrum $\pm1$,
all operators in each row commute, and so do all operators in each column. However, any two operators belonging to different columns {\em and} rows do not commute.  Thus, each operator belongs to two different contexts explicitly displayed in this array.
Furthermore, each operator is the product of the other ones  in the column or in the row to which it belongs, with the sole exception that in the case of the last column each operator is also such a product {\em but times} $-1$. Thus, there is no way that these nine operators behave like real numbers upon multiplication. In other words, if one ascribes to each of the operators whatever realistic values, either +1 or -1, independent of the row or column, one runs into a contradiction with quantum formalism.
The trick used by Anselem {\em et al.} is to treat as subsystem $1$ the polarization degree of freedom of a photon, and as subsystem $2$ the path degree of freedom, as it was the case in \.{Z}ukowski, (1991), Micheler et al. (2000) and Simon et al. (2000). This allowed the construction of six elaborate interferometers equivalent to measurements of all the terms in the inequality (\ref{CABELLO}). The observed value of the left hand side of inequality (\ref{CABELLO}) was for all  twenty tested states close to $5.45$, with the highest measurement error at $0.0032$. After averaging over all states the standard deviation was just $0.0006$, thus the violation of the inequality was as high as by 655 standard deviations. The discrepancy between the ideal quantum value $6$ was due  to imperfections in the complicated interferometers (note, each one had eight exit ports), and the effective observables were slightly deviating from the ideal ones considered in the theoretical reasoning of Cabello. Most recently, a proposal by Klyachko (2008) for a single qutrit contextuality experiment involving only six different measurements was experimentally realized for photons prepared in superposition of three modes (Lapkiewicz, 2011).

%%%%%%%%%%%%%%%%%%%%%%%%%%%JUNE 6, 2011, 8:18 AM %%%%%%%%%%%%%%%%%%%%%%%%%%

\section{Quantum communication}

\label{communication}

Quantum communication ultimately aims at absolute
security and faithful transfer of information, classical or quantum.
Photons are the fastest information carrier, and
due to their very weak coupling to the environment, are
an obvious choice for quantum communication,
especially for long-distances.
Hence, the ability of manipulating the quantum features (such as coherence and
entanglement) of photons is a precious resource.

In this section, we will review several breakthroughs in the field of quantum
communication\footnote{We exclude quantum cryptography, which has been extensively
reviewed by Gisin \textit{et al}. (2002).}: By exploiting entanglement one can
efficiently encode classical messages (Bennett and Wiesner, 1992;
section~\ref{sec:DCoding}), transfer quantum information to a remote location
(Bennett \textit{et al}., 1993; section~\ref{sec:teleportation}), entangle two
remote particles that have no common past (\.{Z}ukowski \textit{et al}., 1993;
section~\ref{sec:swapping}), and purify a large ensemble in a less entangled states
into a smaller ensemble with higher entanglement (Bennett \textit{et
al}., 1996b; Deutsch \textit{et al}., 1996; Gisin, 1996; Horodecki \textit{et
al}., 1996; Pan \textit{et al}., 2001b; section~\ref{sec:purification}).

Needless to say, one of the ultimate dreams is long-distance or even global
($10^{3}$-$10^{4}$ km) quantum communication. As a combination of the ideas of
entanglement purification and swapping, the quantum repeater protocol (Briegel
\textit{et al}., 1998), see section~\ref{sec:repeater}, is an efficient method for
beating decoherence and photon losses in attempts to create long-distance
high-quality entanglement.

In section~\ref{sec:free-space} we discuss steps on the road toward
satellite-based quantum communication and its first step, i.e., free-space
distribution of entangled photon pairs over a distance of 10 km
achieved in 2005 (Resch \textit{et al}., 2005; Peng \textit{et al}., 2005) and, more recently, over
144 km (Ursin \textit{et al}., 2007; Fedrizzi
\textit{et al}., 2009).

\subsection{Quantum dense coding}

\label{sec:DCoding}

One can encode two bits of classical information with two qubits in such a way that each qubit carries a
single bit. To this end, in the case of two qubits represented by polarization
states one could use states  $HH$, $HV$, $VH$ and
$VV$. The idea of quantum dense coding,
introduced by Bennett and Wiesner (1992) is that, by manipulating only
\emph{one} of the two particles in a Bell state, one can also encode two bits of information.

The procedure runs as follows:

\textit{Step 1}. \textit{Sharing maximal entanglement}. A maximally entangled
qubit pair (say, in the state $|\psi^{+}\rangle_{AB}$) is  shared by
Alice and Bob (Fig. \ref{coding}). They agree in advance that $|\psi
^{-}\rangle_{AB}$, $|\phi^{-}\rangle_{AB}$, $|\phi^{+}\rangle_{AB}$, and
$|\psi^{+}\rangle_{AB}$  respectively represent the binary numbers 00,
01, 10, and 11.

\textit{Step 2}. \textit{Coding of the message}. According to the value Bob wants
to transmit to Alice, he performs one out of four possible unitary
transformations (identity operation $\hat{I}$, $\sigma_{x}$, $\sigma_{y}$, and
$\sigma_{z}$) on his qubit $B$ alone. The three non-identity operations
transform, in an one-to-one way, the original state $|\psi^{+}\rangle_{AB}$,
respectively, into $|\phi^{+}\rangle_{AB}$,
$|\phi^{-}\rangle_{AB}$, and $|\psi^{-}\rangle_{AB}$. Once this is done,
Bob sends his qubit to Alice. Note that this possibility
of transforming any of the four basis states to any other by only manipulating {\em one}
of the two qubits holds only for  the maximally entangled states. For product
(and classical) states it is always necessary to have control over both qubits (bits)
to encode two bits in four distinguishable states.

\textit{Step 3. Decoding of  the message}. Upon reception, Alice performs a
Bell-state measurement, distinguishing between the four code-states and thus
allowing her to read out both bits of information.
The quantum dense coding  doubles the information
capacity of the transmission channel: what is actually sent is only one
qubit. This more efficient way of coding information at first glance seems to be at odds
with the Holevo theorem (1973), which states  that  maximally
one bit can be encoded on a \textit{single} qubit. Entanglement,
a property of pairs of qubits, allows to circumvent this theorem and to encode
information entirely in the relative properties of the pair, i.e.
in their correlations.

The first experimental realization of quantum dense coding was reported by the
Innsbruck group (Mattle \textit{et al}., 1996). The preparation of the
polarization entangled photon pairs, the single-qubit operations at Bob's
station, and Alice's Bell-state analyzer can all be done with the SPDC and linear optical techniques
presented in the previous sections. In the Innsbruck experiment, each of the two
$|\psi^{\pm}\rangle_{AB}$ states  could be distinguished, and, they could be distinguished from
$|\phi^{\pm}\rangle_{AB}$. However, , with interferometric Bell-state analysis, there was no possibility to tell which of
the states $\phi^{\pm}$ caused the given detection event.
Thus three different messages could be operationally encoded by manipulating
a single qubit only.\footnote{The states $|\psi^{\pm}\rangle_{AB}$ carried two different
values, while $|\phi^{\pm}\rangle_{AB}$ the third.} Thus  an increase of
channel capacity to $\log_{2}3\simeq1.58$ bits was possible---the
highest value achievable using linear optics and classical communications,
 see Vaidman et al. (1999) and Lutkenhaus et al. (1999). However, the observed
signal-to-noise ratio reduced the actual channel capacity to on average 1.13 bit for the
cases of  successful transfers. Nevertheless the classical and Holevo values of 1 were breached.

\begin{figure}
[ptb]
\begin{center}
\includegraphics[width=0.45\textwidth]
{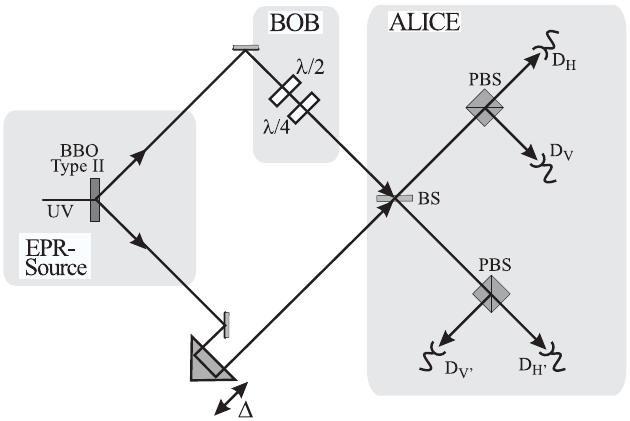}%
\caption{Experimental set-up for quantum dense coding (Mattle \textit{et al}.,
1996).}%
\label{coding}%
\end{center}
\end{figure}

Schuck \textit{et al}. (2006) realized
a complete linear-optical Bell-state analyzer which is able to distinguish all four Bell states , provided the initial pair is entangled in two degrees of freedom, called hyper-entangled, here polarization entanglement plus the
intrinsic time-energy correlation in SPDC (Kwiat
and Weinfurter (1998)). With this ability, the dense coding protocol was implemented
for all four Bell-states in the polarization degree-of-freedom, achieving an overall channel capacity of $1.18(3)$ bits per photon.
%%%%%%%%%%%%%%%%%%%%%%%%OCT-10-13:25%%%%%%%%%%%%%%%%%%%%%%%%%%%%%%%%%%%
Later developments were using the
observation that with hyperentanglement
in at least two degrees-of-freedom, four Bell states
in one of the degrees-of-freedom can be distinguished by local measurements (Walburn et al. (2003)).
Barreiro et al. (2008) exploited pairs of photons simultaneously
entangled in spin and orbital angular momentum
and achieved  a channel capacity of
1.630(6) bits, finally beating the channel capacity limit of 1.58
bits of the  conventional linear-optics implementations.

%%%%%%%%%%%%%%%%%%%%%%%%%%%2011-JUNE-06-10:26AM%%%%%%%%%%%%%%%%

\subsection{Quantum teleportation}

\label{sec:teleportation}

The fascinating possibility of quantum teleportation was discovered by
Bennett, Brassard, Cr\'{e}peau, Jozsa, Peres, and Wootters (1993). Quantum
teleportation is indeed not only a critical ingredient for many more quantum communication protocols and
for quantum computation (Gottesman and Chuang, 1999; Knill, Laflamme, and
Milburn, 2001)---its experimental realization allows new studies of the
fundamentals of quantum theory.

\subsubsection{Theory: qubit teleportation involving an EPR channel and two bit transfer}

\begin{figure}
\begin{center}
\includegraphics[width=0.4\textwidth]
{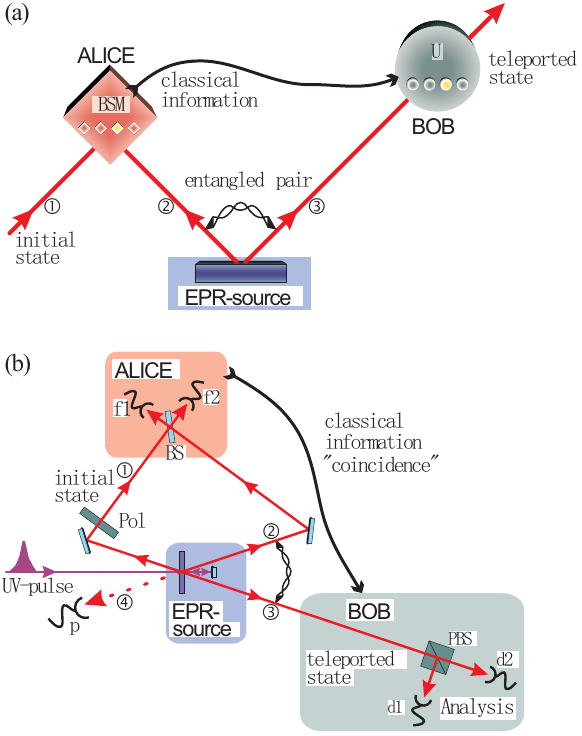}%
\caption{(a) Principle of quantum teleportation. Alice has particle 1, whose qubit
state she wants to teleport to Bob. Alice and Bob also share an ancillary maximally
entangled pair of qubits carried by particles 2 and 3 emitted by an EPR source. Alice
performs a Bell-state measurement jointly on particle 1 and one of the ancillaries. She
sends a two bit classical information informing Bob about the result of her Bell state
measurement. Based on this, he  performs one of four unitary transformations (U)
on his ancillary particle. The transformations are such that Bob's ancilla's qubit
ends up in the state,  which was originally associated with particle 1. (b) Setup of the Innsbruck
teleportation experiment (Bouwmeester \textit{et al}., 1997). A pulse of
ultraviolet light passing through a non-linear crystal creates an ancillary
pair of entangled photons 2 and 3 in a polarization state $|\psi^{-}\rangle_{12}$.
The pulse is reflected, and during its second
passage through the crystal another pair of
photons can be created. One of them plays the role of photon 1. Its polarization is put into a certain state,
teleportation of which is the aim of the experiment. The other one serves as a trigger indicating, by causing a
count at P, that  photon 1  is under way. Alice looks for coincidences after the BS at which
the  photon 1 and one of the ancillaries are superposed. In the case of
receiving a classical information indicating that Alice obtained a coincidence count in
detectors f1 and f2, which implies a collapse into  $|\psi^{-}\rangle_{12}$, Bob knows that his photon
3 is in the original polarization state of photon 1. He can check whether this is indeed so by a polarization
analysis with a PBS and the detectors d1 and d2 behind it.}%
\label{telidea}%
\end{center}
\end{figure}

The idea of quantum teleportation is illustrated in the Fig.~\ref{telidea}(a).
Suppose, particle 1 carries a qubit in the state $|\chi\rangle$ which Alice should
teleport to Bob, that is to transfer it to his particle. Let us consider pure states,
and represent our discussion in terms of qubits  carried by polarization states of photons. We shall assume that
$|\chi\rangle_{1}%
=\alpha|H\rangle_{1}+\beta|V\rangle_{1}$  is the original polarization state of particle 1.
Of course, Alice does not know the state of this qubit.
The trivial idea, i.e., that Alice performs certain measurements on
$|\chi\rangle_{1}$ by which she would obtain all the information
necessary for Bob to reconstruct the state is ruled out: an
experiment on a qubit can give only one bit of information. This only suffice
to determine which state can be ruled out, but is insufficient to
reconstruct the actual state. To this end we need infinitely many
measurements on identical copies of the state. The projection
postulate makes it impossible to fully determine the state of a
single quantum system, or, from another point of view, the no-cloning principle (Wootters and Zurek,
1982) excludes the possibility to create additional high-fidelity replicas of the original state.

%%%%%%%%%%%%%%%%%%%%%%%%OCT-10-19:01pm%%%%%%%%%%%%%%%%%%%%%%%%%%%%%%%%%%%%%

Still, according to Bennett \emph{et al.} there is a way out. Suppose that an
ancillary pair of photons 2 and 3 is shared by Alice and Bob, and that it is in the
polarization entangled state $|\psi^{-}\rangle_{23}$. The entire system,
comprising of the unknown state of particle 1 and the entangled pair, is in a
state $|\chi\rangle_{1}|\psi^{-}\rangle_{23}$. By expanding the state of
particles 1 and 2 (they are in the hand of Alice) in the Bell basis, one gets
\begin{equation}%
\begin{array}
[c]{ccc}%
|\chi\rangle_{1}|\psi^{-}\rangle_{23} & = & \frac{1}{2}\left[  \left\vert
\psi^{-}\right\rangle _{12}\left(  -\alpha\left\vert H\right\rangle _{3}%
-\beta\left\vert V\right\rangle _{3}\right)  \right. \\
&  & +\left\vert \psi^{+}\right\rangle _{12}\left(  -\alpha\left\vert
H\right\rangle _{3}+\beta\left\vert V\right\rangle _{3}\right) \\
&  & +\left\vert \phi^{-}\right\rangle _{12}\left(  \alpha\left\vert
V\right\rangle _{3}+\beta\left\vert H\right\rangle _{3}\right) \\
&  & +\left.  \left\vert \phi^{+}\right\rangle _{12}\left(  \alpha\left\vert
V\right\rangle _{3}-\beta\left\vert H\right\rangle _{3}\right)  \right]  .
\end{array}
\label{newexp}%
\end{equation}
Now, if Alice performs a polarization Bell-state measurement on her two particles then the four possible outcomes are equally likely,
regardless of the unknown state $|\chi\rangle_{1}$. However, once particles 1
and 2 are projected into one of the four entangled states, the polarization state of particle 3 is
instantaneously projected into one of the four pure states appearing in Eq.
(\ref{newexp}). They can be rewritten in the following form
\begin{equation}%
\begin{array}
[c]{cccc}%
-\left\vert \chi\right\rangle _{3}, & -\hat{\sigma_{z}}\left\vert \chi\right\rangle
_{3}, & \hat{\sigma_{x}}\left\vert \chi\right\rangle _{3}, & \hat{\sigma_{y}}\left\vert
\chi\right\rangle _{3},
\end{array}
\label{pure}%
\end{equation}
where the hatted symbols represent  Pauli operators, which act
as unitary transformations. Each of these possible resultant states of Bob's
particle 3 is related in a one-to-one way with the original state $\left\vert
\chi\right\rangle _{1},$ which Alice wanted to teleport. In the case of the
first (singlet) outcome, the state of polarization of particle 3 is the same as the initial
state of particle 1 (except for an irrelevant phase factor), so Bob does not
need to do anything further to finish the transfer the original, in general unknown, state of polarization of particle 1. In the
other three cases, Bob can  apply one of the unitary
transformations of Eq.~(\ref{pure}) to convert the state of particle 3 into
the original one of particle 1.  However, all this is possible only if he additionally receives via a classical
communication channel a two-bit information on  the Bell-state
measurement result  obtained by Alice.\footnote{Note, that Alice and Bob before the teleportation must agree on choice of the basis of her Bell-state measurement.} After Bob's unitary operation, the final state of polarization of particle 3 becomes the new representative of Alice's unknown state,
$\left\vert \chi\right\rangle _{1}$. The original state of Alice's particle 1 is
irrecoverably erased by the Bell-state measurement, as the Bell-state measurement
does not reveal any information on the properties of any of the particles
prior to the measurement. This is why quantum teleportation circumvents the
no-cloning theorem.

%%%%%%%%%%%%%%%%%%%%%%%OCT-10-19:40%%%%%%%%%%%%%%%%%%%%%%%%%%%%%%%%%%%%%%%%%%

The transfer of quantum information from particle 1 to particle 3 can happen
over arbitrary distances. It is not necessary for Alice
to know where Bob is (although they do need to share some reference
frame information in order for the protocol to work). Furthermore, as quantum teleportation is a linear
operation applied to $\left\vert \chi\right\rangle _{1}$, it works for mixed
states, or entangled states, equally well; the initial state $\left\vert
\chi\right\rangle _{1}$ can be completely unknown not only to Alice but to
anyone. Here, a fascinating case is that $\left\vert \chi\right\rangle _{1}$
could even be quantum mechanically completely undefined at the time the
Bell-state measurement takes place. This is the case when, as already remarked
by Bennett \textit{et al}. (1993), particle 1 itself is a member of an
entangled pair, ultimately leading to entanglement swapping (\.{Z}ukowski
\textit{et al}., 1993; Bose \textit{et al}., 1998). Quantum teleportation does
not violate causality: a  transfer of two bits of classical information
is absolutely necessary to conclude the process.

Generally speaking, the basic criteria to
achieve a \textit{bona fide} qubit\footnote{For higher dimensional system this set of conditions
has an extension.} teleportation are: (1) the experimental
scheme without any changes is capable of teleporting any pure or mixed qubit state, this includes the possibility of entanglement swapping, (2) a
fidelity better than the classical one of $2/3$, see Massar and Popescu (1995) can be achieved,
(3) at least in principle, the scheme should be extendable to long distances, (4) the state
to be teleported should be of external nature, that is, it is carried by a particle which plays no role
in the preparation of the quantum part of the teleportation channel (essentially the EPR pair). This
in principle allows to teleport any unknown qubit state delivered by some outside party.

\subsubsection{Experimental quantum teleportation}

Figure \ref{telidea}(b) is a schematic of the Innsbruck experimental setup of
Bouwmeester \emph{at al}. (1997)\footnote{An operational blueprint for such an experiment is first mentioned  in \.Zukowski {\em et al.} (1993). }. A pulse of ultraviolet laser
passing forth and back through a BBO crystal (type II) creates two polarization-entangled EPR pairs.
The pair used as the ancillary one, is labeled here as photons 2 and 3, is distributed to Alice and Bob.
The photon 1 of the other pair passes a polarizer which prepares
it in the initial state to be teleported, and photon 4 is a trigger
indicating that the photon 1 is under way.
After photon 1 is given to Alice, she combines it with her photon 2 and performs the Bell-state analysis.

To demonstrate that teleportation is allowed by Nature it is sufficient to identify one of the four Bell states.
Bell state measurement on photons 1 and 2 is done with the use of a BS. As
explained in Section~\ref{sec:bsm}, if there is a coincidence
detection between the two outputs of the beam splitter, then the
photons are projected to the antisymmetric state
$|\psi^{-}\rangle_{12}$. The Bell-state analysis relies on the interference of two
independently created photons. Therefore, one has to guarantee that behind the BS
the information which photon came from which source is completely erased. This
was done using the methods described in
section~\ref{sec:SWAP} (\.{Z}ukowski \emph{et al}. 1995).
In the experiment, the UV-pump pulse had a duration of 200 fs.
By using narrow bandwidth filters ($\Delta\lambda=4$ nm) in front of
the detectors registering photons 1 and 2, a coherence time of
about 500 fs could be obtained which was sufficiently longer than the pump pulse
duration, so that one could not infer anymore during which passage
through the crystal which of the two photons was created. This generated high
visibility of the multi-photon interference. Furthermore, single-mode fiber
couplers acting as spatial filters were used to guarantee good mode overlap of
the detected photons.

\begin{figure}
[ptb]
\begin{center}
\includegraphics[width=0.42\textwidth]
{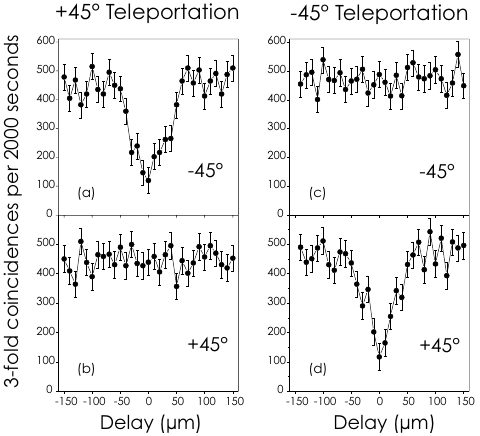}%
\caption{The measured three-fold coincidence rates at d2f1f2 (+45$^{\circ}$) and
d1f1f2 (-45$^{\circ}$) for two cases of the state to be teleported
being polarized under +45$^{\circ}$ (a and b) or  -45$^{\circ}$ (c and d), respectively. The
coincidence rates are plotted as a function of the delay (in $\mu$m) between the
arrival of photon 1 and 2 at Alice's BS [see Fig. \ref{telidea}(b)]. The
three-fold coincidence rates are plotted after subtracting the spurious
three-fold background contribution. The data, together with similar results for
other polarizations, constitute a positive result of a test for  teleportation of an arbitrary state.
(Bouwmeester \textit{et al}., 1997).}%
\label{tep1}%
\end{center}
\end{figure}

To experimentally demonstrate that an arbitrary unknown quantum state can be
teleported, it is sufficient to show that the scheme works for all
mutually orthogonal axes of the polarization (Poincar\'{e}) sphere. The
experimental results for teleportation of photon 1 polarized under
$+45^{\circ}$ ($-45^{\circ}$)\ is shown in the left (right) column of
Fig.~\ref{tep1}. Bouwmeester \textit{et al}.
demonstrated that quantum teleportation works for orthogonal
states $|H\rangle$ and $|V\rangle$ as well as for $|H\rangle+|V\rangle$, $|H\rangle-|V\rangle$ and
$|H\rangle+i|V\rangle$. Thus teleportation was tested for an exhaustive set of mutually unbiased
(in other words, fully complementary) bases of polarization (qubit) states. The average fidelities measured for these states
were 0.81(1), well above the $2/3$ threshold.

%%%%%%%%%%%%%%%%%%%%%%%%%%%%%%%OCT-10-23:20%%%%%%%%%%%%%%%%%%%%%%%%%%%%%%%%%%%%%%%

%EndExpansion

\subsubsection{Teleportation onto freely flying photons}

Most applications of quantum teleportation include the subsequent manipulation
of the teleported photon. Thus a freely propagating output state, which is
teleported with high fidelity, is strongly desired. In the Innsbruck experiment,
however, owing to the probabilistic nature of SPDC, there was also the chance to register a pf1f2 coincidence (Fig.~\ref{telidea}b), seemingly indicating the preparation of a single photon and the identification of a $|\psi^{-}\rangle_{12}$, which could occur unfortunately due to an unwanted event of two-pair emission during the second passage of the UV-pulse, with no emission in the first passage. In such a case, no photon propagates to Bob. Thus in the experiment, a successful teleportation act had to be confirmed by a detection event at Bob's side. Because of this fact,
Braunstein and Kimble (1998) in their comment classified  the experiment as involving a ``post-selection'', and implied that the
fidelity of the process therefore was not sufficient. In their reply (Bouwmeester \textit{et al}. Nature, 1998), they
point out that the situation should be interpreted as reducing the efficiency of teleportation rather than its quality. Possible solutions
(Braunstein and Kimble, 1998; Kok and Braunstein, 2000) could include the
discrimination of one- and two-photon events at detector p [Fig.~\ref{telidea}%
(b)], a quantum non-demolition measurement of the photon number in
mode 3. It should be noted that if one uses a single-photon source
or entangled pairs from a single quantum emitter, e.g., from a
quantum dot (c.f. Stevenson et al. 2006, Akopian et al. 2006), one in principle
does not need to worry about the double pair emission of SPDC
at least up to the antibunching quality of the source. However, due to the
lack of appropriate technology,  such scheme has not been thus far realized.

In the experiment of Pan \textit{et al}., (2003a) such problems were basically removed.
The scheme was such, that a coincident  registration of photons at the  Bell measurement
station was heralding that with a high probability one has a propagating photon carrying the
teleported  state, see Fig.~\ref{telidea}(b). Such a process was called by the authors
``teleportation into freely propagating photons''.  The basic idea of this experimental method is that the entangled
ancillary pair was provided much more frequently than the photon to be
teleported [a similar idea was also used in the teleportation experiment using time-bin entanglement carried
out by Marcikic \textit{et al}. (2003)]. Thus, when a qubit which was to be
teleported arrived the teleportation machinery was almost always ready.
Technically, the main idea was to reduce the number of unwanted f1-f2 coincidence counts.
This was accomplished by attenuating the beam 1 by a factor of $\gamma$, while
leaving the intensity in modes 2-3 unchanged. With such an arrangement a three-fold coincidence f1-f2-p occurs
with probability $\gamma p^{2}$  for a successful
teleportation ($p$ is the probability of having a single
pair creation during a SPDC process). With a significantly lower probability $(\gamma p)^{2}$ one
has a spurious coincidences without a photon at Bob's side. Thus, for a sufficiently low $\gamma$ it is not necessary anymore in a very good approximation to actually detect the photon 3 to be certain that teleportation occurred. The photon 3
give us a a freely
propagating  beam of teleported qubits.

To demonstrate a non-conditional teleportation, a series of neutral filters
were inserted in mode 1, showing that the probability of a successful
teleportation  conditioned on an f1-f2-p three-fold coincidence
increases with decreasing filter transmission $\gamma$ (e.g., the observed probability of success was $0.138\pm
0.002$ for $\gamma=0.05$).
The average fidelity for the unconditional teleportation for three mutually unbiased bases was $\sim0.80(2)$.

%%%%%%%%%%%%%%%%%%%%%%%%%%%%%%%%%%%%%%%6 JUNE 2011 - 11:36 am%%%%%%%%%%%%%%%%%%%%%%%%%%%%%%%%
\subsubsection{Teleportation of a qubit carried by a photon of the ancillary EPR pair}

It is well known that with standard optical devices (passive linear optics plus detectors) one can measure any observable associated with a single photon. Thus if the photon carries two qubits, any two qubit measurement can be performed, including a Bell state measurement, involving states of two different photon ``degrees of freedom", e.g. polarization and path. Thus, as teleportation is from an algebraic point of view a three qubit operation, and as there is no easy solution for a Bell state measurement for two photons, each carrying a qubit, one can resort to the following. One can have a scheme in which a single photon carries two qubits, the qubit to be teleported and one  of the qubits of the EPR maximally entangled pair. This effectively boils down to an emulation of the third particle (sub-system) in the process (for such an emulation in the case of GHZ correlations see Zukowski (1991), an in experiment Michler {\em et al.} (2000a)). However, the fact that one of the particles is emulated does not allow one to teleport a qubit state of an independently arriving particle, and it is difficult to imagine an entanglement swapping process which leaves as a result two spatially separated qubits, previously independently emitted, in a maximally entangled state. Thus the comparative straightforwardness of a Bell state measurement has a trade off: the process is not fully versatile, and in some respects does not mirror the original idea.

In the protocol with the emulation, the quantum state  to be teleported  can prepared
by performing a unitary operation on an additional degree of freedom of
one of EPR  particles of the quantum channel.  A protocol of this kind
has been proposed by Popescu (1995) and was experimentally realized in Rome
(Boschi \textit{et al}., 1998) with a teleportation fidelity of $0.85(1)$. As the protocol does not involve interference
of photons from two separate emissions, and as only one EPR pair is manipulated, it
avoids many difficulties.  Just one SPDC source is needed, and it works with just a one pair-emission process.

%For instance, it is very easy to perform a full
%Bell-state measurement. However, if one aims at teleporting a state of an
%independent qubit, one would have to somehow transfer its state onto
%Alice's particle of the EPR pair---a task equally difficult as full Bell-state analysis and original quantum teleportation.

\begin{figure}
[ptb]
\begin{center}
\includegraphics[width=0.48\textwidth]
{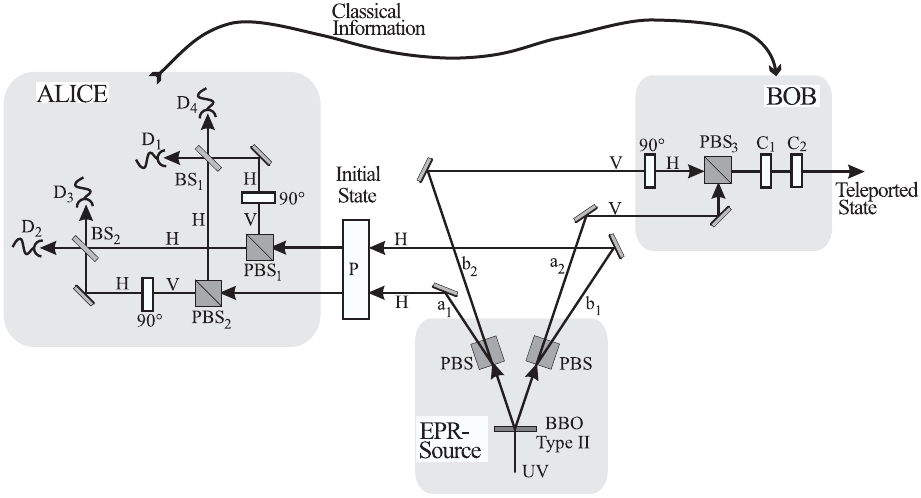}%
\caption{Experimental setup for the \textquotedblleft two-qubit at one sub-system
scheme\textquotedblright\ of quantum teleportation (Boschi \textit{et al}.,
1998).}%
\label{rome}%
\end{center}
\end{figure}
%EndExpansion

The main idea (see the experimental setup in Fig.~\ref{rome}) is to use
the spatial and polarization degrees of freedom of photons. One emulates the
particle which carries the to-be-teleported-qubit with the use of an
additional degree of freedom of an EPR particle sent to Alice.
%\footnote{Compare Zukowski (1991) where such
%an emulation of one of the GHZ entangled particles was proposed}.
The first step
is to produce two photons entangled in their directions of propagation (this will serve as the EPR pair),
i.e., entangled in momentum, but each with a well-defined polarization. Thus
one starts with
\begin{equation}
\frac{1}{\sqrt{2}}\left(  |a_{1}\rangle_{1}|a_{2}\rangle_{2}+|b_{1}\rangle
_{1}|b_{2}\rangle_{2}\right)  |H\rangle_{1}|V\rangle_{2}\,. \label{mom-ent}%
\end{equation}
The area indicated "EPR-source" in Fig.~\ref{rome} shows how this can be
achieved.\footnote{One first generates two-photon polarization-entangled state
$\left\vert \psi^{+}\right\rangle _{12}$. As a PBS transmits (deflects) $H$
($V$) photons, $\left\vert \psi^{+}\right\rangle _{12}$ is then transformed
into momentum entanglement in Eq.~(\ref{mom-ent}), in which photons with label
1 (2) are $H$ ($V$) polarized.} On the way to Alice photon 1 is intercepted by
the Preparer $P$ who changes the polarization from $H$ to an arbitrary quantum
superposition
\begin{equation}
|\chi\rangle_{1}=\alpha|H\rangle_{1}+\beta|V\rangle_{1}.
\end{equation}
This is the quantum state that Alice will transmit to Bob. The Preparer
transforms the polarization in both paths $a_{1}$ and $b_{1}$ in the same way.
The total state $|\Phi\rangle$ of the two photons after his/her action is
\begin{equation}
|\Phi\rangle=\frac{1}{\sqrt{2}}\left(  |a_{1}\rangle_{1}|a_{2}\rangle
_{2}+|b_{1}\rangle_{1}|b_{2}\rangle_{2}\right)  |\chi\rangle_{1}|V\rangle
_{2},
\end{equation}
which is a formal analogue of the initial state in Eq. (\ref{newexp}).

Next, Alice performs a Bell-state-like measurement on the two degrees of freedom of her (single) particle. The
four \textquotedblleft Bell states\textquotedblright\ are represented by the
following correlated polarization-path states of the photon:
\begin{align}
\left\vert \bar{\psi}^{\pm}\right\rangle _{1}  &  =\frac{1}{\sqrt{2}}%
(|a_{1}\rangle_{1}|V\rangle_{1}\pm|b_{1}\rangle_{1}|H\rangle_{1}),\nonumber\\
\left\vert \bar{\phi}^{\pm}\right\rangle _{1}  &  =\frac{1}{\sqrt{2}}%
(|a_{1}\rangle_{1}|V\rangle_{1}\pm|b_{1}\rangle_{1}|H\rangle_{1}).
\end{align}
The measurement of photon 1 with respect to this basis can in principle be
achieved with $100\%$ success rate (A photon detection by $D_{1}$, $D_{2}$,
$D_{3}$, or $D_{4}$ corresponds directly to a projection onto one of the four
Bell states; see Fig.~\ref{rome}).

Since, in terms of the four single-photon Bell states, one has
\begin{align}
|\Phi\rangle &  =\frac{1}{2}[\left\vert \bar{\psi}^{+}\right\rangle _{1}%
(\beta|a_{2}\rangle_{2}+\alpha|b_{2}\rangle_{2})|H\rangle_{2}\nonumber\\
&  +\left\vert \bar{\psi}^{-}\right\rangle _{1}(\alpha|a_{2}\rangle_{2}%
+\beta|b_{2}\rangle_{2})|H\rangle_{2}\nonumber\\
&  +\left\vert \bar{\phi}^{+}\right\rangle _{1}(\alpha|a_{2}\rangle_{2}%
-\beta|b_{2}\rangle_{2})|H\rangle_{2}\nonumber\\
&  +\left\vert \bar{\phi}^{-}\right\rangle _{1}(\beta|a_{2}\rangle_{2}%
-\alpha|b_{2}\rangle_{2})|H\rangle_{2}, \label{rome123}%
\end{align}
the final step of the protocol is that Alice informs Bob which detector
clicked. With this information Bob can reproduce the initial polarization
state by transforming the momentum superposition of photon 2 (see
Eq.~(\ref{rome123})) into a corresponding polarization state, and
applying suitable polarization transformations (following the two bit classical information from Alice). They represent the unitary
corrections necessary to put his photons into the polarization state that was set by the Preparer at  the other EPR
photon.

%%%%%%%%%%%%%%%%%%%%%%%%%%%OCT-16-19:09%%%%%%%%%%%%%%%%%%%%%%%%%%%%%%%%

\subsubsection{Teleportation with various physical systems}

Each teleportation experiment done thus far has  advantages
and disadvantages [for a comparison between various methods, see (Bouwmeester
\textit{et al}., 1999b)]. Quantum teleportation of continuous-variable states
(Furusawa \textit{et al}., 1998; Braunstein and van Loock, 2005) has
the advantage that full Bell-state analysis is possible with linear optics (within
the experimental bandwidth). Yet, it is hard to
extend to a long-distance case. The unavoidable degradation of
squeezed-states sets in during longer-distance transfers. This consequently
leads to a rapid lowering of the quality of squeezed-state entanglement.
Quantum teleportation using nuclear magnetic resonance (Nielsen \textit{et
al}., 1998) or trapped atoms (Riebe \textit{et al.}, 2004; Barrett \textit{et
al.}, 2004) has an obvious advantage in that the input quantum state can be
teleported with an efficiency of 100\%. Yet, it is difficult (if
not impossible) to implement it over long distances.

The Innsbruck teleportation technique with its later improvements enables one to
aim at a long-distance teleportation (Marcikic \textit{et al}., 2003; Ursin
\textit{et al}., 2004) and toward more complicated schemes (Zhao \textit{et al}.,
2004; Zhang \textit{et al}., 2006b). There are other interesting developments. Marcikic \textit{et al}. (2003) realized a teleportation of qubits at telecommunication wavelengths over a
fiber length of 2 km. Adopting Boschi \textit{et al}.'s protocol, Jin \textit{et al}. (2010) emulated free-space
quantum teleportation over 16km.

\subsubsection{More-involved teleportations}
%%%%%%%%%%%%%%%%%%%%%%%%%%%%June 2011-06- 11:56 am%%%%%%%%%%%%%%%%%%%%%%%%%%%

\paragraph{Open-destination quantum teleportation}

The so-called
open-destination teleportation, of Karlsson and Bourennane (1998), is a protocol
allowing to transfer a state to one of several potential recipients. It can be decided who gets the state even after the initial to-be-teleported state, $|\chi\rangle $, is wiped out in a Bell-state measurement.
Such a teleportation scheme was experimentally demonstrated, for
$N=3$, by Zhao \textit{et al}. (2004).

\begin{figure}
[ptb]
\begin{center}
\includegraphics[
height=1.6137in,
width=2.1172in
]%
{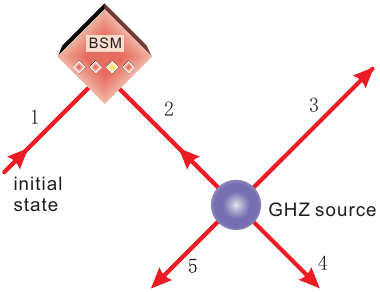}%
\caption{The basic idea of open-destination teleportation (Zhao \textit{et
al}., 2004).}%
\label{opentel}%
\end{center}
\end{figure}

Figure \ref{opentel} shows the basic scheme.
One uses a quantum channel in form of a $N+1$ qubit GHZ state, say for $N=3$ and for polarization qubits:
 \begin{equation}
|\Phi\rangle_{2345}=\frac{1}{\sqrt{2}}\left(  |H\rangle_{2}|H\rangle
_{3}|H\rangle_{4}|H\rangle_{5}+|V\rangle_{2}|V\rangle_{3}|V\rangle
_{4}|V\rangle_{5}\right).  \,
\end{equation}
 and requires, as always, a transfer of two bit classical information. The  polarization state to-be-teleported $|\chi\rangle_{1}=\alpha
|H\rangle_{1}+\beta|V\rangle_{1}$  is first encoded into an $N$-particle
coherent superposition of these GHZ particles. By making a Bell state measurement on, say, photons 1 and 2, one projects the remaining photons into one of four states. E.g. whenever the result of the Bell-state measurement is a $\psi^+$ state, one gets
 \begin{equation}
\frac{1}{\sqrt{2}}\left(  \beta|H\rangle
_{3}|H\rangle_{4}|H\rangle_{5}+\alpha|V\rangle_{3}|V\rangle
_{4}|V\rangle_{5}\right).  \,
\end{equation}
The state $|\chi\rangle$ can be read out
at any of the three  particles, by performing a suitable projection measurement on  $N-1$ of them, here on 2, and a  unitary transformation dependent of the received two bits of data, which is carried out on one of the GHZ particles.\footnote{Either on the remaining one, upon which no measurement is done, or one any one of them before the $N-1$ measurements are done. Further on, we shall follow the first option, as it is simpler.}  Assume that we want to transfer our state to particle 5. To this end upon the receipt of information concerning the result of the BSM, the partner 5 must perform on his particle a $\sigma_x$ transformation which interchanges polarizations $H$ and $V$. The partners 3 and 4 make measurements in the $|\pm\rangle=(|H\rangle\pm|V\rangle)/\sqrt{2}$ basis. The recipient of the state is informed about the measurement results\footnote{The basis of measurement is earlier agreed by the partners of  the protocol.}. Once the recipient gets this additional information,  only if there was just one result associated with a projection to $|-\rangle$, he/she performs the sign flipping  $\sigma_z$ transformation. The state is recovered.

%%%%%%%%%%%%%%%%%%%%%%%%%%%OCT-19-16:14%%%%%%%%%%%%%%%%%%%%%%%%%%%%%%%%%%%%%

In contrast to the original teleportation scheme, after the encoding operation
the destination of teleportation is left open until we perform a polarization
measurement (\textquotedblleft decoding\textquotedblright) on two of the
remaining three photons. This implies that, even though photons 3, 4 and 5 are
far apart, one can still choose which particle should finally carry
the teleported state. No prior agreement on the final destination of the
teleportation is necessary.

In the Zhao \textit{et al}. (2004) experiment the required four-photon GHZ entanglement was generated (conditional upon joint detection) using
the techniques of sections \ref{sec:GHZ-creation}, and the pseudo-single photon
state to be teleported was in form of polarization of an attenuated laser beam containing on average 0.05 photons per pulse. The authors detected
five-fold coincidence with a rate of 12 per hour and
measured fidelities of teleportation from photon 1 to photon 5 and from photon
1 to photon 4. For $+/-$ linear and $R/L$ circular polarization states these were
$\sim0.80(4)$.

%%%%%%%%%%%%%%%%%%%%%%%%%OCT-19-18:26%%%%%%%%%%%%%%%%%%%%%%%%%%%%%%%%%%%%%%

\paragraph{Quantum teleportation of composite two-qubit states}

\begin{figure}
[ptb]
\begin{center}
\includegraphics[width=0.42\textwidth]
{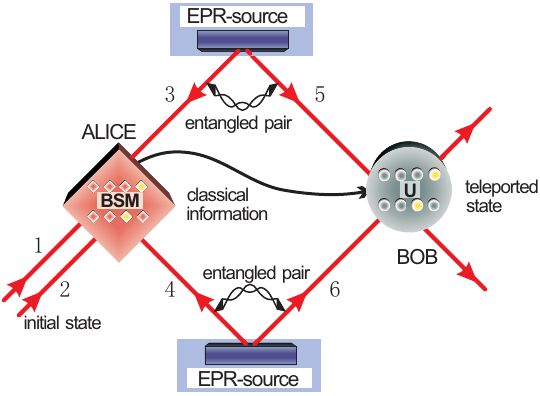}%
\caption{Basic idea of two-qubit quantum teleportation (Zhang \textit{et al}.,
2006c).}%
\label{tele2}%
\end{center}
\end{figure}

Zhang \textit{et al}., (2006b) demonstrated a teleportation of two-qubit states
with a six-photon interferometer. Suppose Alice wants to send an unknown state of a
composite system consisting of qubits 1 and 2:
\begin{align}
\left\vert \chi\right\rangle _{12}  &  =\alpha\left\vert H\right\rangle
_{1}\left\vert H\right\rangle _{2}+\beta\left\vert H\right\rangle
_{1}\left\vert V\right\rangle _{2}\nonumber\\
&  +\gamma\left\vert V\right\rangle _{1}\left\vert H\right\rangle _{2}%
+\delta\left\vert V\right\rangle _{1}\left\vert V\right\rangle _{2}
\label{tq-tele}%
\end{align}
to a distant receiver, Bob (Fig.~\ref{tele2}). Before teleportation Alice and
Bob share two ancillary entangled photon pairs (photon pairs 3-5 and 4-6)
which are both prepared in a Bell state, say, $\left\vert \phi^{+}\right\rangle
=(\left\vert HH\right\rangle +\left\vert VV\right\rangle )/\sqrt{2}$. Following the
standard teleportation protocol, Alice first teleports the state of photon 1
to photon 5 by consuming the entangled pair 3-5. The result of this step is
$\left\vert \chi\right\rangle _{52}$. Similarly, Alice can also teleport
the state of photon 2 to photon 6 by consuming the entangled pair 4-6. After
a successful implementation of the two steps, the original two-qubit state
$\left\vert \chi\right\rangle _{12}$ is teleported to qubits 5 and 6 in
$\left\vert \chi\right\rangle _{56}$.

The teleportation of two-qubit states was realized by teleporting  two
photonic qubits individually. Thus, neither the two original qubits nor the
teleported qubits have to be in the same place. Such a flexibility is desired
in distributed quantum information processing such as quantum telecomputation and quantum state sharing.
The method
can be easily generalized to teleport a state of an $N$-qubit composite system.%

%EndExpansion

Zhang \textit{et al}. (2006b) managed to obtain on average
10$^{5}$ photon pairs per second from each
EPR source. As a result, on average 10 six-photon events per minute were registered.
The two-qubit teleportation protocol was implemented for three different
initial states $\left\vert X\right\rangle _{A}=\left\vert H\right\rangle
_{1}\left\vert V\right\rangle _{2}$, $\left\vert X\right\rangle _{B}%
=(\left\vert H\right\rangle _{1}+\left\vert V\right\rangle _{1})(\left\vert
H\right\rangle _{2}-i\left\vert V\right\rangle _{2})/2$\ and $\left\vert
X\right\rangle _{C}=(\left\vert H\right\rangle _{1}\left\vert V\right\rangle
_{2}-\left\vert V\right\rangle _{1}\left\vert H\right\rangle _{2})/\sqrt
{2}=\left\vert \psi^{-}\right\rangle $. The measured fidelity for $\left\vert
X\right\rangle _{A}$, $\left\vert X\right\rangle _{B}$, and $\left\vert X\right\rangle _{C}$
was $0.86(3)$, $0.75(2)$ and $0.65(3)$, respectively.
All the measured fidelities were well beyond the state
estimation limit of 0.40 for a two-qubit
system (for a derivation of the limit see Hayashi \textit{et al.} (2005)).

\subsection{Entanglement swapping}

\label{sec:swapping}

\subsubsection{Theory}

Entanglement swapping (\.{Z}ukowski \textit{et al}.,
1993) provides a method of entangling two particles that never interacted
or even have no common past. It can also be interpreted as teleportation of
entanglement, i.e., teleportation of undefined states of a particle entangled with another subsystem  (Bennett \textit{et
al}., 1993). We would like to mention that one of the original motivations of
entanglement swapping is the so called \textquotedblleft event-ready
detection\textquotedblright\ of the entangled particles, a concept suggested
by Bell (Bell, 1987; Clauser and Shimony, 1978). Entanglement swapping,
together with entanglement purification, is a key element of the quantum
repeater protocol (Briegel \textit{et al}., 1998; D\"{u}r \textit{et al}.,
1999; see also section~\ref{sec:repeater}) and opens a way to efficiently distribute entanglement for massive particles (Bose \textit{et al.}, 1998).

\begin{figure}
[ptb]
\begin{center}
\includegraphics[width=0.38\textwidth]
{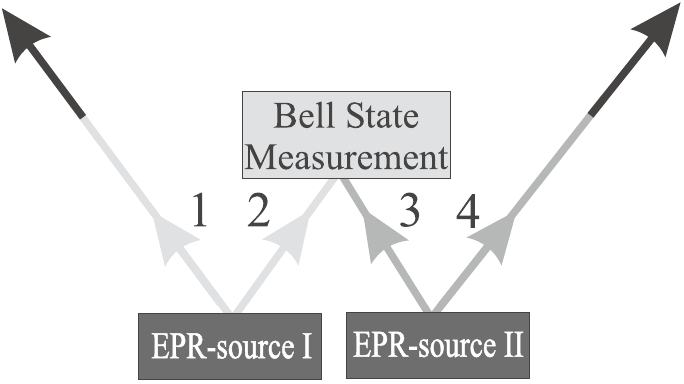}%
\caption{Principle of entanglement swapping.}%
\label{eswap}%
\end{center}
\end{figure}

Consider the arrangement of Fig.~\ref{eswap}. We have two EPR sources. Assume
that each source emits a pair of entangled photons in a state, say,
$\left\vert \psi^{-}\right\rangle $ so that the total state of the four
photons is $\left\vert \Psi\right\rangle _{1234}=\left\vert \psi
^{-}\right\rangle _{12}\left\vert \psi^{-}\right\rangle _{34}$. While pairs
1-2 and 3-4 are entangled, there is no entanglement of any of the photons 1 or
2 with any of the photons 3 or 4.%

Next, one performs a Bell-state measurement on photons 2 and 3. According to the
expansion
\begin{equation}%
\begin{array}
[c]{ccc}%
\left\vert \Psi\right\rangle _{1234} & = & \frac{1}{2}\left(  \left\vert
\psi^{+}\right\rangle _{14}\left\vert \psi^{+}\right\rangle _{23}-\left\vert
\psi^{-}\right\rangle _{14}\left\vert \psi^{-}\right\rangle _{23}\right. \\
&  & \left.  -\left\vert \phi^{+}\right\rangle _{14}\left\vert \phi
^{+}\right\rangle _{23}+\left\vert \phi^{-}\right\rangle _{14}\left\vert
\phi^{-}\right\rangle _{23}\right)  \ ,
\end{array}
\, \label{2state}%
\end{equation}
this measurement always projects photons 1 and 4 also onto a Bell state. For
example, if the result of the Bell-state measurement of photons 2 and 3 is
$\left\vert \psi^{-}\right\rangle $, then the resulting state for photons 1
and 4 is also $\left\vert \psi^{-}\right\rangle $. In all cases photons 1 and
4 emerge entangled despite the fact that they never interacted in the past. In
Fig.~\ref{eswap} entangled particles are indicated by the same degree of darkness of the lines.
Note that particles 1 and 4 become entangled after the Bell-state measurement
on particles 2 and 3. Without knowing which result of the BSM measurement
occurred, however, the state of photons $1$ and $4$ would remain
maximally mixed.

Given an ideal arrangement with sources that emit only a single pair of
entangled photons each, the process of entanglement swapping also gives a
means to generate event-ready entanglement. Namely, as soon as Alice
completes the Bell-state measurement on particles 2 and 3, we know
that photons 1 and 4 are on their way, ready for detection in an
entangled state. In this way one has the possibility to perform an
event-ready test of Bell's inequality (Bell, 1987; \.{Z}ukowski \textit{et al}.,
1993). For a further discussion on event-ready entanglement, see section~\ref{sec:eventready}.%

%%%%%%%%%%%%%%%%%%%%%OCT-19-19:09%%%%%%%%%%%%%%%%%%%%%%%%%%%%%%

\begin{figure}
[ptb]
\begin{center}
\includegraphics[width=0.46\textwidth]
{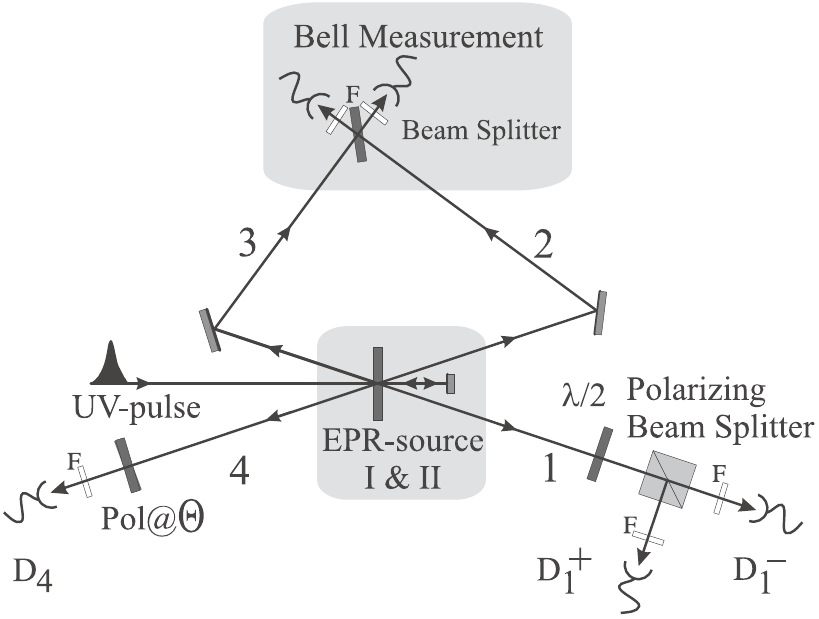}%
\caption{Experimental setup of entanglement swapping (Pan \emph{et al.},
1998). The generation scheme for photon pairs 1-2 and 3-4, and the Bell-state measurement
identifying the $\left\vert \psi^{-}\right\rangle _{23}$ state is identical as in
the Innsbruck teleportation experiment. The Bell-state
measurement on photons 2 and 3 projects the two remaining photons 1 and 4  onto an
entangled state. To analyze their entanglement one looks at coincidences
between detectors D$1^{+}$ and D4, and between detectors D$1^{-}$ and D4, for
different polarization angles $\Theta$. Note that, since the detection of
coincidences between detectors D$1^{+}$ and D4, and D$1^{-}$ and D4 are
conditioned on the detection of the $\left\vert \psi^{-}\right\rangle _{23}$
state (which happens whenever one has a coincidence behind the beamsplitter), one in fact is looking for 4-fold coincidences to signify the event of entanglement swapping. Narrow bandwidth filters (F) are
positioned in front of each detector to make photons from different emissions indistinguishable.}%
\label{swapsetup}%
\end{center}
\end{figure}
%EndExpansion

\subsubsection{First experimental demonstration}

The above scheme was realized using an SPDC source as shown in Fig.~\ref{swapsetup}, by Pan \textit{et al.}
%%%%%%%%%%%%%%%%%%%%%JUNE2011-06-124:57 PM%%%%%%%%
(1998). As in the Innsbruck teleportation experiment, only the anti-symmetric
Bell-state $\left\vert \psi^{-}\right\rangle _{23}$ was detectable in this Bell-state measurement. After such a
detection event, signaled by a coincidence behind the BS, according to the entanglement swapping rules,
photon 1 and 4 is projected into the same entangled state $\left\vert
\psi^{-}\right\rangle _{14}$. This entanglement was tested by analyzing the
polarization correlations between photons 1 and 4 conditioned on coincidences
between the detectors of the Bell-state analyzer. When varying the analysis angle  $\Theta$ for the photons going to
detector $D_4$, the coincidences  with $D_1^+$ and $D_1^-$ should follow two complementary sine curves in
dependence on $\Theta$. The observed sinusoidal
behavior (the interference pattern) of the coincidence rates had a visibility
of $0.65(2)$, which clearly surpasses the 0.5 limit for a classical interference for coincidence measurements. A later
experiment (Pan \textit{et al}., 2001a) achieved a visibility of $\sim
0.84$, which is sufficient for violating a Bell inequality (the threshold is 0.71) for photons 1 and
4. Further advancements were achieved by Jennewein \textit{et al}. (2002)
by implementing  a 2-state Bell-analyzer. Using a fiber coupler an ideal spatial mode-overlap was
obtained. A Bell-inequality, for measurements on photons 1 and 4, was violated
by the factor of $1.211(45)$, i.e., by 4.6 standard deviations.

 A
test of Bell's inequality involving  swapping of entanglement has some appealing
features, aside from being an \textquotedblleft event-ready\textquotedblright%
\ one\footnote{A successful BSM measurement defines the ensemble of photon pairs, $1$ and $4$, which are subject to a Bell test.}.
In addition, it can be  performed in a delayed-choice mode, as suggested by
Peres (2000) and realized by Jennewein \textit{et al}. (2002). In such an
experiment one delays the instant of time at which the Bell-state measurement
is performed on photons 2 and 3. Thus entanglement between photons 1 and 4 in
subensembles associated with a specific result for 2 and 3, is revealed,
\textit{a posteriori}, after they have already been measured and may no longer
exist. Most recently, an experiment with active switching and space-like separation of the
relevant decision was carried out by Ma et al. (2011).

\subsubsection{Other experiments on entanglement swapping}

\label{otherswapping}

Recently, entanglement swapping experiments with
an increased complexity  involving three pairs of entangled photons, have been
demonstrated: multi-stage entanglement swapping (Goebel et al., 2008) and
multi-particle entanglement swapping (Lu et al., 2009).

If one aims to build a quantum repeater (Briegel \textit{et al}., 1998, see also Sec.\ref{sec:repeater})
one has to achieve entanglement swapping with synchronized entangled photon sources among all distributed segments. It thus
requires stable interference between two  independently emitted photons.
Yang \textit{et al}., (2006), Kaltenbaek \textit{et al}., (2006), (2009) successfully
realized the necessary technique using synchronized femto-second (fs) lasers
to solve the above problem. Kaltenbaek \textit{et al}., (2006)\footnote{The original purpose of the experiment was to demonstrate that independently emitted photons do interfere. Thus the team used two independently pumped PDC crystals. The only link between the two pumping lasers was via an electronic pulse synchronization. A recording of a pair of idlers heralded  that two signals were on the way to the BS at which the Hong-Ou-Mandel coincidence dip was observed. The visibility was well surpassing the classical limit. Thus non-classical interference of entirely independent photons was observed. } reported an
active synchronization method: the two independent fs pulsed lasers pumping
the two separate SPDC sources were electronically synchronized to emit pulses
at the same time.  To enable interference the two
photons registered behind the BS cannot be distinguished in any way.
To this end, the now standard methods suggested in  (Zukowski et al.,
1995)  and discussed in  Section IV D were used.

The entangled photons generated via the usual SPDC, as used in the above experiments,
have broadband linewidth (usually on the order of several THz) determined by the phase-matching
condition. Thus there the challenge was to achieve sufficiently sharp synchronization of the photons.
Halder \textit{et al}. (2007) took a different approach to achieve entanglement swapping
by a precise time measurement. The photon detector used in the experiment was a niobium nitride superconducting single photon
detector with a time resolution of 74ps\footnote{Conventional room temperature silicon detectors have a time jitter of $\sim500$ps.}. The photons
were filtered using 10pm-bandwidth filters, which corresponds to a coherence time of 350ps well above the temporal resolution of the detectors.
Hence, ultra-coincidence photon timing could be obtained, and pulsed sources could be replaced by continuous-wave
sources, which do not require any synchronization.\footnote{The experiment is a realization of the original scheme of Zukowski et al. (1993).}

The passive filtering used by Halder \textit{et al}. is however, extremely inefficient (the 10pm-filter transmits $<1\%$ only of all down-converted photons).
Thus a very bright narrow-band entangled photon source is highly desirable.
A recent experiment (Bao \textit{et al}., 2008) realized such a source with a linewidth of 9.6 MHz. Due to the
long coherence time, synchronization for such sources is unnecessary, while
coincidence measurements with time resolution of several ns with current
commercial single-photon detectors will be sufficient to see interference of photons originating from independent sources (work in progress).

Entanglement swapping provides a tool to entangle qubits without direct interaction.
An interesting application  is that we can entangle distant, independent \emph{matter} qubits
through photon-mediated entanglement swapping. Imagine we start with two entangled
atom-photon pairs (Blinov \textit{et al}. 2004; Volz \textit{et al}. 2006). By implementing a Bell state measurement of the two photons, we can project
the two atomic qubits into an maximally entangled state.
%This is a key ingredient in implementation of quantum repeater protocols (see Section~\ref{sec:repeater}) and for
%distributed measurement-based quantum computing (see Section~\ref{sec:oneway}).
Proof-of-principle experiments have been performed by Moehring \textit{et al}. (2007)
who entangled two trapped atomic ions separated one meter apart using entanglement swapping exploiting interference of photons emitted by the ions, and by Yuan \textit{et al}. (2008) in atomic ensembles.
These experiments still suffer from low success probability and imperfect state fidelity. For instance, the
Moehring \textit{et al}. (2007) experiment had a success probability of $3.6\times10^{-9}$ and the fidelity of the states of the entangled
ions was 0.63(3). The ion-ion entanglement fidelity was improved to be 0.81 in a later experiment by Matsukevich et al. (2008).
Together with the high efficiency of the measurement of the quantum state of an ion, this high fidelity allowed
to observe a Bell inequality violation with an efficiency high enough to close the detection loophole.

\subsection{Beating noisy environment}

\label{sec:purification}

So far, significant experimental progress has been achieved in
small-scale realizations of quantum information processing. However, interesting
challenges arise in bringing quantum information processing to technologically
useful scales. This is primarily due to the unavoidable decoherence\footnote{For
general aspects on decoherence, we refer to a review  by Zurek (2003).}
caused by a coupling between the quantum system and the environment.
In quantum communication, it is the noisy quantum channel that degrades
the quality of entanglement between  particles  the further they propagate.
Yet, the implementation of any of the quantum
communication schemes (as reviewed above) over large distances requires that
two distant parties share entangled pairs with high quality. Similarly,
during quantum computation the coherence of a quantum system also decreases
exponentially with an increasing operation time, consequently
leading to failure in the quantum computation. It is therefore necessary to
overcome  decoherence in any realistic large-scale realization
of quantum information processing.

An important tool to overcome the noise in the quantum communication channel is
entanglement distillation, concentration and purification, proposed by Bennett \textit{et al}. (1996a, 1996b,
1996c) and Deutsch \textit{et al}. (1996). A linear-optical implementation
of entanglement purification was suggested and experimentally
demonstrated by Pan \textit{et al}. (2001b, 2003b). Quantum repeater (Briegel
\textit{et al}., 1998; D\"{u}r \textit{et al}., 1999), based on entanglement
purification and entanglement swapping, would provide an efficient way to generate
highly entangled states between two distant locations. Remarkably, the quantum
repeater protocol tolerates general errors on the percent level, which is
reachable using entanglement purification based on linear optics (Pan
\textit{et al}., 2001b, 2003b). A study (D\"{u}r and Briegel, 2003)
shows that entanglement purification can also be used to increase, by several orders of magnitude, the quality
of logical operations between two qubits. In
essence, this implies that the threshold for tolerable errors in quantum
computation is within reach using entanglement purification and linear optics.

\subsubsection{Entanglement distillation and concentration}

Entanglement concentration aims to obtain with a nonzero probability a higher
entanglement from \textit{pure states} with a lower entanglement.
There are two methods to achieve this. The first is the so-called
Procrustean method (Bennett \textit{et al}., 1996a). It requires
that the photon pairs are all in a pure non-maximally entangled
state, say, $\left\vert \Psi \right\rangle_{\text{nonmax}}=\alpha|H\rangle|V\rangle+\beta|V\rangle |H\rangle$,
where $\alpha$ and $\beta$ are two \textit{known} amplitudes.
In this case, the scheme only involves local filtering operations (Gisin, 1996;
Horodecki \textit{et al}., 1996) on single pairs. Second, the Schmidt
decomposition scheme (Bennett \textit{et al}., 1996a) works for
photon pairs that are all in a pure but \textit{unknown}
non-maximally entangled state $\left\vert \Psi\right\rangle
_{\text{nonmax}}$. In practice, this scheme is more difficult to
implement as it requires simultaneous collective measurements on
many photons.

Kwiat \textit{et al}., (2001) used the Procrustean method to demonstrate
experimentally distillation of maximally entangled states
from non-maximally entangled inputs. Using partial polarizers,
they performed a filtering process to maximize the entanglement of
pure polarization-entangled photon pairs generated by SPDC. The method
was also applied  to initial states that were partially mixed. After filtering,
the distilled states show violations of a Bell's inequality, while the
initial states do not have this property. For two special types of two-qubit mixed states,
Verstraete et al. (2001) constructed the optimal local filtering operations for
distilling entanglement from the mixed state, with an experimental demonstration
done by Wang et al. (2006).

The Schmidt decomposition scheme becomes
practically feasible after the proposal of a linear-optical implementation of
entanglement concentration (Zhao \textit{et al}., 2001; Yamamoto \textit{et
al}., 2001). Two independent experiments (Yamamoto \textit{et
al}., 2003; Zhao \textit{et al}., 2003b) were reported for linear-optical
entanglement concentration.

\subsubsection{Entanglement purification}
\label{puri}

\begin{figure}
[ptb]
\begin{center}
\includegraphics[width=0.38\textwidth]
{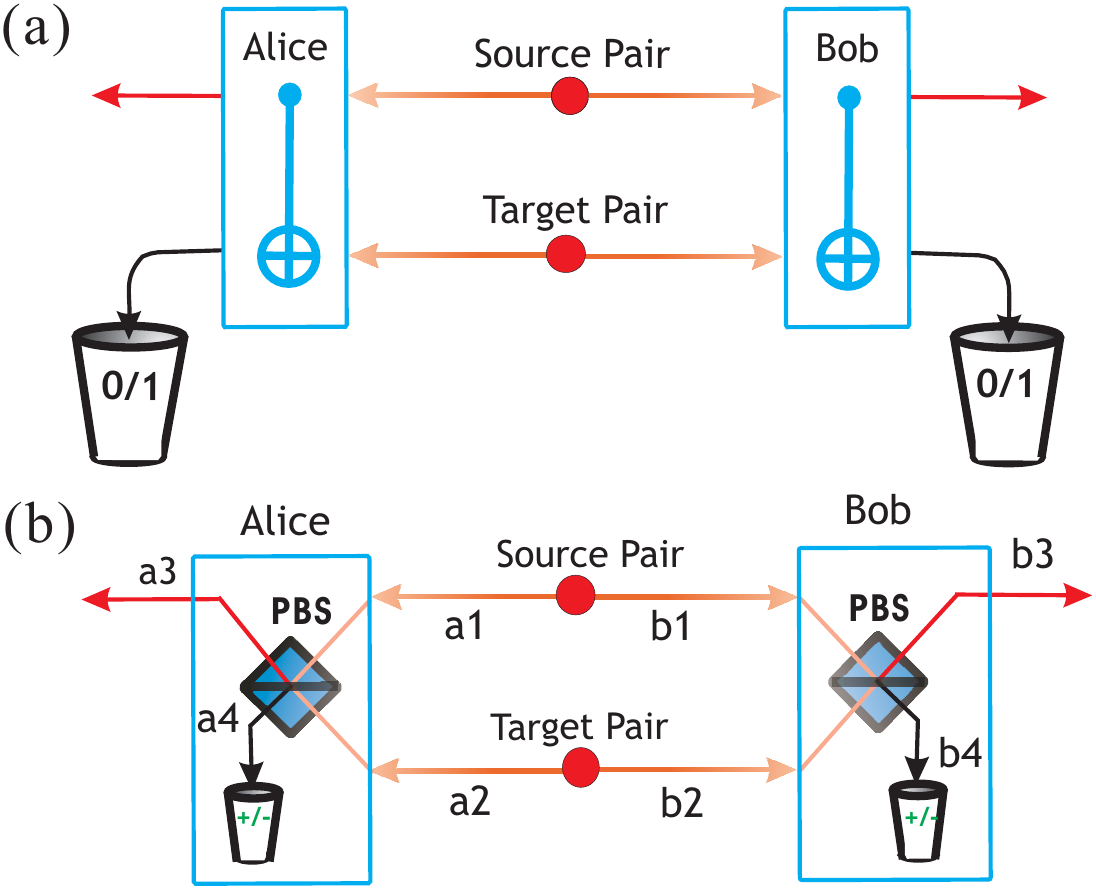}%
\caption{Scheme of entanglement purification by Bennett
  \textit{et al.} (1996a) and Pan \textit{et al}., (2001b).
(a). Two poorly entangled (source and target) pairs are initially shared by Alice and Bob. They both perform a (local)
CNOT operation on the two particles at their hands, measure the particles
belonging to the target pair in the $0/1$ basis and compare the measured results via classical
communication. If the results are the same, then the remaining pair will have a higher degree of
entanglement than the original two pairs. In this case they keep the source pair. In the case of
obtaining opposite results, they discard it. By repeating the same procedure, always starting from
the pairs produced in the former purification step, it is possible to distill pairs of arbitrarily
high entanglement quality (for more details see Bennett \textit{et al.} 1996a).
(b). An alternative and more feasible scheme which does not require a CNOT
operation but only polarizing beam splitters. The PBS transmits horizontal ($|H\rangle$), and
reflects vertical ($|V\rangle$), polarization. By selecting only those events for which there is
one, and only one, photon in each output mode of the PBS, one can project two photons
input from different spatial modes into the subspace spanned by $|H\rangle|H\rangle$ and
$|V\rangle|V\rangle$ (for more details see text).}%
\label{pur-cnot-pbs}%
\end{center}
\end{figure}

The underlying idea of entanglement purification is that, by
using local operations and classical communication (LOCC) only, to extract from multiple copies of imperfect
states (\emph{arbitrary mixed states}) fewer copies of entangled state asymptotically to near-unity fidelity.
Schemes of entanglement purification ware introduced by Bennett
\textit{et al.}, 1996a, 1996c; Deutsch \textit{et al}., 1996, as illustrated
in Fig.~\ref{pur-cnot-pbs}(a). However, a drawback of these theoretical schemes is that they require
CNOT operations. In the context of long-distance
quantum communication, the probability of errors caused by the CNOT operation must be within a few
percent, which, unfortunately, is somewhat beyond the current experimental techniques. A more feasible purification scheme was  proposed
by Pan et al. (2001). They showed that  purification does not have to entirely rely on  CNOT operations. In some cases a simple linear optical element,  a  polarizing
beam splitter, suffices (see Fig. \ref{pur-cnot-pbs}b).

%%%%%%%%%%%%%%%%%%%%%%%%%OCT19-20:42%%%%%%%%%%%%%%%%%%%%%%%%%%%%%

The linear-optical purification scheme [shown in Fig. \ref{pur-cnot-pbs}(b)]
will be presented here using a specific example. Let our
initial state  be
\begin{equation}
\rho_{ab}=F|\phi^{+}\rangle_{ab}\langle\phi^{+}|+(1-F)|\psi^{-}\rangle
_{ab}\langle\psi^{-}|, \label{mixstate}%
\end{equation}
where $|\psi^{-}\rangle_{ab}$ is an unwanted admixture. The subscripts $a$ and $b$ indicate
the particles at Alice's and Bob's locations, respectively.

Alice and Bob share a big number of  pairs described by $\rho_{ab}$. They start by picking at random  two such  pairs. Each of them superimposes their photons on a PBS. An
essential step in the purification scheme is to select those cases for which there
is exactly one photon in each of the four spatial output modes.  We shall refer
to them as \textquotedblleft four-mode cases\textquotedblright.
This corresponds to a projection onto the subspace in which  two photons at
the same experimental location (Alice's or Bob's) have equal polarization. This is similar to the
bilateral CNOT-operation of the original scheme.
Note that the polarizations at  two different locations do not have to be
the same. After performing the purification procedure (selection of
four-mode cases, measurements in modes a4 and b4 in the $+/-$ basis,
and local operations conditional on the measurement results), Alice and Bob
will finally create a new ensemble described by the density operator
\begin{equation}
\rho_{ab}^{\prime}=F^{\prime}|\phi^{+}\rangle_{ab}\langle\phi^{+}%
|+(1-F^{\prime})|\psi^{-}\rangle_{ab}\langle\psi^{-}|,
\end{equation}
with a larger fidelity $F^{\prime}=F^{2}/[F^{2}+(1-F)^{2}]$ (for $F>1/2$) of
pairs in the desired state than before the purification.

Though it seems that only a rather special example, single bit-flip error, has been considered,
the same method actually applies to the arbitrary mixed
states $\rho_{ab}$, provided that they contain a sufficiently large fraction $F>1/2$ of photon
pairs in a maximally entangled state. This works as follows: one can first purify away single
bit-flip errors; phase errors can then be easily transformed into bit-flip errors by a $45^{\circ}$
polarization rotation and treated in a subsequent purification step.

%%%%%%%%%%%%%%%%%%%%%%%%%%OCT-19-20:54%%%%%%%%%%%%%%%%%%%%%%%%%%%%%%%%%%%%%%%%%%%%

\begin{figure}
[ptb]
\begin{center}
\includegraphics[width=0.36\textwidth]
{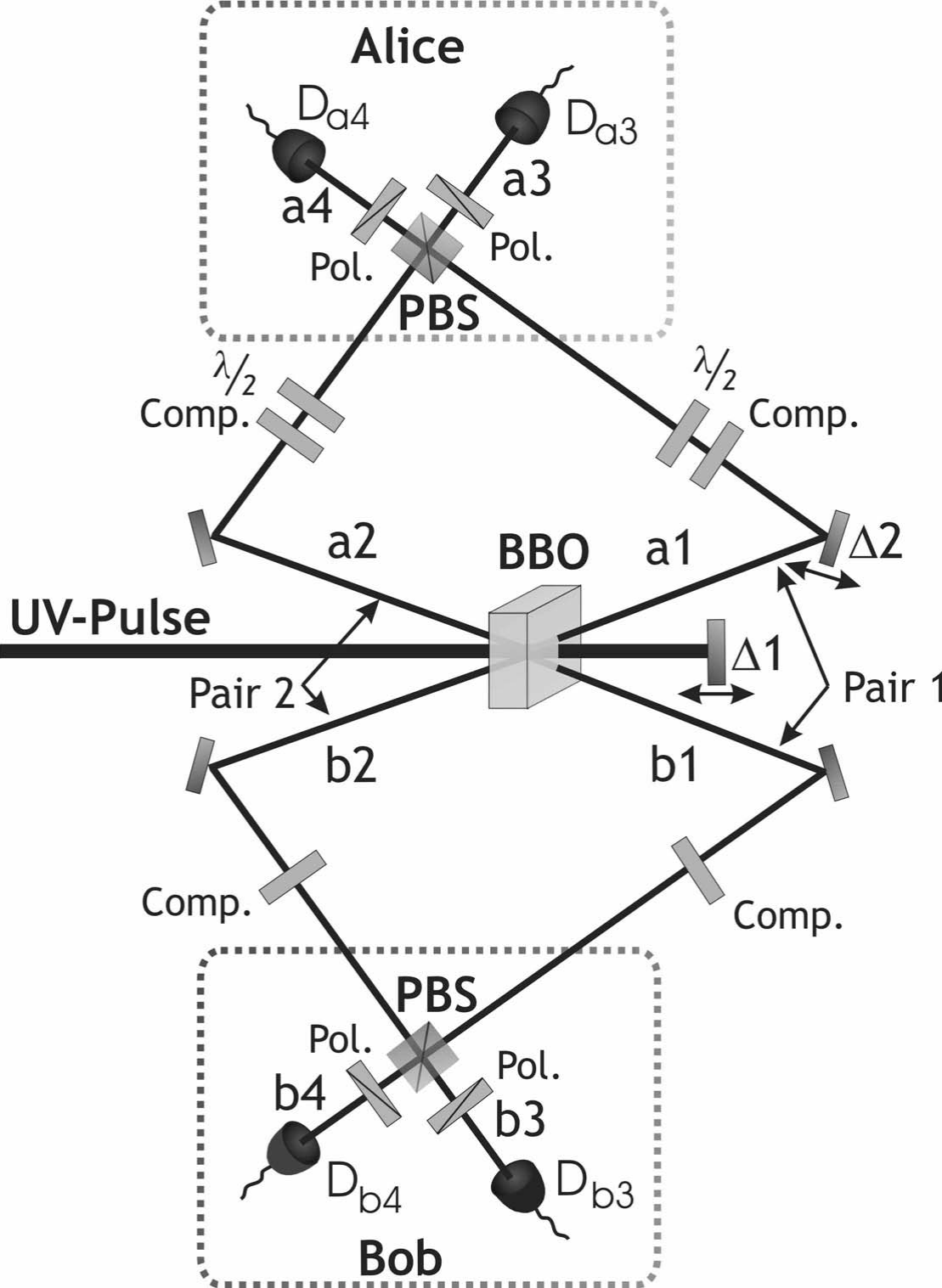}%
\caption{Experimental set-up for entanglement purification (Pan \textit{et
al}., 2003b). UV laser pulses pass through a BBO crystal twice to
produce two polarization-entangled photon pairs, i.e., a pair 1 in a1-b1 and
a pair 2 in a2-b2. Four compensators (Comp.) are used to offset the
birefringent effect caused by the BBO crystal during parametric
down-conversion. The photons in the modes b1 and b2  pass through a
half-wave plate ($\lambda/2$) to simulate a noise that reduces the
entanglement quality. Next, the two pairs are sent via local PBSs.
This results in entanglement purification. Adjusting the positions of the delay
mirrors $\Delta1$ and $\Delta2$ tunes the optical paths in such a way that the photons at local
measurement stations arrive at their PBS simultaneously. Detections of exactly one
photon in each of the four outputs (a3, a4, b3 and b4) behind a $45^{\circ}$
polarizer (Pol.) lead to a successful purification act.}%
\label{pur-setup3d}%
\end{center}
\end{figure}

An experimental demonstration of the entanglement purification scheme has been reported by Pan \textit{et al}.
(2003b). The  setup is shown in Fig.~\ref{pur-setup3d}.
For each run  two pairs of an initial, mixed state (\ref{mixstate}) were prepared with  SPDC and half-wave
plates. Next, the two photons at Alice's (Bob's) side in the mode a1-a2 (b1-b2) were  interfered
at a PBS.
After the four photons' passage through the two PBS, and under the condition  that one detects
one and only one photon polarized along the $\pm$ basis in each of the modes a4 and b4, the two
photons in the mode a3-b3, according to quantum mechanical calculations, have a higher fidelity to be in the pure entangled state.
There is, however, a complication in the actual experiment. Owing to the probabilistic nature
of SPDC, with a probability of the same order of magnitude, two photon pairs can be emitted into
a one mode pair. Fortunately, as pointed out by Simon and Pan (2002), this does not ruin
 the purification protocol. Simply, for the higher order emissions causing a four-mode detection case the
 photons in a3-b3 are projected, due to interference, to the entangled state. The scheme requires a fine stabilization of the phases
between the amplitudes of the four-mode contributions.

%EndExpansion

\begin{figure}
[ptb]
\begin{center}
\includegraphics[width=0.45\textwidth]
{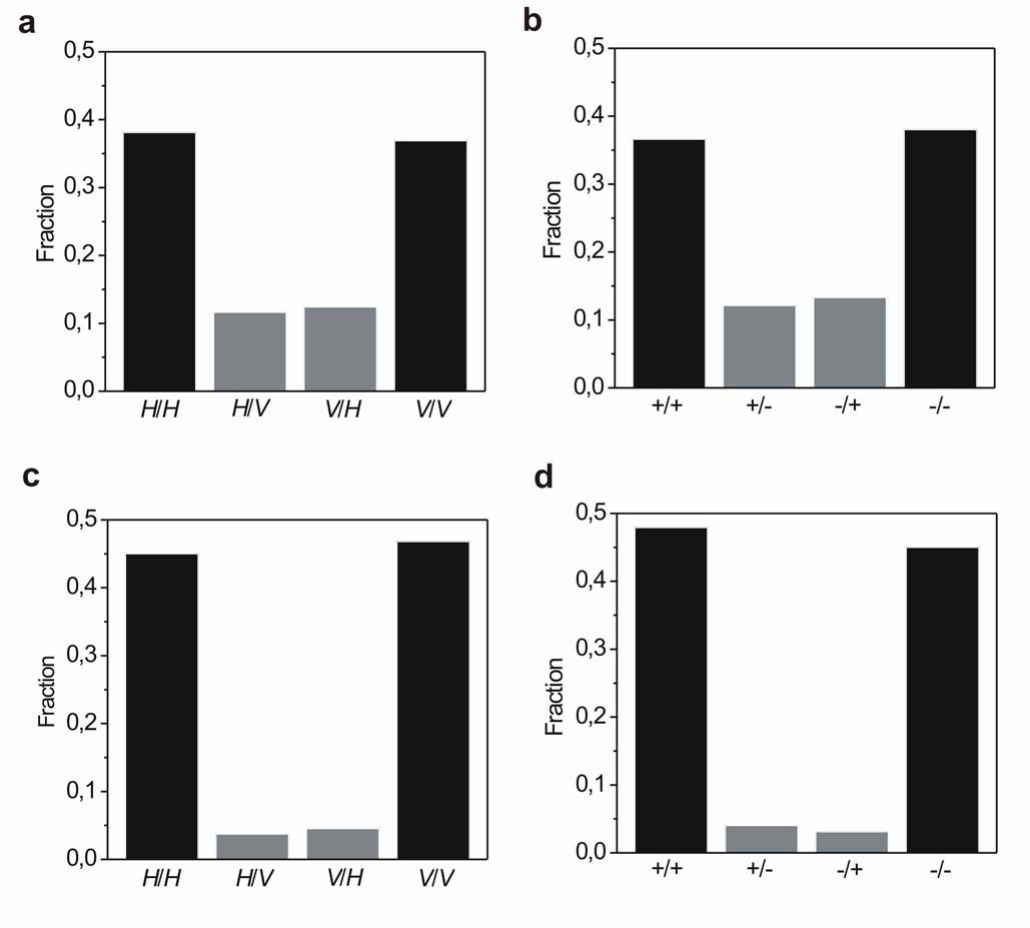}%
\caption{Experimental results (Pan \textit{et
al}., 2003b). (a) and (b) show  measured
fractions,  in the $H/V$ and in the $+/-$ bases, for the original mixed
state. (c) and (d) show measured fractions for the purified state in the
modes a3 and b3, in the same pair of bases. A comparison of the
fractions in (a) and (b), with the results shown in (c) and (d) indicates  entanglement purification.}%
\label{pur-data}%
\end{center}
\end{figure}
%EndExpansion

%Once one has perfect spatial and temporal overlap and fixes the
%relative phase $\phi_{4}$ to zero, one is ready to demonstrate
%entanglement purification.
In the first purification experiment, a mixed
state as in (\ref{mixstate}) with fidelity of $F=0.75$ was prepared. The
preparation was positively tested by measuring fractions both in the
$H/V$ and in the $+/-$ bases, as shown in Figs. \ref{pur-data}(a) and
\ref{pur-data}(b), respectively.

The measured results in Fig.\ref{pur-data}%
(c) and Fig.\ref{pur-data}(d) show a significant improvement of
entanglement fidelity to the value of $F^{\prime}=0.92\pm0.01$ for the
purified sub-ensemble. In a second experiment entanglement purification was
performed for a mixed state with $F=0.80$. After purification, the observed
entanglement fidelity for the sub-ensemble in the modes a3 and b3 was about
$0.94\pm0.01$. For each initial mixed state the purification scheme worked.
The two-photon interference visibility of the original mixed states was $50\%$ and $60\%$,
respectively. In a single purification step it was increased to $84\%$ and $88\%$. This is well above
the threshold to violate  a Bell inequality. In a subsequent development, a Bell experiment with
purified states was performed (Walther
\textit{et al}. (2005a)). The states of initially poorly entangled
photons, below thresholds to violate a Bell-CHSH inequality, were prepared by a
controllable decoherence. After a purification, $S_{\mathrm{Bell}}=2.29\pm0.13$
was measured. This violates the inequality by 2.2 standard deviations.

%%%%%%%%%%%%%%%%%%%%%%%%%%%%%%%%%%%%%%%%

If one estimates the accuracy of local operations at the used PBSs, one obtains theoretical values of fidelity better than $98\%$ for the process itself, or
equivalently an error probability of at most $2\%$. Entanglement
purification with such a high quality is important not only for
quantum communication, but also for quantum computation. With linear
optics the accuracy of single photon operations on polarization and
spatial degrees of freedom can be extremely high (a typical accuracy of
commercial products is $\sim10^{-3}$). These facts, together with the
experimental realization of high-fidelity teleportation (Pan
\textit{et al}., 2003a), imply that the threshold of tolerable
error rates for quantum repeaters could be achieved.

Although the efficiency of this entanglement purification scheme (Pan et al. 2001b) is
in theory $1/4$, the actual success probability in the experiment (Pan et al. 2003b)
was much lower as it relied on the SPDC to probabilistically create two pairs of
entangled photons, thus only a small fraction of photons actually
went through the purification system. In addition to the low efficiency, another weakness of
 this experiment is that due to double pairs emission events in SPDC,  a phase
stabilization was required. This is unfeasible in long-distance quantum communication.
Considering possible applications of entanglement purification,
these two problems can be solved by using entangled photon pairs deterministically generated using
quantum dots or other emitters.

\subsection{Long-distance entanglement distribution}

\label{sec:free-space}

The ultimate goal of quantum communication is to work
at long distances. A summary of the recent experimental progress
is listed in Table \ref{table:LDED} (The distance for the distribution of entanglement
in optical fibers was extended from 50 km (Marcikic \textit{et al}., 2004) to
the order of $\sim100$ km (Takesue \textit{et al}, 2007; H\"{u}bel \textit{et
al}, 2007; Honjo \textit{et al}, 2007)).
\begin{table}[hbt]
\caption{ \label{table:LDED}Summary of some recent experimental progresses on entanglement distribution
over long distances. S refers to the Bell-CHSH parameter. V refers to the visibility of the interference.}
\begin{tabular}{|c|l|l|l|l|}
\hline
~Year & ~Authors & ~Distance & ~S/Visibility & ~Rate~\\
\hline
\multicolumn{5}{|l|}{~\textit{free-space distribution of polarization entanglement}} \\
\hline
2003 & Aspelmeyer \textit{et al}. & 600 m & S=2.41(10) & $>$15 Hz\\
\hline
2005 & Resch \textit{et al}. &  7.8 km & S=2.27(2) & 84 Hz\\
\hline
2005 & Peng \textit{et al}.  & 13 km & S=2.45(9) & 150 Hz\\
\hline
2007 & Ursin \textit{et al}. & 144 km & S=2.508(37) & 20-40 Hz\\
\hline
2009 & Fedrizzi \textit{et al}. & 144 km & S=2.612(114) & 0.071 Hz\\
\hline
\multicolumn{5}{|l|}{~\textit{time-bin entanglement distribution via optical fiber}} \\
\hline
2004 & Marcikic \textit{et al}. &  50 km & S=2.185(12) & 5 Hz\\
\hline
2006 & Takesue \textit{et al}. &  60 km & V=75.8\% & 0.3 Hz \\
\hline
2007 & Honjo \textit{et al}. & 100 km & V=81.6\% & 1.4 Hz \\
\hline
\end{tabular}
\end{table}

For real life applications of fiber-based quantum
communication one has to face several major limiting factors including photon loss
and photon detection noise (mainly dark counts). For quantum key
distribution, the rate for dark counts etc. for a given photon detector is constant,
while the key rate decreases with increasing fiber length. Therefore, the
signal-to-noise ratio decreases exponentially with the length of the fiber. At
a certain fiber length the signal-to-noise ratio is so low that secure keys cannot be generated. A further extension of the distance over which reliable
quantum communication is possible requires detectors of lower noise, fiber
links of lower loss, quantum communication systems of faster working
rates and so on. In summary, the present-day technology puts a strong limitation
on the distance for practical fiber-based quantum communication\footnote{In the context of quantum cryptography, recent
revolutionary progress has been achieved by introducing the idea of decoy
states (Hwang, 2003; Lo, 2004; Wang, 2005).
The decoy state scheme, which is designed such that Alice randomly sends some of her laser pulses
with a lower average photon number, can be used to detect a photon-number-splitting
attack, as Eve has no way to tell which pulses are signal and which decoy.
Thus using classically attenuated laser pulses, one can extend the secure
quantum communication distance from $\sim$30 km, as in the conventional scheme (Waks \textit{et al}., 2002), to $\sim$
100 km with the decoy-state protocol, and still gets higher key generation rates.
Such a scheme was experimentally realized via
optical fiber (Rosenberg \textit{et al}., 2007; Peng \textit{et al}., 2007) and
via free-space links (Schmitt-Manderbach \textit{et al}., 2007).
The decoy-state protocol allows the same security level as in the case
of  true single photon sources. Taking the advantage of ultra low loss fibers
and low-noise superconducting detectors Korneev \textit{et al}., 2007 and
Marsili \textit{et al}., 2008 created a prototype of quantum key distribution working at a distance of
250km.}.  This underlines the necessity of developing  quantum
repeaters.

A promising way to realize long-distance quantum communication is to
exploit satellite-based free-space distribution of single photons, or entangled
photon pairs (Aspelmeyer \textit{et al}., 2003b).
In the scheme, the photonic
quantum states are sent from Earth's surface and reflected from one
satellite to another, and finally sent back to the Earth. Since the effective
thickness of the atmosphere is on the order of 5-10 km, while the outer space
 photon loss and decoherence is negligible, with the help of satellites one
can achieve global free-space quantum communication, provided the quantum
states  survive the passage through the aerosphere.

Along these lines, an important experimental progress has been made in the
free-space distribution of attenuated laser pulses [over 23.4 km, see
(Kurtsiefer \textit{et al}., 2002); over 144 km, see (Schmitt-Manderbach
\textit{et al}., 2007)] and of entangled photon pairs [over 600 m, see
(Aspelmeyer \textit{et al}., 2003a); over 7.8 km, see (Resch \textit{et al}.,
2005); over 13 km, see (Peng \textit{et al}., 2005); and over 144 km, see (Ursin
\textit{et al}., 2007; Fedrizzi \textit{et al}., 2009), see table \ref{table:LDED} for a summary].

%The free-space
%entanglement distribution experiment performed in Hefei (Peng \textit{et al}.,
%2005) confirmed that entanglement can survive even for  photons which  pass, through a noisy ground air,
%distances which are well beyond the effective thickness of the atmosphere.
%The link efficiency of entangled photon pairs achieved in the Hefei experiment is about
%a few percent, which is well beyond the threshold required for satellite-based
%free-space quantum communication.

More recently, a 144-km free-space link was built between two Canary Islands
and used for distribution of one photon of an entangled pair (Ursin \textit{et al}., 2007),
and later, both photons (Fedrizzi \textit{et al}., 2009).
The final
photon states were found to preserve excellent, noise-limited fidelity, even
though they experienced extreme attenuation due to mainly turbulent atmospheric
effects.
The total channel loss of 64 dB corresponded to the estimated
attenuation regime for a two-photon satellite quantum communication scenario.
The entanglement of the received two-photon states was confirmed by violating
the CHSH inequality by more than 5 standard deviations. From a fundamental
point of view, this means that the photons are subject to virtually no
decoherence during their 0.5 ms long flight through air.
For those aiming at a  world-wide quantum communication
this is an encouraging development.
 The photon-pair flight
time of $\sim0.5$ ms represents the longest lifetime of photonic Bell states reported
so far.

\subsection{Quantum memory and quantum repeaters}

\label{sec:repeater}

Above we have shown that entanglement purification enables one to overcome the degradation of the quality of photon entanglement.
Still, a major drawback of schemes for communication between distant
nodes is the exponential scaling of the error probability with the length
of the connecting channels.
 The quantum repeater protocol (Briegel \textit{et al}., 1998;
D\"{u}r \textit{et al}., 1999) provides a blueprint of a general framework to remedy this problem by nesting entanglement purifiaction and swapping steps.
Once constructed it would enable one to establish high-quality long-distance entanglement with
resources increasing only polynomially with transmission distance.

Further, a quantum memory for single photons, with the ability of interconverting between
stationary and flying qubits (see section~\ref{sec:memory}) is a crucial element in the
quantum repeater scheme. There are several candidates for localized qubits.
For instance, one may use atomic internal states to store local information.
Mapping between the atomic and photonic qubits requires a strong coupling
between atoms and photons via high-finesse cavities (Raimond \textit{et al}., 2001;
Leibfried \textit{et al}., 2003; Walther \textit{et al}., 2006) or initial atom-photon entanglement together with entanglement swapping. Below we will focus on the
atomic-ensemble based schemes [Duan \textit{et al}., 2001, 2002;
Chen \textit{et al}., 2007; Jiang \textit{et al}., 2007; Zhao \textit{et al}.,
2007; see also (Sangouard \textit{et al}., 2009) for a review].

We emphasize that quantum memories  have applications not only in long-distance quantum communication,
but they also provide a route to a more efficient multi-photon entanglement (see Sec.\ref{sec:multi}) or
linear optics quantum computing (see Sec.\ref{computing}). So far, the majority of the reported multiphoton
interferometry experiments face the problem of  random  arrivals of
 SPDC photon pairs. Thus, scalability of this  approach is questionable.
Given a quantum repeater, ideally with long storage time, high writing and retrieval
efficiencies\footnote{Extensive efforts still need to be undertaken to make a quantum memory
usable for this purpose, see Sec.\ref{sec:memory}.}, the randomly generated SPDC photon pairs can be stored
and synchronized with the arrival of other photon pairs. This
would, for instance, enable efficient generation of multiphoton
states  in a time which
increase only polynomially with number of involved qubits.

%%%%%%%%%%%%%%%%%%%%%%%%%%%%

\subsubsection{Quantum repeater protocol}

\label{sec:qr}

In classical communication, the problem of exponential attenuation can be
overcome by using repeaters at certain points in the channel. They amplify
the signal and restore it to its original shape. In analogy to
fault-tolerant quantum computing (Nielson and Chuang, 2000; Preskill, 1998), the quantum repeater proposal (Briegel \textit{et
al}., 1998; D\"{u}r \textit{et al}., 1999) is a
cascaded entanglement purification protocol for communication systems.

\begin{figure}[tb]
\begin{center}
\includegraphics[width=0.45\textwidth]
{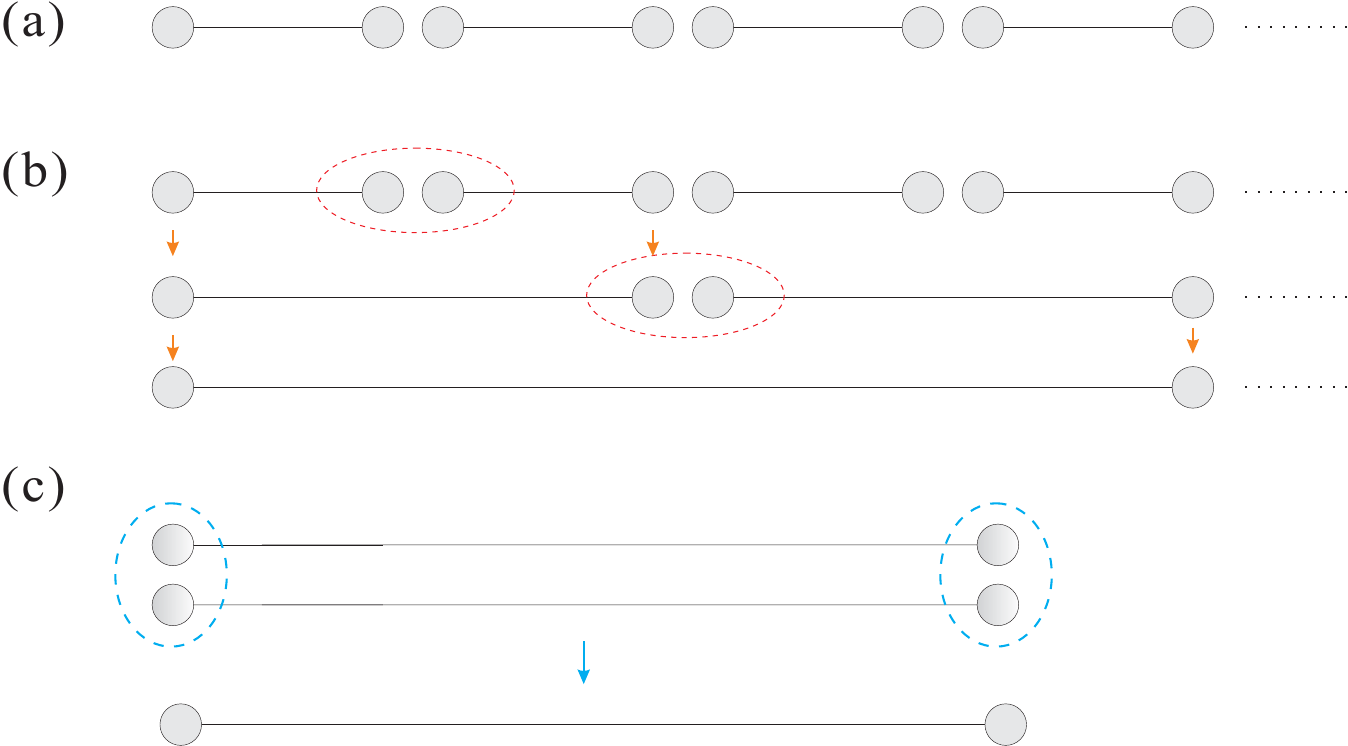}%
\caption{Quantum repeater scheme of Briegel \textit{et al}.,
(1998). (a) Creation of a sequence of entangled pairs. (b)
nested purification protocol which combines the methods of entanglement
swapping and purification, assisted with repeated creation of auxiliary pairs.
(c) Purification of entangled pairs stored in distant locations.}%
\label{coding}%
\end{center}
\end{figure}

The quantum repeater protocol comprises  three elements:

\begin{enumerate}
\item A method for creation of entanglement between particles at distant
nodes, which uses auxiliary particles at intermediate \textquotedblleft
connection points\textquotedblright\ and a \textit{nested purification
protocol}.

\item Entanglement purification, even with imperfect means.

\item A protocol for which the time needed for entanglement creation scales
polynomially, whereas the required material resources per connection point grow
only logarithmically with the distance.
\end{enumerate}

Exemplarily, here we will describe a scheme for the physical
realizations of a quantum repeater which has been proposed by
Duan, Lukin, Cirac and Zoller\footnote{Other
physical implementations include the quantum repeater based on solid-state
photon emitters (Childress \textit{et al}., 2005, 2006) and a hybrid quantum
repeater using bright coherent light and electronic- and nuclear-spins (van
Loock \textit{et al}., 2006).} (DLCZ, 2001, 2002).
They suggested atomic ensembles as local memory
qubits. They  have a collectively enhanced coupling to light, even without the
aid of high-finesse cavities. This  scheme incorporates
entanglement swapping, built-in entanglement purification and quantum memory.%

\begin{figure}
[ptb]
\begin{center}
\includegraphics[width=0.35\textwidth]%
{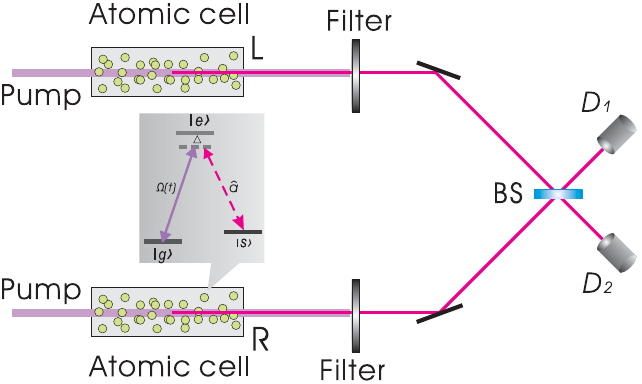}%
\caption{Schematic of a setup for generating entanglement between the two atomic
ensembles L and R in the DLCZ scheme (Duan \textit{et al}., 2001). The inset
shows the relevant level structure of the atoms in the ensemble with
$\left\vert g\right\rangle $, the ground state, $\left\vert s\right\rangle ,$
the metastable state for storing a qubit, and $\left\vert e\right\rangle ,$
the excited state. The transition $\left\vert g\right\rangle \rightarrow
\left\vert e\right\rangle $ is coupled by a classical laser light, with the Rabi
frequency $\Omega$. The forward scattered Stokes light comes from the
transition $\left\vert e\right\rangle \rightarrow\left\vert s\right\rangle $.
An off-resonant coupling with a large detuning $\Delta$ is
assumed.}%
\label{dlczfig}%
\end{center}
\end{figure}

Figure~\ref{dlczfig} is a schematic of a setup for entangling  two atomic
ensembles (optically-thick atomic cells of $N_{a}$ identical atoms) L and R
which are spatially separated within the channel attenuation length. A pair of
metastable lower states $|g\rangle$ and $|s\rangle$ can correspond to
hyperfine or Zeeman sublevels of electronic ground states of alkali atoms. Long
lifetimes for relevant coherences in such systems have been observed both in a room-temperature
dilute atomic gas (Phillips \textit{et al}., 2001), and in a sample of cold
trapped atoms (Liu \textit{et al}., 2001).
%EndExpansion

All the atoms are initially prepared in the ground states $\left\vert
g\right\rangle_i $. A sample is illuminated by a short, off-resonant laser pulse
that induces Raman transitions into  states $\left\vert s\right\rangle_i $.
Particularly important is  the forward-scattered Stokes light (the
signal mode $\hat{a}$). It is uniquely correlated with the excitation of
the symmetric collective atomic mode $\hat{S}\equiv\left(  1/\sqrt{N_{a}%
}\right)  \sum_{i}\left\vert g\right\rangle{_{i}} {_{i}}\left\langle s\right\vert $,
where the summation is taken over all the atoms. The light-atom interaction
generates, after the interaction time $t_{\Delta}$, a two-mode ($\hat{a}$ and
$\hat{S}$) squeezed state
%\footnote{That is, a state $exp(r_c\hat{a^+}\hat{S^+})|\Omega\rangle$ where
%$|\Omega\rangle$ is the vacuum (which for atoms means all of them are in the ground state)
% and $r_c$ is the strength of squeezing parameter.}
 (Braunstein and van Loock, 2005), with the squeezing
parameter $r_{c}$ proportional $t_{\Delta}$. If $t_{\Delta}$ is very small,
the two-mode squeezed state can be written in the perturbative form
\begin{equation}
\left\vert \zeta\right\rangle =\left\vert 0_{a}\right\rangle \left\vert
0_{p}\right\rangle +\sqrt{p_{c}}\hat{S}^{\dagger}\hat{a}^{\dagger}\left\vert
0_{a}\right\rangle \left\vert 0_{p}\right\rangle +O\left(  p_{c}\right)  ,
\label{atomph}%
\end{equation}
where $p_{c}=\tanh^{2}r_{c}\ll1$ is the small excitation probability and
$O\left(  p_{c}\right)  $ represents the terms with more excitations, whose
probabilities are equal or smaller than $p_{c}^{2}$. The Hilbert space vectors $\left\vert
0_{a}\right\rangle $ and $\left\vert 0_{p}\right\rangle $ are, respectively,
the atomic and photonic vacuum states with $\left\vert 0_{a}\right\rangle
\equiv\bigotimes_{i}\left\vert g\right\rangle _{i}$. For a large $N_{a}$, the
collectively enhanced signal-to-noise ratio may strongly  boost the
efficiency of the scheme.

This setup enables one to generate entanglement between two distant
ensembles L and R using the configuration shown in Fig.~\ref{dlczfig}. If
two laser pulses excite both ensembles simultaneously, the whole system
is described by the state $\left\vert \zeta\right\rangle _{L}\otimes\left\vert
\zeta\right\rangle _{R}$, where $\left\vert \zeta\right\rangle _{L}$ and
$\left\vert \zeta\right\rangle _{R}$ are given by Eq.~(\ref{atomph}). The subscripts $L$ and $R$ denote the respective cells (in (\ref{atomph}) one should add such subscripts to all mathematical objects). The forward
scattered Stokes signal from both ensembles is combined at the BS and a
photodetector click in either D1 \textit{or} D2 measures the combined
radiation from two samples, $\hat{a}_{+}^{\dagger}\hat{a}_{+}$ or $\hat{a}%
_{-}^{\dagger}\hat{a}_{-}$ with $\hat{a}_{\pm}=\left(  \hat{a}_{L}\pm
e^{i\varphi}\hat{a}_{R}\right)  /\sqrt{2}$. The symbol $\varphi$ denotes an unknown
difference of the phase shifts in the two channels. Depending  on which
detector clicks, one applies $\hat{a}_{+}$ or $\hat{a}_{-}$ to the whole
state $\left\vert \zeta\right\rangle _{L}\otimes\left\vert \zeta\right\rangle
_{R}$. The resulting projected state of the ensembles L and R is nearly maximally
entangled. It reads (we neglect $O(p_{c})$ terms)
\begin{equation}
\left\vert \Psi_{\varphi}\right\rangle _{LR}^{\pm}=\left(  \hat{S}%
_{L}^{\dagger}\pm e^{i\varphi}\hat{S}_{R}^{\dagger}\right)  /\sqrt
{2}\left\vert 0_{a}\right\rangle _{L}\left\vert 0_{a}\right\rangle _{R}.
\label{dlczent}%
\end{equation}
For each round the probability for getting a click is given by $p_{c}$. thus,  we
need to repeat the process about $1/p_{c}$ times to warrant a successful preparation of entanglement. The average preparation time is given by $T_{0}\sim
t_{\Delta}/p_{c}$.

The entanglement generation (as well as entanglement connection) in the DLCZ
scheme is based on single-photon interference at
photodetectors\footnote{Such a method was first proposed to entangle single atoms
(Cabrillo \textit{et al}., 1999; Bose \textit{et al}., 1999).}, which requires a stable
long-distance interferometric stability. The fluctuations of the relative phase $\varphi$
caused by the environment would wash out the coherence (i.e., entanglement) in
Eq.~(\ref{dlczent}).
For instance, to maintain path length phase stability
at the level of $\lambda/10$ ($\lambda$: wavelength) for single photons,
typically of $\lambda\sim1$ $\mu$m, generated from atomic ensembles (Eisaman
\textit{et al.}, 2005) requires a precise control of timing jitter at a
sub-femtosecond level, which is almost  experimentally impossible
(Holman \textit{et al}., 2005). For more detailed analysis on phase-stability
problem of the DLCZ scheme, we refer to (Chen \textit{et al}., 2007).

The phase-stability problem can be overcome by interfering two
photons, one coming from each remote ion or atom in a cavity (Bose \textit{et
al}., 1999; Browne \textit{et al}., 2003; Feng \textit{et al}., 2003; Simon
and Irvine, 2003), which was experimentally implemented by Moehring \textit{et al}.,
(2007) and Maunz \textit{et al}., (2007). A robust implementation of a quantum repeater using
atomic ensembles was proposed by Chen \textit{et al}., (2007; Jiang \textit{et al}.,
(2007); and Zhao \textit{et al}., (2007). With the help of two-photon interference it
eliminates the stringent requirement of long-distance phase stabilization.

Though the DLCZ scheme does not meet all the criteria for long-distance quantum
communication, it provides a promising approach to a fully controllable
single-photon source based on atomic ensembles, which seems to be much easier
for experimental demonstrations. Let us summarize the basic ideas behind it.
The atomic ensemble generates a correlated state in Eq.~(\ref{atomph}), which
is an exact analog of the SPDC radiation. By measuring the forward signal mode with a
single-photon detector, under the condition that the detector clicks, the collective
atomic mode is projected to a single-excitation state. Such excitations can be
stored for a reasonably long time in metastable states
(the so-called ground-state manifold) of the atoms.
On demand the single-atomic excitation can be transferred to a single photon
(still within the storage time) with a method described
in the next section. This is with fully controllable properties: the emitted
single-photon pulse is directed  forward; the emission time is
controllable by the repumping time; and the pulse shape is controllable by
varying the time dependence of the Rabi frequency of the repumping pulse.

So far, significant advances have been achieved along these lines.
For a \textit{partial} list,\footnote{For a comprehensive review,
see (Sangouard \textit{et al}., 2009).} let us mention the
following: controllable generation, storage and retrieval of single
photons with tunable frequency, timing and bandwidth (Chou
\textit{et al}., 2004; Eisaman \textit{et al.}, 2004, 2005;
Chaneli\`{e}re \textit{et al.}, 2005); a deterministic single-photon
source using measurement-based feedback protocol (Matsukevich
\textit{et al}., 2006b; Laurat \textit{et al}., 2006; Chen
\textit{et al}., 2006b); conditional control of two atomic memories
(Felinto \textit{et al}., 2006); entanglement of two atomic
ensembles (Metsukevich \textit{et al}., 2006a) and its distribution
between two quantum nodes located 3 meters apart (Chou \textit{et
al}., 2007); mapping photonic entanglement into and out of an
atomic-ensemble-based quantum memory (Choi \textit{et al}., 2008);
optimal control of light pulse storage and retrieval (Novikova
\textit{et al}., 2007); the Hong-Ou-Mandel interference of photon
pairs from two independent ensembles (Chaneli\'{e}re \textit{et
al}., 2007). A quantum repeater node following the robust protocol
(Chen \textit{et al}., 2007; Zhao \textit{et al}., 2007) was
experimentally demonstrated by Yuan \textit{et al}. (2008).
These experiments
are currently limited by the relatively short coherence time ($\sim20~\mu$s ) of the memory qubits and the low conversion efficiency
($\sim15$\%) between photonic and atomic states. We refer to
Kimble (2008) for a more in-depth review on this topic.

\subsubsection{Quantum state transfer between matter and photons}

\label{sec:memory}

The technique of quantum state transfer between matter and photons
is indispensable for both long-distance quantum communication and
large-scale optical quantum computing (see section~\ref{computing}).
In such applications the matter itself should be endowed with a long
storage time. This makes atoms strong candidates for localized
photonic information carriers. The early proposals (Cirac
\textit{et al}., 1997; van Enk, Cirac, and Zoller, 1997) along these
lines use the strong coupling of photons and single atoms in
high-finesse cavities.

The basic idea of quantum light memory is in transferring a photonic
state to the excitations of atomic internal states. In such a way it
can be stored. After some controllable time, it should be possible
to transfer back the excitations to photons restoring the original
quantum state. The experimentally challenging technology at the
interface of photons and single atoms motivated search for
alternative routes to matter-light quantum interfaces. Along this
line, theoretical ideas on quantum light memory have been proposed
(Kozhekin \textit{et al}., 2000; Lukin \textit{et al}., 2000;
Fleischhauer and Lukin, 2000, 2002; Duan \textit{et al}., 2001;
Duan, Cirac, and Zoller, 2002; Chen \textit{et al}., 2007), and the
relevant experimental advances (Kash \textit{et al}., 1999; Phillips
\textit{et al}., 2001; Liu \textit{et al}., 2001; Schori \textit{et
al.}, 2002; Julsgaard \textit{et al}., 2004; van der Wal \textit{et
al}., 2003; Bajcsy \textit{et al}., 2003; Matsukevich and Kuzmich;
2004; H\`{e}tet \textit{et al}., 2008) have been
reported.\footnote{For comprehensive reviews, see (Sangouard
\textit{et al}., 2009), and (Hammerer, S\o rensen, and Polzik,
2008).}

\begin{figure}
[ptb]
\begin{center}
\includegraphics[width=0.38\textwidth]%
{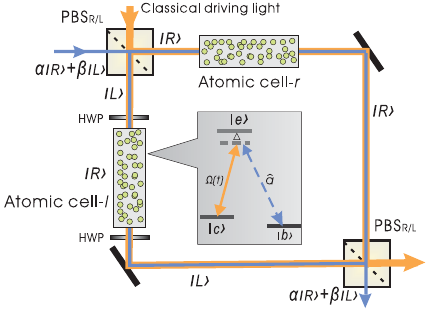}%
\caption{Quantum memory for photonic polarization qubits. Two
identical ensembles are identically driven by a classical field,
which is equally right- and left-circularly polarized. Classical and
quantized light fields are fed into the first PBS$_{R/L}$
(\textquotedblleft rotated\textquotedblright\ PBS which reflects
left-circular photons and transmits right-circular photons) and
leave at two different outputs of the second PBS$_{R/L}$. Two
half-wave plates (HWP), enabling the transformation $\left\vert
R\right\rangle \leftrightarrow \left\vert L\right\rangle $, are
placed along the $\left\vert L\right\rangle $-output of the first
PBS$_{R/L}$. As the atomic cell-$r$ (cell-$l$) works as a quantum
memory for single photons with right-circular (left-circular)
polarization, via the adiabatic transfer method, the whole setup is
therefore a quantum memory of any single-photon polarization states.
The inset shows the relevant level structure of the atoms, with the
ground state $\left\vert b\right\rangle $, the storage metastable
state $\left\vert c\right\rangle $, and the excited state
$\left\vert e\right\rangle $ (Chen \textit{et
al}., 2007).}%
\label{polmemory}%
\end{center}
\end{figure}

The atomic-ensemble-based quantum memory consists of a coherently
driven atomic ensemble ($N\gg1$ atoms) of large optical thickness
with a level structure shown in the inset of Fig.~\ref{polmemory}.
The $\left\vert c\right\rangle $-$\left\vert e\right\rangle $
transition\ is coherently driven by a classical field of Rabi
frequency $\Omega(t)$, and the $\left\vert b\right\rangle
$-$\left\vert e\right\rangle $ transition is coupled to a quantized
single-mode (the multimode case is similar) light field (described
by an annihilation operator $\hat{a}$). The coupling constant is
denoted by $g$. Under the two-photon resonance (i.e., the two
detunings for the two transitions shown in the inset of
Fig.~\ref{polmemory} are both equal to $\Delta$), the classical
driving field can induce transparency for the quantized light field
and a substantial group-velocity reduction, and even the complete
stopping of the light (for reviews, see Lukin and Imamo\v{g}lu,
2001; Lukin, 2003; Fleischhauer \textit{et al}., 2005). The
Hamiltonian of the whole system ($N$ atoms plus the quantized light
field), in a frame rotating at the optical frequency, reads
$H=\hbar\Omega(t)\hat
{S}_{ec}+\hbar g\sqrt{N}\hat{a}\hat{S}_{eb}+H.c.$, where $\hat{S}_{ec}%
=\sum_{i}^{N}\left\vert e\right\rangle _{ii}\left\langle
c\right\vert $,
$\hat{S}_{eb}=\frac{1}{\sqrt{N}}\sum_{i}^{N}\left\vert
e\right\rangle _{ii}\left\langle b\right\vert, $ and $H.c.$ denotes
Hermitian conjugate of the previous expression. This Hamiltonian has
a its zero-energy eigenstates, the so-called \textquotedblleft dark
states\textquotedblright. When the atom number is much larger than
the photon number, the dark states represent elementary excitations
of bosonic quasiparticles, i.e., the dark-state polaritons. For more
details on this concept, see (Lukin \textit{et al}., 2000;
Fleischhauer
and Lukin, 2000, 2002)%

%EndExpansion

By adiabatically changing $\Omega(t)$ between the two limiting cases
($\Omega(t)\ll g\sqrt{N}$, or $\Omega(t)\gg g\sqrt{N}$) one can
coherently map dark-state polariton states onto either purely
atom-like states where the photons are stored, or purely photon-like
states, which corresponds to the release of the stored photons. In
principle, a quantum memory based on the adiabatic transfer method
is reversible, preserves pulse shape of the stored photons
(Fleischhauer and Lukin, 2000; 2002) and may have an efficiency very
close to unity. As there is no excited atomic state in the
dark-state subspace, the storage time can be very long.

The original quantum memory was proposed for storing a coherent
superposition of photon-number states. However, two atomic ensembles
can be entangled by storage of two entangled light fields (Lukin
\textit{et al}., 2000). Recent experiments achieved single quantum
excitation memory times of 1 ms using cold atomic ensembles (Zhao
\textit{et al}., 2009a), 6 ms using atomic rubidium confined in a
one-dimensional optical lattice (Zhao \textit{et al}., 2009b), and
0.1s using quantum memory confined in an optical lattice with laser
compensation of the lattice light shifts (Radnaev et al. 2010).

\section{Photonic quantum computing}

\label{computing}

As we have seen in the above section, the photon, thanks to its high
transmittance through air and glass fibers and its extremely long
decoherence time, has arguably been the best
candidate for quantum communication. However, things
become trickier when we come to the field of
quantum computation. The weak interaction between photons, which is
of  a significant benefit in quantum communication, turns to be a drawback
where non-trivial two-qubit quantum gates are essential. For a long time
it seemed  obvious that linear optical two-photon
gates can be done only in a non-deterministic fashion and thus quantum
computing cannot be scalable. However,  in 2001
Knill, Laflamme and Milburn (KLM) proved that scalable optical quantum
computing is possible using only single-photon sources, linear optical
elements, and photon-number resolving detectors. The KLM scheme subsequently
spurred new experiments demonstrating probabilistic controlled
two-photon gates. Despite of KLM's effort, the resource overhead
required for optical quantum computing is absolutely daunting.
Several improvements of this protocol, particularly those based on cluster
states or error encoding, have dramatically reduced this worrying resource
overhead, and started to bridge the gap between the theoretical scalability and
practical implementations.

We have witnessed considerable theoretical and experimental progress
in optical quantum computing in these years. This topic has  been
reviewed earlier by Kok \textit{et al}. (2007), O'Brien (2007),
O'Brien \textit{et al}. (2009), and Ralph \textit{et al}. (2009).
This section serves as  a supplement to these previous
reviews. Thus, we will skip some theoretical details and
mainly focus on recent experimental advances.

\subsection{Linear-optical two-qubit logic gates}

\label{sec:LO-gate}

Knill \textit{et al}. (2001) showed that the success rate of
the logic gates can be arbitrarily close to one by using more
ancilla photons and detectors. A similar conclusion has been independently
obtained by Koashi \textit{et al}., (2001) using entangled ancilla photons.
A novel aspect of this protocol is that, despite the lack of the
photon-photon interaction, quantum measurements with photon number
resolving detectors can induce effective
nonlinearity sufficient for the realization of two-qubit gates.
The original KLM scheme was only very recently implemented in a sophisticated setup up
using polarization encoding, and Sagnac-interferometers for increased stability, by Okamoto
\textit{et al}., (2010).

Further improvements reduced the complexity and
improved the efficiency of the original scheme by introducing certain assumptions
and restrictions, enabling a series of experiments and  demonstrations.
Hofmann and Takeuchi, (2002) and independently Ralph \textit{et al}., (2002a, 2002b) developed quantum gates under the restriction of what is here called a two, or four-mode case, that is the successful operation of the gate can be verified if the two photons involved are detected in certain outputs (this is also called "detection in coincidence basis" or "conditioned detection").  Essentially a single two-photon interference is enough, together with a state-dependent filtering, to perform probabilistic CNOT-operations. The restriction does not allow further operations on the two photons involved and thus limits the depth of calculations, however, the simplicity of the gate makes it a very useful and reliable tool if no further joint operations on the two photons are required. The original proposal used dual-rail encoding and was first implemented by O'Brien \textit{et al}., (2003). An even simpler set-up becomes possible with polarization encoding (Kiesel \textit{et al}., 2005b; Langford \textit{et al}., 2005; Okamoto \textit{et al}., 2005) which in turn could be already applied, e.g., to observe cluster states for one-way quantum computing (see below,  Kiesel \textit{et al}., 2005a). More recently, Politi \textit{et al}. (2008) reported a high-fidelity silica-on-silicon integrated optical
realizations of key quantum photonic circuits.  Laing \textit{et al}. (2010) reported a two-photon quantum
interference visibility of $99.5(4)\%$, a {\small CNOT} gate (the obtained
average fidelity of logical basis was $96.9(2)\%$) and a path-entangled
two-photon state (with fidelity of $>92\%$).  Crespi \textit{et al}. (2011) reported the first probabilistic logic gates on integrated circuits also for polarization qubits.

In a separate development, Koashi \textit{et al}., 2001, and Pittman \textit{et al}., 2001, showed that by using entangled pairs of photons as ancilla, the success of the gate operation can be inferred by the detection of photons in ancilla-outputs. This enables one to perform the gate operation in a nondestructive manner. Such gates, assisted by entangled or unentangled photon pairs, were
reported using four or five photons, see Gasparoni \textit{et al}., (2004), Zhao \textit{et al}., (2005b), Bao \textit{et al}., (2007), Tokunaga \textit{et al}., (2008)
and Gao \textit{et al}., (2010b).

\begin{figure}
[bt]
\begin{center}
\includegraphics[width=0.42\textwidth
]%
{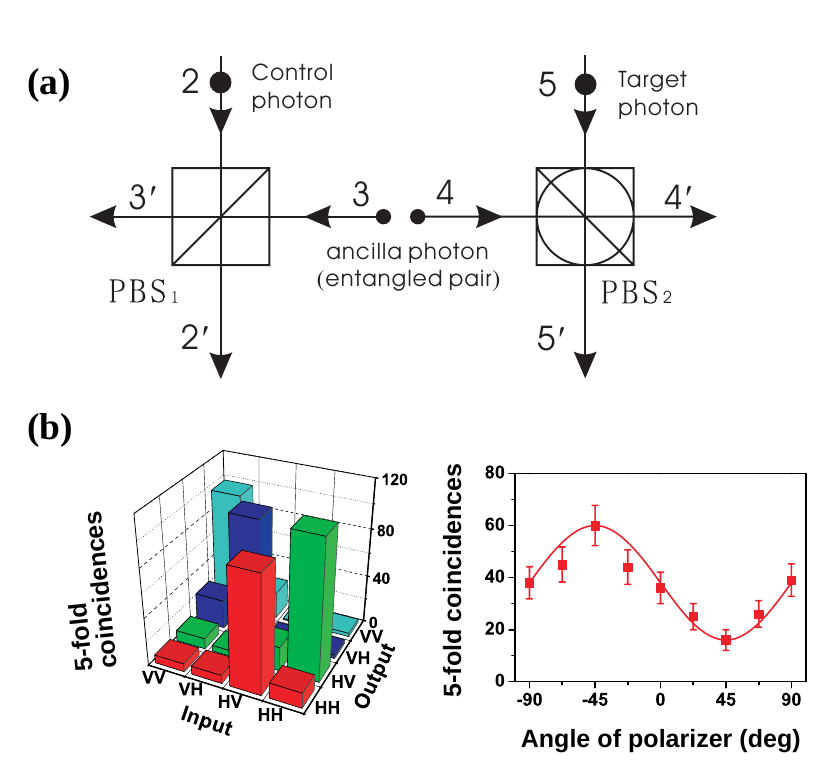}%
\caption{(a) A nondestructive {\small CNOT} gate constituting of a PBS, half-wave
plates (HWP) and using an ancilla entangled photon pair $|\psi^{-}\rangle_{34}$
(Pittman \textit{et al}., 2001). (b) Experimental results of two-photon CNOT gate (Zhao et al. 2005).}%
\label{zhao-cnot}%
\end{center}
\end{figure}

In this section, we present, for pedagogical purposes, the working
principle and a proof-of-principle demonstration (Zhao
\textit{et al}., 2005b) of a nondestructive {\small CNOT} gate for two
independent photons. As suggested by Pittman \textit{et al}., (2001),
the success of the gate can be detected by measurement of  ancilla
photons and this information is classically feed-forwardable. This is
an important feature for both circuit-model, and one-way model of
scalable optical quantum computing.

Suppose that one aims to perform a {\small CNOT} gate on an arbitrary two-qubit
state $\left\vert \chi\right\rangle _{25}$ [Eq.~(\ref{tq-tele})]  using
an ancilla entangled photon pair in the Bell state $|\psi^{-}\rangle_{34}$.
Note that PBS$_{2}$ in Fig.~\ref{zhao-cnot}(a) transmits $|H+V\rangle$
while reflects $|H-V\rangle$ polarization. The output state of the
whole apparatus is (Pittman \textit{et al}., 2001)
\begin{align}
\left\vert \chi\right\rangle _{25}|\psi^{-}\rangle_{34}  &  \rightarrow
\frac{1}{4}[|V^{\prime}\rangle_{3^{\prime}}|V\rangle_{4^{\prime}%
}(\mathrm{CNOT}_{2^{\prime}5^{\prime}}\left\vert \chi\right\rangle
_{2^{\prime}5^{\prime}})\nonumber\\
&  +|H^{\prime}\rangle_{3^{\prime}}|V\rangle_{4^{\prime}}(\hat{z}_{5^{\prime}%
}\mathrm{CNOT}_{2^{\prime}5^{\prime}}\left\vert \chi\right\rangle _{2^{\prime
}5^{\prime}})\nonumber\\
&  +|H^{\prime}\rangle_{3^{\prime}}|H\rangle_{4^{\prime}}(\hat{x}_{5^{\prime}%
}\hat{z}_{5^{\prime}}\mathrm{CNOT}_{2^{\prime}5^{\prime}}\left\vert
\chi\right\rangle _{2^{\prime}5^{\prime}})\nonumber\\
&  +|V^{\prime}\rangle_{3^{\prime}}|H\rangle_{4^{\prime}}(\hat{x}_{5^{\prime}%
}\mathrm{CNOT}_{2^{\prime}5^{\prime}}\left\vert \chi\right\rangle _{2^{\prime
}5^{\prime}})]\nonumber\\
&  +\frac{\sqrt{3}}{2}|...\rangle_{\mathrm{not}\text{\textrm{ four-mode cases}%
}}\text{ .} \label{lo-cnot}%
\end{align}
Consider a detection of a pair of photons at the output modes $3^{\prime}$ and $4^{\prime}$, only one photon
at each output (the four mode case). Depending on the registered polarizations, up to a specific unitary transformation, a non-destructive {\small CNOT} gate operation is then performed on $\left\vert
\chi\right\rangle _{2^{\prime}5^{\prime}}$.

%%%%%%%%%%%%%%%%%%%%%%%%%%%%%%%%%%%%OCT-06-9:35am%%%%%%%%%%%%%%%%%%%%%%%%
\subsection{Cluster-state quantum computing}

\label{sec:oneway}

Another significant step is the discovery of \textquotedblleft cluster-state quantum
computing\textquotedblright\ (Raussendorf and Briegel, 2001; Raussendorf
\textit{et al}., 2003; Briegel \textit{et al}., 2009), which is based on the
preparation of highly entangled multi-qubit states, the so-called
\textquotedblleft cluster states\textquotedblright\ (Briegel and Raussendorf,
2001) and adaptive one-qubit measurements. Besides its
thought-provoking theoretical structure, this model also brings a number of practical
advantages for physical realization of quantum computation.
In scenarios in which quantum gates can be performed directly in,
at best, a non-deterministic fashion, the one-way model is particularly useful.
Linear optical cluster-state quantum computation
is the most prominent example. For existing,
short surveys of the topic, see (O'Brien, 2007; Briegel \textit{et al}., 2009).

Cluster states can be created by a controllable Ising-type interaction
(Briegel and Raussendorf, 2001; Raussendorf and Briegel, 2001;
Raussendorf \textit{et al}., 2003). It was recently shown that an
efficient preparation of cluster states is possible with probabilistic
two-qubit controlled phase flip gates (Duan and Raussendorf, 2005; Chen
\textit{et al}., 2006a). Few-photon
cluster states were created in several recent experiments (Zhang
\textit{et al}., 2006a; Walther \textit{et al}., 2005b; Kiesel \textit{et
al}., 2005a; Lu \textit{et al}., 2007; Tokunaga \textit{et al}., 2008).

\subsubsection{Constructing photonic cluster states}

\label{sec:LO-oneway}

By combing the one-way model with linear optical quantum computing, recent
theoretical proposals  require much less resources and effectively replace the
original KLM scheme (see Nielsen, 2004; Browne and Rudolph, 2005; Bodiya and Duan, 2006; Chen
\textit{et al}., 2008; an efficient parity-encoded optical quantum
computing model by Gilchrist \textit{et al}., 2007). Nielsen showed that efficiency can be greatly enhanced by building photonic
cluster states using easy non-deterministic gates. The resource overhead (Bell states, operations, etc.)
for a reliable entangling gate in Nielsen's scheme is $\sim10^3$, and thus about two orders of magnitude
less than the original KLM protocol. Furthermore, by introducing two linear-optical fusion operations, Browne and Rudolph (2005) achieved a
greater degree of efficiency ($\sim10^2$) and in a simpler scheme than the previous proposals. Matter qubits can also be constructed into
cluster states using linear optics and photon interference, as proposed by
Barrett and Kok (2005). Here we focus on a linear-optical architecture for one-way
quantum computing.

\begin{figure}
[ptb]
\begin{center}
\includegraphics[width=0.38\textwidth]
{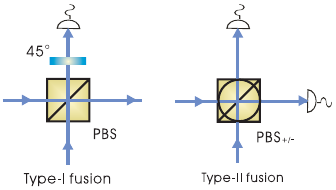}%
\caption{Nondeterministic photonic qubit fusion gates. The type-I fusion gate combines two
input single photons on a PBS and one of the outputs is measured in the $H\pm V$ basis. The type-II fusion
gate combines two photons on a $45^\circ$ rotated PBS (that is, both inputs and outputs are rotated
using a HWP), and both outputs are detected. See text for more details (Browne and Rudolph, 2005).}%
\label{fusion}%
\end{center}
\end{figure}

\begin{figure}
[ptb]
\begin{center}
\includegraphics[width=0.46\textwidth]
{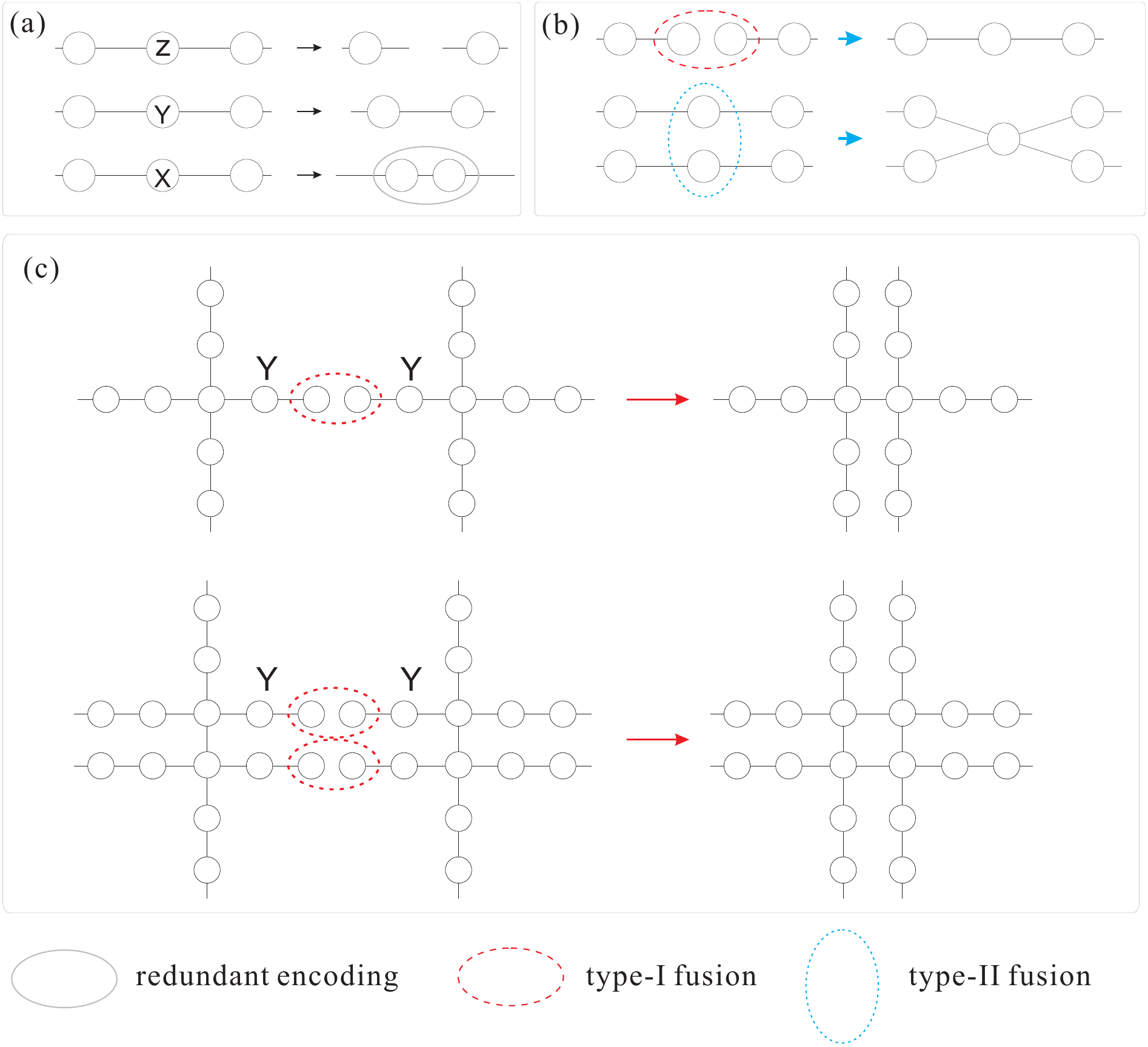}%
\caption{Construction of two-dimensional cluster states using photon fusion (Browne and Rudolph, 2005; Duan and Raussendorf, 2005).
(a) Certain measurements on a cluster qubit will leave
the remaining qubits in a new cluster state with a different
layout. (b) The effect of type-I and type-II fusion
operations on successful connection of two linear cluster states.
(c) One method of efficient construction of two-dimensional cluster states (see also Duan and Raussendorf, 2005)}%
\label{fusion}%
\end{center}
\end{figure}

\paragraph{A linear-optical architecture}

\label{sec:architecture}

The Browne-Rudolph scheme has an
 important practical advantage. It requires stable interferometry over
only the coherence length of the photons, i.e., the interferometer path lengths differences
only need to be kept constant to within tens of microns, not fractions of a wavelength.
Its two basic operations are the
type-I and type-II \textquotedblleft two-qubit fusion\textquotedblright%
\ operations (Fig.~\ref{fusion}). The physical mechanism of type-I fusion
is again two-photon interference at the PBS, which is used as a parity
check (Pan and Zeilinger, 1998; Pan \textit{et al}., 2001b; Cerf \textit{et
al}., 1998; Pittman \textit{et al}., 2001). Starting from a supply of
2-qubit polarization cluster states $\frac{1}{2}(|HH\rangle+|VH\rangle+|HV\rangle
-|VV\rangle)$ (which are equivalent to a Bell state and can be
created via the methods described in section \ref{sec:eventready}), the type-I
fusion operation allows one to efficiently generate arbitrarily long linear
cluster states. If the Type-I fusion is applied to the end-qubits of linear
(i.e., one-dimensional) clusters of lengths $n$ and $m$, successful outcomes (with a probability of $50\%$)
generate a linear cluster of length $(n+m-1)$. The type-I
fusion operation fails (also with a probability of $50\%$) when either zero or two photons
of either polarization are detected. The failure outcomes have the effect of
measuring both input qubits in the $\sigma_{z}$-eigenbasis, which leaves the remaining qubits in a cluster state of the same layout as before
the measurement, but now with all the bonds broken to the measured qubit.

Browne and Rudolph (2005) also showed that one can finally prepare a square-lattice
cluster of $N$ qubits with a temporal overhead scaling logarithmically with
$N$, and with an operational overhead (i.e., number of fusion operations)
scaling as $\sim N\ln N$ (Chen \textit{et al}., 2008). The described protocol
is a linear-optical realization of the Duan-Raussendorf proposal (2005), but
combines the advantages of the Browne-Rudolph scheme, whose overall efficiency
is thereby demonstrated. A crucial element for a realistic realization is
quantum memory for polarization qubits, which was
discussed in section~\ref{sec:memory}.

\paragraph{Event-ready entangler}

\label{sec:eventready}

In the above linear-optical architecture, two-photon entangled pairs are the
basic resources. In the case of SPDC, one usually does not know when a pair is emitted.
Only a firing of  photon detectors informs one that a spontaneous emission act happened.
However, most schemes of optical quantum computation, including the nondestructive CNOT gates,
scalable fusion of cluster states, require that the photon pairs
should be created in an ``event-ready" (or heralded) way.

Rudolph proposed a way to generate one pair of
event-ready entangled photon from four single photons, with a method that requires only
linear optics and photon-number-discriminating detectors. This was experimentally simulated using four photons from SPDC by Zhang \textit{et al}.,
(2008a). Another scheme of generating triggered photon pairs, which does not need
true single photon sources but totally tests upon SPDC, was proposed by Sliwa \textit{et al}. (2003). Following this proposal,
in two experiments, Wagenknecht \textit{et al}. (2010) and Barz \textit{et al}. (2010), have demonstrated heralded generation of
photon states, that are maximally entangled in polarization.

Probably a more promising realization of a triggered entangled photon source will come from
the biexciton (two electron-hole pairs) radiative decay in a self-assembled quantum dot.
This was demonstrated in Stevenson \textit{et al}. (2006). A quantum dot can emit a single pair
of entangled photons on demand, with a probability near close to one. However, it has a very low extraction efficiency. Very recently, Dousse \textit{et al}. (2010)
used carefully fabricated cavity to increase the collection efficiency, and created a
source of polarization entangled photon pairs with a state fidelity of 0.67 and a rate of 0.12 per an excitation pulse.

\subsection{Few-photon quantum computing experiments}

In recent years, we have also witnessed a number of proof-of-principle demonstrations
of quantum computing involving several photons and linear optics (experimental
realizations of photonic CNOT gates have been discussed in section \ref{sec:LO-gate}). For example,
Mohseni \textit{et al}. (2003) and Tame \textit{et al}. (2007)
demonstrated the two-qubit Deutsch-Josza algorithm in
a circuit and  a one-way model. The Grover's search algorithm (Grover, 1997) has been
realized by Kwiat \textit{et al}. (2000) by designing an optical circuit, and later
by Walther \textit{et al}. (2005b), Prevedel \textit{et al}. (2007b),
Chen \textit{et al}. (2007), and Vallone \textit{et al}. (2008), who used four-qubit cluster states.

\begin{figure}
[ptb]
\begin{center}
\includegraphics[width=0.48\textwidth]%
{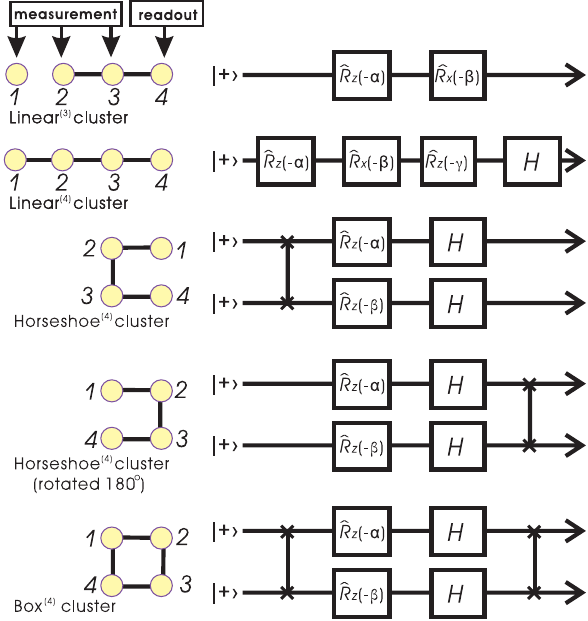}%
\caption{Few-qubit cluster states and the quantum circuits they implement. For
each three- and four-qubit cluster, its quantum state ($|\Phi_{\text{lin}%
3}\rangle$, $|\Phi_{\text{lin}4}\rangle$, $|\Phi_{\sqsubset4}\rangle$,
$|\Phi_{\sqsupset4}\rangle$, or $|\Phi_{\square4}\rangle$) and the computation
carried out in the one-way quantum computer model is shown (Walther
\textit{et al}., 2005).}%
\label{ow-exp}%
\end{center}
\end{figure}

In the experiment of Walther \textit{et al}. (2005b),  via a SPDC a four-photon
polarization-entangled cluster state was created (with a fidelity of $F=0.63\pm0.02$)
\begin{align}
|\Phi_{\text{c}}\rangle &  =\frac{1}{2}(\left\vert H\right\rangle
_{1}\left\vert H\right\rangle _{2}\left\vert H\right\rangle _{3}\left\vert
H\right\rangle _{4}+\left\vert H\right\rangle _{1}\left\vert H\right\rangle
_{2}\left\vert V\right\rangle _{3}\left\vert V\right\rangle _{4}\nonumber\\
&  +\left\vert V\right\rangle _{1}\left\vert V\right\rangle _{2}\left\vert
H\right\rangle _{3}\left\vert H\right\rangle _{4}-\left\vert V\right\rangle
_{1}\left\vert V\right\rangle _{2}\left\vert V\right\rangle _{3}\left\vert
V\right\rangle _{4}).%
\end{align}
With local unitary operations $|\Phi_{\text{c}}\rangle$ can be arranged to
various cluster shapes $|\Phi_{\text{lin}4}\rangle$, $|\Phi
_{\sqsubset4}\rangle$, $|\Phi_{\sqsupset4}\rangle$, and $|\Phi_{\square4}\rangle$
(see Fig.~\ref{ow-exp}).
Using the cluster state $|\Phi_{\text{c}}\rangle$, a universal
set of quantum logic operations, single-qubit rotations and non-trivial
two-qubit gates were  demonstrated (see Fig.~\ref{ow-exp}).
In addition, Walther \textit{et al}. also implemented
a two-qubit Grover's quantum search algorithm using
the box cluster $|\Phi_{\square4}\rangle$. The measured probability of
the quantum computer to determine the correct outcome was about 90\%.
A drawback of the experiment is that no feed-forward
was used. This reduced the success rate of the computation by a factor of
two, for every measurement. An improved experiment done
by Prevedel \textit{et al}. (2007b) incorporated active fast feed-forward,
so that the earlier measurement outcomes could change
the setting of a future measurement in real time.
Particularly, the computational step (i.e., the
individual feed-forward cycle) could be operated in less than 150 ns using
electro-optical modulators.

Shor's quantum algorithm provides a way of factoring large integers in polynomial
time, a task for which no efficient classical method is known. Recently, a compiled version
of Shor's algorithm has been demonstrated using four photonic qubits
by Lu \textit{et al}. (2007) and Lanyon \textit{et al}. (2007).
Aiming to solve the easiest case of Shor's algorithm (15=3$\times$5),
these two experiments designed a simplified linear optical
network to implement the quantum circuits of the modular exponential
execution. The results yielded a high success probability
($\thicksim0.99$) of factoring. It is notable that
in these optical experiments during the computation genuine multi-particle entanglement and
multi-path interference were observed, which did not appear in
the previous implementations using nuclear magnetic resonance (Vandersypen
\textit{et al}. 2001). Later optical implementations of Shor's algorithm later
moved into an integrated photonic chip device (Politi \textit{et al}. 2009).

Another interesting avenue of the ongoing multi-photon experiments is to exploit
the preliminary photonic quantum computers for simulation of many-body physics, a powerful
application of quantum computers proposed in the early 1980s.
Following proposals by Han \textit{et al} (2007) and Pachos (2007), four- and six-qubit graph states
were created to mimic the ground state of the Kitaev spin-lattice model
(Kitaev, 2003). Using the photonic graph states to simulate
the creation and braiding of the anyons in the Kitaev model, a phase shift of $\pi$ related to the
anyon braiding was observed, confirming the predictions for the fractional statistics
of Abelian $1/2$ anyons (Lu \textit{et al}. 2009; Pachos \textit{et al}. 2009).
Recently, Lanyon \textit{et al}. (2010) employed a photonic quantum computer to solve
a quantum chemistry problem: calculation of the energies of
the hydrogen molecule. In their experiment,
the iterative phase estimation quantum algorithm was performed in full. With an assistance of
a classical computer, it was possible to calculate the
complete energy spectrum, up to 20 bits of precision. Ma \textit{et al.} (2011)
used polarization states of four photons to simulate a frustrated Heisenberg
spin system: a spin-1/2 tetramer.

%%%%%%%%%%%%%%%%%%%%%%%%%%%%%%%%%%06-OCT-17:47%%%%%%%%%%%%%%%%%%%%%%%%%%%%%%%%%%%%%%
\subsection{Toward scalable optical quantum computing}

While in small-scale tests of optical quantum computing we have witnessed a progress,
serious problems exist in scaling up this technique. Here we briefly discuss
the key challenges and the ongoing efforts in designing fault-tolerant
architectures, fighting against experimental noise, and improving
single-photon sources and detectors.

Quantum computers will be very susceptible to noise which rapidly destroys the fragile quantum
information. Much effort has been devoted to the understanding of the scalability
under realistic noise models.
The \emph{threshold theorem} has established that if the noise is
below some value, an arbitrarily long quantum computation can be realized. Several
fault-tolerant architectures for optical quantum computing have been proposed.
Nielsen and Dawson (2005) have shown that the thresholds are respectively $<3\times10^{-3}$
and $<10^{-4}$ for photon loss and depolarizing noise (see also
Dawson \textit{et al}. 2006). In addition, there are proposals
of one-way quantum computing in decoherence-free
subspaces (Tame \textit{et al}., 2007; Jiang \textit{et al}., 2009) and
topological cluster-state quantum computing (Raussendorf \textit{et al}. 2006;
2007; Raussendorf and Harrington, 2007).
The latter proposal achieved a high error threshold of $0.75\times10^{-3}$, the
highest known for a local architecture. For photon loss alone, Ralph \textit{et al}. (2005) and Varnava
\textit{et al}. (2006) and (2008) have designed loss-tolerant quantum computer
schemes within the circuit-model and one-way model; the latter scheme can tolerate an
overall optical loss, including source inefficiency and collection loss, up to an
impressive 33$\%$. Gong et al. (2010) proposed a new scheme where the efficiency threshold for loss tolerance
requires the product of source and detector efficiencies to be $>50\%$. Despite of the progress,
one should note that when the losses are high,
the resource requirements become impractically high. Moreover, unfortunately, these loss-tolerant
codes tend to amplify the depolarizing errors (e.g. bit flips and phase flips);
the tradeoff has been discussed and new schemes have been designed which tolerates both errors in Rohde et al. (2007).

Some basic quantum error correction codes have been tested in optical
experiments. O'Brien \textit{et al}. (2005)  demonstrated a two-qubit code
for correction of a $Z$-measurement error. With a continuous variable encoding,
Aoki \textit{et al}. (2009) realized a nine-qubit Shor's code, which is able to correct an
arbitrary single-qubit error. Decoherence-free subspaces, a type of passive error-preventing codes,
have been experimentally realized using two photons by Kwiat \textit{et al}. (2000) and four photons
by Bourennane \textit{et al}. (2004b). Furthermore, decoherence-free subspace approach  was applied in an optical
demonstration of the Deutsch-Jozsa algorithm (Mohseni \textit{et al}., 2003), and for reliable measurement-based
one-way information transfer (Prevedel \textit{et al}., 2007a).
To fight against the qubit-loss error, an especially serious problem for photonic qubits,
Lu \textit{et al}. (2008) demonstrated a four-qubit Grassl erasure correction code
(for the theoretical proposals see Grassl (1997) and Ralph (2005)), and a tree-shaped graph state (Vanara \textit{et al}. 2006).
The tested method is applicable in both the quantum circuit
model and in the one-way model.

Despite the progress, the fault-tolerant thresholds are
still well beyond what is achievable with today's technology. Optical quantum
computing makes critical use of sources of on-demand single-photons
which are indistinguishable and can be collected efficiently. The majority of
experimental demonstrations so far have relied on the SPDC photons,
which suffers from undesired higher-order photon emission
(Weinhold \textit{et al}. 2008, this has been considered as the major
source of error for most experiments reviewed here, see Section~\ref{sec:SWAP}), large bandwidth, and the probabilistic
manner of photon pair emission. New generations of
single-photon sources have been  developed: they are based on
solid-state devices, atoms, molecules and ions
etc. (see Lounis and Orrit, (2005) for a recent review). These new
single-photon emitters include single quantum dots (Michler \textit{et
al.}, 2000b; Santori \textit{et al.}, 2002; Pelton \textit{et al.},
2002; Bennett \textit{et al}., 2005; Englund \textit{et al}., 2007; Strauf
\textit{et al}., 2007; Shields, 2007), nitrogen-vacancy color centers
(Kurtsiefer \textit{et al.}, 2000; Brouri \textit{et al.}, 2000), neutral atoms (Kuhn
\textit{et al.}, 2002; McKeever \textit{et al.}, 2004; Darqui\'{e} \textit{et
al.}, 2005), ions (Keller \textit{et al.}, 2004), and molecules (Brunel
\textit{et al}., 1999; Lounis and Moerner, 2000). Solid state
sources of single photons hold the promise of a ready integration, and much
experimental effort has been recently devoted to improving the single-photon
quality, collection efficiency, and interference of photons from remote
independent quantum dots. Other controllable single-photon sources
can be devised using trapped single-atoms in high-finesse optical cavities, which are
spectrally narrow and have a well-defined spatial mode. However, there are problems associated with low out-coupling efficiency.
Single photons emitted by atomic ensembles, as we discussed
in section \ref{sec:repeater}, are another promising source.
This type of single photon source naturally enjoys a very narrow ($\sim$MHz)
linewidth and good indistinguishability. However, in addition to the low photon extraction efficiency, much work needs to be done to
improve the retrieval efficiency (converting atomic collective excitations into photons) up to unity.

To meet the stringent demands of scalable optical quantum computing,
one faces yet another challenge:  new single-photon
detectors that have near-unity efficiencies, high repetition rates,
low dark count rates, and the ability to resolve
the photon number. Currently, mostly used room temperature silicon single-photon
detectors can be operated at 10MHz with a peak efficiency of 65\%,
a dark count rate of about 100 Hz, and a timing jitter
of typically 500ps; work is in progress to
improve these parameters (see \textit{e.g.} Kardyna \textit{et al}., 2008).
Significant progress (Rosenberg \textit{et
al}., 2005; Lita, Miller and Nam, 2008; Divochiy \textit{et al}., 2008) has
been made on superconducting detectors capable of resolving photon-number,
with an ultra low dark count rate (0.15 Hz at the
wavelength of 1.3 $\mu$m), and high efficiency up to 95\%. We refer the Reader to
a review by Hadfield, (2009) for more details.

Finally, it is worth mentioning that chip-scale waveguide quantum circuits
have been created recently, and used to demonstrate high-visibility Hong-Ou-Mandel
interference, {\small CNOT} gates, an instance of Shor's algorithm
(Politi \textit{et al}. 2008, 2009), and quantum walk (Peruzzo \textit{et al.}, 2010). Current silica waveguide circuits have dimensions of about
one centimeter per logic gate. This is a step toward integrated optics
architecture for improved performance, miniaturization,
and scalability. An open challenge is to integrate such devices with single-photon
sources and detectors.

\section{Concluding Remarks}

We have reviewed the principles and experimental techniques for manipulation of multi-photon entangled states, which have enabled a series of pioneering experiments in the field of quantum information. A number of important applications have been highlighted: Laboratory tests demonstrating the contradiction between quantum mechanics and local realism  performed with entangled photons, that is, the Bell and Greenberger-Horne-Zeilinger experiments (Weihs \textit{et al}. 1998, Pan \textit{et al}. 2001). Quantum teleportation--the transfer and reconstruction of quantum states over arbitrary distances became an experimental reality with four-photon interferometries (Bouwmeester \textit{et al}. 1997, Marcikic \textit{et al}. 2003). A variation of teleportation, entanglement swapping (Pan \textit{et al}. 1998), which together with entanglement purification (Pan \textit{et al}. 2003) and quantum memory (e.g. Yuan \textit{et al}. 2008) are the essential components of quantum repeaters necessary for quantum networking and long distance quantum communication. Proof-of-principle demonstrations of linear optics quantum computing (e.g. O'Brien \textit{et al}. 2003, Walther \textit{et al}. 2005, Lu \textit{et al}. 2008, Lanyon \textit{et al}. 2010) and super-resolving phase measurement (e.g. Nagata \textit{et al}. 2007) with multi-photon devices.

The ultimate goal are long-distance quantum communication and scalable optical quantum computing. However, many technological challenges remain. Parametric down-conversion (Kwiat \textit{et al}. 1995; White \textit{et al}. 1999) has been serving as the main workhorse for the multi-photon experiments reviewed here; up to eight entangled photons have been observed (Yao \textit{et al}. 2007a). However, due to its intrinsic limitations, there is a bottleneck with regard to the attainable brightness and fidelity of multi-photon states based on it. This calls for the development of a next generation of more reliable and scalable single photon sources (Lounis and Orrit, 2005). Other challenges include efficient coupling and detection of single photons and quantum memories for photons with long storage time and high retrieval efficiency etc. Continuing effort is devoted in this direction and encouraging results have been obtained. For instance, single photons and entangled photons have been generated from self-assembled quantum dots embedded in a microcavity, with extraction efficiency up to $80\%$ (Dousse \textit{et al}. 2010). In the case of a long-lived quantum memory based on atomic ensembles  storage times were reported to reach  up to 8 ms (Zhao \textit{et al}. 2009), with potential improvements when using cavities. Armed with these new techniques, the control of multi-photon states will reach a higher level. In any event, we expect that the techniques reviewed in this article should be applicable in the future experiments.

\begin{acknowledgments}
We would like to thank Y.-J. Wei, Y.-M. He, Y. He, Z.-E. Su, Z.-S. Yuan, B. Zhao, X.-H. Bao, M. Zhang and N.-L. Liu for valuable
comments and helps. This work was supported by the National Natural Science Foundation of China,
the National Fundamental Research Program (under grant No. 2011CB921300) and the
Chinese Academy of Sciences. M.\.{Z}. was supported by Professorial
Subsidy of FNP and the MNiI Grant No 1 P03B 04927, MNiSW Grant No. N202
208538, and EU programs QAP and Q-ESSENCE. He acknowledges  Austrian-Polish and German-Polish exchange programmes, and finally CAS visting professorship. H.W. acknowledges the support from DFG (MAP) and BMBF (QPENS and QuORep) and the EU programs QAP and Q-ESSENCE. A.Z. acknowledges the supports
from EPSRC, QIPIRC, FWF and EC under the Integrated Project Qubit Application.
\end{acknowledgments}

\appendix

\section{The two-photon states produced by SPDC}

\emph{Crystal-field interaction.--}In the interaction Hamiltonian of the
electromagnetic field with an atom or a molecule, the dominating part is
$\hat{H}_{a-f}\sim\hat{\bm{\mu}}_{e}\cdot\bm{E}(\bm{x},t)$, i.e., scalar
product of the dipole moment, $\bm{\mu}$, of the atoms or molecules with the
local electric field. Since the electric polarization, $\bm{p}(\bm{x},t)$, of
a medium is given by the mean dipole moment of the atoms or molecules per unit
volume, the principal term of the field-crystal interaction Hamiltonian,
$H_{int}$, is proportional to $\int_{V}\bm{p}(\bm{x},t)\cdot
\bm{E}(\bm{x},t)d^{3}x,\label{Hnl1}$ where $V$ is the volume of the crystal.
One can assume that $\bm{E}(\bm{x},t)$ interacts with $\bm{p}(\bm{x},t)$ only
in the point $\bm{x}$, thus the $i$-th component of polarization is in the
most general case given by
\begin{equation}%
\begin{array}
[c]{rcl}%
p_{i}(\bm{x},t) & = & \sum_{j=1}^{3}\chi_{ij}^{(1)}(\bm{x})E_{j}(\bm{x},t)\\
& + & \sum_{j,k=1}^{3}\chi_{ijk}^{(2)}(\bm{x})E_{j}(\bm{x},t)E_{k}%
(\bm{x},t)+\cdots,
\end{array}
\end{equation}
where $\chi_{ij}^{(1)}$ are $\chi_{ijk}^{(2)}$ are the (macroscopic)
polarizability tensors. For any crystal with centro-symmetric structure the
quadratic term of the polarizability vanishes. Thus, as we shall see, the SPDC effect exists
only for birefringent media having a nonzero value of $\chi^{(2)}$. If one
assumes that $\chi_{ijk}^{(2)}(\bm{x})$ has the same value for all points
within the crystal, one gets
\begin{align*}
H_{int}  &  \sim\int_{V}\bm{p}(\bm{x},t)\cdot\bm{E}(\bm{x},t)d^{3}x\\
&  =\int_{V}\bm{p}^{:lin}(\bm{x },t)\cdot\bm{E}(\bm{x},t)d^{3}x\\
&  +\int_{V}\bm{p}^{:nl}(\bm{x},t)\cdot\bm{E}(\bm{x},t)d^{3}x,
\end{align*}
where $\bm{p}^{:lin}$ ($\bm{p}^{:nl}$) is the linear (nonlinear) term of
polarization. The nonlinear part of the Hamiltonian is
\begin{equation}
H^{NL}\sim\int_{V}\sum_{ijk}\chi_{ijk}^{(2)}E_{i}(\bm{x},t)E_{j}%
(\bm{x},t)E_{k}(\bm{x},t)d^{3}x. \label{Hnl2}%
\end{equation}

The quantized electric field can be expressed
(in the interaction picture) as
\begin{align}
\bm{E}(\bm{x},t)  &  =\sum_{\lambda=1}^{2}\int d^{3}k\frac{i}{\sqrt
{2\omega(2\pi)^{3}}}\nonumber\\
&  \times\hat{\epsilon}(\bm{k},\lambda)a(\bm{k},\lambda)e^{i(\bm{k}\cdot
\bm{x}-\omega t)}+h.c.\nonumber\\
&  =\bm{E}^{(+)}(\bm{x},t)+\bm{E}^{(-)}(\bm{x},t), \label{poleE}%
\end{align}
where $\bm{E}^{(-)}(\bm{x},t)=[\bm{E}^{(+)}(\bm{x},t)]^{\dagger}$, and the
summation is over two orthogonal linear polarizations, $h.c.$ denotes the
hermitian conjugate of the previous term, and $\hat{\epsilon}(\bm{k},\lambda)$
is a unit vector defining the linear polarization. The symbol
$a(\bm{k},\lambda)$ denotes the annihilation operator of a monochromatic
photon with wave vector $\bm{k}$, and polarization $\hat{\epsilon
}(\bm{k},\lambda)$. The principal commutation rules for such creation and
annihilation operators are given by\footnote{These new operators are linked
with the ones discussed earlier (see section~\ref{sec2-lo}) by the relation
$a^{l}=\sum_{\lambda}\int d^{3}kg_{\lambda\bm{k}}^{l}a(\bm{k},\lambda)$.}
$[a(\bm{k},\lambda),a^{\dagger}(\bm{k}^{\prime},\lambda^{\prime}%
)]=\delta_{\lambda,\lambda^{\prime}}\delta^{(3)}(\bm{k}-\bm{k}^{\prime})$,
$[a^{\dagger}(\bm{k},\lambda),a^{\dagger}(\bm{k}^{\prime},\lambda^{\prime
})]=0$ and $[a(\bm{k},\lambda),a(\bm{k}^{\prime},\lambda^{\prime})]=0$.

\emph{The relevant terms in the Hamiltonian.--}One can neglect the depletion
of the laser field and assume that the total field is $\bm{E}^{Laser}%
(\bm{x},t)+\bm{E}(\bm{x},t),$ where $\bm{E}^{Laser}$ is a classical field. The
quantum field $\bm{E}$ describes the emitted photons. The down conversion
takes place, thanks to only the terms in (\ref{Hnl2}) of the form
\begin{equation}
\int_{V}\sum_{ijk}\chi_{ijk}^{(2)}E_{i}^{Laser}E_{j}^{(-)}E_{k}^{(-)}d^{3}x.
\label{hn14}%
\end{equation}
Simply, only $E^{(-)}$ contains the creation operators, and thus acting on the
vacuum state $|\Omega\rangle$ can give rise to a two-photon state. Thus, we
forget about all other terms and analyze only $H^{NL}$ in the form of
(\ref{hn14}) plus its Hermitian conjugate.

Let us describe the laser field as a monochromatic plane wave $\hat{z}{E_{0}%
}(e^{i(\bm{k_0}\cdot\bm{x}-\omega_{0}t-\phi)}+c.c.)$, where $E_{0}$ is the
field amplitude.\footnote{Since an arbitrary electromagnetic field is a
superposition of the plane waves, starting with this trivial case it is very
easy to get the general description.} Then, from (\ref{hn14}), one gets for
$H^{NL}$
\begin{align}
&  \sim\int_{V}d^{3}x\sum_{jk}\{\chi_{3jk}^{(2)}E_{0}[e^{i(\bm{k}_{0}%
\cdot\bm{x}-\omega_{0}t-\phi)}+c.c]\nonumber\\
&  \times\sum_{\lambda}\int d^{3}kf(\omega)\hat{\epsilon}_{j}(\bm{k},\lambda
)a^{\dagger}(\bm{k},\lambda)e^{-i(\bm{k}\cdot\bm{x}-\omega t)}\\
&  \times\sum_{\lambda^{\prime}}\int d^{3}k^{\prime}f(\omega^{\prime}%
)\hat{\epsilon}_{k}(\bm{k}^{\prime},\lambda)a^{\dagger}(\bm{k}^{\prime
},\lambda)e^{-i(\bm{k'}\cdot\bm{x}-\omega^{\prime}t)}+\mbox{h.c.},
\nonumber\label{Hnl5}%
\end{align}
with $f(\omega)$ being a factor dependent on $\omega$. Its specific structure
is irrelevant here. Extracting only those elements of the above expressions
which contain $\bm{x}$ and $t$, one sees that their overall contribution to
$H^{NL}$ is given by $\Delta(\pm\bm{k}_{0}-\bm{k}-\bm{k}^{\prime}%
)e^{-i(\pm\omega_{o}-\omega-\omega^{\prime})t}$, where $\Delta_{\pm}%
(...)=\int_{V}d^{3}xe^{i(\pm\bm{k}_{0}-\bm{k}-\bm{k}^{\prime})\cdot\bm{x}}$.
The terms with the time dependent factors, $e^{i(\omega_{o}+\omega
+\omega^{\prime})t}$, average out in any time integration (see below), and thus we can
drop them. If we assume that our crystal is a cube $L\times L\times L$, then
for $L\rightarrow\infty$, $\Delta_{\pm}$ approaches $\delta(\pm\bm{k}_{0}%
-\bm{k}-\bm{k}^{\prime})$. Thus, emission of the photon pairs is possible only
for the directions for which the condition $\bm{k}_{0}\approx
\bm{k}+\bm{k}^{\prime}$ is met. Finally one has
\begin{align}
&  H^{NL}\sim\sum_{\lambda,\lambda^{\prime}}\int d^{3}k\int d^{3}k^{\prime
}\Delta(\bm{k}_{0}-\bm{k}-\bm{k}^{\prime})A_{\lambda,\lambda^{\prime}}%
^{eff}\nonumber\\
&  \times e^{-i(\omega_{0}-\omega-\omega^{\prime})t}a^{\dagger}(\bm{k},\lambda
)a^{\dagger}(\bm{k}^{\prime},\lambda^{\prime})+\mbox{h.c.},
\end{align}
%where $\Delta(\bm{k}_{0}-\bm{k}-\bm{k}^{\prime})=\Delta_{+}(\bm{k}_{0}%
%-\bm{k}-\bm{k}^{\prime})+\Delta_{-}(-\bm{k}_{0}-\bm{k}-\bm{k}^{\prime})$ and
where $A_{\lambda,\lambda^{\prime}}^{eff}=\sum_{j,k}E_{0}\chi_{3jk}^{(2)}%
\hat{\epsilon}_{j}(\bm{k},\lambda)\hat{\epsilon}_{k}(\bm{k}^{\prime}%
,\lambda^{\prime})$ is the effective strength of the laser-crystal coupling.
Henceforth, we shall replace the symbol $A_{\lambda,\lambda^{\prime}}^{eff}$
by $F_{o}(\bm{k}_{0})$.

\emph{The state of photons emitted in the SPDC process.--}The pump-crystal
coupling is weak. The evolution of the state $\left\vert \Psi_{D}%
(t)\right\rangle $ (in the interaction picture) is given by $i\hbar\frac
{d}{dt}|\Psi_{D}(t)\rangle=H^{NL}(t)|\Psi_{D}(t)\rangle$. In the first order
in the perturbation expansion
\begin{equation}%
\begin{array}
[c]{rcl}
& |\Psi_{D}(t)\rangle\simeq|\Psi_{D}(t_{0})\rangle+\frac{1}{i\hbar}\int
_{t_{0}}^{t}H^{NL}(t^{\prime})|\Psi_{D}(t_{0})\rangle dt^{\prime}. &
\end{array}
\label{disturb}%
\end{equation}
Put $t_{0}=-\infty$, and take the vacuum state (no photons) $|\Omega\rangle$
as the initial state $|\Psi(0)\rangle$. Only in the term with the
integral one can find creation of pairs of photons. For $t\rightarrow\infty$ it contains an integral of
the following form: $\int_{-\infty}^{+\infty}dt^{\prime}e^{it^{\prime}%
(\omega+\omega^{\prime}-\omega_{0})}$ which is $2\pi\delta(\omega
+\omega^{\prime}-\omega_{0})$. Thus, the two-photon component of the state, at
$t=\infty$, is effectively given by
\begin{align}
&  \sum_{\lambda,\lambda^{\prime}}\int d^{3}k\int d^{3}k^{^{\prime}}%
F_{o}(\bm{k}_{0})\Delta(\bm{k}_{0}-\bm{k}-\bm{k}^{\prime})\nonumber\\
&  \times\delta(\omega+\omega^{\prime}-\omega_{0})a^{\dagger}(\bm{k},\lambda
)a^{\dagger}(\bm{k}^{\prime},\lambda^{\prime})|\Omega\rangle,
\label{STATE-PDC}%
\end{align}
and the frequencies of the emissions satisfy the relation\footnote{One should
add here a note that in reality this relation is not absolutely sharp. The
molecular polarization was treated here phenomenologically. Still, once a more
refined model is used the relationship is sharp enough, so that the deviations
from perfect equality are beyond the resolution of the present measuring
setups.} $\omega_{0}=\omega+\omega^{\prime}$.

\emph{Directions of emissions.--}Since $\omega=|\bm{k}|\frac{c}{n(\omega
,\lambda)}$, where $\frac{c}{n(\omega,\lambda)}=c(\omega,\lambda)$ is the
speed of light in the given medium, which depends on frequency and
polarization, the condition for frequencies becomes $|\bm{k}_{0}|c(\omega
_{0})\simeq|\bm{k}|c(\omega,\lambda)+|\bm{k}^{\prime}|c(\omega^{\prime
},\lambda^{\prime})$. This, together with $\bm{k}_{0}\simeq\bm{k}_{s}%
+\bm{k}_{i}$ fixes the possible emission directions, frequencies and
polarizations.\footnote{If one has $\omega\simeq\omega^{\prime}$ then we have
a frequency degenerate SPDC, and if $\hat{k}\simeq\hat{k}^{\prime}$, then we
have a co-linear one.}

\emph{Time correlations.--}The probability of a detection of a photon, of,
e.g., the horizontal polarization $H$, at a detector situated at point
$\bm{x}$ and at time $t$, is proportional to $\eta\mathrm{Tr}[\varrho
(t)E_{H}^{(-)}(\bm{x},t)E_{H}^{(+)}(\bm{x},t)],\label{pro10}$ where $\eta$ is
the coefficient which characterizes the quantum efficiency of the detection
process, $\varrho$ is the density operator, $E_{H}$ is the horizontal
component of the field in the detector. For the above relation to be true, we
also assume that only photons of a specified direction of the wave vector enter
 via the aperture of the detector. For a pure state, this reduces to
$p(\bm{x},t,H)\simeq\langle\psi|E_{H}^{(-)}E_{H}^{(+)}|\psi\rangle$. The
probability of a joint detection of two photons, of polarization $H$, at the
locations $\bm{x}_{1}$ and $\bm{x}_{2}$, and at the moments of time $t$ and
$t^{^{\prime}}$, is proportional to
\begin{align}
p(\bm{x}_{1},t;\bm{x}_{2},t^{^{\prime}})  &  \sim\langle\psi|E_{H}%
^{(-)}(\bm{x}_{1},t)E_{H}^{(-)}(\bm{x}_{2},t^{^{\prime}})\nonumber\\
&  \times E_{H}^{(+)}(\bm{x}_{2},t^{^{\prime}})E_{H}^{(+)}(\bm{x}_{1}%
,t)|\psi\rangle. \label{pro11}%
\end{align}
If the detectors are very far away from each other, and from the crystal, then
the photon field reaching them can be treated as free-evolving. We put into
(\ref{pro11}) the photon state (\ref{STATE-PDC}). Let $t=t_{1}$ and
$t^{\prime}=t_{2}$, and $|\psi\rangle=|\psi(t=\infty)\rangle$, then
(\ref{pro11}) can be written down as
\begin{align}
p(\bm{x}_{1},t|\bm{x}_{2},t^{\prime})  &  \simeq\langle\psi|E_{H}%
^{(-)}(\bm{x}_{1},t)E_{H}^{(-)}(\bm{x}_{2},t^{\prime})\nonumber\\
&  \times E_{H}^{(+)}(\bm{x}_{1},t)E_{H}^{(+)}(\bm{x}_{2},t^{\prime}%
)|\psi\rangle. \label{pro12}%
\end{align}

To simplify the description, let us replace the annihilation and creation
operators, which were used above, with new operators $a_{i}(\omega)$ and their
conjugates, which describe \textquotedblleft unidirectional\textquotedblright%
\ excitations of the photon field [i.e., we assume that the detectors see only
the photons of a specified direction of propagation, a good assumption if the
detectors are far from the crystal, and the apertures are narrow, see (Fearn
and Loudon, 1987)]. The index $i$ defines the direction (fixed) of the wave
vector. The new operators satisfy commutation relation, which are a
modification of those given above to the current specific case $[a_{i}%
(\omega),a_{j}^{\dagger}(\omega^{\prime})]=\delta_{ij}\delta(\omega
-\omega^{\prime})$, $[a_{i}(\omega),a_{j}(\omega^{\prime})]=0$. If we choose
just two propagation directions that fulfill the phase matching conditions,
then effectively one can put
\begin{equation}
E_{H}^{(+)}(\bm{x}_{i},t)=\int d\omega e^{-i\omega t}f_{i}(\omega)a_{i}%
(\omega)
\end{equation}
with $i=1,2$, and where $f_{1}$ and $f_{2}$ are the frequency response
functions of the filter-detector system.

We assume that the maxima of the functions agree with the frequencies given by
the phase matching conditions. Introducing a unit operator $\hat{I}=\sum
_{i=0}^{\infty}|b_{i}\rangle\langle b_{i}|$, where $|b_{i}\rangle$ is the
basis states, into (\ref{pro12}), we obtain
\begin{align}
p(\bm{x}_{1},t|\bm{x}_{2},t^{\prime})  &  \simeq\langle\psi|E_{H}%
^{(-)}(\bm{x}_{1},t)E_{H}^{(-)}(\bm{x}_{2},t^{\prime})\nonumber\\
&  \times\hat{I}E_{H}^{(+)}(\bm{x}_{1},t)E_{H}^{(+)}(\bm{x}_{2},t^{\prime
})|\psi\rangle. \label{pro13}%
\end{align}
Since $E_{H}^{(+)}$ contains only the annihilation operators, they transform
the two-photon state $|\Psi\rangle$ into the vacuum state. Thus,
Eq.~(\ref{pro13}) can be put as (Mollow, 1973)
\begin{equation}
p(\bm{x}_{1},t|\bm{x}_{2},t^{\prime})\simeq\langle\psi|E_{H}^{(-)}%
E_{H}^{^{\prime}(-)}|\Omega\rangle\langle\Omega|E_{H}^{(+)}E_{H}^{^{\prime
}(+)}|\psi\rangle, \label{pro14}%
\end{equation}
where the primed expressions pertain to the moment of time $t^{\prime}$ and
the position $\bm{x}_{2}$. Thus we have $p(\bm{x}_{1},t|\bm{x}_{2},t^{\prime
})\simeq|A_{12}(t,t^{\prime})|^{2}$, where $A_{12}(t,t^{\prime})=\langle
\Omega|E_{H}^{(+)}(\bm{x}_{1},t)E_{H}^{^{\prime}(+)}(\bm{x}_{2},t^{\prime
})|\psi\rangle$. With the use of the new creation operators, the state
$|\Psi\rangle$ can be approximated by
\begin{equation}
|\Omega\rangle+\int d\omega_{1}\int d\omega_{2}F_{o}\delta(\omega-\omega
_{1}-\omega_{2})a_{1}^{\dagger}(\omega_{1})a_{2}^{\dagger}(\omega_{2}%
)|\Omega\rangle.
\end{equation}
Therefore, one gets the following formula for the detection amplitude
\begin{align}
A_{12}(t,t^{\prime})  &  =\langle\Omega|\int d\omega^{\prime}e^{-i\omega
^{\prime}t^{\prime}}f_{2}(\omega^{\prime})a_{2}(\omega^{\prime})\nonumber\\
&  \times\int d\omega e^{-i\omega t}f_{1}(\omega)a_{1}(\omega)\int d\omega
_{1}\int d\omega_{2}\nonumber\\
&  \times F_{o}\delta(\omega_{0}-\omega_{1}-\omega_{2})a_{2}^{\dagger}%
(\omega_{2})a_{1}^{\dagger}(\omega_{1})|\Omega\rangle.
\end{align}
Since the creation and annihilation operators for different modes commute, and
since one can use $\langle\Omega|a_{i}(\omega^{\prime})a_{j}^{\dagger}%
(\omega)|\Omega\rangle=\delta_{ij}\delta(\omega^{\prime}-\omega)$, we get
\begin{equation}
A_{12}(t,t^{\prime})=F_{o}e^{-i\omega_{0}t^{\prime}}\int d\omega
e^{-i\omega(t-t^{\prime})}f_{2}(\omega_{0}-\omega)f_{1}(\omega),
\label{AMPLITUDE}%
\end{equation}
and we have
\begin{align}
p(\bm{x}_{1},t|\bm{x}_{2},t^{\prime})  &  \sim|A_{12}(t,t^{\prime}%
)|^{2}\nonumber\\
&  \simeq|\int d\omega e^{-i\omega(t-t^{\prime})}f_{2}(\omega_{0}-\omega
)f_{1}(\omega)|^{2}, \label{pro16}%
\end{align}
i.e., the probability depends on the difference of the detection times.

For instance, assume that: $f_{1}=f_{2}=f$, and that they are Gaussian,
$f(\omega)=Ce^{-\frac{(\omega_{c}-\omega)^{2}}{{\sigma}^{2}}}$, with the
central frequency $\omega_{c}=\omega_{0}/2$. Then we have $f_{1}(\omega
)=f_{2}(\omega_{0}-\omega)=f(\omega)$. The probability of detection of two
photons at the moments $t$ and $t^{\prime}$ reads
\begin{align}
&  p(\bm{x}_{1},t|\bm{x}_{2},t^{\prime})\nonumber\\
&  \sim|\int d\omega e^{-i\omega(t-t^{\prime})}C^{2}e^{-2\frac{{(\omega
_{c}-\omega)}^{2}}{{\sigma}^{2}}}|^{2}\sim e^{-\frac{\sigma^{2}}%
{2}(t-t^{\prime})^{2}}. \label{pro15}%
\end{align}
As $\sigma\rightarrow\infty$ the expression (\ref{pro15}) approaches
$\delta(t-t^{\prime})$. We have a perfect time correlation. For a realistic case
of final bandwidths, the degree of time correlation of the detection of the
SPDC photons depends entirely on the frequency response of the detectors (plus
interference filters, if any, before them).

\emph{The output state of pulsed pumped SPDC.--}Since the pump pulse is a
superposition of monochromatic waves, the output state for this case is an
integral of the monochromatic case over the momentum profile of the pulse:
$|\psi_{pulse}\rangle=\int d^{3}\bm{k}_{0}|\psi(F_{o}(\bm{k}_{0})\rangle$,
where $|\psi(F_{o}(\bm{k}_{0})\rangle$ is the state for the monochromatic case
with wave vector $\bm{k}$ and field amplitude $F_{o}(\bm{k}_{0})$. Since the
frequency of the pulse and the wave vector are not strictly defined, if the
pulse is too short the SPDC photons are less tightly correlated directionally.

The two photon state coming out of an SPDC can be approximated by
\begin{align}
|\Psi\rangle &  =\int d\omega_{0}F_{o}(\omega_{0})\int d\omega_{1}\int
d\omega_{2}\nonumber\\
&  \times\delta(\omega_{0}-\omega_{1}-\omega_{2})a_{1}^{\dagger}(\omega
_{1})a_{2}^{\dagger}(\omega_{2})|\Omega\rangle, \label{PULSED}%
\end{align}
where we have replaced the effective pump amplitude by the spectral
decomposition of the laser pulse $F_{o}(\omega_{0})$.

\emph{Two-photon detection amplitude: the pulsed pump case.--}If we have a
pulsed pump we have to integrate the amplitude (\ref{AMPLITUDE}) over the
frequency content of the pump (just like it is in the case of the state
\ref{PULSED}):
\begin{eqnarray}
& A (t,t^{\prime})  & \nonumber\\
&=\int d\omega_{o}F_{o}(\omega_{o})e^{-\omega_{o}t^{\prime
}}\int d\omega e^{-i\omega(t-t^{\prime})}f_{2}(\omega_{o}-\omega)f_{1}(\omega
)&\nonumber\\
&  =\int dt_{p}F_{o}(t_{p})f_{1}(t-t_{p})f_{2}%
(t^{\prime}-t_{p}).&\nonumber\\
\end{eqnarray}
where, e.g., $F_{o}(t)$ is the Fourier transformation (time profile) of
$F_{o}(\omega)$. Namely, the time correlation of the detections is defined by
the resolution of the respective filters, while the events happen at times
dictated by the pulse. This clearly visible in the case of no filters, and broad band radiation. The (unphysical) limiting case
is reached by replacing  $f$'s by $\delta(t-t_p)$ and $\delta(t'-t_p)$. This gives $F(t)\delta(t'-t)$.

\section*{Reference}

%\bibitem {Aerts}
Aerts S., P.G. Kwiat, J.-A. Larsson, M. Zukowski, 1999, Phys. Rev. Lett. \textbf{ 83}, 2872.

%\bibitem{aoki09}
Aoki, T., Go Takahashi, Tadashi Kajiya, Jun-ichi Yoshikawa, Samuel L. Braunstein, Peter
van Loock and Akira Furusawa, Nature Physics 5, 541 (2009)

%\bibitem {Akopian-dot}
Akopian, N., N.H. Lindner, E. Poem, Y. Berlatzky, J.
Avron, and D. Gershoni, B. D. Gerardot, and P. M. Petroff, 2006, Phys. Rev.
Lett. \textbf{96}, 130501.

%\bibitem {Alber}
Alber, G., T. Beth, M. Horodecki, P. Horodecki, R. Horodecki,
M. Rotteler, H. Weinfurter, R. Werner, and A. Zeilinger, 2001, \textit{Quantum
Information: An Introduction to Basic Theoretical Concepts and Experiments}
(Springer-Verlag, Berlin/Heidelberg).

%\bibitem {shihalley}
Alley C.O., and Y.H. Shih, 1987, \emph{Proc. of 2nd Int.
Symp. Foundations of Quantum Mechanics}, ed. M. Namiki (Physical Society of
Japan, Tokyo).

%\bibitem{ultrabright}
Altepeter, J.B., E.R. Jeffrey, and P.G. Kwiat, 2005, Opt. Exp. \textbf{13}, 8951.

%\bibitem{six-singlet}
Amselem E., M. Radmark, M. Bourennane, A. Cabello, 2009, Phys. Rev. Lett. \textbf{103}  160405.

%\bibitem {Ardehali}
Ardehali, M., 1992, Phys. Rev. A \textbf{46}, 5375.

%\bibitem {Aspect99}
Aspect, A., 1999, Nature (London) \textbf{390}, 189.

%\bibitem[Aspect \textit{et al}., 1982b]{aspect82b}
Aspect, A., J. Dalibard, and G. Roger, 1982a, Phys. Rev. Lett. \textbf{49}, 1804.

%\bibitem[Aspect \textit{et al}., 1981]{aspect81}
Aspect, A., P. Grangier, and
G. Roger, 1981, Phys. Rev. Lett. \textbf{47}, 460.

%\bibitem[Aspect \textit{et al}., 1982a]{aspect82a}
Aspect, A., P. Grangier, and
G. Roger, 1982b, Phys. Rev. Lett. \textbf{49}, 91.

%\bibitem {Aspelmeyer}
Aspelmeyer, M., H. R. B\"{o}hm, T. Gyatso, T. Jennewein,
R. Kaltenbaek, M. Lindenthal, G. Molina-Terriza, A. Poppe, K. Resch, M.
Taraba, R. Ursin, P. Walther, and A. Zeilinger, 2003a, Science \textbf{301}, 621.

%\bibitem {Aspelmeyer03}
Aspelmeyer, M., T. Jennewein, M. Pfenningbauer, W.R.
Leeb, and A. Zeilinger, 2003b, IEEE J. Sel. Top. Quantum Electron. \textbf{9}, 1541.

%\bibitem {Lukin03}
Bajcsy, M., A.S. Zibrov, and M.D. Lukin, 2003, Nature
(London) \textbf{426}, 638.

%\bibitem {Bao2007}
Bao, X.-H., T.-Y. Chen, Q. Zhang, J. Yang, H. Zhang, T.
Yang, and J.-W. Pan, 2007, Phys. Rev. Lett. \textbf{98}, 170502.

%\bibitem {Bao08}
Bao, X.-H., Y. Qian, J. Yang, H. Zhang, Z.-B. Chen, T. Yang,
and J.-W. Pan, 2008, Phys. Rev. Lett. \textbf{101}, 190501.

%\bibitem {mattini-hyper}
Barbieri, M., C. Cinelli, P. Mataloni, and F. De
Martini, 2005, Phys. Rev. A \textbf{72}, 052110.

%\bibitem{barbieri09}
Barbieri M., T. J. Weinhold, B. P. Lanyon, A. Gilchrist, K. J. Resch, M. P.
Almeida and A. G. White, J. Mod. Opt. 56, 209 (2009).

%\bibitem {barnett89}
Barnett, S.M., J.~Jeffers, A.~Gatti, and R.~Loudon, 1989,
Phys. Rev. A, \textbf{57} 2134.

%\bibitem{barz10}
Barz, S., G. Cronenberg, A. Zeilinger and P. Walther, (2010), Nature Photonics 4, 553.

%\bibitem {hyperKwiat}
Barreiro, J.T., N.K. Langford, N.A. Peters, and P.G.
Kwiat, 2005, Phys. Rev. Lett. \textbf{95}, 260501.

%\bibitem {Barreiro08}
Barreiro, J.T., T.-C. Wei, and P.G. Kwiat, 2008, Nature
Phys. \textbf{4}, 282.

%\bibitem {tele-ion}
Barrett, M.D., J. Chiaverini, T. Schaetz, J. Britton, W.M.
Itano, J.D. Jost, E. Knill, C. Langer, D. Leibfried, R. Ozeri, D.J. Wineland,
2004, Nature (London) \textbf{429}, 737.

%\bibitem {BKok}
Barrett, S.D. and P. Kok, 2005, Phys. Rev. A \textbf{71}, 060310(R).

%\bibitem {Belinskii}
Belinskii, A.V. and D.N. Klyshko, 1993, Phys. Usp.
\textbf{36}, 653.

%\bibitem[Bell, 1964]{Bell}
Bell, J.S., 1964, Physics (Long Island City, N.Y.)
\textbf{1}, 195.

%\bibitem {Bell66}
Bell, J.S., 1966, Rev. Mod. Phys. \textbf{38}, 447.

%\bibitem {Bellbook}
Bell, J.S., 1987, \textit{Speakable and Unspeakable in
Quantum Mechanics} (Cambridge University Press, New York).

%\bibitem {Benhelm}
Benhelm, J., G. Kirchmair, C.F. Roos, and R. Blatt, 2008,
Nature Phys. \textbf{4}, 463.

%\bibitem {Bennett-QD}
Bennett, A.J. , D.C. Unitt, P. Atkinson, D.A. Ritchie,
and A.J. Shields, 2005, Opt. Express \textbf{13}, 50.

%\bibitem {BB84}
Bennett, C.H. and G. Brassard, 1984, in \textit{Proceedings of
the IEEE International Conference on Computers, Systems and Singal Proceeding,
Bangalore, India} (IEEE, New York), 175.

%\bibitem {six-person}
Bennett, C.H., G. Brassard, C. Cr\'{e}peau, R. Jozsa, A.
Peres, and W.K. Wootters, 1993, Phys. Rev. Lett. \textbf{70}, 1895.

%\bibitem {bennett96b}
Bennett, C.H., H.J. Bernstein, S. Popescu, and B.
Schumacher, 1996a, Phys. Rev. A \textbf{53}, 2046.

%\bibitem {Pur-Bennett}
Bennett, C.H., G. Brassard, S. Popescu, B. Schumacher,
J.A. Smolin, and W.K. Wootters, 1996b, Phys. Rev. Lett. \textbf{76}, 722 -
Erratum: 1997, Phys. Rev. Lett. \textbf{78}, 2031.

%\bibitem {QEC-Bennett96}
Bennett, C.H., D.P. DiVincenzo, J.A. Smolin, and W.K.
Wootters, 1996c, Phys. Rev. A \textbf{54}, 3824.

%\bibitem {Bennett-DiV}
Bennett, C.H. and D.P. DiVincenzo, 2000, Nature (London)
\textbf{404}, 247.

%\bibitem {denseCod92}
Bennett, C.H. and S.J. Wiesner, 1992, Phys. Rev. Lett.
\textbf{69}, 2881.

%\bibitem {Bennett-RSP}
Bennett, C.H., D.P. DiVincenzo, P.W. Shor, J.A. Smolin,
B.M. Terhal, and W.K. Wootters, 2001, Phys. Rev. Lett. \textbf{87}, 077902.

%\bibitem{bernstein}
Bernstein, H.J., Greenberger, DM; Horne, MA, Zeilinger, A,
Phys. Rev. A 47, 78-84 (1993).

%\bibitem{beugnon06}
Beugnon, J., Jones, M. P. A., Dingjan, J., Darquie, B., Messin, G., Browaeys, A., and
Grangier, P., (2006) Nature 440, 779.

%\bibitem {BB1975}
Bialynicki-Birula, I. and Z Bialynicka-Birula, 1975,
\textit{Quantum Electrodynamics} (Pergamon, Oxford).

%%\bibitem {BlattW}
Blatt, R. and D. Wineland, 2008, Nature (London)
%\textbf{453}, 1008.

%\bibitem {atom-photon04}
Blinov, B.B., D.L. Moehring, L.-M. Duan, and C.
Monroe, 2004, Nature (London) \textbf{428}, 153.

%\bibitem {Bodiya}
Bodiya, T.P. and L.-M. Duan, 2006, Phys. Rev. Lett.
\textbf{97}, 143601.

%\bibitem {Bohm}
Bohm, D., 1951, \textit{Quantum Theory} (Prentice Hall,
Englewood Cliffs, NJ).

%\bibitem {hardy97}
Boschi, D., S. Branca, F. De Martini, and L. Hardy, 1997,
Phys. Rev. Lett. \textbf{79}, 2755.

%\bibitem {TeleEXP98}
Boschi,\ D., S. Branca, F. De Martini, L. Hardy, and S.
Popescu, 1998, Phys. Rev. Lett. \textbf{80}, 1121.

%\bibitem {bose-decay}
Bose, S., P.L. Knight, M.B. Plenio, and V. Vedral, 1999,
Phys. Rev. Lett. \textbf{83}, 5158.

%\bibitem {bose98}
Bose, S., V. Vedral, and P.L. Knight, 1998, Phys. Rev. A
\textbf{57}, 822.

%\bibitem {Boto}
Boto, A.N., P. Kok, D.S. Abrams, S.L. Braunstein, C.P.
Williams, and J.P. Dowling, 2000, Phys. Rev. Lett. \textbf{85}, 2733.

%\bibitem{Bourennane-witness}
Bourennane M., M. Eibl, Ch. Kurtsiefer, S. Gaertner,
H. Weinfurter, O. G\"{u}hne, Ph. Hyllus, D. Bru{\ss}, M. Lewenstein, A. Sanpera, 2004a, Phys. Rev. Lett. \textbf{92}, 087902.

%\bibitem {dfs-Bourennane}
Bourennane, M., M. Eibl, S. Gaertner, C. Kurtsiefer,
A. Cabello, and H. Weinfurter, 2004b, Phys. Rev. Lett. \textbf{92}, 107901.

%\bibitem {TeleEXP97}
Bouwmeester, D., J.-W. Pan, K. Mattle, M. Eibl, H.
Weinfurter, and A. Zeilinger, 1997, Nature (London) \textbf{390}, 575.
%%\bibitem {ereject}Bouwmeester, D., 2001, Phys. Rev. A \textbf{63}, 040301(R).

%\bibitem {TeleEXP97reply}
Bouwmeester, D., J.-W. Pan, M. Daniell, H.
Weinfurter, M. \.{Z}ukowski, and A. Zeilinger, 1998, Nature (London)
\textbf{394}, 841.

%\bibitem {GHZ-3p}
Bouwmeester, D., J.-W. Pan, M. Daniell, H. Weinfurter, and A.
Zeilinger, 1999a, Phys. Rev. Lett. \textbf{82}, 1345.

%\bibitem {dik99}
Bouwmeester, D., J.-W. Pan, H. Weinfurter, and A. Zeilinger,
1999b, J. Mod. Opt. \textbf{47}, 279.

%\bibitem {BEZ}
Bouwmeester, D., A. Ekert, and A. Zeilinger (eds.), 2001,
\textit{The Physics of Quantum Information} (Springer-Verlag, Berlin/Heidelberg).

%\bibitem {highNOON}
Bouwmeester, D., 2004, Nature (London) \textbf{429}, 139.

%%\bibitem {nonl-R08}
Branciard, C., N. Brunner, N. Gisin, C. Kurtsiefer, A.
%Lamas-Linares, A. Ling, and V. Scarani, 2008, quant-ph/0801.2241.

%\bibitem{Braig}
 Braig, C., P. Zarda, C. Kurtsiefer and H. Weinfurter, 2002, App. Phys. B 76, 113

%\bibitem {nonl-R-G}
Branciard, C., A. Ling, N. Gisin, C. Kurtsiefer, A.
Lamas-Linares, and V. Scarani, 2007, Phys. Rev. Lett. \textbf{99}, 210407.

%\bibitem {branning}
Branning, D., W.P. Grice, R. Erdmann, and I.A. Walmsley,
1999, Phys. Rev. Lett \textbf{83}, 955.

%\bibitem {branning2000}
Branning, D., W.P. Grice, R. Erdmann, and I.A.
Walmsley, 2000, Phys. Rev. A \textbf{62}, 013814.

%%\bibitem {NMR-none}
Braunstein, S.L., C.M. Caves, R. Jozsa, N. Linden, S.
%Popescu, and R. Schack, 1999, Phys. Rev. Lett. \textbf{83}, 1054.

%\bibitem {TeleEXP97comment}
Braunstein, S.L. and H.J. Kimble, 1998, Nature
(London) \textbf{394}, 840.

%%\bibitem {bm-bsm}
Braunstein, S.L. and A. Mann, 1995, Phys. Rev. A \textbf{51}, R1727.

%\bibitem {cv-rmp}
Braunstein, S.L. and P. van Loock, 2005, Rev. Mod. Phys.
\textbf{77}, 513.

%\bibitem {Brendel92}
Brendel, J., E. Mohler, and W. Martiennsen, 1992,
Europhys. Lett. \textbf{20}, 575.

%\bibitem {timebin99}
Brendel, J., N. Gisin, W. Tittel, and H. Zbinden, 1999,
Phys. Rev. Lett. \textbf{82}, 2594.

%\bibitem {Briegel09}
Briegel, H.-J., D. E. Browne, W. D\"{u}r, R. Raussendorf,
and M. Van den Nest, 2009, Nature Phys. \textbf{5}, 19.

%\bibitem {Repeater}
Briegel, H.-J., W. D\"{u}r, J.I. Cirac, and P. Zoller,
1998, Phys. Rev. Lett. \textbf{81}, 5932.

%\bibitem {cluster}
Briegel, H.J. and R. Raussendorf, 2001, Phys. Rev. Lett.
\textbf{86}, 910.

%\bibitem {brouri}
Brouri, R., A. Beveratos, J.-Ph. Poizat, and P. Grangier,
2000, Opt. Lett. \textbf{25}, 1294.

%\bibitem {ion-03robust}
Browne, D.E., M.B. Plenio, and S.F. Huelga, 2003, Phys.
Rev. Lett. \textbf{91}, 067901.

%\bibitem {Rudolph05}
Browne, D.E. and T. Rudolph, 2005, Phys. Rev. Lett.
\textbf{95}, 010501.

%%\bibitem {Brukner04}
Brukner, \v{C}., M. \.{Z}ukowski, J.-W. Pan, and A.
%Zeilinger, 2004, Phys. Rev. Lett. \textbf{92}, 127901.

%\bibitem {SP-mole}
Brunel, C., B. Lounis, P. Tamarat, and M. Orrit, 1999, Phys.
Rev. Lett. \textbf{83}, 2722.

%\bibitem {bruss02}
Bru\ss , D., J.I. Cirac, P. Horodecki, F. Hulpke, B. Kraus,
M. Lewenstein, and A. Sanpera, 2002, J. Mod. Opt. \textbf{49}, 1399.

%%\bibitem {bruss}
Bru\ss , D., A. Ekert, S.F. Huelga, J.-W. Pan, and A.
%Zeilinger, 1997, Phil. Trans. R. Soc. Lond. A \textbf{355}, 2259.

%\bibitem {SPDC70}
Burnham, D.C. and D.L. Weinberg, 1970, Phys. Rev. Lett.
\textbf{25}, 84.

%\bibitem{cabello96}
Cabello A. and E. Santos. Phys. Lett. A 214 (1996).

%\bibitem {Cabello-Hardy}
Cabello, A., 2001a, Phys. Rev. Lett. \textbf{86}, 1911.

%\bibitem {Cabello-GHZ}
Cabello, A., 2001b, Phys. Rev. Lett. \textbf{87}, 010403.

%\bibitem {Cabello-reply}
Cabello, A., 2003, Phys. Rev. Lett. \textbf{90}, 258902.

%\bibitem{cabello08}
Cabello, A (2008) Phys. Rev. Lett. 102, 040401

%\bibitem {zoller99}
Cabrillo, C., J.I. Cirac, P. Garc\'{\i}a-Fern\'{a}ndez, and
P. Zoller, 1999, Phys. Rev. A \textbf{59}, 1025.

%\bibitem {QEC-CS}
Calderbank, A.R. and P.W. Shor, 1996, Phys. Rev. A
\textbf{54}, 1098.

%\bibitem {antibunching}
Carmichael, H.J. and D.F. Walls, 1976, J. Phys. B
\textbf{9}, L43; J. Phys. B \textbf{9}, 1199.

%\bibitem {Carteret}
Carteret, H.A., A. Higuchi, and A. Sudbery, 2000, J. Math.
Phys. \textbf{41}, 7932

%\bibitem {cerf98}
Cerf, N.J., C.~Adami, and P.G. Kwiat, 1998, Phys. Rev. A,
\textbf{57} R1477.

%%\bibitem {cerf02}
Cerf, N.J., M. Bourennane, A. Karlsson, and N. Gisin, 2002, Phys. Rev. Lett. \textbf{88}, 127902.

%\bibitem {Chane06}
Chaneliere, T., D.N. Matsukevich, S.D. Jenkins, T.A.B.
Kennedy, M.S. Chapman, and A. Kuzmich, 2006, Phys. Rev. Lett. \textbf{96}, 093604.

%\bibitem {Chane07}
Chaneliere, T., D. N. Matsukevich, S. D. Jenkins, S.-Y.
Lan, R. Zhao, T. A. B. Kennedy, and A. Kuzmich, 2007, Phys. Rev. Lett.
\textbf{98}, 113602.

%\bibitem {storage-single-cold}
Chaneliere, T., D.N. Matsukevich, S.D.
Jenkins, S.-Y. Lan, T.A.B. Kennedy, and A. Kuzmich, 2005, Nature (London)
\textbf{438}, 833.

%\bibitem {chenqin}
Chen, Q., J. Cheng, K.-L. Wang, and J.F. Du, 2006a, Phys.
Rev. A \textbf{73}, 012303.

%\bibitem {shuai}
Chen, S., Y.-A. Chen, T. Strassel, Z.-S. Yuan, B. Zhao, J.
Schmiedmayer, and J.-W. Pan, 2006b, Phys. Rev. Lett. \textbf{97}, 173004.

%\bibitem {third-man}
Chen, Y.-A., A.-N. Zhang, Z. Zhao, X.-Q. Zhou, C.-Y. Lu,
C.-Z. Peng, T. Yang, J.-W. Pan, 2005, Phys. Rev. Lett. \textbf{95}, 200502.

%%\bibitem {Error-rej}
Chen, Y.-A., A.-N. Zhang, Z. Zhao, X.-Q. Zhou, and J.-W.
%Pan, 2006c, Phys. Rev. Lett. \textbf{96}, 220504.

%\bibitem {GHZ-Chen}
Chen, Z.-B., J.-W. Pan, Y.-D. Zhang, \v{C}. Brukner, and A.
Zeilinger, 2003, Phys. Rev. Lett. \textbf{90}, 160408.

%\bibitem {ChenQC06}
Chen, Z.-B., Q. Zhang, X.-H. Bao, J. Schmiedmayer, and
J.-W. Pan, 2006d, Phys. Rev. A \textbf{73}, 050302(R).

%\bibitem {chenzhao}
Chen, Z.-B., B. Zhao, Y.-A. Chen, J. Schmiedmayer, and
J.-W. Pan, 2007, Phys. Rev. A \textbf{76}, 022329.

%\bibitem {chenzhao06b}
Chen, Z.-B., B. Zhao, and J.-W. Pan, 2008, to be published.

%\bibitem {Childress-PRA}
Childress, L., J.M. Taylor, A.S. S\o rensen, and M.D.
Lukin, 2005, Phys. Rev. A \textbf{72}, 052330.

%\bibitem {Childress}
Childress, L., J.M. Taylor,, A.S. S\o rensen, and M.D.
Lukin, 2006, Phys. Rev. Lett. \textbf{96}, 070504.

%%\bibitem {Childs}
Childs, A.M., D.W. Leung, and M.A. Nielsen, 2005, Phys. Rev.
%A \textbf{71}, 032318.

%%\bibitem {ChoLee}
Cho, J. and H.W. Lee, 2005, Phys. Rev. Lett. \textbf{95}, 160501.

%\bibitem {Choi08}
Choi, K.S., H. Deng, J. Laurat, and H.J. Kimble, 2008, Nature
(London) \textbf{452}, 67.

%\bibitem {Chou07}
Chou, C.W., J. Laurat, H. Deng, K.S. Choi, H. de Riedmatten,
D. Felinto, and H.J. Kimble, 2007, Science \textbf{316}, 1316.

%\bibitem {Kimble04PRL}
Chou, C.W., S.V. Polyakov, A. Kuzmich, and H.J. Kimble,
2004, Phys. Rev. Lett. \textbf{92}, 213601.

%\bibitem {Chuang-Nielsen}
Chuang, I.L. and M.A. Nielsen, 1997, J. Mod. Opt.
\textbf{44}, 2455.

%\bibitem {Cinelli}
Cinelli, C., M. Barbieri, R. Perris, P. Mataloni, and F. De
Martini, 2005, Phys. Rev. Lett. \textbf{95}, 240405.

%\bibitem {Cinelli04}
Cinelli, C., G. Di Nepi, F. De Martini, M. Barbieri, and
P. Mataloni, 2004, Phys. Rev. A \textbf{70}, 022321.

%\bibitem {Cirac-Zoller94}
Cirac, J.I. and P. Zoller, 1994, Phys. Rev. A
\textbf{50}, R2799.

%\bibitem {network97}
Cirac, J.I., P. Zoller, H.J. Kimble, and H. Mabuchi, 1997,
Phys. Rev. Lett. \textbf{78}, 3221.

%\bibitem {cirelson}
Cirel'son, B.S., 1980, Lett. Math. Phys. \textbf{4}, 93.

%\bibitem {CH}
Clauser, J.F., M.A. Horne, 1974,
Phys. Rev. D \textbf{10}, 526.
%\bibitem {chsh}Clauser, J.F., M.A. Horne, A. Shimony, and R.A. Holt, 1969,
Phys. Rev. Lett. \textbf{23}, 880.

%\bibitem {clauser}
Clauser, J.F. and A. Shimony, 1978, Rep. Prog. Phys.
\textbf{41}, 1881.

%\bibitem{coffman}
Coffman, V., Kundu, J. and Wootters, W. K. Phys. Rev. A 61, 052306 (2000).

%%\bibitem {qssPRL}
Cleve, R., D. Gottesman, and H.-K. Lo, 1999, Phys. Rev. Lett.
%\textbf{83}, 648.

%%\bibitem {cory}
Cory, D.G., M.D. Price, W. Maas, E. Knill, R. Laflamme, W.H.
%Zurek, T.F. Havel, and S.S. Somaroo, 1998, Phys. Rev. Lett. \textbf{81}, 2152.

%\bibitem {grangier05}
Darqui\'{e}, B., M.P.A. Jones, J. Dingjan, J. Beugnon, S.
Bergamini, Y. Sortais, G. Messin, A. Browaeys, and P. Grangier, 2005, Science
\textbf{309}, 454.

%%\bibitem {Nielsen06}
Dawson, C.M., H.L. Haselgrove, and M.A. Nielsen, 2006a,
%Phys. Rev. Lett. \textbf{96}, 020501.

%\bibitem {Nielsen06pra}
Dawson, C.M., H.L. Haselgrove, and M.A. Nielsen, 2006,
Phys. Rev. A \textbf{73}, 052306.

%%\bibitem {tb93}
De Barros M.R.X. and P.C. Becker, 1993, Opt. Lett. \textbf{18}, 631.

%%\bibitem {garuccio}
De Caro, L. and A. Garuccio, 1994, Phys. Rev. A
%\textbf{50}, R2803.

%%\bibitem {cBSM}
DelRe, E., B. Crosignani, and P. Di Porto, 2000, Phys. Rev.
%Lett. \textbf{84}, 2989.

%\bibitem {Gisin-tb}
de Riedmatten, H., I. Marcikic, W. Tittel, H. Zbinden, and
N. Gisin, 2003, Phys. Rev. A \textbf{67}, 022301.

%\bibitem {Pur-Deutsch}
Deutsch, D., A. Ekert, R. Jozsa, C. Macchiavello, S.
Popescu, and A. Sanpera, 1996, Phys. Rev. Lett. \textbf{77}, 2818 [Erratum:
1998, Phys. Rev. Lett. \textbf{80}, 2022].

%%\bibitem {Devoret}
Devoret, M.H. and J.M. Martinis, 2004, Quantum Inf.
%Processing \textbf{3}, 163.

%\bibitem {Dewdney}
Dewdney, C., P.R. Holland, and A. Kyprianidis, 1987, J.
Phys. A: Math. Gen. \textbf{20}, 4717.

%\bibitem {Walther-antiB}
Diedrich, F. and H. Walther, 1987, Phys. Rev. Lett.
\textbf{58}, 203.

%%\bibitem {Dieks}
Dieks, D., 1982, Phys. Lett. A \textbf{92}, 271.

%\bibitem {femto97}
Di Giuseppe, G., L. Haiberger, F. De Martini, A.V.
Sergienko, 1997, Phys. Rev. A \textbf{56}, R21.

%\bibitem {Dirac}
Dirac, P.A.M., 1927, Proc. R. Soc. London A \textbf{114}, 243.

%\bibitem {DiV}
DiVincenzo, D.P., 2000, Fortschr. Phys. \textbf{48}, 771.

%\bibitem {Divochiy}
Divochiy, A., F. Marsili, D. Bitauld, A. Gaggero, R. Leoni,
F. Mattioli, A. Korneev, V. Seleznev, N. Kaurova, O. Minaeva, G. Gol'tsman,
K.G. Lagoudakis, M. Benkhaoul, F. L\'{e}vy, and A. Fiore, 2008, Nature Photon.
\textbf{2}, 302.

%\bibitem{Ultrabright-Dousse10}
Dousse, A.,  J. Suffczynski,  A. Beveratos,  O. Krebs,  A. Lematre,  I. Sagnes,  J. Bloch,  P. Voisin  P. Senellart, Nature,
\textbf{466}, 217 (2010).

%\bibitem {Dowling98}
Dowling, J.P., 1998, Phys. Rev. A \textbf{57}, 4736.

%%\bibitem {Du08}
Du, S., P. Kolchin, C. Belthangady, G.Y. Yin, and S.E. Harris,
%2008, Phys. Rev. Lett. \textbf{100}, 183603.

%%\bibitem {dsf-duan}
Duan, L.-M. and G.C. Guo, 1997, Phys. Rev. Lett.
%\textbf{79}, 1953.

%\bibitem {DLCZ}
Duan, L.-M., M.D. Lukin, J.I. Cirac, and P. Zoller, 2001,
Nature (London) \textbf{414}, 413.

%\bibitem {DLCZ-pra}
Duan, L.-M., J.I. Cirac, and P. Zoller, 2002, Phys. Rev. A
\textbf{66}, 023818.

%\bibitem {Duan-ow}
Duan, L.-M. and R. Raussendorf, 2005, Phys. Rev. Lett.
\textbf{95}, 080503.

%\bibitem {duer03}
D\"{u}r, W. and H.-J. Briegel, 2003, Phys. Rev. Lett.
\textbf{90}, 067901.

%\bibitem {duer07}
D\"{u}r, W. and H.-J. Briegel, 2007, Rep. Prog. Phys.
\textbf{70}, 1381.

%\bibitem {duer}
D\"{u}r, W., H.-J. Briegel, J.I. Cirac, and P. Zoller, 1999,
Phys. Rev. A \textbf{59}, 169 [Erratum: 1999, Phys. Rev. A \textbf{60}, 725].

%\bibitem {Eberhard}
Eberhard, P.H., 1993, Phys. Rev. A \textbf{47}, R747.

%\bibitem {Edamatsu}
Edamatsu, K., R. Shimizu, and T. Itoh, 2002, Phys. Rev.
Lett. \textbf{89}, 213601.

%\bibitem {eibl}
Eibl, M., S. Gaertner, M. Bourennane, C. Kurtsiefer, M.
\.{Z}ukowski, and H. Weinfurter, 2003, Phys. Rev. Lett. \textbf{90}, 200403.

%\bibitem {Einstein1905}
Einstein, A., 1905, Ann. Phys. \textbf{17}, 132.

%\bibitem {EPR}
Einstein, A., B. Podolsky, and N. Rosen, 1935, Phys. Rev.
\textbf{47}, 777.

%\bibitem {storage-single}
Eisaman, M.D., A. Andr\'{e}, F. Massou, M.
Fleischhauer, A.S. Zibrov, and M.D. Lukin, 2005, Nature (London) \textbf{438}, 837.

%\bibitem {Eisaman04}
Eisaman, M.D., L. Childress, A. Andr\'{e}, F. Massou, A.S.
Zibrov, and M.D. Lukin, 2004, Phys. Rev. Lett. \textbf{93}, 233602.

%\bibitem {N-entangle-Twin}
Eisenberg, H.S., G. Khoury, G.A. Durkin, C. Simon,
and D. Bouwmeester, 2004, Phys. Rev. Lett. \textbf{93}, 193901.

%\bibitem {Ekert91}
Ekert, A.K., 1991, Phys. Rev. Lett. \textbf{67}, 661.

%\bibitem{elitzur93}
Elitzur A.C., Vaidman L., Found. Phys. 23, 987 (1993)

%%\bibitem {Engel}
Engel, H.-A., L.P. Kouwenhoven, D. Loss, and C.M. Marcus,
%2004, Quantum Inf. Processing \textbf{3}, 115.

%\bibitem{Englert2001}
Englert B.-G., C. Kurtsiefer,and H. Weinfurter, Phys. Rev. A, \textbf{63}, 032303 (2001).

%\bibitem {Englund}
Englund, D., A. Faraon, B. Zhang, Y. Yamamoto, and J.
Vu\v{c}kovi\'{c}, 2007, Opt. Express \textbf{15}, 5550.

%\bibitem{fatal04}
Fattal, D., Kyo Inoue, Jelena Vuckovic, Charles Santori, Glenn S. Solomon, and Yoshihisa
Yamamoto, (2004) Phys. Rev. Lett. 92, 037903

%\bibitem {fearn87}
Fearn H. and R.~Loudon, 1987, Opt. Commun. \textbf{64} 485.

%\bibitem {Fed09}
Fedrizzi, A., R. Ursin, T. Herbst, M. Nespoli, R. Prevedel, T.
Scheidl, F. Tiefenbacher, T. Jennewein, and A. Zeilinger, 2009, Nature Phys.
\textbf{5}, 389.

%\bibitem{Fedrizzi}
 Fedrizzi, A, T. Herbst, A. Poppe, T. Jennewein, A. Zeilinger, Opt. Express 15, 15377 (2007)

%\bibitem {Felinto}
Felinto, D., C.W. Chou, J. Laurat, E.W. Schomburg, H. de
Riedmatten, and H.J. Kimble, 2006, Nature Phys. \textbf{2}, 844.

%\bibitem {robust-atom}
Feng, X.-L., Z.-M. Zhang, X.-D. Li, S.-Q. Gong, and
Z.-Z. Xu, 2003, Phys. Rev. Lett. \textbf{90}, 217902.

%\bibitem {Feynman64}
Feynman, R.P., R.B. Leighton, and M. Sands, 1963,
\textit{The Feynman Lectures on Physics III} (Addison-Wesley, Reading, Mass.).

%\bibitem {SP-swap}
Fiorentino, M., T. Kim, and F.N.C. Wong, 2005, Phys. Rev. A
\textbf{72}, 012318.

%\bibitem {SP-CNOT}
Fiorentino, M. and F.N.C. Wong, 2004, Phys. Rev. Lett.
\textbf{93}, 070502.

%\bibitem {EIT-RMP}
Fleischhauer, M., A. Imamo\={g}lu, and J.P. Marangos, 2005,
Rev. Mod. Phys. \textbf{77}, 633.

%\bibitem {mem-DSP}
Fleischhauer, M. and M.D. Lukin, 2000, Phys. Rev. Lett.
\textbf{84}, 5094.

%\bibitem {mem-pra}
Fleischhauer, M. and M.D. Lukin, 2002, Phys. Rev. A
\textbf{65}, 022314.

%%\bibitem {Fowler}
Fowler A.G. and K. Goyal, 2008, arXiv:0805.3202v2.

%\bibitem {Franson89}
Franson, J.D., 1989, Phys. Rev. Lett. \textbf{62}, 2205.

%\bibitem {FC}
Freedman, S.J. and J.F. Clauser, 1972, Phys. Rev. Lett.
\textbf{28}, 938.

%\bibitem{fulconis07}
Fulconis, J., O. Alibart, J. L. O'Brien, W. J. Wadsworth,
and J. G. Rarity, Phys. Rev. Lett. 99, 120501 (2007)

%\bibitem {tele-cv}
Furusawa, A., J.L. S\o rensen, S.L. Braunstein, C.A. Fuchs,
H.J. Kimble, and E.S. Polzik, 1998, Science \textbf{282}, 706.

%\bibitem{Garay07}
Garay-Palmett K., H.J. McGuinness, O. Cohen, et al. Opt. Expr. 15 14870-14886 (2007)

%%\bibitem {Gao09}
Gao, W.-B., A.G. Fowler, R. Raussendorf, X.-C. Yao, H. Lu, P.
%Xu, C.-Y. Lu, C.-Z. Peng, Y. Deng, Z.-B. Chen, and J.-W. Pan, 2009, arXiv:0905.1542v2.

%\bibitem{gao10a}
Gao, W.-B., C.-Y. Lu, X.-C. Yao, P.Xu, O. Guhne, Y.-A. Chen, C.-Z. Peng,
Z.-B. Chen, and J.-W. Pan, Nature Phys. 6,  331 (2010a)

%\bibitem{gao10b}
Gao, W.-B., X.-C. Yao, P.Xu, O. Guhne, C.-Y. Lu, C.-Z. Peng,
Z.-B. Chen, and J.-W. Pan, Phys. Rev. Lett. 104, 020501 (2010b)

%\bibitem {Garg}
Garg A. and N.D. Mermin, 1987, Phys. Rev. A \textbf{35}, 3831.

%\bibitem {CNOT-Vienna}
Gasparoni, S., J.-W. Pan, P. Walther, T. Rudolph, and A.
Zeilinger, 2004, Phys. Rev. Lett. \textbf{93}, 020504.

%\bibitem {Genovese}
Genovese, M., 2005, Phys. Rep. \textbf{413}, 319.

%\bibitem {timepol}
Genovese, M. and C. Novero, 2002, Eur. Phys. J. D
\textbf{21}, 109.

%%\bibitem {gilbert}
Gilbert, G., M. Hamrick, and Y.S. Weinstein, 2006, Phys.
%Rev. A \textbf{73}, 064303.

%\bibitem {Gilchrist}
Gilchrist, A., A.J.F. Hayes, and T.C. Ralph, 2007, Phys.
Rev. A \textbf{75}, 052328.

%\bibitem {Gill2002}
Gill, R.D., G. Weihs, A. Zeilinger, and M. \.{Z}ukowski,
2002, Proc. Nat. Acad. Sci. \textbf{99}, 14632.

%\bibitem {gisin-filter}
Gisin, N., 1996, Phys. Lett. A \textbf{210}, 151.

%\bibitem {gisin98}
Gisin, N. and H. Bechmann-Pasquinucci, 1998, Phys. Lett. A
\textbf{246}, 1.

%\bibitem {gisin-peres}
Gisin N. and A. Peres, 1992, Phys. Lett. A \textbf{162}, 15.

%\bibitem {QC-RMP}
Gisin, N., G. Ribordy, W. Tittel, and H. Zbinden, 2002, Rev.
Mod. Phys. \textbf{74}, 145.

%\bibitem {Glauber}
Glauber, R.J., 1963, Phys. Rev. \textbf{130}, 2529.

%\bibitem {Geobel08}
Goebel, A.M., C. Wagenknecht, Q. Zhang, Y.-A. Chen, K.
Chen, J. Schmiedmayer, and J.-W. Pan, 2008, Phys. Rev. Lett. \textbf{101}, 080403.

%%\bibitem {Gottesman-PhD}
Gottesman, D., 1997, Ph.D. thesis (California
%Institute of Technology, Pasadena, CA).

%\bibitem{Gong10}
Gong, Y.-X., X.-B. Zou, T. C. Ralph, S.-N. Zhu, and G.-C. Guo, Phys. Rev. A 81, 052303 (2010)

%\bibitem {Gottesman-Chuang}
Gottesman, D., and I.L. Chuang, 1999, Nature
(London) \textbf{402}, 390.

%\bibitem{grassl97}
Grassl M, Beth T, Pellizari T (1997) Phys Rev A 56, 33

%\bibitem {Grangier}
Grangier, P., G. Roger, and A. Aspect, 1986, Europhys.
Lett. \textbf{1}, 173.

%\bibitem {GHZ1990}
Greenberger,\ D.M., M.A. Horne, A. Shimony, and A.
Zeilinger, 1990, Am. J. Phys. \textbf{58}, 1131.

%\bibitem {GHZ}
Greenberger,\ D.M., M.A. Horne, and A. Zeilinger, 1989, in
\textit{Bell's Theorem, Quantum Theory, and Conceptions of the Universe},
edited by M. Kafatos (Kluwer Academic, Dordrecht).

%\bibitem {GHZ94}
Greenberger, D.M., M.A. Horne, and A. Zeilinger, 1993, Phys.
Today \textbf{46} (8), 22.

%\bibitem{Grespi}
Crespi, A., R. Ramponi, R. Osselame, L. Sansoni, I. Bongioanni, F. Sciarrino,G.Vallone, P. Mataoni, 2011, arxiv: quant-ph/1105.1454

%\bibitem {femto98}
Grice, W.P., R. Erdmann, I.A. Walmsley, and D. Branning,
1998, Phys. Rev. A \textbf{57}, R2289.

%\bibitem {femto}
Grice, W.P. and I.A. Walmsley, 1997, Phys. Rev. A \textbf{56}, 1627.

%\bibitem{grice01}
Grice, W.P., A.B. U'Ren and I.A. Walmsley (2001), Phys. Rev. A, 64, 063815

%\bibitem {qutrit-qc}
Gr\"{o}blacher, S., T. Jennewein, A. Vaziri, G. Weihs, and
A. Zeilinger, 2006, New J. Phys. \textbf{8}, 75.

%\bibitem {nonLR}
Gr\"{o}blacher, S., T. Paterek, R. Kaltenbaek, \v{C}. Brukner,
M. \.{Z}ukowski, M. Aspelmeyer, and A. Zeilinger, 2007, Nature (London)
\textbf{446}, 871.

%\bibitem {grover}
Grover, L.K., 1997, Phys. Rev. Lett. \textbf{79}, 325.

%\bibitem{guhne07}
G\"uhne, O., Lu CY, Gao WB, Pan JW (2007) Phys. Rev. A. 76, 030305

%\bibitem{ghune2009}
G\"uhne O. , G. Toth, Physics Reports 474, 1, (2009)

%%\bibitem {Guo}
Guo, G.-P. and G.-C. Guo, 2003, Phys. Lett. A \textbf{318}, 337.

%\bibitem {swap-cv}
Halder, M., A. Beveratos, N. Gisin, V. Scarani, C. Simon,
and H. Zbinden, 2007, Nature Phys. \textbf{3}, 692.

%\bibitem{HALDER-2009}
 Halder, M., J. Fulconis, B. Cemlyn, A. Clark, C. Xiong,
W. J. Wadsworth and J. G. Rarity, 2009, Opt. Exp. \textbf{17}, 4670.

%\bibitem {Hammerer}
Hammerer, K., A.S. S\o rensen and E.S. Polzik, 2008, Rev.
Mod. Phys. (to be published); arXiv:0807.3358v3.

%\bibitem{han}
Han, Y.-J., R. Raussendorf, and L.-M. Duan, Phys. Rev.
Lett. 98, 150404 (2007).

%\bibitem {HBT}
Hanbury Brown R. and R.Q. Twiss, 1956, Nature (London)
\textbf{177}, 27.

%\bibitem{hardy92b}
Hardy L., Phys. Rev. Lett. 68, 2981 (1992)

%\bibitem {hardy93}
Hardy, L., 1993, Phys. Rev. Lett. \textbf{71}, 1665.

%\bibitem {Haroche}
Haroche, S., 1995, Ann. N.Y. Acad. Sci. \textbf{755}, 73.

%\bibitem {hayashi}
Hayashi, A., T. Hashimoto, and M. Horibe, 2005, Phys. Rev. A
\textbf{72}, 032325.

%\bibitem {graph-rev}
Hein, M., W. D\"{u}r, J. Eisert, R. Raussendorf, M. Van
den Nest, H.-J. Briegel, 2006, quant-ph/0602096.

%\bibitem {graph}
Hein, M., J. Eisert, and H.J. Briegel, 2004, Phys. Rev. A
\textbf{69}, 062311.

%\bibitem {Herzog1994}
Herzog, T. J., J. G. Rarity, H. Weinfurter, and
A. Zeilinger, 1994, Phys. Rev. Lett. \textbf{72}, 629.

%\bibitem {Hetet}
H\'{e}tet, G., B.C. Buchler, O. Gl\"{o}ckl, M.T.L. Hsu, A.M.
Akulshin, H.-A. Bachor, and P.K. Lam, 2008, Opt. Express \textbf{16}, 7369.

%\bibitem {Hof02}
Hofmann, H.F. and S. Takeuchi, 2002, Phys. Rev. A \textbf{66} 024308.

%\bibitem {Holevo}
Holevo, A.S., 1973, Probl. Inf. Transm. \textbf{9}, 177.

%\bibitem {Holland}
Holland, M.J. and K. Burnett, 1993, Phys. Rev. Lett.
\textbf{71}, 1355.

%\bibitem {jitter}
Holman, K.W., D. D. Hudson, J. Ye, and D.J. Jones, 2005, Opt.
Lett. \textbf{30}, 1225.

%\bibitem {Hong-spdc}
Hong, C.K. and L. Mandel, 1985, Phys. Rev. A \textbf{31}, 2409.

%\bibitem {HOM}
Hong, C.K., Z.Y. Ou, and L. Mandel, 1987, Phys. Rev. Lett.
\textbf{59}, 2044.

%\bibitem {Honjo100}
Honjo, T., H. Takesue, H. Kamada, Y. Nishida, O. Tadanaga,
M. Asobe, and K. Inoue, 2007, Opt. Express \textbf{15}, 13957.

%\bibitem {hornezeilinger85}
M.A.Horne, A.Zeilinger, 1985, Symposium on the Foundations of Modern Physics, Joensuu, P.
Lahti and P. Mittelstaedt (Eds.), World Scientific Publ. (Singapore), 435-9.

%\bibitem {TP-interf89}
Horne, M., A. Shimony, and A. Zeilinger, 1989, Phys.
Rev. Lett. \textbf{62}, 2209.

%\bibitem {TP-interf}
Horne, M., A. Shimony, and A. Zeilinger, 1990, Nature
(London) \textbf{347}, 429.

%\bibitem {Horne86}
Horne, M. and A. Zeilinger, 1986, Ann. N.Y. Acad. Sci.
\textbf{480}, 469.

%\bibitem {horo95}
Horodecki, R., P. Horodecki, and M. Horodecki, 1995, Phys.
Lett. A \textbf{200}, 340.

%\bibitem {horo96}
Horodecki, M., P. Horodecki, and R. Horodecki, 1996, Phys.
Lett. A \textbf{223}, 1.

%\bibitem {horo97}
Horodecki, M., P. Horodecki, and R. Horodecki, 1997, Phys.
Rev. Lett. \textbf{78}, 547.

%\bibitem {horo98}
Horodecki, M., P. Horodecki, and R. Horodecki, 1998, Phys.
Rev. Lett. \textbf{80}, 5239.

%\bibitem {Horo-rev}
Horodecki, R., P. Horodecki, M. Horodecki, and K.
Horodecki, 2009, Rev. Mod. Phys. \textbf{81}, 865.

%\bibitem {bell-s1}
Howell, J.C., A. Lamas-Linares, and D. Bouwmeester, 2002,
Phys. Rev. Lett. \textbf{88}, 030401.

%\bibitem {Noncontext-EXP03}
Huang, Y.-F., C.-F. Li, Y.-S. Zhang, J.-W. Pan, and
G.-C. Guo, 2003, Phys. Rev. Lett. \textbf{90}, 250401.

%\bibitem {Hubel100}
H\"{u}bel, H., M.R. Vanner, T. Lederer, B. Blauensteiner,
T. Lorunser, A. Poppe, and A. Zeilinger, 2007, Opt. Express \textbf{15}, 7853.

%\bibitem {Hwang2003}
Hwang, W.Y., 2003,\ Phys. Rev. Lett. \textbf{91,} 057901.

%\bibitem{irvine2005}
Irvine M. W. T., J. F. Hodelin, C. Simon, and D. Bouwmeester, Phys. Rev.
Lett. \textbf{95}, 030401 (2005).

%\bibitem {Jacobson}
Jacobson, J., G. Bj\"{o}rk, I. Chuang, and Y. Yamamoto,
1995, Phys. Rev. Lett. \textbf{74}, 4835.

%\bibitem{james01}
James, D. F. V., P. G. Kwiat, W. J. Munro, and A. G. White,
Phys. Rev. A \textbf{64}, 052312 (2001).

%\bibitem {qrng-innsbruck}
Jennwein, T., U. Achleitner, Ch. Simon, G. Weihs, H. Weinfurter, A. Zeilinger, 2000, Rev. Sci. Instr. \textbf{41}, 1675.

%\bibitem {nonlocal-swap}
Jennewein, T., G. Weihs, J.-W. Pan, and A. Zeilinger,
2002, Phys. Rev. Lett. \textbf{88}, 017903.

%\bibitem {Jiang}
Jiang, L., J.M. Taylor, and M.D. Lukin, 2007, Phys. Rev. A
\textbf{76}, 012301.

%\bibitem {Jiang09}
Jiang, L., A.M. Rey, O. Romero-Isart, J.J.
Garc\'{\i}a-Ripoll, A. Sanpera, and M.D. Lukin, 2009, Phys. Rev. A
\textbf{79}, 022309.

%\bibitem{16km}
Jin, X.-.M., \textit{et al}. Nature Photonics 4, 376 (2010).

%\bibitem {Julsgaard04}
Julsgaard, B., J. Sherson, J.I. Cirac, J.
Fiur\'{a}\v{s}ek, and E.S. Polzik, 2004, Nature (London) \textbf{432}, 482.

%\bibitem {Kaltenbaek}
Kaltenbaek, R., B. Blauensteiner, M. \.{Z}ukowski, M.
Aspelmeyer, and A. Zeilinger, 2006, Phys. Rev. Lett. \textbf{96}, 240502.

%\bibitem {Kaltenbaek09}
Kaltenbaek, R., R. Prevedel, M. Aspelmeyer, and A.
Zeilinger, 2009, Phys. Rev. A \textbf{79}, 040302(R).

%\bibitem {karlsson}
Karlsson, A. and M. Bourennane, 1998, Phys. Rev. A
\textbf{58}, 4394.

%\bibitem {atomGAS}
Kash, M.M., V.A. Sautenkov, A.S. Zibrov, L. Hollberg, G.R.
Welch, M.D. Lukin, Yu. Rostovtsev, E.S. Fry, and M.O. Scully, 1999, Phys. Rev.
Lett. \textbf{82}, 5229.

%\bibitem{kardynal08}
Kardynal, B. E., Z. L. Yuan and A. J. Shields, Nature Photonics 2, 425 (2008)

%\bibitem {Zeilinger-n}
Kaszlikowski, D., P. Gnaci\'{n}ski, M. \.{Z}ukowski, W.
Miklaszewski, and A. Zeilinger, 2000, Phys. Rev. Lett. \textbf{85}, 4418.

%\bibitem {lange04}
Keller, M., B. Lange, K. Hayasaka, W. Lange, and H. Walther,
2004, Nature (London) \textbf{431}, 1075.

%\bibitem {keller}
Keller, T.E., and M.H. Rubin, 1997, Phys. Rev. A \textbf{56}, 1534.

%\bibitem {keller98}
Keller, T.E., and M.H. Rubin, and Y. Shih, 1998, Phys.
Lett. A \textbf{244}, 507.

%\bibitem {Kiesel}
Kiesel, N., C. Schmid, U. Weber, G. T\'{o}th, O. G\"{u}hne,
R. Ursin, and H. Weinfurter, 2005a, Phys. Rev. Lett. \textbf{95}, 210502.

%\bibitem {Kiesel-b}
Kiesel, N., C. Schmid, U. Weber, R. Ursin, and H.
Weinfurter, 2005b, Phys. Rev. Lett. \textbf{95}, 210505.

%\bibitem{Kiesel-Dicke}
Kiesel, N., C. Schmid, G. T\'{o}th, E. Solano, and H. Weinfurter, 2007,
Phys. Rev. Lett. \textbf{98}, 063604.

%\bibitem {Kim03}
Kim, Y.-H., 2003, Phys. Rev. A \textbf{67}, 040301(R).

%%\bibitem {kim2001}
Kim, Y.-H., M.V. Chekhova, S.P. Kulik, M.H. Rubin, and Y.H.
%Shih, 2001, Phys. Rev. A \textbf{63}, 062301.

%\bibitem {kim00b}
Kim, Y.-H., S.P. Kulik, and Y.H. Shih, 2001, Phys. Rev. Lett.
\textbf{86}, 1370.

%\bibitem{Wong}
Kim, T., M. Fiorentino, and F. N. C. Wong, Phys. Rev.
A 73, 012316 (2006).

%\bibitem {Kimble1977}
Kimble, H.J., M. Dagenais, and L. Mandel, 1977, Phys.
Rev. Lett. \textbf{39}, 691.

%\bibitem {Kimble1978}
Kimble, H.J., M. Dagenais, and L. Mandel, 1978, Phys.
Rev. A \textbf{18}, 201.

%\bibitem{Kimble2008}
Kimble, H.J., 2008, Nature 453, 1023 .

%\bibitem{kitaev}
Kitaev, A.Y. Ann. Phys. (N.Y.) 303, 2 (2003).

%\bibitem{klyachko}
Klyachko, A. A., M.A. Can, u.S. Biniciog,
and A.S. Shumovsky, (2008) Phys. Rev. Lett. 101, 20403.

%\bibitem {klyshkoSPDC}
Klyshko, D.N., 1967, Soviet Phys-JETP Lett. \textbf{6}, 23.

%\bibitem {Kly93}
Klyshko, D.N., 1993, Phys. Lett. A \textbf{172}, 399.

%\bibitem {SPDC}
Klyshko, D.N., 1988, \textit{Photons and Nonlinear Optics}
(Gordon and Breach, New York).

%%\bibitem {Knill01}
Knill, E., R. Laflamme, R. Martinez, and C. Negrevergne,
%2001, Phys. Rev. Lett. \textbf{86}, 5811.

%\bibitem {KLM}
Knill, E., R. Laflamme, and G. Milburn, 2001, Nature (London)
\textbf{409}, 46.

%\bibitem {Koashi}
Koashi, M., T. Yamamoto, and N. Imoto, 2001, Phys. Rev. A,
\textbf{63} 030301.

%\bibitem {KS}
Kochen, S. and E.P. Specker, 1967, J. Math. Mech. \textbf{17}, 59.

%\bibitem {Kok00}
Kok, P. and S.L. Braunstein, 2000, Phys. Rev. A \textbf{61}, 042304.

%\bibitem {Kok02}
Kok, P., H. Lee, and J.P. Dowling, 2002, Phys. Rev. A
\textbf{65}, 052104.

%\bibitem {Kok01}
Kok, P., A.N. Boto, D.S. Abrams, C.P. Williams, S.L.
Braunstein, and J.P. Dowling, 2001, Phys. Rev. A \textbf{63}, 063407.

%\bibitem {Kok2005}
Kok, P., W.J. Munro, K. Nemoto, T.C. Ralph, J.P. Dowling,
and G.J. Milburn, 2007, Rev. Mod. Phys. \textbf{79}, 135.

%%\bibitem {Kol06}
Kolchin, P., S. Du, C. Belthangady, G.Y. Yin, and S.E. Harris,
%2006, Phys. Rev. Lett. \textbf{97}, 113602.

%\bibitem {Korneev-SSPD}
Korneev, A., Yu. Vachtomin, O. Minaeva, A. Divochiy, K.
Smirnov, O. Okunev, G. Gol'tsman, C. Zinoni, N. Chauvin, L. Balet, F. Marsili,
D. Bitauld, B. Alloing, L.H. Li, A. Fiore, L. Lunghi, A. Gerardino, M. Halder,
C. Jorel, H. Zbinden, 2007, IEEE J. Quant. Electron. \textbf{13}, 944.

%\bibitem {mem-kmp}
Kozhekin, A.E., K. M\o lmer, and E. Polzik, 2000, Phys. Rev.
A \textbf{62}, 033809.

%\bibitem {Krenn}
Krenn, G. and A. Zeilinger, 1996, Phys. Rev. A \textbf{54}, 1793.

%\bibitem {SP-Rempe02}
Kuhn, A., M. Hennrich, and G. Rempe, 2002, Phys. Rev.
Lett. \textbf{89}, 067901.

%\bibitem {SP-solid}
Kurtsiefer, C., S. Mayer, P. Zarda, and H. Weinfurter,
2000, Phys. Rev. Lett. \textbf{85}, 290.

%\bibitem {Kurtsiefer02}
Kurtsiefer, C., P. Zarda, M. Halder, H. Weinfurter,
P.M. Gorman, P.R. Tapster, and J.G. Rarity, 2002, Nature (London)
\textbf{419}, 450.

%\bibitem{Wong_pulsed}
Kuzucu, O., and F. N. C. Wong, Phys. Rev. A 77, 032314 (2008)

%\bibitem{kwiat92}
Kwiat, P. G., A. M. Steinberg, and R. Y. Chiao, Phys. Rev. A 45, 7729 (1992).

%\bibitem {hyper-E}
Kwiat, P.G., 1997, J. Mod. Opt. \textbf{44}, 2173.

%\bibitem {distill-01EXP}
Kwiat, P.G., S. Barraza-Lopez, A. Stefanov, and N.
Gisin, 2001, Nature (London) \textbf{409}, 1014.

%\bibitem {dfs-kwiat}
Kwiat, P.G., A.J. Berglund, J.B. Altepeter, and A.G.
White, 2000, Science \textbf{290}, 498.

%\bibitem {kwiat3}
Kwiat, P.G., P.H. Eberhard, A.M. Steinberg, and R.Y. Chiao,
1994, Phys. Rev. A \textbf{49}, 3209.

%\bibitem {Kwiat-SPDC}
Kwiat, P.G., K. Mattle, H. Weinfurter, A. Zeilinger, A.V.
Sergienko, and Y. Shih, 1995, Phys. Rev. Lett. \textbf{75}, 4337.

%\bibitem {kwiat93}
Kwiat, P.G., A.M. Steinberg, and R.Y. Chiao, 1993, Phys.
Rev. A \textbf{47}, R2472.

%\bibitem {kwiat-bsm}
Kwiat, P.G. and H. Weinfurter, 1998, Phys. Rev. A
\textbf{58}, R2623.

%\bibitem {laloe}
Lalo\"{e}, F., 2001, Am. J. Phys. \textbf{69}, 655.

%\bibitem {E-laser}
Lamas-Linares, A., J.C. Howell, and D. Bouwmeester, 2001,
Nature (London) \textbf{412}, 887.

%\bibitem {Landau}
Landau, L.J., 1987, Phys. Lett. A \textbf{120}, 54.

%\bibitem {Langford}
Langford, N.K., T.J. Weinhold, R. Prevedel, K.J. Resch, A.
Gilchrist, J.L. O'Brien, G.J. Pryde, and A.G. White, 2005, Phys. Rev. Lett.
\textbf{95}, 210504.

%\bibitem{lanyon07}
Lanyon B. P., T. J. Weinhold, N. K. Langford, M. Barbieri, D. F. V. James, A.
Gilchrist, and A. G. White, Phys. Rev. Lett. \textbf{99}, 250505 (2007).

%\bibitem{lanyon09a}
Lanyon B. P., M. Barbieri, M. P. Almeida, T. Jennewein, T. C. Ralph, K. J.
Resch, G. J. Pryde, J. L. O'Brien, A. Gilchrist and A. G. White,
Nature Physics 5, 134 (2009a).

%\bibitem{lanyon09b}
Lanyon B. P., and N K Langford (2009b) New J. Phys. 11 013008.

%\bibitem{lanyon10}
Lanyon B. P., J. D. Whitfield, G. G. Gillet, M. E. Goggin, M. P. Almeida, I.
Kassal, J. D. Biamonte, M. Mohseni, B. J. Powell, M. Barbieri, A. Aspuru-Guzik
and A. G. White, Nature Chemistry 2, 106 (2010).

%\bibitem{lapkiewicz}
Lapkiewicz, R., L. Peizhe, S. Christoph, N.K. Langford, S. Ramelow, M. Wiesniak
and A. Zeilinger, (2011) Nature 474, 490.

%\bibitem{laskowski}
Laskowski W., M. Wiesniak M., M. Zukowski, et al., J. Phys. B: Atom. Molec. Opt. Phys. 42, 114004 (2009)

%\bibitem {raurat}
Laurat, J., H. de Riedmatten, D. Felinto, C.-W. Chou, E.
Schomburg, and H.J. Kimble, 2006, Opt. Express \textbf{14}, 6912.

%\bibitem {QBeat}
Legero, T., T. Wilk, M. Hennrich, G. Rempe, and A. Kuhn, 2004,
Phys. Rev. Lett. \textbf{93}, 070503.

%\bibitem {Leggett}
Leggett, A.J., 2003, Found. Phys. \textbf{33}, 1469.

%\bibitem {ion-rmp}
Leibfried, D., R. Blatt, C. Monroe, and D. Wineland, 2003,
Rev. Mod. Phys. \textbf{75}, 281.

%%\bibitem {Leonhardt03}
Leonhardt, U., 2003, Rep. Prog. Phys. \textbf{66}, 1207.

%\bibitem {witness00}
Lewenstein, M., B. Kraus, J.I. Cirac, and P. Horodecki,
2000, Phys. Rev. A \textbf{62}, 052310.

%\bibitem {Lewis}
Lewis, G.N., 1926, Nature (London) \textbf{118}, 874.

%\bibitem{Laing10}
Laing, A., et al. Appl. Phys. Lett. 97, 211109 (2010)

%%\bibitem {dsf-Lidar}
Lidar, D.A., I.L. Chuang, and K.B. Whaley, 1998, Phys.
%Rev. Lett. \textbf{81}, 2594.

%%\bibitem {Lin}
Lin, G.W., X.B. Zou, X.M. Lin, and G.C. Guo, 2009, Europhys.
%Lett. \textbf{86}, 30006.

%%\bibitem {Ling08}
Ling, A., M.P. Peloso, I. Marcikic, V. Scarani, A.
%Lamas-Linares, and C. Kurtsiefer, 2008, Phys. Rev. A \textbf{78}, 020301(R).

%\bibitem {Lita}
Lita, A.E., A.J. Miller, and S.W. Nam, 2008, Opt. Express
\textbf{16}, 3032.

%\bibitem {Liu01}
Liu, C., Z. Dutton, C.H. Behroozi, and L.V. Hau, 2001, Nature
(London) \textbf{409}, 490.

%%\bibitem {llody95}
Lloyd, S., 1995, Phys. Rev. Lett. \textbf{75}, 346.

%%\bibitem {lloyd}
Lloyd, S., 1998, Phys. Rev. A \textbf{57}, R1473.

%%\bibitem {cBSM87}
Lloyd, S., M.S. Shahriar, and J.H. Shapiro, and P. R. Hemmer,
%2001, Phys. Rev. Lett. \textbf{87}, 167903.

%\bibitem {LoISIT2004}
Lo, H.-K., 2004, Proceedings of IEEE ISIT (International
Symposium on Information Theory), p.~137.

%%\bibitem {Lo-Ma-Chen2005}
Lo, H.-K., X.-F. Ma, and K. Chen, 2005, Phys. Rev.
%Lett. \textbf{94,} 230504.

%%\bibitem {Loss}
Loss, D. and D.P. DiVincenzo, 1998, Phys. Rev. A \textbf{57}, 120.

%\bibitem {moerner00}
Lounis, B. and W.E. Moerner, 2000, Nature (London)
\textbf{407}, 491.

%\bibitem {Lounis05}
Lounis B. and M. Orrit, 2005, Rep. Prog. Phys \textbf{68}, 1129.

%\bibitem {lu07a}
Lu, C.-Y., X.-Q. Zhou, O. G\"{u}hne, W.-B. Gao, J. Zhang, Z.-S.
Yuan, A. Goebel, T. Yang, and J.-W. Pan, 2007a, Nature Phys. \textbf{3}, 91.

%\bibitem{lu07b}
Lu C.-Y., D. E. Browne, T. Yang, and J.-W. Pan, 2007b Phys. Rev. Lett. \textbf{99}, 250504
.

%\bibitem {lu08}
Lu, C.-Y., W.-B. Gao, J. Zhang, X.-Q. Zhou, T. Yang, and J.-W.
Pan, 2008, Proc. Natl. Acad. Sci. USA \textbf{105}, 11050.

%\bibitem{lu09a}
Lu C.-Y., W.-B. Gao, O. G\"uhne, X.-Q. Zhou, Z.-B. Chen, and J.-W. Pan, Phys.
Rev. Lett. \textbf{102}, 030502 (2009).

%\bibitem{lu09b}
 Lu C.-Y., T. Yang, and J.-W. Pan, Phys. Rev. Lett. \textbf{103}, 020501 (2009).

%%\bibitem {Lucero}
Lucero, E., M. Hofheinz, M. Ansmann, R.C. Bialczak, N. Katz,
%M. Neeley, A.D. O'Connell, H. Wang, A.N. Cleland, and J.M. Martinis, 2008,
%Phys. Rev. Lett. \textbf{100}, 247001.

%\bibitem {AE-rmp}
Lukin, M.D., 2003, Rev. Mod. Phys. \textbf{75}, 457.

%\bibitem {EIT-Nature}
Lukin, M.D. and A. Imamo\u{g}lu, 2001, Nature (London)
\textbf{413}, 273.

%\bibitem {mem-Lukin}
Lukin, M.D., S.F. Yelin, and M. Fleischhauer, 2000, Phys.
Rev. Lett. \textbf{84}, 4232.

%\bibitem {Lukenhaus}
L\"{u}tkenhaus, N., J. Calsamiglia, and K.-A. Suominen,
1999, Phys. Rev. A \textbf{59}, 3295.

%\bibitem {Lvovsky}
Lvovsky, A.I., 2002, Phys. Rev. Lett. \textbf{88}, 098901.

%%\bibitem {ma}
Ma, L.-S., R. K. Shelton, H. C. Kapteyn, M. M. Murnane, and J.
%Ye,\ 2001, Phys. Rev. A \textbf{64}, 021802.

%\bibitem{heisenberg}
Ma, X.-S., B. Dakic, W. Naylor, A. Zeilinger, P. Walther,
(2011) Nature Physics, doi:10.1038/nphys1919.

%\bibitem {freq-qubit}
Madsen, M.J., D.L. Moehring, P. Maunz, R.N. Kohn, Jr.,
L.-M. Duan, and C. Monroe, 2006, Phys. Rev. Lett. \textbf{97}, 040505.

%%\bibitem {Magyar}
Magyar, G. and L. Mandel, 1963, Nature (London) \textbf{198}, 255.

%\bibitem {angular}
Mair, A., A. Vaziri, G. Weihs, and A. Zeilinger, 2001,
Nature (London) \textbf{412}, 313.

%%\bibitem {Majer}
Majer, J., J.M. Chow, J.M. Gambetta, J. Koch, B.R. Johnson,
%J.A. Schreier, L. Frunzio, D.I. Schuster, A.A. Houck, A. Wallraff, A. Blais,
%M.H. Devoret, S.M. Girvin, and R.J. Schoelkopf, 2007, Nature (London)
%\textbf{449}, 443.

%%\bibitem {jj-rmp}
Makhlin, Y., G. Sch\"{o}n, and A. Shnirman, 2001, Rev. Mod.
%Phys. \textbf{73}, 357.

%\bibitem {Mandel83}
Mandel, L., 1983, Phys. Rev. A \textbf{28}, 929.

%\bibitem {Mandel}
Mandel, L. and E. Wolf, 1995, \textit{Optical Coherence and
Quantum Optics} (Cambridge University Press, Cambridge).

%\bibitem{mandel}
Mandel, L., , (1999) Rev. Mod. Phys. 71, 274.

%\bibitem{mann}
Mann, A, M. Revzen, W. Schleich, (1992) Phys. Rev. A 46, 5363-5366.

%\bibitem {timebin02}
Marcikic, I., H. de Riedmatten, W. Tittel, V. Scarani, H.
Zbinden, and N. Gisin, 2002, Phys. Rev. A \textbf{66}, 062308.

%\bibitem {Tele-telecom}
Marcikic, I., H. de Riedmatten, W. Tittel, H. Zbinden,
N. Gisin, 2003, Nature (London) \textbf{421}, 509.

%\bibitem {Dis-E-fiber}
Marcikic, I., H. de Riedmatten, W. Tittel, H. Zbinden,
M. Legr\'{e} and N. Gisin, 2004, Phys. Rev. Lett. \textbf{93}, 180502.

%%\bibitem {Marcikic06}
Marcikic, I., A. Lamas-Linares, and C. Kurtsiefer, 2006,
%Appl. Phys. Lett. \textbf{89}, 101122.

%\bibitem {Cabello-comment}
Marinatto, L., 2003, Phys. Rev. Lett. \textbf{90}, 258901.

%\bibitem {Marsili-SSPD}
Marsili, F., D. Bitauld, A. Fiore, A. Gaggero, F.
Mattioli, R. Leoni, M. Benkahoul, F. L\'{e}vy, 2008, Opt. Express \textbf{16}, 3191.

%\bibitem {massar}
Massar, S. and S. Popescu, 1995, Phys. Rev. Lett.
\textbf{74}, 1259.

%\bibitem {ensemble-ent-store}
Matsukevich, D.N., T. Chaneli\`{e}re, S.D.
Jenkins, S.-Y. Lan, T.A.B. Kennedy, and A. Kuzmich, 2006a, Phys. Rev. Lett.
\%textbf{96}, 030405.

%\bibitem {Matsukevich06}
Matsukevich, D.N., T. Chaneli\`{e}re, S.D. Jenkins,
S.-Y. Lan, T.A.B. Kennedy, and A. Kuzmich, 2006b, Phys. Rev. Lett.
\textbf{97}, 013601.

%\bibitem {Kuzmich04}
Matsukevich, D.N., and A. Kuzmich, 2004, Science
\textbf{306}, 663.

%\bibitem {Monroe08}
Matsukevich, D.N., P. Maunz, D.L. Moehring, S. Olmschenk,
and C. Monroe, 2008, Phys. Rev. Lett. \textbf{100}, 150404.

%\bibitem {denseCodEXP}
Mattle, K., H. Weinfurter, P.G. Kwiat, and A. Zeilinger,
1996, Phys. Rev. Lett. \textbf{76}, 4656.

%\bibitem {Maunz}
Maunz, P., D.L. Moehring, S. Olmschenk, K.C. Younge, D.N.
Matsukevich, and C. Monroe, 2007, Nature Phys. \textbf{3}, 538.

%\bibitem {Mazurenko-QKD}
Mazurenko, Y., R. Giust, and J.P. Goedgebuer, 1997, Opt.Commun. \textbf{133},87.

%\bibitem {SP-cQED}
McKeever, J., A. Boca, A.D. Boozer, R. Miller, J.R. Buck, A.
Kuzmich, and H.J. Kimble, 2004, Science \textbf{303}, 1992.

%\bibitem {Mermin90}
Mermin, N.D., 1990a, Phys. Today \textbf{43} (6), 9.

%\bibitem {merminPRL}
Mermin, N.D., 1990b, Phys. Rev. Lett. \textbf{65}, 1838.

%\bibitem {merminPRLc}
Mermin, N.D., 1990c, Phys. Rev. Lett. \textbf{65}, 3373.

%\bibitem {mermin93}
Mermin, N.D., 1993, Rev. Mod. Phys. \textbf{65}, 803.

%\bibitem {BSM96}
Michler, M., K. Mattle, H. Weinfurter, and A. Zeilinger, 1996,
Phys. Rev. A \textbf{53}, R1209.

%\bibitem {Noncontext-EXP00}
Michler, M., H. Weinfurter, and M. \.{Z}ukowski,
2000a, Phys. Rev. Lett. \textbf{84}, 5457.

%\bibitem {michler00}
Michler, P., A. Kiraz, C. Becher, W.V. Schoenfeld, P.M.
Petroff, L. Zhang, E. Hu, and A. Imamo\={g}lu, 2000b, Science \textbf{290}, 2282.

%%\bibitem {Migdall}
Migdall, A.L., D. Branning, S. Castelletto, and M. Ware,
%2002, Phys. Rev. A \textbf{66}, 053805.

%%\bibitem {Miller}
Miller, R., T.E. Northup, K.M Birnbaum, A. Boca, A.D. Boozer,
%and H.J. Kimble, 2005, J. Phys. B: At. Mol. Opt. Phys. \textbf{38}, S551.

%\bibitem {Mintert05}
Mintert, F., A.R.R. Carvalho, M. Ku\'{s}, and A.
Buchleitner, 2005, Phys. Rep. \textbf{415}, 207.

%\bibitem {super-phase}
Mitchell, M.W., J.S. Lundeen, and A.M. Steinberg, 2004,
Nature (London) \textbf{429}, 161.

%\bibitem {Monroe07}
Moehring, D.L., P. Maunz, S. Olmschenk, K.C. Younge, D.N.
Matsukevich, L.-M. Duan, and C. Monroe, 2007, Nature (London) \textbf{449}, 68.

%\bibitem {dfs-Mohseni}
Mohseni, M., J.S. Lundeen, K.J. Resch, and A.M.
Steinberg, 2003, Phys. Rev. Lett. \textbf{91}, 187903.

%\bibitem {qutrit-trigger}
Molina-Terriza, G., A. Vaziri, J.
\v{R}eh\'{a}\v{c}ek, Z. Hradil, and A. Zeilinger, 2004, Phys. Rev. Lett.
\textbf{92}, 167903.

%\bibitem {Mollow}
Mollow, B.R., 1973, Phys. Rev. A \textbf{8}, 2684.

%\bibitem {link06}
Mor, T. and N. Yoran, 2006, Phys. Rev. Lett. \textbf{97}, 090501.

%\bibitem {Mosley}
Mosley, P.J., J.S. Lundeen, B.J. Smith, P. Wasylczyk, A.B.
U'Ren, C. Silberhorn, and I.A. Walmsley, 2008, Phys. Rev. Lett. \textbf{100}, 133601.

%\bibitem {Nagata07}
Nagata, T., R. Okamoto, J. O`Brien, K. Sasaki, and S.
Takeuchi, 2007, Science \textbf{316}, 726.

%%\bibitem {Neeman}
Ne'eman, Y., 1986, Found. Phys. \textbf{16}, 361.

%\bibitem {Neves}
Neves, L., G. Lima, J. G. Aguirre G\'{o}mez, C. H. Monken, C.
Saavedra, and S. P\'{a}dua, 2005, Phys. Rev. Lett. \textbf{94}, 100501.

%\bibitem {Nielsen04}
Nielsen, M.A., 2004, Phys. Rev. Lett. \textbf{93}, 040503.

%\bibitem {Nielson-Chuang}
Nielson, M. and I.L. Chuang, 2000, \textit{Quantum
Computation and Quantum Information} (Cambridge University Press, Cambridge).

%\bibitem {Nielsen05pra}
Nielsen, M.A. and C.M. Dawson, 2005, Phys. Rev. A
\textbf{71}, 042323.

%\bibitem {tele-nmr}
Nielsen, M.A., E. Knill, and R. Laflamme, 1998, Nature
(London) \textbf{396}, 52.

%\bibitem {Novikova}
Novikova, I. A.V. Gorshkov, D.F. Phillips, A.S. S\o rensen,
M.D. Lukin, and R.L. Walsworth, 2007, Phys. Rev. Lett. \textbf{98}, 243602.

%\bibitem {allOPT-CNOT}
O'Brien, J.L., G.J. Pryde, A.G. White, T.C. Ralph, and
D. Branning, 2003, Nature (London) \textbf{426}, 264.

%\bibitem{orbien04}
O'Brien, J. L., G. J. Pryde, A. Gilchrist, D. F. V. James, N. K. Langford, T. C.
Ralph, and A. G. White, Phys. Rev. Lett. \textbf{93}, 080502 (2004).

%\bibitem {Obrien07}
O'Brien, J.L., 2007, Science \textbf{318}, 1567.

%\bibitem{orbien09}
O'Brien, J. L., A. Furasawa, J. Vuckovic, Nature Photonics 3, 687 (2009)

%\bibitem {Okamoto05}
Okamoto, R., H.F. Hofmann, S. Takeuchi, and K. Sasaki, 2005,
Phys. Rev. Lett. \textbf{95}, 210506.

%%\bibitem{okamoto09}
Okamoto, R., J.L. O'Brien, H.F. Hofmann, T. Nagata, K.
%Sasaki, and S. Takeuchi, Science \textbf{323}, 483 (2009)

%\bibitem{okamoto}
Okamoto, R., Jeremy L. O'Brien, Holger F. Hofmann, Shigeki Takeuchi, arXiv:1006.4743.

%\bibitem {pixel}
O'Sullivan-Hale, M.N., I.A. Khan, R.W. Boyd, and J.C. Howell,
2005, Phys. Rev. Lett. \textbf{94}, 220501.

%\bibitem {ou88}
Ou, Z.Y. and L. Mandel, 1988a, Phys. Rev. Lett. \textbf{61}, 50.

%\bibitem {ou88b}
Ou, Z.Y. and L. Mandel, 1988b, Phys. Rev. Lett. \textbf{61}, 54.

%\bibitem {Pachos07}
 Pachos J. 2007,  Annals of Physics \textbf{322}, 1254.

%\bibitem{pachos}
Pachos J., W. Wieczorek, Ch. Schmid, N. Kiesel, R. Pohlner, and H.
Weinfurter, New J. Phys. \textbf{11}, 083010 (2009).

%%\bibitem {dsf-Palma}
Palma, G.M., K.-A. Suominen, and A.K. Ekert, 1996, Proc.
%R. Soc. London A \textbf{452}, 567.

%\bibitem{paleari04}
Paleari, F., (2004) Opt. Exp. 12, 2816.

%\bibitem {GHZ-Pan00}
Pan, J.-W., D. Bouwmeester, M. Daniell, H. Weinfurter, and
A. Zeilinger, 2000, Nature (London) \textbf{403}, 515.

%\bibitem {swapPAN98}
Pan, J.-W., D. Bouwmeester, H. Weinfurter, and A.
Zeilinger, 1998, Phys. Rev. Lett. \textbf{80}, 3891.

%\bibitem {fourP01}
Pan, J.-W., M. Daniell, S. Gasparoni, G. Weihs, and A.
Zeilinger, 2001a, Phys. Rev. Lett. \textbf{86}, 004435.

%\bibitem {Tele-free}
Pan, J.-W., S. Gasparoni, M. Aspelmeyer, T. Jennewein, and
A. Zeilinger, 2003a, Nature (London) \textbf{421}, 721.

%\bibitem {PurEXP}
Pan, J.-W., S. Gasparoni, R. Ursin, G. Weihs, and A.
Zeilinger, 2003b, Nature (London) \textbf{423}, 417.

%\bibitem {Pur-LO01}
Pan, J.-W., C. Simon, C. Brukner, and A. Zeilinger, 2001b,
Nature (London) \textbf{410}, 1067.

%\bibitem {GHZ-nan98}
Pan, J.-W. and A. Zeilinger, 1998, Phys. Rev. A
\textbf{57}, 2208.

%\bibitem{Patel}
Patel R.B., A.J. Bennett, I. Farrer, C.A. Nicoll, D.A. Ritchie, and A.J. Shields, 2010, Nature Photonics 4, 632

%\bibitem {nonl-R-AZ}
Paterek, T., A. Fedrizzi, S. Gr\"{o}blacher, T. Jennewein,
M. \.{Z}ukowski, M. Aspelmeyer, and A. Zeilinger, 2007, Phys. Rev. Lett.
\textbf{99}, 210406.

%\bibitem {Paul86}
Paul, H., 1986, Rev. Mod. Phys. \textbf{58}, 209.

%\bibitem {SP-dot02}
Pelton, M., C. Santori, J. Vu\v{c}kovi\'{c}, B.Y. Zhang,
G.S. Solomon, J. Plant, and Y. Yamamoto, 2002, Phys. Rev. Lett. \textbf{89}, 233602.

%\bibitem {Peng}
Peng, C.-Z., T. Yang, X.-H. Bao, J. Zhang, X.-M. Jin, F.-Y.
Feng, B. Yang, J. Yang, J. Yin, Q. Zhang, N. Li, B.-L. Tian, and J.-W. Pan,
2005, Phys. Rev. Lett. \textbf{94}, 150501.

%\bibitem {Peng07}
Peng, C.-Z., J. Zhang, D. Yang, W.-B. Gao, H.-X. Ma, H. Yin,
H.-P. Zeng, T. Yang, X.-B. Wang, and J.-W. Pan, 2007, Phys. Rev. Lett.
\textbf{98}, 010505.

%\bibitem {peres96}
Peres, A., 1996, Phys. Rev. Lett. \textbf{77}, 1413.

%\bibitem {peres}
Peres, A., 2000, J. Mod. Opt. \textbf{47}, 531.

%\bibitem {peres-book}
Peres, A., 2002, \textit{Quantum Theory: Concepts and
Methods} (Kluwer Academic Publishers, New York).

%\bibitem{walk}
Peruzzo, A., \textit{et al.}, (2010) Science 329, 1500.

%\bibitem {max-mix}
Peters, N.A., J.B. Altepeter, D. Branning, E.R. Jeffrey,
T.-C. Wei, and P.G. Kwiat, 2004, Phys. Rev. Lett. \textbf{92}, 133601.

%%\bibitem {mandel67}
Pfleegor, R.L. and L. Mandel, 1967, Phys. Rev.
%\textbf{159}, 1084.

%\bibitem {vapor}
Phillips, D.F., A. Fleischhauer, A. Mair, R.L. Walsworth, and
M.D. Lukin, 2001, Phys. Rev. Lett. \textbf{86}, 783.

%\bibitem{pittman96}
Pittman, T. B., D. V. Strekalov, A. Migdall, M. H. Rubin, A. V. Sergienko, and Y. H. Shih,
(1996) Phys. Rev. Lett. 77, 1917.

%\bibitem {Pitt01}
Pittman, T.B., B.C. Jacobs, and J.D. Franson, 2001, Phys.
Rev. A \textbf{64} 062311.

%\bibitem {PittmanPRA2003}
Pittman, T.B., M.J. Fitch, B.C. Jacobs, and J.D.
Franson, 2003, Phys. Rev. A \textbf{68}, 032316.

%%\bibitem {pitt02a}
Pittman, T.B., B.C. Jacobs, and J.D. Franson, 2002a, Phys.
%Rev. A \textbf{66}, 042303.

%%\bibitem {Pittman02}
Pittman, T.B., B.C. Jacobs, and J.D. Franson, 2002b, Phys.
%Rev. Lett. \textbf{88}, 257902.

%\bibitem {PittmanBI}
Pittman T.B. and J.D. Franson, 2003, Phys. Rev. Lett.
\textbf{90}, 240401.

%\bibitem {rempe04}
T. Legero, T. Wilk, M. Hennrich, G. Rempe, and A. Kuhn,  (2004) Phys. Rev. Lett. 93, 070503.

%\bibitem {Politi}
Politi, A., M.J. Cryan, J.G. Rarity, S. Yu, J.L. O'Brien,
2008, Science \textbf{320}, 646.

%\bibitem{politi09}
Politi, A., J. C. F. Matthews, and J. L. O'Brien
Science \textbf{325}, 1221 (2009)

%\bibitem {popescu95}
Popescu, S., 1995, quant-ph/9501020.

%\bibitem {popescu}
Popescu, S., L. Hardy, and M. \.{Z}ukowski, 1997, Phys. Rev.
A \textbf{56}, R4353.

%%\bibitem {poyatos}
Poyatos, J.F., J.I. Cirac, and P. Zoller, 1997, Phys. Rev.
%Lett. \textbf{78}, 390.

%\bibitem {Preskill}
Preskill, J., 1998, \textit{Lecture Notes on Quantum
Computation} (CIT Physics 219/Computer Science 219) at http://www.theory.caltech.edu/people/preskill/ph229/.

%\bibitem {Prevedel09}
Prevedel, R., G. Cronenberg, M.S. Tame, M. Paternostro,
P. Walther, M.S. Kim, and A. Zeilinger, 2009, Phys. Rev. Lett. \textbf{103}, 020503.

%\bibitem {Prevedel07a}
Prevedel, R., M.S. Tame, A. Stefanov, M. Paternostro,
M.S. Kim, and A. Zeilinger, 2007a, Phys. Rev. Lett. \textbf{99}, 250503.

%\bibitem {Prevedel07}
Prevedel, R., P. Walther, F. Tiefenbacher, P. B\"{o}hi,
R. Kaltenbaek, T. Jennewein, and A. Zeilinger, 2007b, Nature (London)
\textbf{445}, 65.

%\bibitem{radmark}
Radmark, M., M. Zukowski, M. Bourennane, Phys. Rev. Lett. 103, 150501 (2009).

%\bibitem {radloff}
Radloff, W., 1971, Ann. Phys. (Leipzig) \textbf{26}, 178.

%\bibitem{radnaev}
Radnaev, A.G., Y. O. Dudin, R. Zhao, H. H. Jen, S. D. Jenkins, A. Kuzmich and T. A. B. Kennedy Nature Physics 6, 894 (2010)

%\bibitem {qed-rmp}
Raimond, J.M., M. Brune, and S. Haroche, 2001, Rev. Mod.
Phys. \textbf{73}, 565.

%\bibitem {loss-tol}
Ralph, T.C., A.J.F. Hayes, and A.~Gilchrist, 2005, Phys.
Rev. Lett. \textbf{95} 100501.

%\bibitem {Ralph02}
Ralph, T.C., N.K. Langford, T.B. Bell, and A.G. White,
2002a, Phys. Rev. A \textbf{65}, 062324.

%\bibitem {Ralgh02b}
Ralph, T.C., A.G. White, W.J. Munro, and G.J. Milburn,
2002b, Phys. Rev. A \textbf{65}, 012314.

%\bibitem{Ralph09}
Ralph, T.C. and G. J. Pryde (2009) Progress in Optics, vol. 54 pp 209-269 (ed. Emil Wolf) Elsevier, Great Britain.

%%\bibitem{rangarjan}
Rangarajan, R., M. Goggin and P. Kwiat, Opt. Express, \textbf{17} 18920 (2009).

%\bibitem {rarity95}
Rarity, J.G., 1995, Ann. N.Y. Acad. Sci. \textbf{755}, 624.

%\bibitem {usingspdc}
Rarity, J.G. and P.R. Tapster, 1990a, Phys. Rev. Lett.
\textbf{64}, 2495.

%\bibitem {Rarity90b}
Rarity, J.G., P.R. Tapster, E. Jakeman, T. Larchuk, R.A.
Campos, M.C. Teich, and B.E.A. Saleh, 1990b, Phys. Rev. Lett. \textbf{65}, 1348.

%\bibitem {rarity98}
Rarity, J.G., P.R. Tapster, and R. Loudon, 1996, in
\textit{Quantum Interferometry}, edited by F. De Martini, G. Denardo, and Y.
Shih (VCH, Weinheim).

%\bibitem {ow-pra}
Raussendorf, R., D.E. Browne, and H.J. Briegel, 2003, Phys.
Rev. A \textbf{68}, 022312.

%\bibitem {one-way}
Raussendorf, R. and H.J. Briegel, 2001, Phys. Rev. Lett.
\textbf{86}, 5188.

%%\bibitem {one-way-rev}
Raussendorf, R. and H.J. Briegel, 2002, Quant. Inf.
%Comp. \textbf{6}, 433.

%%\bibitem {RHG}
Raussendorf, R. and J. Harrington, 2007, Phys. Rev. Lett.
%\textbf{98}, 190504.

%\bibitem {RGH06}
Raussendorf, R., J. Harrington, and K. Goyal, 2006, Ann. Phys.
\textbf{321}, 2242.

%\bibitem {Raussendorf07}
Raussendorf, R., J. Harrington, and K. Goyal, 2007,
New J. Phys. \textbf{9}, 199.

%\bibitem {reck94}
Reck, M., A.~Zeilinger, H.~J. Bernstein, and P.~Bertani,
1994, Phys. Rev. Lett., \textbf{73} 58.

%\bibitem {free-space-Vienna}
Resch, K.J., M. Lindenthal, B. Blauensteiner, H.R.
B\"{o}hm, A. Fedrizzi, C. Kurtsiefer, A. Poppe, T. Schmitt-Manderbach, M.
Taraba, R. Ursin, P. Walther, H. Weier, H. Weinfurter, and A. Zeilinger, 2005,
Opt. Express \textbf{13}, 202.

%\bibitem {riebe}
Riebe, M., H. H\"{a}fffner, C.F. Roos, W. H\"{a}nsel, J.
Benhelm, G.P.T. Lancaster, T.W. K\"{o}rber, C. Becher, F. Schmidt-Kaler,
D.F.V. James, and R. Blatt, 2004, Nature (London) \textbf{432}, 602.

%\bibitem{Rohde07}
 Rohde, P.P., Ralph, T.C., Munro, W.J., Phys. Rev. A 75, 010302(R) (2007)

%\bibitem{romero}
 Romero, J., et al, New J. Phys., 12, 123007 (2010)

%\bibitem {Rosenberg07}
Rosenberg, D., J.W. Harrington, P.R. Rice, P.A. Hiskett,
C.G. Peterson, R.J. Hughes, A.E. Lita, S.W. Nam, and J.E. Nordholt, 2007,
Phys. Rev. Lett. \textbf{98}, 010503.

%\bibitem {RosenbergPRA}
Rosenberg, D., A.E. Lita, A.J. Miller, and S.W. Nam,
2005, Phys Rev. A \textbf{71}, 061803.

%\bibitem {Rosenfeld}
Rosenfeld, W., F. Hocke, F. Henkel, M. Krug, J. Volz, M.
Weber, and H. Weinfurter, 2008, Phys. Rev. Lett. \textbf{101}, 260403.

%\bibitem {Rosenfeld2009}
Rosenfeld, W., M. Weber, J. Volz, F. Henkel, M. Krug, A. Cabello, M. Zukowski, H. Weinfurter, 2009, Adv. Sci. Lett. \textbf{2}, 469.

%\bibitem {Rowe2001}
Rowe, M.A., D. Kielpinski, V. Meyer, C.A. Sackett, W.M.
Itano, C. Monroe, and D. J. Wineland, 2001, Nature (London) \textbf{409}, 791.

%\bibitem {roy}
Roy, S.M. and V. Singh, 1991, Phys. Rev. Lett. \textbf{67}, 2761.

%\bibitem {Rubin}
Rubin, M.H., D.N. Klyshko, Y.H. Shih, and A.V. Sergienko,
1994, Phys. Rev. A \textbf{50}, 5122.

%\bibitem {ryff}
Ryff, L.C., 1997, Am. J. Phys. \textbf{65}, 1197.

%\bibitem {sign}
Sanaka, K., T. Jennewein, J.-W. Pan, K. Resch, and A.
Zeilinger, 2004, Phys. Rev. Lett. \textbf{92}, 017902.

%\bibitem {Sangouard}
Sangouard, N., C. Simon, H. de Riedmatten, and N. Gisin,
2009, arXiv:0906.2699.

%\bibitem {SP-indist}
Santori, C., D. Fattal, J. Vu\v{c}kovi\'{c}, G.S. Solomon,
and Y. Yamamoto, 2002, Nature (London) \textbf{419}, 594.

%\bibitem{santos91}
Santos, E. (1991) Phys. Rev. Lett. 66, 1388

%\bibitem{santos92}
Santos, E. (1992) Phys. Rev. A 46, 3646

%%\bibitem {SP-dot01}
Santori, C., M. Pelton, G. Solomon, Y. Dale, and Y.
%Yamamoto, 2001, Phys. Rev. Lett. \textbf{86}, 1502.

%%\bibitem {Scarani-rmp}
Scarani, V., S. Iblisdir, N. Gisin, and A. Ac\'{\i}n,
%2005, Rev. Mod. Phys. \textbf{77}, 1225.

%%\bibitem {Scarani}
Scarani, V. and N. Gisin, 2001, Phys. Rev. Lett.
%\textbf{87}, 117901.

%%\bibitem {Scheel}
Scheel, S., W.J. Munro, J. Eisert, K. Nemoto, and P. Kok,
%2006, Phys. Rev. A \textbf{73}, 034301.

%%\bibitem {Schlosshauer}
Schlosshauer, M., 2004, Rev. Mod. Phys. \textbf{76}, 1267.

%%\bibitem {HW07}
Schmid, C., N Kiesel, W Wieczorek, and H Weinfurter, 2007, New
%J. Phys. \textbf{9}, 236.

%%\bibitem {cnot-ion}
Schmidt-Kaler, F., H. H\"{a}ffner, M. Riebe, S. Gulde,
%G.P.T. Lancaster, T. Deuschle, C. Becher, C.F. Roos, J. Eschner, and R. Blatt,
%2003, Nature (London) \textbf{422}, 408.

%\bibitem {Schmitt07decoy}
Schmitt-Manderbach, T., H. Weier, M. F\"{u}rst, R.
Ursin, F. Tiefenbacher, T. Scheidl, J. Perdigues, Z. Sodnik, C. Kurtsiefer,
J.G. Rarity, A. Zeilinger, and H. Weinfurter, 2007, Phys. Rev. Lett.
\textbf{98}, 010504.

%\bibitem {Schori02}
Schori, C., B. Julsgaard, J.L. S\o rensen, and E.S. Polzik,
2002, Phys. Rev. Lett. \textbf{89}, 057903.

%\bibitem {Schrodinger}
Schr\"{o}dinger, E., 1935, Naturwissenschaften
\textbf{23}, 807; \textbf{23}, 823; \textbf{23}, 844; the English translation
appears in Wheeler J.A. and W.H. Zurek, 1983, \emph{Quantum Theory and
Measurement} (Princeton University Press, New York).

%\bibitem {Schrodinger2}
Schr\"{o}dinger, E., 1935, Proceedings of the Cambridge Philosophical Society, 31, (1935), 555-563

%\bibitem {Schuck06}
Schuck, C., G. Huber, C. Kurtsiefer, and H. Weinfurter,
2006, Phys. Rev. Lett. \textbf{96}, 190501.

%%\bibitem {cBSM83}
Scully, M.O., B.-G. Englert, and C.J. Bednar, 1999, Phys.
%Rev. Lett. \textbf{83}, 4433.

%\bibitem {eraser}
Scully, M.O., B.-G. Englert, and H. Walther, 1991, Nature
(London) \textbf{351}, 111.

%\bibitem {Scully-book}
Scully, M.O. and M.S. Zubairy, 1997, \textit{Quantum
Optics} (Cambridge University Press, Cambridge).

%%\bibitem {SeevinckPRL}
Seevinck, M. and G. Svetlichny, 2002, Phys. Rev. Lett.
%89, 060401.

%\bibitem{selleri}
Selleri and Zeilinger (1988) Found. Phys. 18, 1141.

%\bibitem {sergienko}
Sergienko, A.V., M. Atat\"{u}re, Z. Walton, G. Jaeger,
B.E.A. Saleh, and M.C. Teich, 1999, Phys. Rev. A \textbf{60}, R2622.

%\bibitem {sq-fiber}
Shelby, R.M., M.D. Levenson, S.H. Perlmutter, R.G. DeVoe,
and D.F. Walls, 1986, Phys. Rev. Lett. \textbf{57}, 691.

%\bibitem {Shields}
Shields, A.J., 2007, Nature Photon. \textbf{1}, 215.

%\bibitem {Shih2003}
Shih, Y.H., 2003, Rep. Prog. Phys. \textbf{66}, 1009.

%\bibitem {Shih}
Shih, Y.H. and C.O. Alley, 1988, Phys. Rev. Lett. \textbf{61}, 2921.

%\bibitem {Shor95}
Shor, P.W., 1995, Phys. Rev. A \textbf{52}, R2493.

%\bibitem{Sliwa03}
Sliwa, C., and Banaszek, K., 2003, Phys. Rev. A. 67, 030101.
%%\bibitem {Sillanpaa}Sillanp\"{a}\"{a}, M.A., J.I. Park, and R.W. Simmonds,
%2007, Nature (London) \textbf{449}, 438.

%\bibitem {E-laserTH}
Simon, C. and D. Bouwmeester, 2003, Phys. Rev. Lett.
\textbf{91}, 053601.

%\bibitem {Simon07}
Simon, C., H. de Riedmatten, M. Afzelius, N. Sangouard, H.
Zbinden, and N. Gisin, 2007, Phys. Rev. Lett. \textbf{98}, 190503.

%\bibitem {robust-ion}
Simon, C. and Irvine, W.T.M. 2003, Phys. Rev. Lett.
\textbf{91}, 110405.

%\bibitem {Pur-LOPRL}
Simon, C. and J.-W. Pan, 2002, Phys. Rev. Lett.
\textbf{89}, 257901.

%\bibitem {Simon00}
Simon, C., M. \.{Z}ukowski, H. Weinfurter, and A. Zeilinger,
2000, Phys. Rev. Lett. \textbf{85}, 1783.

%\bibitem{SleatorWeinfurter1995}
Sleator, T., and H. Weinfurter, 1995, Phys. Rev. Lett. \textbf{74}, 4087.

%\bibitem {sq-1985}
Slusher, R.E., L.W. Hollberg, B. Yurke, J.C. Mertz, and J.F.
Valley, 1985, Phys. Rev. Lett. \textbf{55}, 2409.

%\bibitem {Smithey}
Smithey, D.T., M. Beck, M. Belsley, and M.G. Raymer, 1992,
Phys. Rev. Lett. \textbf{69}, 2650.

%%\bibitem {KLM-pol}
Spedalieri, F.M., H. Lee, and J.P. Dowling, 2006, Phys. Rev.
%A \textbf{73}, 012334.

%%\bibitem {Steane96}
Steane, A.M., 1996, Phys. Rev. Lett. \textbf{77}, 793.

%%\bibitem {Steane98}
Steane, A., 1998, Rep. Prog. Phys. \textbf{61} 117.

%\bibitem{specker60}
Specker, E.P., Dialectica 14, 239 (1960).

%\bibitem {Stefanov02}
Stefanov, A., H. Zbinden, N. Gisin, and A. Suarez, 2002,
Phys. Rev. Lett. \textbf{88}, 120404.

%\bibitem {Stefanov03}
Stefanov, A., H. Zbinden, N. Gisin, and A. Suarez, 2003,
Phys. Rev. A \textbf{67}, 042115.

%\bibitem {Stevenson}
Stevenson, R.M., R.J. Young, P. Atkinson, K. Cooper, D.A.
Ritchie, and A.J. Shields, 2006, Nature (London) \textbf{439}, 179.

%\bibitem {Strauf}
Strauf, S., N.G. Stoltz, M.T. Rakher, L.A. Coldren, P.M.
Petroff, and D. Bouwmeester, 2007, Nature Photon. \textbf{1}, 704.

%\bibitem{Sun06}
 Sun, F. W., B. H. Liu, Y. F. Huang, Z. Y. Ou,
and G. C. Guo, Phys. Rev. A 74, 033812 (2006)

%\bibitem {Sun-QKD}
Sun, P.C., Y. Mazurenko, and Y. Fainman, 1995, Opt.Lett. \textbf{20},1062.

%\bibitem {Takesue100}
Takesue, H., S.W. Nam, Q. Zhang, R.H. Hadfield, T. Honjo,
K. Tamaki, and Y. Yamamoto, 2007, Nature Photon. \textbf{1}, 343.

%\bibitem {tame07a}
Tame, M.S., M. Paternostro, and M.S. Kim, 2007, New J. Phys.
\textbf{9}, 201.

%\bibitem{tame07b}
Tame, M.S., R. Prevedel, M. Paternostro, P. Bohi, M.S. Kim and A. Zeilinger. Phys.
Rev. Lett. \textbf{98}, 140501 (2007).

%\bibitem {Taylor}
Taylor, G.I., 1909, Proc. Camb. Phil. Soc. Math. Phys. Sci.
\textbf{15}, 114.

%\bibitem {Terhal}
Terhal, B., 2000, Phys. Lett. A \textbf{271}, 319.

%\bibitem {Tittel98}
Tittel, W., J. Brendel, H. Zbinden, and N. Gisin, 1998,
Phys. Rev. Lett. \textbf{81}, 3563.

%\bibitem {Tittel-Weihs}
Tittel, W. and G. Weihs, 2001, Quant. Inf. Comp.
\textbf{1}(2), 3.

%\bibitem {Thom06}
Thompson, J.K., J. Simon, H. Loh, and V. Vuletic, 2006,
Science \textbf{313}, 74.

%\bibitem {Tokunaga08}
Tokunaga, Y., S. Kuwashiro, T. Yamamoto, M. Koashi, and N. Imoto, 2008, Phys. Rev. Lett. \textbf{100}, 210501.

%\bibitem{torgerson95}
Torgerson, J.R., D. Branning, C.H. Monken, and L. Mandel. Phys. Lett. A 204, 323 (1995)

%\bibitem {Torres}
Torres, J.P., A. Alexandrescu, and L. Torner, 2003, Phys.
Rev. A \textbf{68}, 050301.

%\bibitem {Torres05}
Torres, J.P., F. Maci\`{a}, S. Carrasco, and L. Torner,
2005, Opt. Lett. \textbf{30}, 314.

%%\bibitem {Uffink}
Uffink, J., 2002, Phys. Rev. Lett. \textbf{88}, 230406.

%\bibitem {Uren}
U'Ren, A.B., K. Banaszek, and I.A. Walmsley, 2003, Quantum Inf.
Comp. \textbf{3}, 480.

%\bibitem {Tele-Danube}
Ursin, R., T. Jennewein, M. Aspelmeyer, R. Kaltenbaek,
M. Lindenthal, P. Walther, and A. Zeilinger, 2004, Nature (London)
\textbf{430}, 849.

%\bibitem {Ursin07}
Ursin, R., F. Tiefenbacher, T. Schmitt-Manderbach, H. Weier,
T. Scheidl, M. Lindenthal, B. Blauensteiner, T. Jennewein, J. Perdigues, P.
Trojek, B. Oemer, M. Fuerst, M. Meyenburg, J. Rarity, Z. Sodnik, C. Barbieri,
H. Weinfurter, and A. Zeilinger, 2007, Nature Phys. \textbf{3}, 481.

%\bibitem {Vaidman99}
Vaidman, L. and N. Yoran, 1999, Phys. Rev. A \textbf{59}, 116.

%\bibitem{vallone}
Vallone G, E. Pomarico, F. De Martini, Phys. Rev. A \textbf{78}, 042335 (2008).

%\bibitem{vandersypen}
Vandersypen, L. M. K. et al., Nature (London) 414, 883
(2001);

%\bibitem {memory-Science}
van der Wal, C.H., M.D. Eisaman, A. Andre, R.L.
Walsworth, D.F. Phillips, A.S. Zibrov, M.D. Lukin, 2003, Science \textbf{301}, 196.

%\bibitem {vanEnk97}
van Enk, S.J., J.I. Cirac, and P. Zoller, 1997, Phys. Rev.
Lett. \textbf{78}, 4293.

%\bibitem {bell3}
van Houwelingen, J.A.W., N. Brunner, A. Beveratos, H. Zbinden,
and N. Gisin, 2006, Phys. Rev. Lett. \textbf{96}, 130502.

%\bibitem {loock}
van Loock, P., T.D. Ladd, K. Sanaka, F. Yamaguchi, K. Nemoto,
W.J. Munro, and Y. Yamamoto, 2006, Phys. Rev. Lett. \textbf{96}, 240501.

%\bibitem {Varnava}
Varnava, M., D.E. Browne, and T. Rudolph, 2006, Phys. Rev.
Lett. \textbf{97}, 120501.

%\bibitem{varnava08}
Varnava M., D.E. Browne and T. Rudolph, Phys. Rev. Lett. \textbf{100},
060502 (2008).

%\bibitem {qutrit-OAM}
Vaziri, A., J.-W. Pan, T. Jennewein, G. Weihs, and A.
Zeilinger, 2003, Phys. Rev. Lett. \textbf{91}, 227902.

%\bibitem {qutrit-02}
Vaziri, A., G. Weihs, and A. Zeilinger, 2002, Phys. Rev.
Lett. \textbf{89}, 240401.

%\bibitem{verstraete01}
Verstraete, F., J. Dehaene, and B. DeMoor, Phys. Rev. A
64, 010101(R) (2001).

%\bibitem{vertesi10}
Vertesi T. et al. Phys. Rev. Lett. 104, 060401 (2010)

%\bibitem {cBSM85}
Vitali, D., M. Fortunato, and P. Tombesi, 2000, Phys. Rev.
Lett. \textbf{85}, 445.

%\bibitem {Volz}
Volz, J., M. Weber, D. Schlenk, W. Rosenfeld, J. Vrana, K.
Saucke, C. Kurtsiefer, and H. Weinfurter, 2006, Phys. Rev. Lett. \textbf{96}, 030404.

%\bibitem{wagenknecht10}
Wagenknecht, C., C.-M. Li, A. Reingruber, X.-H. Bao, A. Goebel, Y.-A. Chen, Q. Zhang, K. Chen, and J.-W. Pan, 2010, Nature Photonics, 4, 549.

%\bibitem {Waks02}
Waks, E., A. Zeevi, Y. Yamamoto, 2002, Phys. Rev. A
\textbf{65}, 052310.

%\bibitem {walborn}
Walborn, S.P., S. P\'{a}dua, and C.H. Monken, 2003b, Phys.
Rev. A \textbf{68}, 042313.

%\bibitem {Walls-Milburn}
Walls, D.F. and G.J. Milburn, 1994, \textit{Quantum
Optics} (Springer-Verlag, Berlin).

%\bibitem {deBroglie}
Walther, P., J.-W. Pan, M. Aspelmeyer, R. Ursin, S.
Gasparoni, and A. Zeilinger, 2004, Nature (London) \textbf{429}, 158.

%\bibitem {Pur-nonlocal}
Walther, P., K.J. Resch, C. Brukner, A.M. Steinberg,
J.-W. Pan, and A. Zeilinger, 2005a, Phys. Rev. Lett. \textbf{94}, 040504.

%\bibitem {one-way-EXP}
Walther, P., K.J. Resch, T. Rudolph, E. Schenck, H.
Weinfurter, V. Vedral, M. Aspelmeyer, and A. Zeilinger, 2005b, Nature (London)
\textbf{434}, 169.

%\bibitem {C-QED}
Walther, H., B.T.H. Varcoe, B.-G. Englert, and T. Becker,
2006, Rep. Prog. Phys. \textbf{69}, 1325.

%%\bibitem {Walther-bsm}
Walther, P. and A. Zeilinger, 2005, Phys. Rev. A
%\textbf{72}, 010302.

%\bibitem {Walton03}
Walton, Z.D., M.C. Booth, A.V. Sergienko, B.E.A. Saleh, and
M.C. Teich, 2003, Phys. Rev. A \textbf{67}, 053810.

%\bibitem {Walton04}
Walton, Z.D., A.V. Sergienko, B.E.A. Saleh, M.C. Teich,
2004, Phys. Rev. A \textbf{70}, 052317.

%\bibitem {Wang2005}
Wang, X.-B., 2005,\ Phys. Rev. Lett. \textbf{94,} 230503.

%\bibitem{wang2006}
Wang, Z-W., Xiang-Fa Zhou, Yun-Feng Huang, Yong-Sheng Zhang, Xi-Feng Ren, Guang
-Can Guo, (2006) Phys. Rev. Lett. 96, 220505.

%%\bibitem {WZM}
Wang, L.J., X.Y. Zou, and L. Mandel, 1991, Phys. Rev. A
%\textbf{44}, 4614.

%%\bibitem {QZM92}
Wang, L.J., X.Y. Zou, and L. Mandel, 1992, J. Opt. Soc. Am. B
%\textbf{9}, 605.

%\bibitem {Weber}
Weber, B., H. P. Specht, T. M\"{u}ller, J. Bochmann, M.
M\"{u}cke, D. L. Moehring, and G. Rempe, 2009, Phys. Rev. Lett. \textbf{102}, 030501.

%\bibitem {Weihs1998}
Weihs, G., T. Jennewein, C. Simon, H. Weinfurter, and A.
Zeilinger, 1998, Phys. Rev. Lett. \textbf{81}, 5039.

%\bibitem {weihs96}
Weihs, G., M.~Reck, H.~Weinfurter, and A.~Zeilinger, 1996,
Opt. Lett., \textbf{21} 302.

%\bibitem{weinhold08}
Weinhold T. J., A. Gilchrist, K. J. Resch, A. C. Doherty, J. L. O'Brien, G. J.
Pryde, and A. G. White, arXiv 0808.0794 (2008).

%\bibitem {WeinfurterBSM}
Weinfurter, H., 1994, Europhys. Lett. \textbf{25}, 559.

%\bibitem {WZ2001}
Weinfurter, H. and M. \.{Z}ukowski, 2001, Phys. Rev. A
\textbf{64}, 010102.

%\bibitem {werner}
Werner, R.F., 1989, Phys. Rev. A \textbf{40}, 4277.

%\bibitem {Werner-rev}
Werner R.F. and M.M. Wolf, 2001, Quantum Inf. Comput.
\textbf{1} (3), 1.

%\bibitem {nonmax-Bell}
White, A.G., D.F.V. James, P.H. Eberhard, and P.G.
Kwiat, 1999, Phys. Rev. Lett. \textbf{83}, 3103.

%\bibitem{White01}
White, A. G., James, D. F. V., Munro, W. J. and Kwiat, P. G. Phys. Rev. A 65, 012301
(2001).

%\bibitem{wieczorek08}
Wieczorek, W., Ch. Schmid, N. Kiesel, R. Pohlner , O. Guhne, and H.
Weinfurter, Phys. Rev. Lett. \textbf{101}, 010503 (2008).

%\bibitem{Wieczorek09}
Wieczorek, W., R. Krischek, N. Kiesel, P. Michelberger,
G. T\'{o}th, and H. Weinfurter, 2009a, Phys. Rev. Lett. \textbf{103}, 020504.

%\bibitem{Wieczorek09-projection}
Wieczorek, W., N. Kiesel, C. Schmid, and H. Weinfurter, 2009b, Phys. Rev. A \textbf{79}, 022311.

%\bibitem {Wiesner}
Wiesner, S., 1983, SIGACT News \textbf{15}, 78.

%\bibitem {WZ82}
Wootters, W.K. and W.H. Zurek, 1982, Nature (London)
\textbf{299}, 802.

%\bibitem{wootters98}
Wootters, W. K. (1998) Phys. Rev. Lett. 80, 2245

%\bibitem {sq-1986}
Wu, L.-A., H.J. Kimble, J.L. Hall, and H. Wu, 1986, Phys.
Rev. Lett. \textbf{57}, 2520.

%%\bibitem {Yamamoto08}
Yamamoto, T., K. Hayashi, \c{S}.K. \"{O}zdemir, M.
%Koashi, and N. Imoto, 2008, Nature Photon. \textbf{2}, 488.

%\bibitem {conc-Y2001}
Yamamoto, T., M. Koashi, and N. Imoto, 2001, Phys. Rev. A
\textbf{64}, 012304.

%\bibitem {Conc-EXP}
Yamamoto, T., M. Koashi, \c{S}.K. \"{O}zdemir, and N.
Imoto, 2003, Nature (London) \textbf{421}, 343.

%\bibitem {Yang}
Yang, T., Q. Zhang, J. Zhang, J. Yin, Z. Zhao, M. \.{Z}ukowski,
Z.-B. Chen, and J.-W. Pan, 2005, Phys. Rev. Lett. \textbf{95}, 240406.

%\bibitem {Yang06}
Yang, T., Q. Zhang, T.-Y. Chen, S. Lu, J. Yin, J.-W. Pan,
Z.-Y. Wei, J.-R. Tian, and J. Zhang, 2006, Phys. Rev. Lett. \textbf{96}, 110501.

%\bibitem{Yao2011}
Yao, X.-C. et al. manuscript in preparation (2011).

%\bibitem {link03}
Yoran N. and B. Reznik, 2003, Phys. Rev. Lett. \textbf{91}, 037903.

%%\bibitem {Young}
Young, R.J, R.M. Stevenson, P. Atkinson, K. Cooper, D.A.
%Ritchie, and A.J Shields, 2006, New J. Phys. \textbf{8}, 29.

%%\bibitem {Sixia-N}
Yu, S., Z.-B. Chen, J.-W. Pan, and Y.-D. Zhang, 2003, Phys.
%Rev. Lett. \textbf{90}, 080401.

%\bibitem {Yuan07}
Yuan, Z.-S., Y.-A. Chen, S. Chen, B. Zhao, M. Koch, T.
Strassel, Y. Zhao, G.-J. Zhu, J. Schmiedmayer, and J.-W. Pan, 2007, Phys. Rev.
Lett. \textbf{98}, 180503.

%\bibitem {Yuan08}
Yuan, Z.-S., Y.-A. Chen, B. Zhao, S. Chen, J. Schmiedmayer,
and J.-W. Pan, 2008, Nature (London) \textbf{454}, 1098.

%\bibitem {YS}
Yurke B. and D. Stoler, 1992a, Phys. Rev. Lett. \textbf{68}, 1251.

%\bibitem {YSpra}
Yurke B. and D. Stoler, 1992b, Phys. Rev. A \textbf{46}, 2229.

%%\bibitem {Zanardi98}
Zanardi, P., 1999, Phys. Rev. A \textbf{60}, R729.

%%\bibitem {dsf-Zanardi}
Zanardi, P. and M. Rasetti, 1997, Phys. Rev. Lett.
%\textbf{79}, 3306.

%\bibitem {Zbinden}
Zbinden, H., J. Brendel, N. Gisin, and W. Tittel, 2001,
Phys. Rev. A \textbf{63}, 022111.

%\bibitem {Zeilinger81}
Zeilinger, A., 1981, Am. J. Phys. \textbf{49}, 882.

%\bibitem {Zeilinger93}
Zeilinger, A., H. J. Bernstein, D. M. Greenberger, M. A.
Horne, and M. \.{Z}ukowski, 1993, in \textit{Quantum Control and Measurement},
edited by H. Ezawa and Y. Murayama (North Holland, Amsterdam), p. 9.

%\bibitem {GHZ-gen97}
Zeilinger, A., M.A. Horne, H. Weinfurter, and M.
\.{Z}ukowski, 1997, Phys. Rev. Lett. \textbf{78}, 3031.

%\bibitem {HappyPhoton}
Zeilinger, A., G. Weihs, T. Jennewein, and M.
Aspelmeyer, 2005, Nature (London) \textbf{433}, 230.

%\bibitem {zeldovich}
Zel'dovich, Ya.B. and D.N. Klyshko, 1969, JETP Lett.
\textbf{9}, 40.

%\bibitem {An-Ning}
Zhang, A.-N., C.-Y. Lu, X.-Q. Zhou, Y.-A. Chen, Z. Zhao, T.
Yang, and J.-W. Pan, 2006a, Phys. Rev. A \textbf{73}, 022330.

%\bibitem {zhangbo}
Zhang, Q., X.-H. Bao, C.-Y. Lu, X.-Q. Zhou, T. Yang, T.
Rudolph, and J.-W. Pan, 2008a, Phys. Rev. A \textbf{77}, 062316.

%\bibitem {qzhang}
Zhang, Q., A. Goebel, C. Wagenknecht, Y.-A. Chen, B. Zhao, T.
Yang, A. Mair, J. Schmiedmayer, and J.-W. Pan, 2006b, Nature Phys. \textbf{2}, 678.

%%\bibitem {Zhang100}
Zhang, Q., H. Takesue, S.W. Nam, C. Langrock, X. Xie, B.
%Baek, M.M. Fejer, and Y. Yamamoto, 2008b, Opt. Express \textbf{16}, 5776.

%%\bibitem {dfs-zhang}
Zhang, Q., J. Yin, T.-Y. Chen, S. Lu, J. Zhang, X.-Q. Li,
%T. Yang, X.-B. Wang, and J.-W. Pan, 2006d, Phys. Rev. A \textbf{73}, 020301(R).

%\bibitem {ZhaoB09}
Zhao, B., Y.-A. Chen, X.-H. Bao, T. Strassel, C.-S. Chuu,
X.-M. Jin, J. Schmiedmayer, Z.-S. Yuan, S. Chen, and J.-W. Pan, 2009a, Nature
Phys. \textbf{5}, 95.

%\bibitem {zhaobo06}
Zhao, B., Z.-B. Chen, Y.-A. Chen, J. Schmiedmayer, and
J.-W. Pan, 2007, Phys. Rev. Lett. \textbf{98}, 240502.

%\bibitem {ZhaoR09}
Zhao, R., Y.O. Dudin, S.D. Jenkins, C.J. Campbell, D.N.
Matsukevich, T.A.B. Kennedy, and A. Kuzmich, 2009b, Nature Phys. \textbf{5}, 100.

%\bibitem {zhao01}
Zhao, Z., J.-W. Pan, and M.S. Zhan, 2001, Phys. Rev. A
\textbf{64}, 014301.

%\bibitem {Zhao-GHZ}
Zhao, Z., T. Yang, Y.-A. Chen, A.-N. Zhang, M.
\.{Z}ukowski, and J.-W. Pan, 2003a, Phys. Rev. Lett. \textbf{91}, 180401.

%\bibitem {Conc-EXP-Zhao}
Zhao, Z., T. Yang, Y.-A. Chen, A.-N. Zhang, and J.-W.
Pan, 2003b, Phys. Rev. Lett. \textbf{90}, 207901.

%\bibitem {fiveP}
Zhao, Z., Y.-A. Chen, A.-N. Zhang, T. Yang, H.J. Briegel, and
J.-W. Pan, 2004, Nature (London) \textbf{430}, 54.

%\bibitem {CNOT-Hefei}
Zhao, Z., A.-N. Zhang, Y.-A. Chen, H. Zhang J.-F. Du, T.
Yang, and J.-W. Pan, 2005, Phys. Rev. Lett. \textbf{94}, 030501.

%%\bibitem {ZWM}
Zou, X.Y., L.J. Wang, and L. Mandel, 1991, Phys. Rev. Lett.
%\textbf{67}, 318.

%\bibitem{zukowski91}
\.{Z}ukowski, M., 1991, Phys. Lett. A \textbf{157}, 198.

%\bibitem {z93}
\.{Z}ukowski, M., 1993, Phys. Lett. A \textbf{177} 290.

%\bibitem {Zukowski00}
\.{Z}ukowski, M., 2000, Phys. Rev. A \textbf{61}, 022109.

%\bibitem {zb02}
\.{Z}ukowski, M. and \v{C}. Brukner, 2002, Phys. Rev. Lett.
\textbf{88}, 210401.

%\bibitem {z97}
\.{Z}ukowski, M. and D. Kaszlikowski, 1997, Phys. Rev. A
\textbf{56}, R1682.

%\bibitem {Zuk88}
\.{Z}ukowski, M. and J. Pykacz, 1988, Phys. Lett. A
\textbf{127}, 1.

%\bibitem {Nport}
\.{Z}ukowski, M., A. Zeilinger, and M.A. Horne, 1997, Phys.
Rev. A \textbf{55}, 2464.

%\bibitem {swapTH}
.{Z}ukowski, M., A. Zeilinger, M.A. Horne, and A.K. Ekert,
1993, Phys. Rev. Lett. \textbf{71}, 4287.

%\bibitem {thirdman}
\.{Z}ukowski, M., A. Zeilinger, M.A. Horne, and H.
Weinfurter, 1998, Acta Phys. Pol. \textbf{93}, 187.

%\bibitem {Marek99}
\.{Z}ukowski, M., A. Zeilinger, M.A. Horne, and H.
Weinfurter, 1999, Int. J. Theor. Phys. \textbf{38}, 501.

%\bibitem {marek-ny}
\.{Z}ukowski, M., A. Zeilinger, and H. Weinfurter, 1995,
Ann. N.Y. Acad. Sci. \textbf{755}, 91.

%\bibitem {zurek-rmp}
Zurek, W.H., 2003, Rev. Mod. Phys. \textbf{75}, 715.

\end{document}